\documentclass[12pt]{article}
\usepackage{threeparttable}
\usepackage{tikz}
\usepackage{amsmath}
\usepackage{graphicx}
\usepackage{enumerate}
\usepackage[authoryear]{natbib}
\usepackage{url} %
\usepackage{latexsym}
\usepackage{epsfig}
\usepackage{subfigure}
\usepackage{threeparttable}
\usepackage{topcapt}
\usepackage{amssymb}
\usepackage{amsthm}
\newtheorem{assumption}{Assumption}
\usepackage{array}
\usepackage{color}
\usepackage{float}
\usepackage[bookmarks=false,hypertexnames=false]{hyperref}\hypersetup{colorlinks=true, linkcolor=blue, filecolor=gray, urlcolor=blue, citecolor=blue}

\usepackage{mathrsfs}
\usepackage{indentfirst}
\usepackage{psfrag}
\usepackage{multirow}
\usepackage{booktabs}
\usepackage{longtable}

\usepackage{bm}
\usepackage{rotating}
\usepackage{bookmark}
\usepackage{algorithm}
\usepackage{algpseudocode}

\usepackage{hyperref}
\hypersetup{hidelinks,
	colorlinks=true,
	allcolors=black,
	pdfstartview=Fit,
	breaklinks=true}
\usepackage{xr}
\makeatletter

\newcommand*{\addFileDependency}[1]{%
\typeout{(#1)}%

\@addtofilelist{#1}

\IfFileExists{#1}{}{\typeout{No file #1.}}
}\makeatother

\newcommand*{\myexternaldocument}[1]{%
\externaldocument{#1}%
\addFileDependency{#1.tex}%
\addFileDependency{#1.aux}%
}

\myexternaldocument{mean_LLF_sample_spliting20240101supp}
\newcommand{\blind}{1}

\newtheorem{thm}{Theorem}%[section]
\newtheorem{lemma}{Lemma}%[subsection]

\newtheorem{remark}{Remark}

\newtheorem{prop}{Proposition}   %%%%% \newtheorem{algorithm}{Algorithm}
\renewcommand{\baselinestretch}{1}
\def\IR{\rm I \kern-0.20em R}

\renewcommand{\theequation}{\arabic{equation}}
\renewcommand{\baselinestretch}{1.5}

\begin{document}

\def\spacingset#1{\renewcommand{\baselinestretch}%
{#1}\small\normalsize} \spacingset{1}

\if1\blind
{

  \title{\bf Robust Inference for Multiple Predictive Regressions with an Application on Bond Risk Premia
}
  \author{Xiaosai Liao\hspace{.2cm}\\
    Institute of Chinese Financial Studies, \\
    Southwestern University of Finance and Economics\\
    and \\
    Xinjue Li \\
    School of Economics, Yunnan University\\
    and \\
    Qingliang Fan\thanks{ Corresponding author.  ELB 903, The Chinese University of Hong Kong, Shatin, Hong Kong.  E-mail:  michaelqfan@cuhk.edu.hk.} \\
    Department of Economics, The Chinese University of Hong Kong \\
    }
  \maketitle
} \fi

\if0\blind
{
  \bigskip
  \bigskip
  \bigskip
  \begin{center}
    {\LARGE\bf Title}
\end{center}
  \medskip
} \fi

\bigskip
\begin{abstract}
We propose a robust hypothesis testing procedure for the predictability of multiple predictors that could be highly persistent. Our method improves the popular extended instrumental variable (IVX) testing \citep{Phillips2013PredictiveRU,Kostakisetal2015} in that, besides addressing the two bias effects found in \cite{HosseinkouchackDemetrescu2021}, we find and deal with the variance-enlargement effect. We show that two types of higher-order terms induce these distortion effects in the test statistic, leading to significant over-rejection for one-sided tests and tests in multiple predictive regressions. Our improved IVX-based test includes three steps to tackle all the issues above regarding finite sample bias and variance terms. Thus, the test statistics perform well in size control, while its power performance is comparable with the original IVX. Monte Carlo simulations and an empirical study on the predictability of bond risk premia are provided to demonstrate the effectiveness of the newly proposed approach.
\end{abstract}

\noindent%
\vfill

\newpage
\spacingset{1.9} % DON'T change the spacing!

\section{Introduction}
\label{4}

Testing the predictability of asset returns is the focal point of financial studies for different stakeholders, including academia, practitioners, and policymakers. However, the commonly used statistical inference suffers from size distortions due to the high persistence of (some) predictors and the contemporaneous correlations between the innovations of predictors and dependent variables \citep{CampbellYogo2006}. The most popular method to solve this problem is the IVX \citep{Phillips2013PredictiveRU,Kostakisetal2015}, which aims to solve the over-rejection problem of the test statistics based on OLS.

Unfortunately, the IVX-based test \citep{Phillips2013PredictiveRU,Kostakisetal2015} still has size distortion, which is magnified in empirically relevant finite samples, one-sided hypotheses, and multiple predictors cases. In Section \ref{section2}, we show explicitly that the root of the oversized IVX statistic is not only the {\bf Deformation Effect} (DE) and the {\bf Displacement Effect} (DiE) \citep{HosseinkouchackDemetrescu2021,DemetrescuRodrigues2020} but also the {\bf Variance Enlargement Effect} (VEE). In closely related literature, \cite{HosseinkouchackDemetrescu2021} provided a structured approach to control the small sample noncentrality of the IVX $t$-statistic for a given instrumental variable. \cite{DemetrescuRodrigues2020} combined the residual-augmented regression and the recursive demean method to remove the size distortion of IVX caused by higher-order terms. However, \cite{HosseinkouchackDemetrescu2021} did not consider the inference in multiple predictive
regressions, while the recursive demean technique applied by \cite{DemetrescuRodrigues2020} would cause the inference to perform worse than our test in terms of power. Additionally, the inference by \cite{HosseinkouchackDemetrescu2021} and \cite{DemetrescuRodrigues2020} still suffers size distortion, as evidenced by their simulations. \cite{XuGuo2022} applied the Lagrange-multipliers principle to the joint test in multiple predictive models. Nevertheless, \cite{XuGuo2022} ignored the higher-order terms that generate the three effects mentioned above, and as a result, their test still suffered size distortions in many cases.

To our knowledge, no existing literature deals with the VEE in IVX test statistics in a single or multiple predictive regression. We show that, as the number of predictors increases, the VEE makes the IVX test statistic grow even bigger compared to its theoretical value. Therefore, ignoring the VEE leads to serious size distortion problems in multiple predictive regressions.

\subsection{Main Results and Contributions}

Our new three-step procedure (detailed in Section \ref{section4}) improves the size control of IVX while keeping its good power performance:
\begin{enumerate}
\item We entirely eliminate the DE using a novel sample splitting technique without loss of effective sample size. To do so, we use split (evenly from the middle time point) subsamples and obtain three IVX test statistics based on the full sample and two subsamples. The new test statistic is the weighted average of these three test statistics, effectively circumvents the higher-order term causing the DE.
\item We considerably mitigate the DiE, which can cause the finite sample mean of the test statistic to deviate from the expected value of its asymptotic distribution. By subtracting the median of this displacement from the test statistic obtained in the first step, the location of the new test statistic and its asymptotic distribution become very close, thus reducing this displacement effect.
\item The VEE can cause the sample standard deviation of the test statistic obtained in the second step to be greater than its  asymptotic standard deviation. We divide the test statistic in the previous step by the ratio of the two (sample and the asymptotic) standard deviations. The standard deviation of the new test statistic and its asymptotic distribution are very close,  thereby alleviating this effect.
\end{enumerate}
Under some mild conditions detailed in Section \ref{section2}, the test statistics converge to a standard normal or Chi-square distribution. For more robust inference in the finite sample, we combine the three steps for size control and the Lagrange-multipliers principle by \cite{XuGuo2022}. The easy-to-implement procedure is summarized in Algorithm \ref{algorithm1}. Intensive simulation studies in the main text and Appendix are provided to demonstrate the effectiveness of the newly proposed approach.

In the empirical research, we revisit the predictability of the bond risk premia. Using forward rates, macroeconomic principal components, and their linear combinations as predictors of bond risk premia, we show that some of the predictors (especially the highly persistent ones) proposed by OLS or IVX \citep{Phillips2013PredictiveRU,Kostakisetal2015} exhibit no predicting power on bond excess returns based on our test. Additionally, the effective predictors selected by our proposed test have strong economic implications, and these findings are notably different from the IVX test in \cite{Kostakisetal2015}. The empirical results confirm that our newly developed test statistics are more reliable tools to test the predictability of financial returns. Our test should be helpful to practitioners.

The main contributions of the paper are summarized below.
 \begin{enumerate}
 \item We address the distortion effects induced by the higher-order terms of IVX test statistics in a unified approach integrating univariate and multivariate models. We discuss the newly discovered VEE and show its accumulative nature in the finite sample as the number of predictors increases, which explains the lingering size distortion issue in the financial return predictability test literature. Our new test eliminates the DE and significantly reduces the DiE and VEE.
 \item The proposed inference procedure performs very well in size control while keeping the good power performance of the original IVX test. Compared to the literature, the proposed method has unique advantages for one-sided tests in the univariate model and joint and marginal tests in the multivariate model.
 \item From the empirical study perspective, our test sheds light on the significant variables to forecast one-year-ahead excess returns on bonds. Based on our test, the relatively longer-term forward rates are found to have little predicting power for the short-term bond excess return. The evidence also supports the market segmentation theory, which indicates that the long-term forward rates can not forecast the short-term bond excess return.
 \end{enumerate}

\subsection{Brief Literature Review}
There is a rich literature on the over-rejection problem in the inference of predictive regression with highly persistent predictors. Early literature such as Bonferroni's method by \cite{CampbellYogo2006} and the conditional likelihood method by \cite{JanssonMoreira2006} focused on the inference for the univariate model. \cite{Choietal2016} introduced the Cauchy type instrumental variable approach for the test in univariate regression. \cite{Zhuetal2014} and \cite{Yangetal2021} proposed the weighted empirical likelihood approach with a good size control to test predictability in univariate and bivariate predictive models. To conduct inference in multiple predictive models, \cite{Elliott2011} and \cite{BreitungDemetrescu2015} proposed the variable addition approach, but the power performance is unsatisfactory. \cite{HARVEY2021198} used the quasi-GLS demeaned predictor to develop easy-to-implement tests for return predictability in univariate models, displaying attractive finite sample size control and power. The IVX approach is proposed by \cite{MagdalinosPhillips2009} and is also studied by \cite{Kostakisetal2015},  \cite{PhillipsLee2016}, \cite{Yangetal2019},  \cite{DemetrescuRodrigues2020},  \cite{HosseinkouchackDemetrescu2021} and \cite{Liu2022RobustIW}. In the studies of forecasting bond risk premia, \cite{Bauer2018}  proposed a bootstrap procedure to test the predictability of bond risk premia with highly persistent predictors. However, it still tends to over-reject and  does not offer the asymptotic property of the bootstrap procedure. \cite{FAN2021269} proposed an augmented factor model and applied their method to forecast the excess return of U.S. government bonds.

The rest of this paper is organized as follows.  Section \ref{section2} introduces the higher-order terms causing the distortion effects of IVX test statistics.  Section \ref{section4} provides the procedures to construct the test statistics and  presents the asymptotic theories for the proposed estimators and the test statistics.  Section \ref{section5} reports the Monte Carlo simulation results. Section \ref{section6} presents the empirical application. Section \ref{section7} concludes the paper. The online Appendix contains detailed proofs of the theoretical results, extra simulations, and additional empirical results.

Throughout this paper, SD, WD refer to strong and weak dependence, respectively. The symbols $\Rightarrow$, $\xrightarrow{d}$ and $\xrightarrow{p}$ are used to represent weak convergence and convergence in distribution and in probability, respectively. All limits are for $T\rightarrow \infty$ in the limit theories, and $O_p (1)$ is stochastically asymptotically bounded while $o_p(1)$ is asymptotically negligible. For any matrix $A$, $A^{1/2}$ is a matrix defined as $A^{1/2} = Q_L \operatorname{diag}( \lambda_1^{1/2}, \lambda_2^{1/2},\cdots,\lambda_K^{1/2})Q_L^\top$, in which  $Q_L$ is the matrix whose $i$th column is the eigenvectors of $A$, and $\lambda_i$, $i=1,2,\cdots,K$ are the eigenvalues of $A$. It implies that $A^{1/2}A^{1/2}=A$. Also, we define $A^{-1/2}=(A^{1/2})^{-1}$.  $0_K$ is a zero vector with dimension K and $\operatorname{I_K}$ is a $K\times K$ identity matrix.

\section{IVX-based Test and the Higher-order Terms}\label{section2}
We are interested in testing the predictability of $y_t$, such as bond or stock returns, in a unified framework with univariate or multivariate predictors. Without loss of generality, we discuss the following multiple predictive regression model
\begin{align}\label{mulpredModel}
y_t = \mu + x_{t-1}^\top \beta + u_t,
\end{align}
where $\beta=\left(\beta_1,\beta_2,\cdots,\beta_K \right)^\top$.  The model is also studied by \cite{Phillips2013PredictiveRU,Kostakisetal2015}.
The highly persistent variables $x_{t-1}=\left(x_{1,t-1},x_{2,t-1} \cdots,x_{K,t-1}\right)^\top$ are modeled as follows
\begin{align}\label{mulgtuA1}
x_{i,t} = \rho_i   x_{i,t-1}   +v_{i,t},
\end{align}
where $x_{i,0}=o_p(\sqrt{T})$, $\rho_i=1+c_i/T^\alpha$, $c=\operatorname{diag}(c_1,c_2,\cdots,c_K)$,  $v_t=(v_{1,t},v_{2,t},\cdots,v_{K,t})^\top$, $t=1,\cdots,T$, $i=1,\cdots,K$
and $\alpha=0$ or 1, $K$ is the dimension of $x_{t-1}$. We assume no cointegration relationship exists among $x_t$.
Two types of persistence with different values of $c_i$ and $\alpha$ are considered: (1) Strong Dependence [SD]: $\alpha=1$ and $c_i\leq 0$ is a constant; (2) Weak Dependence [WD]: $\alpha=0$  and $|1+c_i|<1$.   \footnote{ \textcolor{black}{We also permit $x_{i,t-1}$ with $i=1,2,\cdots,K$ to have mixed types of persistence; that is to say, $\alpha$ could be a function of $i$. In this setting, the theoretical results for the test statistics are almost the same.}}
Define $\mathcal{F}_t=\sigma\{(u_j,v_j^\top)^\top, j \leq t\}$, which is the information set available at time $t$. Here, following \cite{Kostakisetal2015}, a general weakly dependent innovation structure of the linear process on $(u_t,v_t^\top)^\top$ in equations (\ref{mulpredModel}) and (\ref{mulgtuA1}) is imposed as follows.

\begin{assumption} \label{Assumption A.1}
Define $\epsilon_t=(u_t, \varepsilon_t^\top)^\top$ as martingale difference sequence\footnote{For SD predictors, $u_t$ could be extended from m.d.s. to the linear process.} with respect to the natural filtration $\mathcal{F}_t$, which satisfies
\begin{align}
E\left(\epsilon_t \epsilon_t^\top|\mathcal{F}_t \right) = \Sigma_t,\,\, a.s.,\quad \text{and} \quad \sup_{t} E\|\epsilon_t\|^{4+\nu} <\infty
\end{align}
for some $\nu>0$, where $\Sigma_t$ is a positive definite matrix. Assume $v_t$ is a stationary linear process given by
\[
v_{t}=\sum_{j=0}^\infty F_{j} \varepsilon_{t-j},
\]
Here, $F_{0}=I_K$, \textcolor{black}{$K$ is the dimension of $x_t$} \textcolor{black}{($K\geq 1$)}, and $(F_{j})_{j\geq 0}$ is a sequence of constant matrices such that $\sum_{j=0}^\infty F_{j}$ is of full rank and $\sum_{j=0}^\infty j\|F_{j}\|< \infty$. The process $\epsilon_t$ is strictly stationary ergodic satisfying
\begin{align}
\lim_{m \rightarrow \infty} \left\| Cov\left[vec\left( \epsilon_m\epsilon_m^\top\right),vec\left(\epsilon_0\epsilon_0^\top \right) \right]\right\|=0
\end{align}
\end{assumption}

\begin{remark}
As specified by \cite{Kostakisetal2015}, the sequence $(u_t)_{t=1}^T$ permits the following GARCH(p,q) representation:
\begin{align}\label{garchset}
u_t = h_t \eta_t,\quad
 h_t^2= \varphi_0+ \sum_{j=1}^p \varphi_j h_{t-j}^2 +\sum_{j=1}^q \bar{ \varphi}_j u_{t-j}^2
\end{align}
where $(\eta_t)$ is an i.i.d. (0,1) sequence. $\varphi_0$ is a positive constant. And $\varphi_j$ and $\bar{\varphi}_j$ are non-negative constants satisfying $\sum_{j=1}^p \varphi_j +\sum_{j=1}^q \bar{ \varphi}_j<1$.
\end{remark}

 Under Assumption \ref{Assumption A.1},
 the functional central limit theorem (FCLT) for $(u_t,v_t^\top)^\top$ holds by \cite{Phillips1992AsymptoticsFL},
\begin{align}\label{limit}
&\frac{1}{\sqrt{T}} \sum_{t=1}^{\lfloor rT \rfloor} \left(u_t, v_t^\top\right)^\top
\Rightarrow [B_{u}(r),B_v(r)^\top]^\top
=BM \left(\begin{array}{cc}
  \Sigma_{uu} & \Sigma_{u v} \\
  \Sigma_{uv}^\top & \Omega_{vv}
  \end{array}
  \right),
\end{align}
where $[B_u(r), B_v(r)^\top]^\top$ is a vector of Brownian motions and
$\Sigma_{u v}=\sum_{h=-\infty}^{\infty}E\left( u_{t}v_{t+h}\right)= F_x(1) E\left(u_t \varepsilon_{t}\right)$, $\Sigma_{u u}=E\left(u_{t} u_{t}^\top\right)$ and $\Omega_{vv}= \sum_{h=-\infty}^\infty E(v_tv_{t+h}^\top) = F_x(1)E(v_tv_t^\top)F_x(1)^\top= F_x(1)\Sigma_{vv}F_x(1)^\top$.
Under Assumption \ref{Assumption A.1}, the following result holds by equation (\ref{limit}), \cite{Phillips1987} and \cite{Kostakisetal2015},
\begin{align}\label{cenlim}
T^{-1 / 2} x_{\lfloor rT \rfloor}  \Rightarrow J_x^c(r),\quad \text{for} \quad 0\leq r \leq 1
\end{align}
for SD predictors, where $J_x^c(r)=\int_0^r e^{(r-s)c}d B_v(s)$.

Following \cite{Phillips2013PredictiveRU}, the standard IVX is set to be less persistent than SD predictors to obtain the asymptotic mixture normal distributed estimator $\hat{\beta}_{ivx}$. Specifically, we have
\begin{align}\label{defdeftwo2}
 \hat{\beta}_{ivx} & = \left(\sum\limits_{t=1}^{T} \bar{z}_{t-1} x_{t-1}^\top \right)^{-1}\sum\limits_{t=1}^{T} \bar{z}_{t-1}y_t ,
\end{align}
where $\bar{z}_{t-1}=z_{t-1}- \frac{1}{T}\sum\limits_{t=1}^{T} z_{t-1}$, $z_t=(z_{1,t},z_{2,t}\cdots,z_{K,t})^\top$  and
\begin{align}\label{mulivz}
z_{i,t} = \rho_z z_{i,t-1} + \Delta x_{i,t}, \quad i=1,2,\cdots,K,
\end{align}
where $\rho_z = 1+ c_z/T^\delta $, $c_z<0$ and $1/2<\delta<1$.

In the following, we expand on Proposition 1 of \cite{HosseinkouchackDemetrescu2021}, which focuses on univariate models, and point out two higher-order terms in multivariate models.  We define the random vector ${t_{ivx}}
 \equiv \left( \sum_{t=1}^T \bar{z}_{t-1} \bar{z}_{t-1}^\top \hat{u}_t^2 \right)^{-1/2} \sum_{t=1}^T \bar{z}_{t-1} u_t$ to analyze the DE, DiE and VEE of  the original IVX \citep{Kostakisetal2015,Phillips2013PredictiveRU} in a unified (univariate or multivariate) model,\footnote{When $K=1$, under the null hypothesis $H_0:\beta=0$, ${t_{ivx}}$ becomes the t-test statistic defined by \cite{HosseinkouchackDemetrescu2021}. The three effects exist in univariate models. } since the original IVX test statistic $ Q_{ivx}={t_{ivx}^\top} H_{ivx}^\top (H_{ivx}H_{ivx}^\top )^{-1} H_{ivx} {t_{ivx}}$  is the function of  ${t_{ivx}}$ under the null hypothesis $H_0:R\beta =r_J$,  where $R$ is a $J\times K$ predetermined matrix with the rank $J$, $r_J$ is a  predetermined vector with dimension $J$, and $H_{ivx}  = R\left(\sum\limits_{t=1}^{T} \bar{z}_{t-1}x_{t-1}^\top \right)^{-1}
\left(\sum\limits_{t=1}^{T} \bar{z}_{t-1}\bar{z}_{t-1}^\top \hat{u}_t^2\right)^{1/2}$.

To show the following proposition, we define the normalized instrumental variable $z_{t-1}^*=\Omega_{zz}^{-1/2}z_{t-1}$, where  $\Omega_{zz}= \Sigma_{uu} \Omega_{vv}/(-2c_z)$ for SD predictors and $\Omega_{zz}= \operatorname{E}(x_{t-1}x_{t-1}^\top u_t^2)$ for WD predictors.
\begin{prop}\label{mulpropfdie3}
Under Assumption \ref{Assumption A.1}, the following results hold for SD predictors.
\begin{align}
{t_{ivx}}
&= \left( \frac{1}{T^{1+\delta}} \sum_{t=1}^T \bar{z}_{t-1} \bar{z}_{t-1}^\top \hat{u}_t^2 \right)^{-1/2}\frac{1}{T^{1 / 2+\delta / 2}} \sum_{t=1}^T \bar{z}_{t-1} u_t \\
&=\left( \frac{1}{T^{1+\delta}} \sum_{t=1}^T \bar{z}_{t-1}^* (\bar{z}_{t-1}^*)^\top \hat{u}_t^2 \right)^{-1/2}\frac{1}{T^{1 / 2+\delta / 2}} \sum_{t=1}^T \bar{z}_{t-1}^* u_t \nonumber
&={Z_T}+{B_T}+{C_T}+ o_p\left[T^{(\delta- 1)/2}\right]. \nonumber
\end{align}
\begin{align}
&{Z_T} = \frac{1}{T^{1 / 2+\delta / 2}} \sum_{t=1}^T z_{t-1}^* u_t \xrightarrow{P} \operatorname{N}(0_K,\operatorname{I_K} ).\nonumber\\
\label{dkmul76gh2}
&{B_T}= \varpi_b
 \frac{1}{T^{1 / 2+\delta / 2}} \sum_{t=1}^T z_{t-1}^* u_t\xrightarrow{P} 0.\\
\label{dkmul76gh}
 & T^{(1-\delta)/2}\operatorname{E}\left({B_T} \right)  \rightarrow
 - \frac{K+1}{2}{\rho_{u v^*}}/ \sqrt{-2 c_z},
\end{align}
\begin{align}\label{multpop1th3}
T^{(1 -\delta) / 2} {C_T}&=\left( \frac{1}{T^{1+\delta}} \sum_{t=1}^T \bar{z}_{t-1} \bar{z}_{t-1}^\top \hat{u}_t^2 \right)^{-1/2}\frac{1}{T^{1 / 2+\delta }} \sum_{t=1}^T z_{t-1} \frac{1}{\sqrt{T}}\sum_{t=1}^T u_t \\
&\Rightarrow -(-c_z/2)^{-1/2} \Sigma_{vv}^{-1/2} \frac{B_u(1)J_x^c(1)}{\sqrt{\Sigma_{uu} }} \nonumber
\end{align}
where ${\rho_{u v^*}} = \Sigma_{vv}^{-1/2} \Sigma_{vu}/ \sqrt{\Sigma_{uu}}$,  $\varpi_b = -\frac{1}{2} \left[\frac{1}{T^{1+\delta }} \sum_{t=1}^T z_{t-1}^* (z_{t-1}^*)^\top u_t^2- \operatorname{I}_K \right]$, $0_K$ is a zero vector with dimension K and $\operatorname{I_K}$ is a $K\times K$ identity matrix.
\end{prop}
\begin{remark}
When $K=1$, ${\rho_{u v^*}}$ becomes the correlation coefficient between $u_t$ and $v_t$. Hence Proposition \ref{mulpropfdie3} is the generalization of Proposition 1 of \cite{HosseinkouchackDemetrescu2021}.
\end{remark}

\subsection{The DE, DiE and VEE}
Three critical points are drawn from Proposition \ref{mulpropfdie3}. First, similar to  \cite{HosseinkouchackDemetrescu2021}, the higher-order term ${B_T}$ comes from the correlation between the terms $ \sum_{t=1}^T \bar{z}_{t-1} \bar{z}_{t-1}^\top \hat{u}_t^2 $  and $  \sum_{t=1}^T \bar{z}_{t-1} u_t$  in the random vector ${t_{ivx}}$. Meanwhile, the higher-order term ${C_T}$ comes from $\sum\nolimits_{t=1}^T z_{t-1} \sum\nolimits_{t=1}^T u_t$ in ${t_{ivx}}$.

\textcolor{black}{Second, with SD predictors, the higher-order term ${B_T}$ leads to not only the DiE shown in equation (\ref{dkmul76gh})  but also the VEE shown in equation (\ref{dkmul76gh2}).} The DiE causes the center of the finite sample distribution of ${t_{ivx}}$ to deviate from zero to $T^{-(1-\delta)/2}\lim_{T\rightarrow\infty}T^{(1-\delta)/2}\operatorname{E}\left({B_T} \right)$, \textcolor{black}{which makes the one-sided test of the original IVX to suffer from size distortion \citep{HosseinkouchackDemetrescu2021}.} The VEE causes the covariance matrix of ${t_{ivx}}$ in the finite sample to be enlarged by the higher-order term $B_T$ so that the test statistics $Q_{ivx}$ is enlarged, which is shown in Theorem \ref{thmnew1}. Specifically, equation (\ref{dkmul76gh2}) shows that the covariance matrix of ${t_{ivx}}$ in the finite sample is approximately
\begin{align}
 \operatorname{Var}\left({t_{ivx}}\right) = \operatorname{Var}\left({Z_T} \right)+ \operatorname{Var}\left({B_T} \right) + o(1) =\operatorname{I_K} +\varpi_b  \varpi_b^\top  + o(1).
\end{align}
Therefore, the covariance matrix of $(\operatorname{I_K} +\hat{\varpi}_b  \hat{\varpi}_b^\top  )^{-1/2}{t_{ivx}}$ in the finite sample is closer to $\operatorname{I_K}$ than that of  ${t_{ivx}}$, where $\hat{\varpi}_b= -\frac{1}{2} \left[  \sum_{t=1}^T z_{t-1}^{**} (z_{t-1}^{**})^\top u_t^2- \operatorname{I}_K \right]$ and $z_{t-1}^{**} = \left(\sum_{t=1}^T z_{t-1}z_{t-1}^\top \hat{u}_t^2 \right)^{-1/2}z_{t-1}$. As a result, the finite sample distribution of the new IVX test statistic  $$ \tilde{Q}_{ivx}={t_{ivx}^\top} H_{ivx}^\top \left[H_{ivx} (\operatorname{I_K} +\hat{\varpi}_b  \hat{\varpi}_b^\top) H_{ivx}^\top \right]^{-1} H_{ivx} {t_{ivx}}$$ based on $(\operatorname{I_K} +\hat{\varpi}_b  \hat{\varpi}_b^\top  )^{-1/2}{t_{ivx}}$ is much closer to $\chi_J^2$ than that of the original IVX test statistic $Q_{ivx}$ based on ${t_{ivx}}$. 

The following theorem shows that $Q_{ivx} - \tilde{Q}_{ivx} $ is positive, and VEE aggravates as K increases.
\begin{thm}\label{thmnew1}
 Under Assumption \ref{Assumption A.1} and   the null hypothesis $H_0:R\beta =r_J$, it follows that
\begin{align*}
 Q_{ivx} - \tilde{Q}_{ivx} & =  {t_{ivx}^\top} H_{ivx}^\top \left(H_{ivx}   H_{ivx}^\top \right)^{-1}
 H_{ivx}  \hat{\varpi}_b  \hat{\varpi}_b^\top  H_{ivx}^\top
\left[H_{ivx} (\operatorname{I_K} +\hat{\varpi}_b  \hat{\varpi}_b^\top) H_{ivx}^\top \right]^{-1} H_{ivx} {t_{ivx}}  \\
&=  {t_{ivx}^\top} H_{ivx}^\top \left(H_{ivx}   H_{ivx}^\top \right)^{-1}
 H_{ivx}  \hat{\varpi}_b  \hat{\varpi}_b^\top  H_{ivx}^\top
\left(H_{ivx}  H_{ivx}^\top \right)^{-1} H_{ivx} {t_{ivx}} +  o_p\left[T^{-(1-\delta)/2}\right] \\
&= \sum_{j=1}^K    (H_{j}^{ivx})^2 +  o_p\left[T^{-(1-\delta)/2}\right]>0,
\end{align*}
where   $H^{ivx}=(H_1^{ivx},H_2^{ivx},\cdots,H_K^{ivx})^\top = \hat{\varpi}_b^\top  H_{ivx}^\top
\left(H_{ivx}  H_{ivx}^\top \right)^{-1} H_{ivx} {t_{ivx}}$.
\end{thm}
By Theorem \ref{thmnew1}, as the number of predictors $K$ grows, VEE becomes more severe since $Q_{ivx} - \tilde{Q}_{ivx} $ grows bigger. This causes the original IVX to suffer severe size distortion in multiple predictive models.

Third, equation (\ref{multpop1th3}) shows that the higher-order term ${C_T}$ leads to the DE, meaning that the finite sample distribution of ${t_{ivx}}$ deviates significantly from the  normal distribution.

In summary, the DiE, VEE, and DE induced by the higher-order terms ${B_T}$ and ${C_T}$ in Proposition \ref{mulpropfdie3} explain why the IVX test \citep{Kostakisetal2015,Phillips2013PredictiveRU} suffers size distortions for one-sided tests and tests in multivariate predictive regression.
A straight-forward idea to eliminate the effects of ${B_T}$ and ${C_T}$ is to subtract their consistent estimators from ${t_{ivx}}$.
However, it is not an easy task to eliminate the effect of ${C_T}$ since it arises from $\sum \nolimits_{t=1}^T z_{t-1}\sum \nolimits_{t=1}^T u_t$ which could not be estimated consistently.
%\

\section{The Improved IVX Test}\label{section4}

\subsection{A Sample Splitting Procedure to Eliminate the DE}
To eliminate the size distortion induced by the higher-order term ${C_T}$ in Proposition \ref{mulpropfdie3}, we apply a sample splitting procedure to remove the term $\sum\nolimits_{t=1}^{T} z_{t-1} \sum\nolimits_{t=1}^{T} u_t$ in test statistics, which is the source of DE. Intuitively, we split the full sample into two subsamples and construct the IVX estimators based on the full sample and the two subsamples. By setting appropriate weights, the new estimator, which is the weighted sum of the three estimators (full sample, two subsamples), does not contain $\sum\nolimits_{t=1}^{T} z_{t-1} \sum\nolimits_{t=1}^{T} u_t$. As a result, the DE is eliminated by the following steps.

We split the full sample into two subsamples $\{(y_t,x_{t-1})\}_{t=1}^{T_0}$ and $\{(y_t,x_{t-1})\}_{t=T_0}^{T}$, where $T_0=\lfloor \lambda T\rfloor$ and $0<\lambda<1$ is a constant. In practice, we suggest evenly split to subsamples such that $\lambda=0.5$, whose reason is further specified by Remark \ref{red7ngn2} in subsection \ref{subsection3.2}.
First, we define the IVX estimators based on two subsamples as follows.
\begin{align}
\label{muldef2new}
 \hat{\beta}_a & = \left(\sum\limits_{t=1}^{T_0} \bar{z}_{t-1}^a x_{t-1}^\top \right)^{-1}\sum\limits_{t=1}^{T_0} \bar{z}_{t-1}^a y_t,\\ \label{muldef3new}
 \hat{\beta}_b & = \left(\sum\limits_{t=T_0+1}^{T} \bar{z}_{t-1}^b x_{t-1}^\top \right)^{-1}\sum\limits_{t=T_0+1}^{T} \bar{z}_{t-1}^b y_t,
\end{align}
where $\bar{z}_{t-1}^a= z_{t-1}- \frac{1}{T_0}\sum\nolimits_{t=1}^{T_0} z_{t-1}$, $\bar{z}_{t-1}^b= z_{t-1}- \frac{1}{T-T_0}\sum\nolimits_{t=T_0+1}^{T} z_{t-1}$.
Second,  we apply equations (\ref{mulpredModel}), (\ref{defdeftwo2}), (\ref{muldef2new}) and (\ref{muldef3new}) to obtain the following equations.
\begin{small}
\begin{align}\label{muldeftwo2}
\left[\sum\limits_{t=1}^{T} \left( z_{t-1}- \frac{1}{T}\sum\limits_{t=1}^{T} z_{t-1} \right) x_{t-1}^\top \right](\hat{\beta}_{ivx}-\beta) & = \sum\limits_{t=1}^{T} z_{t-1}u_t - \frac{1}{T}\sum\limits_{t=1}^{T} z_{t-1} \sum\limits_{t=1}^{T} u_t , \\ \label{muldeftwo3}
\left[\sum\limits_{t=1}^{T_0} \left( z_{t-1}- \frac{1}{T_0}\sum\limits_{t=1}^{T_0} z_{t-1} \right) x_{t-1}^\top \right](\hat{\beta}_a-\beta) &= \sum\limits_{t=1}^{T_0} z_{t-1} u_t- \frac{1}{T_0}\sum\limits_{t=1}^{T_0} z_{t-1} \sum\limits_{t=1}^{T_0} u_t ,\\ \label{muldeftwo4}
\left[\sum\limits_{t=T_0+1}^{T} \left( z_{t-1}- \frac{\sum_{t=T_0+1}^{T} z_{t-1} }{T-T_0} \right) x_{t-1}^\top \right] (\hat{\beta}_b-\beta) & = \sum\limits_{t=T_0+1}^{T} z_{t-1} u_t - \frac{\sum\limits_{t=T_0+1}^{T} z_{t-1}}{T-T_0} \sum\limits_{t=T_0+1}^{T} u_t.
\end{align}
\end{small}
By equations (\ref{muldeftwo3}) and (\ref{muldeftwo4}), the following equations hold.
\begin{footnotesize}
\begin{align}
\label{mul8new}
S_a\left[\sum\limits_{t=1}^{T_0} \left( z_{t-1}- \frac{1}{T_0}\sum\limits_{t=1}^{T_0} z_{t-1} \right) {x}_{t-1}^\top \right](\hat{\beta}_a-\beta) &= S_a \sum\limits_{t=1}^{T_0} z_{t-1} u_t- \frac{1}{T}\sum\limits_{t=1}^{T} z_{t-1} \sum\limits_{t=1}^{T_0} u_t ,\\ \label{mul9new}
S_b\left[\sum\limits_{t=T_0+1}^{T} \left( z_{t-1}- \frac{1}{T-T_0}\sum\limits_{t=T_0+1}^{T} z_{t-1} \right) {x}_{t-1}^\top \right] (\hat{\beta}_b-\beta)& = S_b \sum\limits_{t=T_0+1}^{T} z_{t-1} u_t - \frac{1}{T}\sum\limits_{t=1}^{T} z_{t-1} \sum\limits_{t=T_0+1}^{T} u_t,
\end{align}
\end{footnotesize}
where
\begin{align}\label{7hj4e45d}
S_a &= \frac{1}{T}\sum\limits_{t=1}^{T} z_{t-1}\left(\frac{1}{T_0}\sum\limits_{t=1}^{T_0} z_{t-1}^\top \right) \left(\frac{1}{T_0}\sum\limits_{t=1}^{T_0} z_{t-1}^\top \frac{1}{T_0}\sum\limits_{t=1}^{T_0} z_{t-1}\right)^{-1},\\
S_b &= \frac{1}{T}\sum\limits_{t=1}^{T} z_{t-1}\left(\frac{1}{T-T_0}\sum\limits_{t=T_0+1}^{T} z_{t-1}^\top \right) \left(\frac{1}{T-T_0}\sum\limits_{t=T_0+1}^{T} z_{t-1}^\top \frac{1}{T-T_0}\sum\limits_{t=T_0+1}^{T} z_{t-1}\right)^{-1}. \nonumber
\end{align}
By subtracting the sum of equations (\ref{mul8new}) and (\ref{mul9new}) from equation (\ref{muldeftwo2}),  \textcolor{black}{the term $\sum\nolimits_{t=1}^{T} z_{t-1} \sum\nolimits_{t=1}^{T} u_t$ of $\hat{\beta}_{ivx}$ in equation (\ref{defdeftwo2}) is removed as follows. }
\begin{align}\label{mulkde2con}
\begin{split}
& W_1 \hat{\beta}_{ivx}-W_2\hat{\beta}_a -W_3 \hat{\beta}_b - (W_1-W_2-W_3)\beta \\
& = (\operatorname{I_K}-S_a) \sum\limits_{t=1}^{T_0} z_{t-1}u_t + (\operatorname{I_K}-S_b) \sum\limits_{t=T_0+1}^{T} z_{t-1}u_t
= \sum\limits_{t=1}^{T_0} \tilde{z}_{t-1}u_t,
\end{split}
\end{align}
where $W_1=\sum\nolimits_{t=1}^{T} \bar{z}_{t-1} x_{t-1}^\top $, $W_2=S_a \sum\nolimits_{t=1}^{T_0} \bar{z}_{t-1}^a x_{t-1}^\top$,
 $W_3=S_b \sum\nolimits_{t=T_0+1}^{T} \bar{z}_{t-1}^b x_{t-1}^\top$ and
\begin{align}\label{mult3inti2st}
\tilde{z}_{t-1} =
\begin{cases}
(\operatorname{I_K}-S_a)z_{t-1},\quad 1\leq t\leq T_0;\\
(\operatorname{I_K}-S_b)z_{t-1},\quad T_0+1 \leq t \leq T.
\end{cases}
\end{align}
Third, we could utilize another instrumental variable estimator with the IV $\tilde{z}_{t-1}$. Define the IV estimator $\hat{\beta}_l $ as follows.
\begin{align}\label{mulkde2con2}
\hat{\beta}_l  \equiv (W_1-W_2-W_3)^{-1}(W_1 \hat{\beta}_{ivx}-W_2\hat{\beta}_a -W_3 \hat{\beta}_b).
\end{align}
By equations (\ref{mulkde2con}) and (\ref{mulkde2con2}) and that $W_1-W_2-W_3 = \sum\limits_{t=1}^{T} \tilde{z}_{t-1}x_{t-1}$, it follows that
\begin{align}\label{kdad53g}
\hat{\beta}_l- \beta &= \left(\sum\limits_{t=1}^{T} \tilde{z}_{t-1}x_{t-1}^\top \right)^{-1} \sum\limits_{t=1}^{T} \tilde{z}_{t-1} u_t.
\end{align}
The key and desirable property of the new instrumental variable $\tilde{z}_{t-1}$ is
\begin{align}\label{mulkjeyi4ns}
\sum\nolimits_{t=1}^{T} \tilde{z}_{t-1}=0.
\end{align}
 Thus $\hat{\beta}_l$ could be deemed the instrumental variable estimator using $\tilde{z}_{t-1}$ as IV,
\begin{align*}
\hat{\beta}_l  &= \left(\sum\limits_{t=1}^{T} \tilde{z}_{t-1}x_{t-1}^\top \right)^{-1} \sum\limits_{t=1}^{T} \tilde{z}_{t-1} y_t \\
 &= \left[\sum\limits_{t=1}^{T} \left( \tilde{z}_{t-1}-\frac{1}{T} \sum\limits_{t=1}^{T} \tilde{z}_{t-1} \right)x_{t-1}^\top \right]^{-1} \sum\limits_{t=1}^{T} \left( \tilde{z}_{t-1}-\frac{1}{T} \sum\limits_{t=1}^{T} \tilde{z}_{t-1} \right) y_t \\
  &= \left[\sum\limits_{t=1}^{T} \left( \tilde{z}_{t-1}-\frac{1}{T} \sum\limits_{t=1}^{T} \tilde{z}_{t-1} \right)x_{t-1}^\top \right]^{-1} \sum\limits_{t=1}^{T} \left( \tilde{z}_{t-1}-\frac{1}{T} \sum\limits_{t=1}^{T} \tilde{z}_{t-1} \right) u_t +\beta.
\end{align*}
Therefore, the higher-order term ${C_T}$ vanishes,
since the term $\sum\limits_{t=1}^{T} z_{t-1} \sum\limits_{t=1}^{T} u_t$ disappears in $\hat{\beta}_l- \beta$. 
\begin{thm}\label{multh1m}
Under Assumption \ref{Assumption A.1}, for SD and WD predictors, it follows that
$$
 D_T(\hat{\beta}_l- \beta)\xrightarrow{d} \operatorname{MN}\left[0,\Sigma_{zx}^{-1}\Sigma_{zz}\left(\Sigma_{zx}^{-1}\right)^\top \right],
$$
where $D_T=T^{(1+\delta)/2}$ for SD predictors and $D_T=\sqrt{T}$ for WD predictors and
 $\Sigma_{zz} = \lambda\left(\operatorname{I_K} - \tilde{S}_a\right)\Omega_{zz}\left(\operatorname{I_K} - \tilde{S}_a\right)^\top +(1-\lambda)\left(1-\tilde{S}_b\right)\Omega_{zz} \left(\operatorname{I_K} - \tilde{S}_b\right)^\top
$, where \quad  $\Sigma_{zx}=  -c_z^{-1}(1-\tilde{S}_a) \int_0^{\lambda} dJ_x^c(r)\, J_x^c(r)^\top
-c_z^{-1}(1-\tilde{S}_b)\int_{\lambda}^{1} dJ_x^c(r) \, J_x^c(r)^\top -c_z^{-1}\left[ 1- \lambda\tilde{S}_a -(1-\lambda)\tilde{S}_b\right]\operatorname{E}(v_tv_t^\top) $ for SD predictors and
$\Sigma_{zx}=\left[1- \lambda\tilde{S}_a -(1-\lambda)\tilde{S}_b \right] \operatorname{E}\left( x_{t-1}x_{t-1}^\top\right)$ for WD predictors
and
\begin{align}\label{sasb69}
&S_a \Rightarrow \tilde{S}_a =
\begin{cases}
 J_x^c(1) J_x^c(\lambda)^\top \left[ J_x^c(\lambda)^\top J_x^c(\lambda) \right]^{-1},\quad \text{SD}; \\
  B_v(1) B_v(\lambda)^\top \left[ B_v(\lambda)^\top B_v(\lambda) \right]^{-1},\quad \text{WD};
\end{cases}\\
&S_b \Rightarrow \tilde{S}_b =
\begin{cases}
 J_x^c(1) \left[J_x^c(1) - J_x^c(\lambda) \right]^\top \left\{ \left[J_x^c(1) - J_x^c(\lambda) \right]^\top \left[J_x^c(1) - J_x^c(\lambda) \right] \right\}^{-1},\quad \text{SD}; \\
  B_v(1) \left[B_v(1) - B_v(\lambda) \right]^\top \left\{ \left[B_v(1) - B_v(\lambda) \right]^\top \left[B_v(1) - B_v(\lambda) \right] \right\}^{-1},\quad \text{WD}.
\end{cases}\nonumber
\end{align}
\end{thm}
Then the test statistic ${Q_l}$ is constructed for the null hypothesis $H_0:R\beta =r_J$.
\begin{equation}\label{walde2tes}
{Q_l} \equiv \left(R \tilde{\beta}_l -r_J \right)^\top \left\{R \operatorname{\widehat{Avar}}(\hat{\beta}_l ) R^\top \right\}^{-1} \left(R \tilde{\beta}_l -r_J \right),
\end{equation}
where $\operatorname{\widehat{Avar}}(\hat{\beta}_l )= H_lH_l^\top$
\footnote{We use the heteroskedasticity consistent estimator similar to that of \citep{MacKinnon1985SomeHC}.} and
$H_l = \left( \sum_{t=1}^T \tilde{z}_{t-1} x_{t-1}^\top \right)^{-1}\left(\frac{T}{T-2K-1} \sum_{t=1}^T \tilde{z}_{t-1} \tilde{z}_{t-1}^\top \hat{u}_t^2 \right)^{1/2}$.
\begin{remark}
Following the literature, $u_t$ is estimated by OLS, which has a  smaller variance than that of the IV estimator for $u_t$. Besides the higher-order terms $B_T$ and $C_T$,  \cite{XuGuo2022} argue that the estimation error of $u_t$  also causes the size distortion. Moreover, the variance of unconstrained OLS estimator $u_t$ increases with $K$.  To avoid size distortion induced by the uncertainty of  the estimators of $u_t$, we apply the Lagrange-multipliers principle \citep{XuGuo2022} to obtain the constrained OLS estimator $\hat{u}_t$ as follows.
\begin{align*}
(\hat{\mu}_s,\hat{\beta}_s)^\top  = \arg\, \min_{\mu,\beta} \left( y_t - \mu - x_{t-1}^\top \beta\right)^2,
s.t. \quad R\beta =r_J,
\end{align*}
and $\hat{u}_t= y_t - \hat{\mu}_s - x_{t-1}^\top \hat{\beta}_s$. Since the procedure to obtain $\hat{u}_t$ uses the information of the null hypothesis $H_0:R\beta =r_J$, the estimator $\hat{u}_t$ is more efficient than the unconstrained  OLS estimator. As a result, the size performance of ${Q_l}$ is much better than the case in which the Lagrange-multipliers principle is not applied.
\end{remark}
Moreover, we construct the t-test statistic ${Q_l^t}$ when $J=1$ for right side test $H_0:\beta_i=0$ vs $H_a:\beta_i>0$ and left side test $H_0:\beta_i=0$ vs $H_a:\beta_i<0$.
\begin{align}
{Q_l^t} \equiv \frac{R \hat{\beta}_l -r_J}{ \left[ R\operatorname{\widehat{Avar}}(\hat{\beta}_l )R^\top\right]^{1/2}},
\end{align}
Note that $Q_l=(Q_l^t)^2$ when $J=1$. 
\begin{thm}\label{t7sf8f1}
Under Assumption \ref{Assumption A.1} and the null hypothesis $H_0:R\beta =r_J$, one can show that the limiting distribution of the test statistics ${Q_l^t}$ and ${Q_l}$ are the standard normal and the $\chi^2$-distribution with $J$ degrees of freedom, respectively.
\end{thm}
Then, the local power of ${Q_l^t}$ with $J=1$ is shown as follows.
\begin{thm}\label{mulkeythe2}
Under Assumption \ref{Assumption A.1}, for SD and WD predictors with $J=1$, it follows that
$$
{Q_l^t} \xrightarrow{d} \operatorname{N}(0,1)+ R\Sigma_{zx}^{-1}\Sigma_{zz}\left(\Sigma_{zx}^{-1}\right)^\top R^\top   b,
$$
where $R\beta - r_J=b_\beta/D_T$ and $b_\beta$ is a constant.
\end{thm}
Theorem \ref{mulkeythe2} shows that the convergence rate to the local power of ${Q_l^t}$ is the same as IVX \citep{Kostakisetal2015,Phillips2013PredictiveRU} and the local power of ${Q_l^t}$ is comparable with that of  IVX \citep{Kostakisetal2015,Phillips2013PredictiveRU}. A similar conclusion could be drawn for the test statistic ${Q_l}$.
\subsection{Methods to Reduce DiE and VEE}\label{subsection3.2}
The asymptotically $\chi^2$-distributed test statistic ${Q_l}$ still suffers size distortion due to the DiE and VEE in ${Q_l}$. To analyze these effects, we define the random vector ${t_l}$ as follows.
$${t_l} \equiv   \left(\frac{T}{T-2K-1} \frac{1}{T^{1+\delta}} \sum_{t=1}^T \tilde{z}_{t-1} \tilde{z}_{t-1}^\top \hat{u}_t^2 \right)^{-1/2}\frac{1}{T^{1 / 2+\delta / 2}} \sum_{t=1}^T \tilde{z}_{t-1} u_t. $$
Note that
\begin{align}\label{kda4g42}
\hat{\beta}_l- \beta &= \left(\sum\limits_{t=1}^{T} \tilde{z}_{t-1}x_{t-1}^\top \right)^{-1}
\left(\frac{T}{T-2K-1}  \sum\limits_{t=1}^{T} \tilde{z}_{t-1}\tilde{z}_{t-1}^\top \hat{u}_t^2\right)^{1/2} {t_l}.
\end{align}
Thus by equations (\ref{walde2tes}) and (\ref{kda4g42}), $ Q_l={{t_l}^\top} H_l^\top (H_l H_l^\top )^{-1} H_l {t_l}$  is the function of  ${t_l}$ under the null hypothesis $H_0:R\beta =r_J$.    Therefore, we can analyze and reduce DiE and VEE in ${Q_l}$ by analyzing and reducing those in ${t_l}$.

To analyze the higher-order term of \textcolor{black}{ ${t_l}$} , we define \textcolor{black}{the random vector} based on the first subsample $\{(y_t,x_{t-1})\}_{t=1}^{T_0}$ as
$$
{t_a^p} =\left( \frac{1}{T^{1+\delta}} \sum_{t=1}^{T_0} z_{t-1} z_{t-1}^\top \hat{u}_t^2 \right)^{-1/2}\frac{1}{T^{1 / 2+\delta / 2}} \sum_{t=1}^{T_0} z_{t-1} u_t
$$
and \textcolor{black}{the random vector} based on the second subsample $\{(y_t,x_{t-1})\}_{t=T_0+1}^{T}$ as
$$
{t_b^p} = \left( \frac{1}{T^{1+\delta}} \sum_{t=T_0+1}^{T} z_{t-1} z_{t-1}^\top \hat{u}_t^2 \right)^{-1/2}\frac{1}{T^{1 / 2+\delta / 2}} \sum_{t=T_0+1}^{T} z_{t-1} u_t.
$$
The random vectors ${t_a^p}$ and ${t_b^p}$ are used to show the property of the higher-order terms of ${t_l}$. Since ${t_a^p}$ does not contain $\sum\nolimits_{t=1}^T z_{t-1} \sum\nolimits_{t=1}^T u_t$, it only contains the higher-order terms ${B_T^{p_a}}$ similar to ${B_T}$. Therefore,
\begin{align}\label{mulkee3y1}
{t_a^p} ={Z_T^{p_a}} +{B_T^{p_a}}+o_p\left(T^{\delta / 2-1 / 2}\right),
\end{align}
where
$
{Z_T^{p_a}} \xrightarrow{d} \operatorname{N}(0_K,\operatorname{I_K}),\quad \sqrt{T_0(1-\rho_z^2)} \operatorname{E}({B_T^{p_a}})\xrightarrow{P} -{\rho_{u v^*}},\quad {B_T^{p_a}}\xrightarrow{P}0,
$
and
\begin{align}\label{mulkee3y2}
{t_b^p} ={Z_T^{p_b}} +{B_T^{p_b}}+o_p\left(T^{\delta / 2-1 / 2}\right),
\end{align}
where
$
{Z_T^{p_b}} \xrightarrow{d} \operatorname{N}(0_K,1),\quad \sqrt{(T-T_0)(1-\rho_z^2)} \operatorname{E}({B_T^{p_b}})\xrightarrow{P} -{\rho_{u v^*}},\quad {B_T^{p_b}}\xrightarrow{P}0
$.
Also, notice that under the null hypothesis $H_0:R\beta =r_J$, it follows that
\begin{align}\label{mulcom4r}
 {t_l} = W_a {t_a^p} + W_b {t_b^p},
\end{align}
where
$
W_a = \left( \frac{1}{T^{1+\delta}} \sum_{t=1}^{T} \tilde{z}_{t-1} \tilde{z}_{t-1}^\top \hat{u}_t^2 \right)^{-1/2}
(\operatorname{I_K}-S_a)
\left( \frac{1}{T^{1+\delta}} \sum_{t=1}^{T_0} z_{t-1} z_{t-1}^\top \hat{u}_t^2 \right)^{1/2}
$
 and
$$
W_b = \left( \frac{1}{T^{1+\delta}} \sum_{t=1}^{T} \tilde{z}_{t-1} \tilde{z}_{t-1}^\top \hat{u}_t^2 \right)^{-1/2}
(\operatorname{I_K}-S_b)
\left( \frac{1}{T^{1+\delta}} \sum_{t=T_0+1}^{T} z_{t-1} z_{t-1}^\top \hat{u}_t^2 \right)^{1/2}.
$$

We can apply equation (\ref{mulcom4r}) to construct the higher-order terms of ${t_l}$ by the higher-order terms of ${t_a^p}$ and ${t_b^p}$ shown in equations (\ref{mulkee3y1}) and (\ref{mulkee3y2}).
Define the normalized instrumental variable $\tilde{z}_{t-1}^*=  (\Sigma_{zz})^{-1/2}\tilde{z}_{t-1}$. Then the  following proposition holds.
\begin{prop}\label{mulpropp2}
Under Assumption \ref{Assumption A.1}, for SD predictors, the following equation holds as $T \rightarrow \infty$,
$$
{t_l}={Z_T^l}+{B_T^l}+o_p\left(T^{\delta / 2-1 / 2}\right),
$$
where
$
{Z_T^l} = (\Sigma_{zz})^{-1/2}\sum\nolimits_{t=1}^T \tilde{z}_{t-1}  u_t \stackrel{d}{\rightarrow} \operatorname{N}(0_K,\operatorname{I_K}), \quad \text {and }
$
${B_T^l }\rightarrow 0. \quad$ And
 \begin{align}\label{mulpop1th12}
{B_T^l} &= \varpi_l {Z_T^l}\\
& \label{mulpop1th332} = W_a {B_T^{p_a}}+ W_b {B_T^{p_b}} + o_p\left[T^{-(1-\delta)/2}\right]
 \end{align}
 where
 $
 \varpi_l = -\frac{1}{2}\left\{ \sum\nolimits_{t=1}^T \tilde{z}_{t-1}^* (\tilde{z}_{t-1}^*)^\top \hat{u}_t^2 -\operatorname{I_K}\right\}
 $
 and
 $$
 T^{(1 -\delta) / 2} {B_T^l} ={R_T^l} + W_l\, T^{-(1 -\delta) / 2}\operatorname{plim}_{T\rightarrow \infty} \; T^{(1 -\delta) / 2} \operatorname{E}\left({B_T}\right),
 $$
where ${R_T^l}= W_a {R_{1,T}^l}+W_b{ R_{2,T}^l}$, $ {R_{1,T}^l}=T^{(1 -\delta) / 2}\left[ {B_T^{p_a}}- \operatorname{E}({B_T^{p_a}}) \right]$ and $ {R_{2,T}^l}=T^{(1 -\delta) / 2}\left[ {B_T^{p_b}}- \operatorname{E}({B_T^{p_b}}) \right]$ satisfying
 $
\operatorname{E}\left( {R_{1,T}^l} \right) = \operatorname{E}\left( {R_{2,T}^l} \right) =0
 $
 and
 $
 W_l = W_a/\sqrt{\lambda} + W_b/\sqrt{1-\lambda}
 $.
\end{prop}
\begin{remark}\label{red8dma}
The term ${R_T^l}$ could be regarded as the ``residual term" of ${B_T^l}$ since it is the weighted sum of two terms whose expectation is zero.  Therefore, $W_l\, T^{-(1 -\delta) / 2} \operatorname{plim}_{T\rightarrow \infty} \; T^{(1 -\delta) / 2} \operatorname{E}\left({B_T}\right)$ is the  main source of DiE induced by ${B_T^l}$  rather than the term ${R_T^l}$.
\end{remark}
Proposition \ref{mulpropp2} shows that the higher-order term ${B_T^l}$ leads to not only the DiE but also the VEE in multivariate models. Due to the DiE, the center of the distribution of ${t_l}$ in the finite sample deviates from zero vector, while the covariance matrix of ${t_l}$ in the finite sample is closer to $\operatorname{I_K}+\varpi_l\varpi_l^\top$ rather than  $\operatorname{I_K}$.

First, we try to reduce the DiE.
Although Remark \ref{red8dma} shows that  the higher order term $B_T^l$'s term
$W_l \,T^{-(1 -\delta) / 2} \operatorname{plim}\nolimits_{T\rightarrow \infty} \; T^{(1 -\delta) / 2} \operatorname{E} ({B_T})$ is the main source of size distortion, the operation that subtracting ${t_l}$ by
$W_l\,T^{-(1 -\delta) / 2}  \operatorname{plim}\nolimits_{T\rightarrow \infty} \; T^{(1 -\delta) / 2}  \operatorname{E}\left({B_T}\right)$
could not eliminate the size distortion caused by the higher order term ${B_T^l}$. The main reason is as follows. $W_l$ converges to a function of Brownian Motion rather than a constant. So the covariance matrix of the random vector  $W_l   \operatorname{plim}_{T\rightarrow \infty} \; T^{(1 -\delta) / 2} \operatorname{E}\left({B_T}\right)$  does not vanish even when the sample size goes to infinity.  As a result,  the finite sample covariance matrix of $W_l \, T^{-(1 -\delta) / 2}\operatorname{plim}_{T\rightarrow \infty} \; T^{(1 -\delta) / 2} \operatorname{E}\left({B_T}\right)$  is non-negligible since its rate $T^{-(1 -\delta) / 2}$ converging to zero is very slow. This non-negligible finite sample covariance matrix of $W_l \, T^{-(1 -\delta) / 2}\operatorname{plim}_{T\rightarrow \infty} \; T^{(1 -\delta) / 2} \operatorname{E}\left({B_T}\right)$  leads   that   the finite sample covariance matrix of ${t_l}-W_l\,T^{(1 -\delta) / 2}  \operatorname{plim}_{T\rightarrow \infty} \; T^{(1 -\delta) / 2} \operatorname{E}\left({B_T}\right)$ significantly deviates from  its asymptotic covariance matrix $\operatorname{I_K}$. This deviation causes the size distortion of the test statistic constructed by ${t_l}-W_l T^{-(1 -\delta) / 2}  \operatorname{plim}_{T\rightarrow \infty} \; T^{(1 -\delta) / 2} \operatorname{E}\left({B_T}\right)$.

To reduce the size distortion arise from $W_l \operatorname{plim}_{T\rightarrow \infty} \; T^{(1 -\delta) / 2} \operatorname{E}\left({B_T}\right)$,  we subtract ${t_l}$ by the estimator of $ T^{-(1 -\delta) / 2} \operatorname{plim}_{T\rightarrow \infty} \; T^{(1 -\delta) / 2} \operatorname{E}\left({B_T}\right)$, which is equal to $ - T^{-(1 -\delta) / 2} \frac{K+1}{2}{\hat{\rho}_{u v^*}}/ \sqrt{-2 c_z}$.
We construct the random vector ${\tilde{t}_n}$ as follows.
\begin{align}\label{dsjkih4}
{\tilde{t}_n} &\equiv    {t_l} + T^{-(1 -\delta) / 2} \frac{K+1}{2}{\hat{\rho}_{u v^*}}/ \sqrt{-2 c_z},
\end{align}
where $ {\hat{\rho}_{u v^*}}=\hat{\Sigma}_{vv}^{-1/2} \hat{\Sigma}_{vu} \hat{\Sigma}_{uu}^{-1/2}$ is the consistent estimator of $\rho_{u v^*}$. Note that $ \hat{\Sigma}_{vu}$, $\hat{\Sigma}_{vv}^{-1/2}$ and $\hat{\Sigma}_{uu}$ are the consistent estimators of $\Sigma_{vu}$, $\Sigma_{vv}^{-1/2}$ and $\Sigma_{uu}$, respectively.

In the following, we illustrate why the DiE is reduced in ${\tilde{t}_n}$ in the following aspects. First, by equation (\ref{dsjkih4}) and Proposition \ref{mulpropp2}, it is straightforward that
\begin{align}\label{multntied1}
{\tilde{t}_n} ={Z_T^l}+{B_T^n}+o_p\left(T^{\delta / 2-1 / 2}\right),
\end{align}
where ${B_T^n}= {B_T^l}- T^{-(1 -\delta) / 2} \operatorname{plim}_{T\rightarrow \infty} \; T^{(1 -\delta) / 2} \operatorname{E}\left({B_T}\right) \xrightarrow{P} 0$ and thus
\begin{align}\label{multntied2}
 T^{(1 -\delta) / 2} {B_T^n} ={R_T^l} + (W_l-\operatorname{I_K}) \operatorname{plim}_{T\rightarrow \infty} \; T^{(1 -\delta) / 2} \operatorname{E}\left({B_T}\right).
\end{align}
Second, we show that the distorting effect of ${B_T^n}$ for the size is much  smaller than that of ${B_T^l}$ in the  following aspects. By Proposition \ref{mulpropp2}, it follows that $W_l-\left[ \lambda(1-\lambda) \right]^{-1/2}\operatorname{I_K}$ is negative definite matrix and $\left[ \lambda(1-\lambda) \right]^{-1/2}\geq 2$. Therefore,  the size distortion induced by  $(W_l-\operatorname{I_K}) \operatorname{plim}_{T\rightarrow \infty} \; T^{(1 -\delta) / 2} \operatorname{E}\left({B_T}\right)$ is  smaller  than that of $W_l \operatorname{plim}_{T\rightarrow \infty} \; T^{(1 -\delta) / 2} \operatorname{E}\left({B_T}\right)$. As a result, the size distortion induced by ${B_T^n}$  is smaller than ${B_T^l}$.

\begin{remark}\label{red7ngn1}
The intuition of the size distortion induced by ${B_T^n}$ is smaller than that of ${B_T^l}$ can be easier to see using an example of a univariate model. First,  $\operatorname{P}(W_l-1>0)\rightarrow 1/2$ when $K=1$, which implies that the median of $W_l-1$ is approximately 0. Moreover, $(W_l-1)^2
 \in \left[0,1+ 2\lambda^{-1/2}(1-\lambda)^{-1/2} +\lambda^{-1}(1-\lambda)^{-1}\right]$ such that $W_l-1$ is bounded. It implies that ${B_T^n}$  is much smaller than  ${B_T^l}$ generally, which means DiE is reduced significantly in ${\tilde{t}_n}$.
\end{remark}

\begin{remark}\label{red7ngn2}
\textcolor{black}{Following Remark \ref{red7ngn1}, $(W_l-1)^2$ reaches the minimum upper bound when $\lambda=0.5$. So we set $\lambda=0.5$ to minimize the DiE of ${\tilde{t}_n}$. Simulation studies in Section \ref{section5} and the appendix support the choice of $\lambda=0.5$ in the finite sample.}
\end{remark}

However, VEE still exists in ${\tilde{t}_n}$ since
\begin{align}\label{mulVEE34G}
\operatorname{Var}\left({\tilde{t}_n}\right)& = \operatorname{Var}\left({Z_T^l} \right)+ \operatorname{Var}\left({B_T^n}\right) + o(1) = \operatorname{Var}\left({Z_T^l} \right)+ \operatorname{Var}\left({B_T^l}\right) + o(1) \\
&=\operatorname{I_K} +\varpi_l\varpi_l^\top + o(1).  \nonumber
\end{align}
Equation (\ref{mulVEE34G}) holds since $\operatorname{Var}( {B_T^n})= \operatorname{Var}\left[{B_T^l}- T^{-(1 -\delta) / 2} \operatorname{plim}_{T\rightarrow \infty} \; T^{(1 -\delta) / 2} \operatorname{E}\left({B_T}\right) \right]= \operatorname{Var}\left({B_T^l}\right) + o(1)$ induced by equation (\ref{dsjkih4}).
To eliminate the VEE,  the random vector ${\tilde{t}_m}$ is constructed as follows.
\begin{align}\label{dsjkih5}
{\tilde{t}_m} \equiv  \left(\operatorname{I_K}+\hat{\varpi}_l\hat{\varpi}_l^\top \right)^{-1/2}  {\tilde{t}_n} \xrightarrow{d} \operatorname{N}\left(0,\operatorname{I_K}\right),
\end{align}
where $\hat{\varpi}_l = -\frac{1}{2}\left\{ \sum\nolimits_{t=1}^T \tilde{z}_{t-1}^{**} (\tilde{z}_{t-1}^{**} )^\top \hat{u}_t^2 -\operatorname{I_K}\right\}$,
$\tilde{z}_{t-1}^{**} =({\hat{\Sigma}}_{zz})^{-1/2}\tilde{z}_{t-1}$ and
$$
{\hat{\Sigma}}_{zz} = \left(\operatorname{I_K} - S_a\right)\frac{\sum_{t=1}^{T_0}\Delta x_{t-1}\Delta x_{t-1}^\top }{1-\rho_z^2}\left(\operatorname{I_K} - S_a\right)^\top + \left(1-S_b\right)\frac{\sum_{t=T_0+1}^T \Delta x_{t-1}\Delta x_{t-1}^\top }{1-\rho_z^2} \left(\operatorname{I_K} - S_b\right)^\top.
$$
By equations (\ref{mulpredModel}), (\ref{kda4g42}), (\ref{dsjkih4}) and (\ref{dsjkih5}), the estimator for $\beta$ is defined as
\begin{align}\label{dsjkih6}
 \tilde{\beta}_m &\equiv H_l\left(\operatorname{I_K}+\hat{\varpi}_l\hat{\varpi}_l^\top \right)^{1/2} {\tilde{t}_m} + \beta= \hat{\beta}_{l}+ {\tilde{B}_m} T^{-(1 -\delta) / 2} \frac{K+1}{2}{\hat{\rho}_{u v^*}}/ \sqrt{-2 c_z},
\end{align}
where and
${\tilde{B}_m}\equiv   \left( \sum_{t=1}^T \tilde{z}_{t-1} x_{t-1}^\top \right)^{-1}\left( \sum_{t=1}^T \tilde{z}_{t-1} \tilde{z}_{t-1}^\top \hat{u}_t^2 \right)^{1/2}\left(\operatorname{I_K}+ \hat{\varpi}_l \hat{\varpi}_l^\top \right)^{1/2}.$
We apply the following estimator for the asymptotic distribution of $\hat{\beta}_m $ to reduce VEE of IVX.
\begin{align}
\operatorname{\widehat{Avar}}(\tilde{\beta}_m ) \equiv H_l\left(\operatorname{I_K}+\hat{\varpi}_l\hat{\varpi}_l^\top\right)  H_l^\top.
\end{align}
Then we define the test statistics for $H_0:R\beta =r_J$.
\begin{align}\label{gqmd3e}
{\tilde{Q}_m} \equiv \left(R \tilde{\beta}_m -r_J \right)^\top \left[R\operatorname{\widehat{Avar}}(\tilde{\beta}_m )R^\top\right]^{-1} \left(R \tilde{\beta}_m -r_J \right)
\end{align}
\begin{prop}\label{thmpropnew1}
 Under Assumption \ref{Assumption A.1} and   the null hypothesis $H_0:R\beta =r_J$, for SD predictors, it follows that
\begin{align*}
 Q_l - {\tilde{Q}_m}
&=  {t_l^\top} H_l^\top R^\top \left(R H_l   H_l^\top R^\top \right)^{-1}
R  H_l  \hat{\varpi}_l  \hat{\varpi}_l^\top  H_l^\top R^\top
\left(R H_l  H_l^\top R^\top \right)^{-1} R H_l {t_l} + o_p\left[T^{-(1-\delta)/2}\right]>0.
\end{align*}
\end{prop}
Thus far, the DE is eliminated for SD predictors in the finite sample, and the DiE and VEE are reduced significantly in ${\tilde{Q}_m}$.

On the other hand,  the asymptotic property of $\tilde{\beta}_m $ and ${\tilde{Q}_m}$ is the same as that of $\tilde{\beta}_l$ and ${\tilde{Q}_l}$ by equation (\ref{dsjkih6}), which is shown as follows.

\begin{thm}\label{multh1m23}
Under Assumption \ref{Assumption A.1}, for SD and WD predictors, it follows that
$$
 D_T(\tilde{\beta}_m- \beta)=D_T(\hat{\beta}_l- \beta)+o_p(1)\xrightarrow{d}
 \operatorname{MN}\left[0,\Sigma_{zx}^{-1}\Sigma_{zz}\left(\Sigma_{zx}^{-1}\right)^\top \right].
$$
\end{thm}
By Theorem \ref{multh1m23}, the asymptotic distribution of the test statistic ${\tilde{Q}_m}$ is the same as that of ${Q_l}$, which is shown as follows.
\begin{thm}\label{thm8f1}
Under Assumption \ref{Assumption A.1} and the null hypothesis $H_0:R\beta =r_J$, one can show that the limiting distribution of ${\tilde{Q}_m}$ is the $\chi^2$-distribution with $J$ degrees of freedom.
\end{thm}

Although ${\tilde{Q}_m}$ performs well in terms of size with SD predictors in the finite sample, it suffers size distortion with WD predictors since the terms to reduce the DiE and VEE for the case with SD predictors are redundant terms with WD predictors. To avoid size distortion with WD predictors, we construct the test statistics by the following steps. First, the weight is introduced ${W_z} =\operatorname{diag}(w_{z_1},w_{z_2},\cdots,w_{z_K})$ such that $w_{z_i} \equiv exp\left[-T(1-\hat{\rho}_i)^2 /K \right]=1+O_p(T^{-1})$ for SD predictors\footnote{We put $K$ in $w_i$ such that the weight $w_{z_i}$ is closer to one in the finite sample with more predictors. Therefore, the size control is improved since the DiE and the VEE are reduced sufficiently.} while $w_{z_i} \equiv exp\left[-T(1-\hat{\rho}_i)^2 /K \right]= o_p(T^{-1})$ for WD predictors,
where $\hat{\rho}_i$ is the consistent estimator of $\rho_i$ for the predictor $x_{i,t}$.
Then the proposed estimators and the test statistics are given as follows.
\begin{align}\label{uusjkih4}
{t_n} &\equiv
 {t_l} + {W_z} T^{-(1 -\delta) / 2} \frac{K+1}{2}{\hat{\rho}_{u\tilde{v}^*}}/ \sqrt{-2 c_z},\\
\label{uusjkih6}
 \hat{\beta}_m  & \equiv \left( \sum_{t=1}^T \tilde{z}_{t-1} x_{t-1}^\top \right)^{-1}\left( \sum_{t=1}^T \tilde{z}_{t-1} \tilde{z}_{t-1}^\top \hat{u}_t^2 \right)^{1/2}\left(\operatorname{I_K}+ {W_z}\hat{\varpi}_l\hat{\varpi}_l^\top{W_z}^\top \right)^{1/2} {t_m} + \beta\\
 & = \hat{\beta}_{l}+ {B_m} {W_z} T^{-(1 -\delta) / 2} \frac{K+1}{2}{\hat{\rho}_{u v^*}}/ \sqrt{-2 c_z}, \nonumber\\
 {Q_m} &\equiv \left(R \hat{\beta}_m -r_J \right)^\top \left[R\operatorname{\widehat{Avar}}(\hat{\beta}_m )R^\top\right]^{-1} \left(R \hat{\beta}_m -r_J \right),\label{qmeqn}
\end{align}
where ${B_m}\equiv \left( \sum_{t=1}^T \tilde{z}_{t-1} x_{t-1}^\top \right)^{-1}\left( \sum_{t=1}^T \tilde{z}_{t-1} \tilde{z}_{t-1}^\top \hat{u}_t^2 \right)^{1/2}\left(\operatorname{I_K}+ {W_z}\hat{\varpi}_l\hat{\varpi}_l^\top{W_z}^\top \right)^{1/2}$
and
$\operatorname{\widehat{Avar}}(\hat{\beta}_m ) \equiv H_l \left(\operatorname{I_K}+{W_z}\hat{\varpi}_l\hat{\varpi}_l^\top {W_z}^\top \right)H_l^\top$.

Since  $w_{z_i}=1+O_p(T^{-1})$ for SD predictors and $w_{z_i}= o_p(T^{-1})$ for SD predictors, we have  ${Q_m}={\tilde{Q}_m}+O_p(T^{-1})$.  By this result and Proposition \ref{thmpropnew1}, we have the following proposition.
\begin{prop}\label{thmpropnew2}
 Under Assumption \ref{Assumption A.1} and the null hypothesis $H_0:R\beta =r_J$, for both SD and WD predictors, it follows that
\begin{align*}
  Q_l - {Q_m}  =
\begin{cases}
 Q_l - {\tilde{Q}_m} +O_p(T^{-1})\\
 {t_l^\top} H_l^\top \left(H_l   H_l^\top \right)^{-1}
 H_l  \hat{\varpi}_l  \hat{\varpi}_l^\top  H_l^\top
\left(H_l  H_l^\top \right)^{-1} H_l {t_l}
+ O_p(T^{-1})>0,\; \text{SD};\\
O_p(T^{-1}),\quad \text{WD};
\end{cases}
\end{align*}
\end{prop}
By Proposition \ref{thmpropnew2}, the finite sample property of ${Q_m}$ with SD predictors is similar to that of ${\tilde{Q}_m}$ rather than that of ${Q_l}$. Thus ${Q_m}$ with SD predictors still has the nice finite sample property  of ${\tilde{Q}_m}$ discussed earlier. Meanwhile, the finite sample property of ${Q_m}$ with WD predictors is similar to that of ${Q_l}$ rather than that of ${\tilde{Q}_m}$. Thus  ${Q_m}$  is free of the redundant terms of ${\tilde{Q}_m}$ with WD predictors. Therefore, ${Q_m}$ is free of size distortion for both joint and marginal tests in multiple models with both SD and WD predictors.

Next, we present the one-sided test for $H_0:\beta_i=0$, that is $H_0:R\beta =r_J$ constructed by setting the rank $J$ of matrix $R$ to be one and $r_J=0$. So we construct the t-test statistic ${Q_m^t}$ for right side test $H_0:\beta_i=0$ vs $H_a:\beta_i>0$ and left side test $H_0:\beta_i=0$ vs $H_a:\beta_i<0$.
\begin{align}
{Q_m^t} \equiv \frac{R \hat{\beta}_m -r_J }{ \left[ R\operatorname{\widehat{Avar}}(\hat{\beta}_m )R^\top\right]^{1/2}}.
\end{align}
Note that $Q_m = (Q_m^t)^2$ when $J=1$. By similar arguments for the nice property of ${Q_m}$,  ${Q_m^t}$ performs well in terms of size for one-sided tests in both univariate and multiple models.

By equation (\ref{dsjkih5}) and that $\hat{\varpi}_l=o_p(1)$, we have
\begin{align}\label{uusjkih5}
{t_m}  &\equiv \left(\operatorname{I_K}+{W_z}\hat{\varpi}_l\hat{\varpi}_l^\top{W_z}^\top \right)^{-1/2} {t_n}\xrightarrow{d} \operatorname{N}\left(0,\operatorname{I_K}\right).
\end{align}
Therefore, the limiting distribution of the t-test statistic ${Q_m^t}$ with $J=1$ and Wald type test statistic ${Q_m}$ with $J\geq 1$ under the null hypothesis is stated in the following theorem with its detailed proof given in the Appendix.\footnote{Our procedure works for univariate models as well since ${Q_m^t}$ is equal to the t-test statistic for univariate model when $J=K=1$.}
\begin{thm}\label{thm4}
Under Assumption \ref{Assumption A.1} and the null hypothesis $H_0:R\beta =r_J$, the limiting distribution of the t-test statistic ${Q_m^t}$ with $J=1$ and those of Wald type test statistic ${Q_m}$ are the standard normal distribution and the $\chi^2$-distribution with $J$ degrees of freedom, respectively.
\end{thm}

We summarize our procedure in Algorithm \ref{algorithm1} below.
\begin{algorithm}[H]
\footnotesize
\caption{\label{algorithm1}The instructions for the construction of the test statistics ${Q_m^t}$ and ${Q_m}$}
\hspace*{0.01in}
\begin{algorithmic}[1]
\State Construct the IV $z_{t-1}$ by equation (\ref{mulivz}).
\State Construct the IV estimators $\hat{\beta}_{ivx}$, $\hat{\beta}_a$ and $\hat{\beta}_b$ using the full sample and two subsamples by equations (\ref{defdeftwo2}), (\ref{muldef2new}) and (\ref{muldef3new}).
\State Eliminate the DE: Construct the weighted IV estimator $\hat{\beta}_l \equiv (W_1-W_2-W_3)^{-1}(W_1 \hat{\beta}_{ivx}-W_2\hat{\beta}_a -W_3 \hat{\beta}_b)$, where $W_1=\sum\nolimits_{t=1}^{T} \bar{z}_{t-1} x_{t-1}^\top $, $W_2=S_a \sum\nolimits_{t=1}^{T_0} \left( z_{t-1}- \frac{1}{T_0}\sum\nolimits_{t=1}^{T_0} z_{t-1} \right) x_{t-1}^\top$ and
$W_3=S_b \sum\nolimits_{t=T_0+1}^{T} \left( z_{t-1}- \frac{1}{T-T_0}\sum\nolimits_{t=T_0+1}^{T} z_{t-1} \right) x_{t-1}^\top$.
\State Control size for both SD and WD predictors:
\begin{itemize}
 \item Reduce the DiE: Construct the estimator $\hat{\beta}_m = \hat{\beta}_{l}+ {B_m} {W_z} T^{-(1 -\delta) / 2} \frac{K+1}{2}{\hat{\rho}_{u v^*}}/ \sqrt{-2 c_z}$, where
${B_m}\equiv \left( \sum_{t=1}^T \tilde{z}_{t-1} x_{t-1}^\top \right)^{-1}\left( \sum_{t=1}^T \tilde{z}_{t-1} \tilde{z}_{t-1}^\top \hat{u}_t^2 \right)^{1/2}\left(\operatorname{I_K}+ {W_z}\hat{\varpi}_l\hat{\varpi}_l^\top{W_z}^\top \right)^{1/2}$.
 \item Reduce the VEE: Construct the estimator for the asymptotic covariance matrix of as $\operatorname{\widehat{Avar}}(\hat{\beta}_m ) \equiv H_l \left(\operatorname{I_K}+{W_z}\hat{\varpi}_l\hat{\varpi}_l^\top {W_z}^\top \right) H_l^\top$, where
$
H_l = \left( \sum_{t=1}^T \tilde{z}_{t-1} x_{t-1}^\top \right)^{-1}\left(\frac{T}{T-2K-1} \sum_{t=1}^T \tilde{z}_{t-1} \tilde{z}_{t-1}^\top \hat{u}_t^2 \right)^{1/2}
$.
\item \textcolor{black}{Reduce the size distortion induced by the variance of the estimator of $u_t$: Use Lagrange-multiplier principle to obtain $\hat{u}_t= y_t - \hat{\mu}_s - x_{t-1}^\top \hat{\beta}_s$ by the constrained OLS estimation
$(\hat{\mu}_s,\hat{\beta}_s)^\top  = \arg\, \min_{\mu,\beta} \left( y_t - \mu - x_{t-1}^\top \beta\right)^2,
\ s.t. \quad R\beta =r_J$.}
\end{itemize}
\State Construct the test statistic ${Q_m} \equiv \left(R \hat{\beta}_m -r_J \right)^\top \left[R\operatorname{\widehat{Avar}}(\hat{\beta}_m )R^\top\right]^{-1} \left(R \hat{\beta}_m -r_J \right)$ with SD and WD predictors for $H_0:R\beta =r_J$.
\If{Focus on the one-sided marginal test $H_0:\beta_i=0$ vs $H_a:\beta_i>0$ and $H_0:\beta_i=0$ vs $H_a:\beta_i<0$ }
 \State Construct the t-test statistic ${Q_m^t} \equiv \frac{R \hat{\beta}_m -r_J}{ \left[ R\operatorname{\widehat{Avar}}(\hat{\beta}_m )R^\top\right]^{1/2}}$.
\EndIf
\end{algorithmic}
\end{algorithm}
\section{Monte Carlo Simulations}\label{section5}

We demonstrate the effectiveness of the proposed procedure using a multivariate case.
We report the results for 1) the joint test $H_0:\beta=0$ vs $H_a:\beta\neq 0$; 2) the marginal test (two-sided) $H_0:\beta_i=0$ vs $H_a:\beta_i \neq 0$; 3) the marginal test (one-sided) $H_0:\beta_i=0$ vs $H_a:\beta_i > 0$ and $H_0:\beta_i=0$ vs $H_a:\beta_i < 0$.

The DGP of $y_t$ and $x_{t-1}$ is equation (\ref{mulpredModel}), (\ref{mulgtuA1}) and (\ref{garchset}), in which $\mu=1$ and $\varphi_0=1$. The DGP of the innovations is as follows: $v_{i,t}=\gamma_i \eta_t + \check{v}_{i,t}$, where $(u_t,\check{v}_{i,t})^\top \;\sim \; i.i.d. \, \operatorname{N}(0_K,\operatorname{I}_{K\times K})$.
Thus the contemporaneous correlation coefficient between $\eta_t$ and $v_{i,t}$ is
$\gamma_i (1+\gamma_i^2)^{-1/2}$,  which is the source of  the size distortion of the standard IVX test statistics \citep{Kostakisetal2015,Phillips2013PredictiveRU} and  DiE and VEE shown in Proposition \ref{mulpropfdie3}.
We  report the simulation results for a GARCH model with $\lambda=0.5$, $\varphi_1=0.1$, $\bar{ \varphi}_1=0.85$ and the sample $T=750$ with the nominal size $5\%$. \footnote{To save space, we did not report the simulation results for GARCH model with the sample size $T=250$ and $T=500$ and ARCH and i.i.d. model with the sample size $T=250$, $T=500$ and $T=750$ in the paper, which is similar to the result of GARCH model with the sample size $T=750$. The codes and results are available upon request.} Simulation is repeated $10,000$ times for each setting.
Set $\beta = \frac{b}{(1+K)/2}$ in Table \ref{joint3} and $\delta=0.95$ and $c_z=-4-K$ for all Tables.   \footnote{\textcolor{black}{This setting (choice of $c_z$) is not exactly the same as the IVX of \cite{Phillips2013PredictiveRU}, in that it makes the instrumental variable $z_{t-1}$ become less persistent as the number of predictors K grows bigger. Therefore, in this setting, we can see the advantages of our test even when the predictors  are less persistent, but the VEE is nonetheless significant (see  discussions after Proposition \ref{mulpropfdie3}).}}

First, the results for the size and power performances of the proposed test statistics $Q_l$ and $Q_m$ for joint test $H_0:\beta=0$ are shown in Table \ref{joint3} in which $K=2,3,\cdots,10$ and $\beta= \frac{b}{(1+K)/2}$. Second, the size and power performances of the proposed test statistics $Q_l^t$ and $Q_m^t$ for two-sided marginal test $H_0:\beta_{i}=0$ vs $H_a:\beta_{i}\neq 0$ and right side marginal test $H_0:\beta_{i}=0$ vs $H_a:\beta_{i}> 0$ are shown in Panel A and Panel B of Table \ref{mar}, in which $K=10$.
The first column of Panel A and Panel B of Table \ref{mar} and the third column of Table \ref{joint3} are the size results while the others are power results. In these models, we set $(\rho_1,\rho_2,\cdots,\rho_{K})^\top=(\bm{\rho})_K$, where $\bm{\rho}= (0.996,0.993,1,0.987,0.967,0.95,0.9,0.98,0.92,0.94)^\top$.  We show the simulation results for $(\gamma_1,\gamma_2,\cdots,\gamma_K)^\top=(\Gamma)_K$, $\Gamma=(-3,2,1,3,1,0.833,0.667,0.5,0.333,0.167)^\top$ and thus the contemporaneous correlation coefficient $\left[\gamma_1 (1+\gamma_1^2)^{-1/2},\cdots,\gamma_K (1+\gamma_K^2)^{-1/2} \right]^\top$ between $u_t$ and $v_t$ are $(\tilde{\Gamma})_K$ and
$ \tilde{\Gamma}=(-0.949,0.894,0.707,0.949,0.707,0.64,0.555,0.447,0.316,0.164)^\top$.

We have the following findings from Tables  \ref{mar} and \ref{joint3}. First, for the joint test $H_0:\beta=0$, the proposed test statistic ${Q_l}$ still suffers size distortion with multiple predictors ($K\geq 3$) and the size distortion grows bigger as the number of predictors K grows bigger. Meanwhile, the proposed test statistic ${Q_m}$ is almost free of size distortion in different settings. It is because ${Q_l}$ still suffers from DiE and VEE while ${Q_m}$ does not. Second, the size performance of ${Q_m^t}$ is better than that of ${Q_l^t}$ although ${Q_m^t}$ suffers small-scale size distortion for two-sided marginal test $H_0:\beta_{i}=0$ vs $H_a:\beta_{i}\neq 0$ and right side marginal test $H_0:\beta_{i}=0$ vs $H_a:\beta_{i}> 0$. Specifically, the size performances of ${Q_l^t}$ and ${Q_m^t}$ are comparable for right side marginal test $H_0:\beta_{i}=0$, while the size performance of ${Q_m^t}$ is much better than that of ${Q_l^t}$ for two-sided marginal test $H_0:\beta_{i}=0$ vs $H_a:\beta_{i}\neq 0$. Third, the power performances of ${Q_l^t}$ and ${Q_m^t}$ are comparable and quite well.

We also conduct other experiments (a univariate case and other settings of multivariate cases, e.g., for different $\lambda$'s) for the robustness of our conclusions. These results are reported in the Appendix. We obtain a similar conclusion there.

\begin{table}[H]
 \centering
 \caption{Result (\%) for Marginal Test with $\lambda=0.5$ and $K=10$}
 \resizebox{\textwidth}{!}{
  \begin{tabular}{c|c|ccccccccc|ccccccccc }
  \hline
  \multicolumn{2}{c|}{ Tests} & \multicolumn{9}{c|}{Panel A: $H_0:\beta_i=0$ vs $H_0:\beta_i \neq 0$} & \multicolumn{9}{c}{Panel B: $H_0:\beta_i=0$ vs $H_0:\beta_i > 0$}\\
  \hline
  \multicolumn{2}{c|}{$\beta$} & 0  & 0.05 & 0.1 & 0.15 & 0.2 & 0.25 & 0.3 & 0.35 & 0.4 & 0  & 0.05 & 0.1 & 0.15 & 0.2 & 0.25 & 0.3 & 0.35 & 0.4 \\
  \hline
  \multirow{10}[0]{*}{${Q_l^t}$}
   & i=1 & 7.3 & 59.8 & 83.7 & 89.7 & 93.1 & 93.9 & 94.8 & 95.4 & 96.0 & 8.1 & 67.3 & 86.8 & 91.6 & 94.2 & 94.8 & 95.8 & 96.1 & 96.8 \\
   & i=2 & 6.8 & 39.4 & 74.1 & 86.6 & 90.8 & 92.4 & 94.6 & 94.8 & 95.5 & 5.9 & 49.3 & 79.4 & 89.0 & 92.5 & 93.6 & 95.4 & 95.7 & 96.1 \\
   & i=3 & 7.0 & 40.8 & 69.3 & 80.2 & 85.9 & 87.0 & 88.8 & 89.7 & 91.0 & 5.5 & 48.3 & 74.2 & 83.7 & 88.1 & 89.0 & 90.6 & 91.7 & 92.4 \\
   & i=4 & 6.8 & 48.2 & 82.0 & 90.4 & 94.0 & 95.2 & 95.9 & 96.6 & 96.5 & 5.5 & 57.4 & 85.5 & 92.3 & 95.0 & 96.1 & 96.4 & 97.2 & 97.0 \\
   & i=5 & 5.9 & 27.6 & 65.9 & 84.8 & 90.9 & 93.3 & 94.6 & 95.8 & 96.3 & 5.7 & 37.3 & 73.3 & 88.2 & 92.9 & 94.7 & 95.5 & 96.6 & 97.0 \\
   & i=6 & 5.7 & 22.8 & 60.0 & 81.0 & 89.3 & 93.2 & 94.3 & 95.7 & 96.5 & 5.5 & 32.2 & 68.4 & 85.5 & 91.5 & 94.7 & 95.5 & 96.5 & 97.1 \\
   & i=7 & 5.4 & 15.8 & 45.7 & 71.0 & 84.5 & 90.3 & 93.1 & 94.8 & 95.6 & 4.9 & 23.7 & 56.2 & 77.8 & 88.0 & 92.4 & 94.5 & 95.8 & 96.3 \\
   & i=8 & 6.7 & 30.7 & 68.7 & 84.7 & 90.5 & 93.1 & 94.8 & 95.5 & 95.8 & 6.2 & 39.8 & 75.8 & 87.9 & 92.3 & 94.3 & 95.6 & 96.2 & 96.5 \\
   & i=9 & 5.6 & 18.1 & 48.8 & 73.7 & 86.1 & 91.3 & 93.9 & 94.8 & 95.9 & 5.4 & 26.8 & 59.0 & 80.3 & 89.5 & 93.2 & 95.0 & 95.6 & 96.7 \\
   & i=10 & 6.1 & 20.7 & 54.7 & 76.6 & 87.8 & 91.9 & 94.0 & 95.3 & 96.3 & 5.7 & 29.8 & 64.1 & 82.3 & 90.7 & 93.6 & 95.2 & 96.2 & 96.9 \\
  \hline
 \multirow{10}[0]{*}{${Q_m^t}$}
   & i=1 & 6.3 & 58.5 & 82.8 & 89.4 & 92.7 & 93.6 & 94.7 & 95.2 & 95.7  & 7.8 & 65.9 & 86.4 & 91.3 & 94.1 & 94.6 & 95.6 & 95.9 & 96.5 \\
   & i=2 & 6.0 & 37.7 & 72.8 & 86.1 & 90.2 & 92.0 & 94.3 & 94.6 & 95.3  & 5.5 & 47.5 & 78.4 & 88.5 & 92.1 & 93.3 & 95.3 & 95.5 & 96.0 \\
   & i=3 & 6.3 & 38.3 & 67.8 & 79.3 & 85.2 & 86.6 & 88.5 & 89.3 & 90.6  & 5.1 & 46.8 & 73.2 & 82.8 & 87.7 & 88.6 & 90.2 & 91.2 & 92.2 \\
   & i=4 & 6.1 & 50.0 & 82.3 & 90.3 & 93.7 & 95.0 & 95.8 & 96.4 & 96.4  & 6.4 & 59.4 & 85.4 & 92.2 & 94.8 & 95.9 & 96.3 & 97.1 & 97.0 \\
   & i=5 & 5.4 & 29.0 & 66.4 & 84.5 & 91.0 & 93.0 & 94.5 & 95.7 & 96.2  & 6.6 & 39.1 & 74.0 & 88.1 & 92.7 & 94.4 & 95.5 & 96.4 & 96.9 \\
   & i=6 & 5.2 & 24.1 & 61.2 & 81.1 & 89.3 & 93.2 & 94.3 & 95.6 & 96.4  & 6.5 & 34.2 & 69.6 & 85.7 & 91.7 & 94.6 & 95.4 & 96.4 & 97.0 \\
   & i=7 & 5.4 & 17.7 & 47.9 & 72.3 & 84.7 & 90.4 & 93.1 & 94.9 & 95.5  & 6.0 & 26.2 & 58.3 & 78.8 & 88.6 & 92.6 & 94.5 & 95.8 & 96.4 \\
   & i=8 & 5.8 & 30.0 & 67.8 & 84.4 & 89.9 & 92.8 & 94.7 & 95.3 & 95.6  & 6.1 & 39.6 & 74.9 & 87.5 & 91.9 & 94.1 & 95.6 & 96.0 & 96.4 \\
   & i=9 & 5.4 & 18.3 & 48.5 & 73.5 & 85.5 & 90.9 & 93.5 & 94.6 & 95.7  & 5.7 & 27.4 & 59.3 & 79.8 & 89.1 & 93.0 & 94.8 & 95.6 & 96.6 \\
   & i=10 & 5.5 & 19.6 & 52.8 & 75.5 & 86.9 & 91.5 & 93.5 & 94.9 & 96.1  & 5.5 & 29.0 & 62.9 & 81.2 & 90.0 & 93.3 & 95.0 & 95.9 & 96.8 \\
   \hline
 \end{tabular}%
  }
 \label{mar}%

\end{table}%
To sum up, the proposed test statistics ${Q_m}$ and ${Q_m^t}$ with $\lambda=0.5$ perform quite well in terms of both size and power in multivariate models. Thus, we recommend that empirical researchers apply our test statistics ${Q_m}$ and ${Q_m^t}$ and set $\lambda=0.5$.

\begin{table}[H]
 \centering
 \caption{Result (\%) for Joint Test $H_0:\beta=0$ with $\lambda=0.5$}
   \makebox[\textwidth][c]{
  \scalebox{0.75}{
    \begin{tabular}{c|ccccccccc|ccccccccc}
    \hline
    Statistics & \multicolumn{9}{c|}{${Q_l}$}    & \multicolumn{9}{c}{${Q_m}$} \\
    \hline
    $b$     & 0     & 0.02  & 0.04  & 0.06  & 0.08  & 0.1   & 0.12  & 0.14  & 0.16  & 0     & 0.02  & 0.04  & 0.06  & 0.08  &
    0.1   & 0.12  & 0.14  & 0.16 \\
    \hline
    K=2   & 6.0   & 23.2  & 60.9  & 79.6  & 88.7  & 92.2  & 94.3  & 95.9  & 96.6  & 4.9   & 19.0  & 55.3  & 76.8  & 87.2  & 91.3  & 93.6  & 95.5  & 96.4  \\
    K=3   & 6.6   & 15.8  & 42.3  & 67.7  & 82.4  & 90.3  & 93.6  & 96.1  & 96.9  & 4.4   & 12.4  & 37.5  & 63.7  & 80.1  & 88.9  & 92.7  & 95.6  & 96.4  \\
    K=4   & 7.1   & 17.2  & 51.8  & 76.9  & 89.8  & 95.6  & 97.7  & 98.6  & 99.3  & 4.6   & 19.1  & 55.6  & 79.0  & 90.8  & 96.3  & 97.7  & 98.7  & 99.3  \\
    K=5   & 6.9   & 16.6  & 46.6  & 74.4  & 87.3  & 95.5  & 97.4  & 98.9  & 99.6  & 4.8   & 18.0  & 51.3  & 77.7  & 89.0  & 96.1  & 97.7  & 99.0  & 99.6  \\
    K=6   & 7.1   & 14.3  & 41.4  & 67.8  & 86.0  & 93.7  & 97.5  & 98.4  & 99.2  & 4.5   & 15.6  & 45.8  & 72.8  & 88.3  & 94.9  & 97.9  & 98.5  & 99.3  \\
    K=7   & 6.4   & 12.6  & 34.0  & 61.9  & 79.7  & 90.1  & 95.2  & 98.0  & 99.2  & 4.5   & 13.5  & 39.5  & 67.3  & 84.1  & 91.9  & 96.2  & 98.3  & 99.3  \\
    K=8   & 7.0   & 12.6  & 31.8  & 57.3  & 77.0  & 89.1  & 95.1  & 97.9  & 98.8  & 4.8   & 12.9  & 36.1  & 63.3  & 81.2  & 91.3  & 96.2  & 98.3  & 99.0  \\
    K=9   & 7.3   & 10.6  & 25.8  & 49.3  & 72.3  & 85.5  & 92.2  & 96.6  & 98.2  & 5.1   & 11.0  & 30.6  & 56.0  & 77.0  & 88.9  & 94.2  & 97.4  & 98.8  \\
    K=10  & 7.5   & 9.6   & 21.8  & 43.5  & 64.5  & 79.1  & 88.4  & 94.1  & 97.2  & 4.4   & 9.8   & 25.6  & 50.2  & 70.5  & 83.8  & 91.4  & 95.5  & 97.6  \\
    \hline
    \end{tabular}%
    }
    }
 \label{joint3}%

\end{table}%

\section{Robust Inference for Bond Risk Premia}\label{section6}
Prediction of bond risk premia is crucial for monetary policy and  investment decisions. Following the literature \citep{2005Bond}, we use the excess log return of n-year U.S. discount bond rx(n), where $n=2,3,4,5$, as the left-hand side variable. We reexamine the predictability of five bond forward rates F1-F5 \citep{2005Bond} and the first eight macroeconomic principal components M1-M8 (the $\hat f_i$, $i=1,\dots , 8$ in \citealp{ludvigson2009macro}) constructed by 132 macro variables in a multiple predictive regression model. Specifically, the forward rate F1 is the log price of the one-year discount bond; and the forward rate Fn of n-year (n=2,\dots 5) bond is the log price difference between (n-1)-year bond and n-year bond.\footnote{The dataset used in this section is from the Fama-Bliss database available on CRSP, and the macroeconomic factors are from the \href{https://www.sydneyludvigson.com/s/Updated\_LN\_Macro\_Factors\_2023FEB.xlsx}{Link of \cite{ludvigson2009macro}}.}  Our main sample period is 1964:01-2003:12, which is the same used in the aforementioned literature. We present the test statistics and the corresponding p-values in parenthesis marked with $\ast$, $\ast\ast$, $\ast\ast\ast$ implying the rejection of the null hypothesis at the 10\%, 5\%, and 1\% levels. We also conduct empirical studies on A1)  reexamining the three linear combinations of F1-F5 and M1-M8, which include CP \citep{2005Bond} constructed by F1-F5 and {LN1}, {LN2} (the F5, F6 in the original paper of \citealp{ludvigson2009macro}) constructed by M1-M8; and A2) evaluating the predictability during the pandemic period 2020:01-2022:12. Due to space limitations, we put the more details of these two studies in Appendix \ref{section7.3}.

\begin{table}
  \centering
  \caption{Persistence $\rho_i$ of Predictors}
  \makebox[\textwidth][c]{
  \scalebox{0.75}{
   \begin{threeparttable}
    \begin{tabular}{ccccccccccccccccc}
    \hline
    \multicolumn{17}{c}{Sample studied in \cite{ludvigson2009macro} (1964:01-2003:12)} \\
        \hline
    \multicolumn{1}{l}{Variables} & F1  & F2  & F3  & F4  & F5  & CP  & M1  & M2  & M3  & M4  & M5  & M6  & M7  & M8  & {LN1}  & {LN2} \\
    $\rho_i$ & 0.996 & 0.997 & 0.997 & 0.996 & 0.997 & 0.980 & 0.760 & 0.306 & 0.563 & 0.458 & 0.581 & 0.554 & 0.185 & 0.203 & 0.793 & 0.900 \\
    \hline
    \multicolumn{17}{c}{Sample during the Covid-19 epidemic period (2020:01-2022:12)} \\
        \hline
    \multicolumn{1}{l}{Variables} & F1  & F2  & F3  & F4  & F5  & CP  & M1  & M2  & M3  & M4  & M5  & M6  & M7  & M8  & {LN1}  & {LN2} \\
    $\rho_i$ & 0.995 & 0.993 & 0.991 & 0.992 & 0.992 & 0.990 & 0.307 & 0.257 & -0.22 & 0.531 & 0.373 & 0.312 & 0.092 & 0.337 & 0.326 & 0.371 \\
    \hline
    \end{tabular}%

    \end{threeparttable}}}
    \vspace{-0.5cm}
  \label{tab:Example one table1}
\end{table}

The predictors' persistence parameters $\rho_i$'s are reported in Table \ref{tab:Example one table1}. It shows that F1-F5 and CP are likely the SD predictors, while the rest of the variables are likely to be WD predictors. Therefore, applying our robust testing procedure is necessary to avoid the spurious predictability induced by the size distortion discussed earlier. We show the testing results using F1-F5 and M1-M8 as regressors  in Table \ref{tab:Example two table4}. This is helpful for practitioners to discover which component of  CP, the {LN1} and {LN2} plays a vital part in predicting the bond excess returns.

We discuss the main results as follows. First, our method shows less significance (of predictability) than the original IVX \citep{Kostakisetal2015,Phillips2013PredictiveRU} and \cite{ludvigson2009macro}. This aligns with the theoretical results in Section \ref{section4} that our method improves the size control performance of the original IVX by reducing the distortion effects of the higher-order terms.  Our approach does not detect some predictability of F1-F5 and M1-M8 found by IVX and test statistics based on OLS. At the same time, all predictability of F1-F5 and M1-M8 found by our method are also detected by IVX and test statistics based on OLS except for the predictability of F5 for rx(5). Our test also finds that the linear combinations predictors CP, {LN1} and {LN2} have predicting power for the one-year bond excess returns at all maturities in the main sample period (with higher p-values than the IVX). During the pandemic period of 2020:01-2022:12, our test does not reject the null that CP, {LN1} and {LN2} have no predictability at all maturities, contrary to the OLS and IVX tests, where the former almost rejects all null hypotheses, and the latter finds CP to have predictability. The details are shown in Table \ref{tab:Example one table2} of the Appendix.

\begin{table}[H]
  \centering
  \caption{P Value (\%) of Inference for  F1-F5 and M1-M8 (1964:01-2003:12)}
 \resizebox{\textwidth}{!}{
   %\begin{threeparttable}
     \begin{tabular}{l|lll|lll|lll|lll}
    \hline
     $y_t$ & \multicolumn{3}{c|}{$rx(2)$} & \multicolumn{3}{c|}{$rx(3)$} & \multicolumn{3}{c|}{$rx(4)$} & \multicolumn{3}{c}{$rx(5)$} \\
     \hline
      Predictors  & $Q_m$  & IVX & OLS & $Q_m$  & IVX & OLS & $Q_m$  & IVX & OLS & $Q_m$  & IVX & OLS \\
    \hline
    {F1}
        & 0.0*** & 0.0*** & 0.0*** & 0.0*** & 0.0*** & 0.0*** & {16.2} & 0.0*** & 0.0*** & {35.0} & 0.0*** & 0.1*** \\

    {F2}
        & 0.0*** & 0.2*** & 0.0*** & 87.1 & 64.2 & 52.1 & 49.7 & 59.0 & 41.6 & 33.4 & 69.7 & 49.1 \\

    {F3}
        & {23.4} & 1.42** & 10.4 & 7.8* & 0.0*** & 0.0*** & {88.2} & 0.0*** & 39.5 & {27.0} & 1.1*** & 20.7 \\

    {F4}
        & 92.5 & 72.3 & 40.4 & 87.2 & 72.1 & 67.0 & 0.3*** & 8.4* & 0.0*** & 20.1 & 33.8 & 0.7*** \\

    {F5}
        & 14.4 & 90.4 & 31.6 & 27.9 & 54.8 & 94.4 & 63.3 & 84.8 & 19.2 &  0.2***  & 15.2 & 0.0*** \\

    {M1}
        & 0.0*** & 0.6*** & 0.0*** & 0.5*** & 0.0*** & 0.0*** & 1.4** & 0.4*** & 0.0*** & 0.9*** & 0.4*** & 0.0*** \\

    {M2}
        & 12.8 & 28.9 & 38.2 & 26.0 & 92.4 & 20.1 & 37.3 & 97.3 & 24.1 & 49.2 & 86.6 & 21.8 \\

    {M3}
        & 3.3** & 0.3*** & 0.0*** & 9.4* & 0.4*** & 0.0*** & {10.7} & 1.7** & 0.0*** & 9.9* & 2.2** & 0.0*** \\

    {M4}
        & {0.3} & 0.2*** & 2.6** & {50.5} & 0.9*** & 8.5* & 48.0 & 0.1*** & 1.6** & {34.4} & 0.6*** & 5.1** \\

    {M5}
        & 6.8* & 4.8** & 0.2*** & 10.4 & 15.9 & 0.3*** &  5.6*  & 21.6 & 0.0*** & 10.4 & 6.3** & 0.0*** \\

    {M6}
        & 58.7 & 24.8 & 48.2 & 89.8 & 16.5 & 56.0 & 73.2 & 87.2 & 95.1 & 77.5 & 20.0 & 73.7 \\

    {M7}
        & 0.3** & 0.2** & 0.0*** & 0.3*** & 0.0*** & 0.0*** & 0.7*** & 0.1*** & 0.0*** & 2.2** & 0.0*** & 0.0*** \\

    {M8}
        & 13.0 & 17.0 & 5.0** & 21.8 & 54.3 & 15.6 & 25.4 & 98.9 & 40.5 & 18.9 & 84.4 & 33.3 \\
    \hline
    \end{tabular}%
    }
   \begin{tablenotes}
         \scriptsize
    \item    {\bf{Notes}}: $Q_m$ is our test, IVX is the test in \cite{Phillips2013PredictiveRU}, and OLS is the test in \cite{ludvigson2009macro}.
    \end{tablenotes}
  \label{tab:Example two table4}
\end{table}%
\vspace{-0.5cm}

From the economic perspective, our empirical result in Table \ref{tab:Example two table4} supports the market segmentation theory
 \citep{2013Flow,foley2016impact,greenwood2018asset}. The market segmentation theory points out that short and long-term securities are not perfectly substitutable, and the short-term bond risk premia could only be predicted by the short-term forward rates and, similarly, the long-term forward rates for long-term bond risk premia. In the main sample, our findings parallel the studies of \citep{ludvigson2009macro} that CP, {LN1} and {LN2} can predict the expected one-year excess return on bonds (see the Appendix for details). However, during the pandemic, the bond return is more likely to be driven by aggressive monetary policies and epidemic shocks \citep{goldstein2021covid, levine2021did}, which  hamper the predicting power of CP, {LN1} and {LN2}. Our test can detect the spurious predictability of CP, {LN1} and {LN2} in forecasting excess bond returns under different macroeconomic scenarios, while IVX and OLS do not.

\section{Conclusion}\label{section7}
This paper improves the popular IVX testing \citep{Kostakisetal2015,Phillips2013PredictiveRU} by addressing not only DE and DiE, pointed out by  \cite{HosseinkouchackDemetrescu2021}, but also VEE which cannot be ignored, especially in multiple predictive regressions. We introduce a three-step method to deal with all the aforementioned issues regarding finite sample bias and variance-inflating terms. As a result, the size performance of the proposed method for the one-sided test and the test in multiple predictive regressions is improved significantly, while the power performance is comparable with the original IVX. Numerical simulations demonstrate the effectiveness of the newly proposed approach. Moreover, an empirical study of the predictability of bond returns shows that the original IVX rejects the null more often than our method, which supports the theoretical results that our procedure reduces the size distortion induced by the higher-order terms of IVX.

\newpage
\setcounter{section}{0}
\renewcommand\thesection{\Alph{section}}
\setcounter{equation}{0}  % reset counter
\renewcommand{\theequation}{A\arabic{equation}}
\setcounter{table}{0}
\renewcommand{\thetable}{B\arabic{table}}
\setcounter{figure}{0}
\renewcommand{\thefigure}{B\arabic{figure}}

\begin{center}
{\large\bf SUPPLEMENTARY MATERIAL OF ``ROBUST INFERENCE FOR MULTIPLE PREDICTIVE REGRESSIONS WITH AN APPLICATION ON BOND RISK PREMIA''}
\end{center}
The Appendices include the following parts: Section \ref{app:a} contains all technical proofs and additional discussions complementary to the main text. Section \ref{app:b} collects extra simulation results. Section \ref{section7.3} provides further details of the empirical exercise.
\vspace{-1em}

\newcounter{counter}[section]

%%%%%%%%%%%%%%%%%%%%%%%%%%%%%%%%%%%%%%%%%%%%%%%%%%%%%%%%%%%%%%%%%%%%%%%%%%%%%%

\section{Proofs}\label{app:a}

To help the readers navigate the testing procedure in the main text, we first explain the names of important test statistics.
\begin{description}

\item[Explanation about test statistics' subscripts in the main text:]  \textcolor{black}{Here we use $G$ as the general notation of a test statistic. The subscript  `ivx' in the random vector (or scalar) $G_{ivx}$ is to show $G_{ivx}$ is constructed by the method of the original IVX \citep{Kostakisetal2015}. The subscript  `a' in the random vector (or scalar) $G_a$ is to show $G_a$ is constructed by time series in the duration of the first subsample, while the subscript `b'  in the random vector (or scalar) $G_a$ is to show $G_a$ is constructed by time series in the duration of the second subsample. The subscript  `l' in the random vector (or scalar) $G_l$ is to show $G_l$ is constructed after the procedure that DE is eliminated. The subscript  `n' in the random vector (or scalar) $G_l$ is to show $G_l$ is constructed after the procedure that DE is eliminated and DiE is reduced. The subscript  `m' in the random vector (or scalar) $G_m$ is to show $G_m$ is constructed after the procedure that DE is eliminated and both DiE and VEE are reduced.}

\end{description}

Then, we give the proofs of the theoretical results. Then, we give the proofs of the theoretical results. The equation numbers refer to those in the main text. The new numbered equations in this appendix have the prefix A before the numbers, such as `A1'.
\begin{proof}[Proof of equation (\ref{mulkjeyi4ns})]
By equation (\ref{mult3inti2st}) and the definitions of $\tilde{S}_a$ and $\tilde{S}_b$ in equation (\ref{7hj4e45d}), it is known that
\begin{align}
\sum\nolimits_{t=1}^{T} \tilde{z}_{t-1}& = (\operatorname{I_K}-S_a) \sum\nolimits_{t=1}^{T_0} z_{t-1}+ (\operatorname{I_K}-S_b) \sum\nolimits_{t=T_0+1}^{T} z_{t-1} \\
& = \sum\nolimits_{t=1}^{T} z_{t-1} -S_a \sum\nolimits_{t=1}^{T_0} z_{t-1} - S_b \sum\nolimits_{t=T_0+1}^{T} z_{t-1} \nonumber\\
& = \sum\nolimits_{t=1}^{T} z_{t-1} - \lambda \sum\nolimits_{t=1}^{T } z_{t-1} - (1-\lambda) \sum\nolimits_{t=1}^{T} z_{t-1}=0 \nonumber
\end{align}
\end{proof}

\begin{lemma}\label{mulleea3pp}
Under Assumption \ref{Assumption A.1}, for SD and WD predictors, it follows that
\begin{align}
D_T^{-1} \sum\limits_{t=1}^{T} \tilde{z}_{t-1} u_t \xrightarrow{d} \operatorname{MN}\left(0,\Sigma_{zz} \right)
\end{align}
\end{lemma}
\begin{proof}[Proof of Lemma \ref{mulleea3pp}]
First, we consider the case with SD predictors. To begin with, we first prove that $\left(\sum_{t=1}^{T_0} z_{t-1}^\top u_t, \sum_{t=T_0+1}^{T} z_{t-1}^\top u_t\right)^\top$ follows asymptotically normal distribution. To this aim, we need to prove that
 $$
 \kappa_1^\top \frac{1}{T^{(1+\delta)/2}}\sum\limits_{t=1}^{T_0} z_{t-1}u_t + \kappa_2^\top \frac{1}{T^{(1+\delta)/2}} \sum\limits_{t=T_0+1}^{T} z_{t-1}u_t \xrightarrow{d}
 \operatorname{N}\left[0,  \lambda \kappa_1^\top\Omega_{zz}\kappa_1 +  (1-\lambda) \kappa_2^\top\Omega_{zz}\kappa_2 \right].
 $$
 for any real vector $\kappa_1$ and $\kappa_2$ with dimension K  and use the Cramer-Wold device. Define
 \begin{align}\label{mulkappiv}
 \breve{z}_{t-1} =
 \begin{cases}
 \kappa_1^\top z_{t-1},\; 1\leq t\leq T_0\\
 \kappa_2^\top z_{t-1},\; T_0+1\leq t\leq T
 \end{cases}.
 \end{align}
 Then
 $\kappa_1^\top \frac{1}{T^{(1+\delta)/2}}\sum\limits_{t=1}^{T_0} z_{t-1}u_t + \kappa_2^\top \frac{1}{T^{(1+\delta)/2}} \sum\limits_{t=T_0+1}^{T} z_{t-1}u_t = \frac{1}{T^{(1+\delta)/2}}\sum\limits_{t=1}^{T_0} \breve{z}_{t-1} u_t$. Note that $\{ \frac{1}{T^{(1+\delta)/2}} \breve{z}_{t-1} u_t\}_{t=1}^T$ is a martingale difference sequence. So we need to verify the Lindeberg condition specified by Corollary 3.1 of \cite{HallHeyde1980} as follows
 \begin{align*}
 \sum_{t=1}^T E_{\mathcal{F}_{k-1}}\left(\left\|\frac{1}{T^{(1+\delta)/2}} \breve{z}_{t-1} u_t\right\|^2 {1}\left\{\left\|\frac{1}{T^{(1+\delta)/2}} \breve{z}_{t-1} u_t \right\|>\nu_\kappa \right\}\right) \xrightarrow{P} 0 \quad \forall \nu_\kappa>0
 \end{align*}
 In the following part, we will prove this equation.
By equation (\ref{mulkappiv}) and Cauchy-Schwarz inequality, it follows that
 \begin{align}\label{mullindb1}
 &\sum_{t=1}^T E_{\mathcal{F}_{k-1}}\left(\left\|\frac{1}{T^{(1+\delta)/2}} \breve{z}_{t-1} u_t\right\|^2 {1}\left\{\left\|\frac{1}{T^{(1+\delta)/2}} \breve{z}_{t-1} u_t \right\|>\nu_\kappa \right\}\right)\\
 &\leq  \|\kappa_1\|^2 \sum_{t=1}^{T_0} \frac{1}{T^{1+\delta}} \left\|z_{t-1} \right\|^2 E_{\mathcal{F}_{k-1}}\left( u_t^2 {1}\left\{\left\|\frac{1}{T^{(1+\delta)/2}} z_{t-1} u_t \right\|>\nu_\kappa/\|\kappa_1\|  \right\}\right)     \nonumber\\
 &+ \|\kappa_2\|^2  \sum_{t=T_0+1}^{T} \frac{1}{T^{1+\delta}} \left\|z_{t-1} \right\|^2 E_{\mathcal{F}_{k-1}}\left( u_t^2 {1}\left\{\left\|\frac{1}{T^{(1+\delta)/2}} z_{t-1} u_t \right\|>\nu_\kappa/\|\kappa_2\|  \right\}\right)     \nonumber
 \end{align}
 It is straightforward that both events $\left\{\left\|\frac{1}{T^{(1+\delta)/2}} z_{t-1} u_t \right\|>\nu_\kappa/\|\kappa_1\| \right\}$ and $\left\{\left\|\frac{1}{T^{(1+\delta)/2}} z_{t-1} u_t \right\|>\nu_\kappa/\|\kappa_2\| \right\}$ implies $\left\{\left\|\frac{1}{T^{(1+\delta)/2}} z_{t-1} u_t \right\|>\nu_\kappa/\max(\|\kappa_1\| ,\|\kappa_2\| )\right\}$. So we have
\begin{align}\label{mullindb2}
{1}\left\{\left\|\frac{1}{T^{(1+\delta)/2}} z_{t-1} u_t \right\|>\nu_\kappa/\|\kappa_2\|  \right\}\leq
 {1}\left\{\left\|\frac{1}{T^{(1+\delta)/2}} z_{t-1} u_t \right\|>\nu_\kappa/\max(\|\kappa_1\| ,\|\kappa_2\| ) \right\},\\\label{mullindb3}
 {1}\left\{\left\|\frac{1}{T^{(1+\delta)/2}} z_{t-1} u_t \right\|>\nu_\kappa/\|\kappa_2\|  \right\}\leq
 {1}\left\{\left\|\frac{1}{T^{(1+\delta)/2}} z_{t-1} u_t \right\|>\nu_\kappa/\max(\|\kappa_1\| ,\|\kappa_2\| ) \right\}.
\end{align}
And
\begin{align}\label{mullindb4}
\|\kappa_1\|^2 \leq \max(\|\kappa_1\|^2,\|\kappa_2\|^2),\quad \|\kappa_2\|^2\leq \max(\|\kappa_1\|^2,\|\kappa_2\|^2).
\end{align}
By equations (\ref{mullindb1}), (\ref{mullindb2}), (\ref{mullindb3}) and (\ref{mullindb4}), it follows that
\begin{align}\label{mullindb5}
 &\sum_{t=1}^T E_{\mathcal{F}_{k-1}}\left(\left\|\frac{1}{T^{(1+\delta)/2}} \breve{z}_{t-1} u_t\right\|^2 {1}\left\{\left\|\frac{1}{T^{(1+\delta)/2}} \breve{z}_{t-1} u_t \right\|>\nu_\kappa \right\}\right)\\
 &\leq \max(\|\kappa_1\|^2, \|\kappa_2\|^2) \sum_{t=1}^{T} \frac{1}{T^{1+\delta}} \left\|z_{t-1} \right\|^2 E_{\mathcal{F}_{k-1}}\left( u_t^2 {1}\left\{\left\|\frac{1}{T^{(1+\delta)/2}} z_{t-1} u_t \right\|>\nu_\kappa/\max(\|\kappa_1\| ,\|\kappa_2\| ) \right\}\right) \nonumber
 \end{align}
By part (ii) of Lemma 5.2 of the proof of the Online Technical Supplement to \cite{Phillips2013PredictiveRU}, we have
\begin{align}\label{mullindb6}
 \sum_{t=1}^{T} \frac{1}{T^{1+\delta}} \left\|z_{t-1} \right\|^2 E_{\mathcal{F}_{k-1}}\left( u_t^2 {1}\left\{\left\|\frac{1}{T^{(1+\delta)/2}} z_{t-1} u_t \right\|>\nu_\kappa/\max(\|\kappa_1\| ,\|\kappa_2\| ) \right\}\right)
 \xrightarrow{P} 0
 \end{align}
By equations (\ref{mullindb5}) and (\ref{mullindb6}), the following Lindeberg condition holds.
\begin{align}\label{mullindb7}
 \sum_{t=1}^T E_{\mathcal{F}_{k-1}}\left(\left\|\frac{1}{T^{(1+\delta)/2}} \breve{z}_{t-1} u_t\right\|^2 {1}\left\{\left\|\frac{1}{T^{(1+\delta)/2}} \breve{z}_{t-1} u_t \right\|>\nu_\kappa \right\}\right)\xrightarrow{P} 0,\quad
 \forall \nu_\kappa>0
 \end{align}

Next, we verify the stability condition for $\{ \frac{1}{T^{(1+\delta)/2}} \breve{z}_{t-1} u_t\}_{t=1}^T$. By equation (\ref{mulivz}), it is known that
\begin{align*}
 \frac{1}{T} \sum_{t=1}^{T_0} z_{t-1}z_{t-1}^\top  &= \rho_z^2 \frac{1}{T} \sum_{t=2}^{T_0}z_{t-2}z_{t-2}^\top + \rho_z \frac{1}{T} \sum_{t=1}^{T_0}\Delta x_{t-1} \Delta x_{t-1}^\top  \\
 &+ \rho_z \frac{1}{T} \sum_{t=2}^{T_0} z_{t-2} \Delta x_{t-1}^\top
 + \frac{1}{T} \sum_{t=2}^{T_0}  \Delta x_{t-1} z_{t-2}^\top
\end{align*}
Thus
\begin{align}\label{mullindb8}
 (1-\rho_z^2)\frac{1}{T} \sum_{t=1}^{T_0} z_{t-1} z_{t-1}^\top &= -\rho_z^2 \frac{1}{T} z_{T_0-1} z_{T_0-1}^\top + \rho_z\frac{1}{T} \sum_{t=1}^{T_0}\Delta x_{t-1} \Delta x_{t-1}^\top \\
 & + \rho_z \frac{1}{T} \sum_{t=1}^{T_0} z_{t-2} \Delta x_{t-1}^\top
 + \frac{1}{T} \sum_{t=1}^{T_0} \Delta x_{t-1} z_{t-2}^\top  \nonumber\\
 &= \lambda \frac{1}{T_0} \sum_{t=1}^{T_0}\Delta x_{t-1} \Delta x_{t-1}^\top + o_p(1). \nonumber
\end{align}
The last step of the above equation holds since $\frac{1}{T^{\delta/2}}z_{T_0-1}=O_p(1)$, $\frac{1}{T^{(1+\delta)/2}} \sum_{t=1}^{T_0} z_{t-2} \Delta x_{t-1}^\top =O_p(1)$ and $\frac{1}{T^{(1+\delta)/2}} \sum_{t=1}^{T_0}  \Delta x_{t-1} z_{t-2}^\top =O_p(1)$ by Online Appendix of \cite{Kostakisetal2015} and $T_0=\lambda T$.
Similarly,
\begin{align}\label{mullindb9}
 (1-\rho_z^2)\frac{1}{T} \sum_{t=T_0+1}^{T} z_{t-1}  z_{t-1}^\top =(1-\lambda) \frac{1}{T-T_0} \sum_{t=T_0+1}^{T}\Delta x_{t-1}  \Delta x_{t-1}^\top + o_p(1).
\end{align}
By equations (\ref{mulkappiv}), (\ref{mullindb8}) and (\ref{mullindb9}), the following stability condition holds.
\begin{align}\label{mullindb10}
\frac{1}{T^{1+\delta}}\sum\limits_{t=1}^{T_0} \breve{z}_{t-1} \breve{z}_{t-1}^\top u_t^2 &=\Sigma_{uu} \frac{1}{T^{1+\delta}}\sum\limits_{t=1}^{T_0} \breve{z}_{t-1}  \breve{z}_{t-1}^\top + o_p(1)  \\
&\xrightarrow{P}  \lambda  \kappa_1^\top \Omega_{zz} \kappa_1 +   (1-\lambda)\kappa_2^\top \Omega_{zz} \kappa_2 \nonumber
\end{align}
By equations (\ref{mullindb7}) and (\ref{mullindb10}), Corollary 3.1 of \cite{HallHeyde1980} yields the following results.
\begin{align}\label{mullindb11}
 \frac{1}{T^{(1+\delta)/2}}\sum\limits_{t=1}^{T_0} \breve{z}_{t-1}u_t& =\kappa_1^\top \frac{1}{T^{(1+\delta)/2}}\sum\limits_{t=1}^{T_0} z_{t-1}u_t + \kappa_2^\top \frac{1}{T^{(1+\delta)/2}} \sum\limits_{t=T_0+1}^{T} z_{t-1}u_t \\
 &\xrightarrow{d}
 \operatorname{N}\left[0, \lambda  \kappa_1^\top \Omega_{zz} \kappa_1 +   (1-\lambda)\kappa_2^\top \Omega_{zz} \kappa_2\right] \nonumber
\end{align}
for any real vectors $\kappa_1$ and $\kappa_2$. By the Cramer-Wold device and equation (\ref{mullindb11}), for SD predictors, we have
\begin{align}\label{mullindb12}
\left(\frac{1}{T^{(1+\delta)/2}}\sum_{t=1}^{T_0} z_{t-1}u_t, \frac{1}{T^{(1+\delta)/2}}\sum_{t=T_0+1}^{T} z_{t-1}u_t\right)^\top \xrightarrow{d}
\operatorname{N}\left\{0, \operatorname{diag}\left[ \lambda\Omega_{zz} , (1-\lambda)\Omega_{zz}\right]\right\}.
\end{align}
Similarly, for WD predictors, we also have
\begin{align}\label{mullindb13}
\left(\frac{1}{\sqrt{T} }\sum_{t=1}^{T_0} z_{t-1}u_t, \frac{1}{\sqrt{T}}\sum_{t=T_0+1}^{T} z_{t-1}u_t\right)^\top \xrightarrow{d}
\operatorname{N}\left\{  {0}, \operatorname{diag}\left[ \lambda\Omega_{zz} , (1-\lambda)\Omega_{zz}\right]\right\}.
\end{align}
By equation (\ref{sasb69}), it follows that
\begin{align}
\left(\operatorname{I_K} -S_a,\operatorname{I_K} -S_b \right) \Rightarrow \left(\operatorname{I_K} -\tilde{S}_a,\operatorname{I_K} -\tilde{S}_b \right),
\end{align}
where the joint convergence is guaranteed by that both $\tilde{S}_a$ and $\tilde{S}_b$ are the function of $B_v(\cdot)$. And by the proof of part (i) of Proposition A1 and Lemma 3.2 of \cite{PhillipsMagdalinos2009}, the joint convergence of $\left(D_T^{-1}\sum_{t=1}^{T_0} z_{t-1}u_t, D_T^{-1} \sum_{t=T_0+1}^{T} z_{t-1}u_t\right)^\top $ and $B_v(\cdot)$ applies. As a result, considering that $\left(\operatorname{I_K} -\tilde{S}_a,\operatorname{I_K} -\tilde{S}_b \right)$ is the function of $B_v(\cdot)$, the joint convergence  of $\left(1-S_a,1-S_b \right)$ and $\left(D_T^{-1}\sum_{t=1}^{T_0} z_{t-1}u_t, D_T^{-1} \sum_{t=T_0+1}^{T} z_{t-1}u_t\right)^\top $ holds.

By equations (\ref{mullindb12}) and (\ref{mullindb13}) and the continuous mapping theorem and the joint convergence, the following equation holds for both SD and WD predictors.
\begin{align}
D_T^{-1}\sum_{t=1}^{T} \tilde{z}_{t-1}u_t &= \left(\operatorname{I_K} -S_a,\operatorname{I_K} -S_b \right)\left(D_T^{-1}\sum_{t=1}^{T_0} z_{t-1}u_t, D_T^{-1}\sum_{t=T_0+1}^{T} z_{t-1}u_t\right)^\top \xrightarrow{d} \operatorname{MN}\left(0,\Sigma_{zz} \right). \nonumber
\end{align}
\end{proof}

\begin{lemma}\label{mulopv7as}
Under Assumption \ref{Assumption A.1}, for SD and WD predictors, we have
$$D_T^{-2}\sum_{t=1}^T \tilde{z}_{t-1} \tilde{z}_{t-1}^\top \hat{u}_t^2\Rightarrow \Sigma_{zz}.$$
\end{lemma}
\begin{proof}[Proof of Lemma \ref{mulopv7as}]
Following the similar procedure for the stability condition shown in equation (\ref{mullindb10}) for $\{ \frac{1}{T^{(1+\delta)/2}} \breve{z}_{t-1} u_t\}_{t=1}^T$ of Proof for Lemma \ref{mulleea3pp}, Lemma \ref{mulopv7as} is proved.

\end{proof}

\begin{lemma}\label{porlem2}
Under Assumption A.1 and  the condition $1/2<\delta<1$, by Taylor expansion of matrix \citep{Taylor2016},  we have
\begin{align}\label{app3d3}
 &\left( \frac{1}{T^{1+\delta }} \sum_{t=1}^T z_{t-1}^*(z_{t-1}^*)^\top \hat{u}_t^2 \right)^{-1/2} =\left( \frac{1}{T^{1+\delta }} \sum_{t=1}^T z_{t-1}^*(z_{t-1}^*)^\top u_t^2 \right)^{-1/2}  +o_p\left[ T^{-(1-\delta)/2} \right] \\
 &=
 \left[\operatorname{I_K}+\left(\frac{1}{T^{1+\delta }} \sum_{t=1}^T z_{t-1}^*(z_{t-1}^*)^\top u_t^2 -\operatorname{I_K} \right) \right]^{-1/2} +o_p\left[ T^{-(1-\delta)/2} \right] \nonumber\\
 & = \operatorname{I_K} -\frac{1}{2} \left(\frac{1}{T^{1+\delta }} \sum_{t=1}^T z_{t-1}^*(z_{t-1}^*)^\top u_t^2 -\operatorname{I_K}\right)+o_p \left(\frac{1}{T^{1+\delta }} \sum_{t=1}^T z_{t-1}^*(z_{t-1}^*)^\top u_t^2 -\operatorname{I_K}\right)+o_p\left[ T^{-(1-\delta)/2} \right] \nonumber \\
  & = \operatorname{I_K} -\frac{1}{2} \left(\frac{1}{T^{1+\delta }} \sum_{t=1}^T z_{t-1}^*(z_{t-1}^*)^\top u_t^2 -\operatorname{I_K}\right)+o_p\left[ T^{-(1-\delta)/2} \right].\nonumber
\end{align}
And
\begin{align}\label{df_app3d3}
 &\left( \frac{1}{T^{1+\delta }} \sum_{t=1}^T \tilde{z}_{t-1}^*(\tilde{z}_{t-1}^*)^\top \hat{u}_t^2 \right)^{-1/2} =\left( \frac{1}{T^{1+\delta }} \sum_{t=1}^T \tilde{z}_{t-1}^*(\tilde{z}_{t-1}^*)^\top u_t^2 \right)^{-1/2}  +o_p\left[ T^{-(1-\delta)/2} \right] \\
 &=
 \left[\operatorname{I_K}+\left(\frac{1}{T^{1+\delta }} \sum_{t=1}^T \tilde{z}_{t-1}^*(\tilde{z}_{t-1}^*)^\top u_t^2 -\operatorname{I_K} \right) \right]^{-1/2} +o_p\left[ T^{-(1-\delta)/2} \right] \nonumber\\
 & = \operatorname{I_K} -\frac{1}{2} \left(\frac{1}{T^{1+\delta }} \sum_{t=1}^T \tilde{z}_{t-1}^*(\tilde{z}_{t-1}^*)^\top u_t^2 -\operatorname{I_K}\right)+o_p \left(\frac{1}{T^{1+\delta }} \sum_{t=1}^T \tilde{z}_{t-1}^*(\tilde{z}_{t-1}^*)^\top u_t^2 -\operatorname{I_K}\right)+o_p\left[ T^{-(1-\delta)/2} \right].\nonumber\\
   & = \operatorname{I_K} -\frac{1}{2} \left(\frac{1}{T^{1+\delta }} \sum_{t=1}^T \tilde{z}_{t-1}^*(\tilde{z}_{t-1}^*)^\top u_t^2 -\operatorname{I_K}\right)+o_p\left[ T^{-(1-\delta)/2} \right].\nonumber
\end{align}

\end{lemma}
\begin{proof}[Proof Lemma \ref{porlem2}]
By online appendix of \cite{Kostakisetal2015}, it follows that
\begin{align}\label{gs87djh1}
\frac{1}{T^{1+\delta }} \sum_{t=1}^T z_{t-1}^*(z_{t-1}^*)^\top \hat{u}_t^2
& =\Omega_{zz}^{-1/2} \frac{1}{T^{1+\delta }} \sum_{t=1}^T z_{t-1}z_{t-1}^\top \hat{u}_t^2 \Omega_{zz}^{-1/2} \\
&\xrightarrow{P} \Omega_{zz}^{-1/2} \Sigma_{vv}/(-2c_z)\Sigma_{uu} \Omega_{zz}^{-1/2} = \Omega_{zz}^{-1/2} \Omega_{zz}^{1/2}\Omega_{zz}^{1/2} \Omega_{zz}^{-1/2} = \operatorname{I_K}. \nonumber
\end{align}
Thus
\begin{align}\label{gs87djh2}
\left(\frac{1}{T^{1+\delta }} \sum_{t=1}^T z_{t-1}^*(z_{t-1}^*)^\top \hat{u}_t^2 - \operatorname{I_K} \right)
\left(\frac{1}{T^{1+\delta }} \sum_{t=1}^T z_{t-1}^*(z_{t-1}^*)^\top \hat{u}_t^2 - \operatorname{I_K} \right)
\xrightarrow{P} 0.
\end{align}
By equation (\ref{gs87djh2}) and continuous mapping theorem, the equation about determinant  holds as follows.
\begin{align}\label{gs87djh3}
\operatorname{det}\left|\left(\frac{1}{T^{1+\delta }} \sum_{t=1}^T z_{t-1}^*(z_{t-1}^*)^\top \hat{u}_t^2 - \operatorname{I_K} \right)
\left(\frac{1}{T^{1+\delta }} \sum_{t=1}^T z_{t-1}^*(z_{t-1}^*)^\top \hat{u}_t^2 - \operatorname{I_K} \right) \right| \xrightarrow{P} 0.
\end{align}
Additionally,
\begin{align}\label{gs87djh4}
\operatorname{det}\left|\left(\frac{1}{T^{1+\delta }} \sum_{t=1}^T z_{t-1}^*(z_{t-1}^*)^\top \hat{u}_t^2 - \operatorname{I_K} \right)
\left(\frac{1}{T^{1+\delta }} \sum_{t=1}^T z_{t-1}^*(z_{t-1}^*)^\top \hat{u}_t^2 - \operatorname{I_K} \right) \right| = \lambda_1 \lambda_2 \cdots \lambda_K,
\end{align}
where $\lambda_i>0$, $i=1,2,\cdots,K$ is the singular value of $\frac{1}{T^{1+\delta }} \sum_{t=1}^T z_{t-1}^*(z_{t-1}^*)^\top \hat{u}_t^2 - \operatorname{I_K} $. By equations (\ref{gs87djh3}) and (\ref{gs87djh4}) and  that $\frac{1}{T^{1+\delta }} \sum_{t=1}^T z_{t-1}^*(z_{t-1}^*)^\top \hat{u}_t^2 - \operatorname{I_K}=o_p(1)$, it follows that
\begin{align}\label{gs87djh5}
\max_{i=1,2,\cdots,K} \lambda_i \xrightarrow{P} 0.
\end{align}
On the other hand, following the similar procedure of  \cite{HosseinkouchackDemetrescu2021}, we have
\begin{align}\label{app2ood3}
 \left( \frac{1}{T^{1+\delta }} \sum_{t=1}^T z_{t-1}^*(z_{t-1}^*)^\top \hat{u}_t^2 \right)^{-1/2} =\left( \frac{1}{T^{1+\delta }} \sum_{t=1}^T z_{t-1}^*(z_{t-1}^*)^\top u_t^2 \right)^{-1/2}  +o_p\left[ T^{-(1-\delta)/2} \right].
\end{align}
By Lemma \ref{porlem2} and (\ref{app2ood3}), equation (\ref{app3d3}) holds. Also, by the similar procedure, equation (\ref{df_app3d3}) is proved.

\end{proof}

\begin{proof}[Proof of Proposition \ref{mulpropfdie3}]
We consider the higher order terms induced by the SD predictors.
By equation (\ref{mulivz}), it follows that
\begin{align}
z_{t-1}^*=\Omega_{zz}^{-1/2}z_{t-1} =\Omega_{zz}^{-1/2}\left(\rho_z z_{t-1} + \Delta x_{t-1} \right)
= \rho_z z_{t-2}^* + \Delta x_{t-1}^*,
\end{align}
where $\Delta x_{t-1}^*= x_{t-1}^* - {x}_{t-2}^* $. We define $v_t^*= \Omega_{zz}^{-1/2}v_t$ and thus
$\Delta x_{t-1}^* =v_t^* +\frac{c_z}{T^\delta} {x}_{t-2}^* $ and
\begin{align}\label{mulhj7che1}
\frac{1}{T} \sum_{t=1}^Tv_t^*(v_t^*)^\top \xrightarrow{P} \operatorname{E}\left[ v_t^*(v_t^*)^\top \right]= - 2c_z/\Sigma_{uu}\operatorname{I_K},\\
\frac{1}{T} \sum_{t=1}^T(v_t^*)^\top v_t^* \xrightarrow{P} \operatorname{E}\left[ (v_t^*)^\top v_t^* \right]= - 2c_z/\Sigma_{uu}  K. \nonumber
\end{align}

Then by equation (\ref{app3d3}) in Lemma \ref{porlem2}, we have
\begin{align}\label{khgpl7hi1}
 \left[ \frac{1}{T^{1+\delta}} \sum_{t=1}^T z_{t-1}^* (z_{t-1}^*)^\top u_t^2 \right]^{-1/2}\frac{1}{T^{1 / 2+\delta / 2}} \sum_{t=1}^T z_{t-1}^* u_t ={Z_T}+{B_T}+o_p\left[T^{-(1-\delta)/2}\right]
\end{align}
where
$$
{Z_T}= \frac{1}{T^{1 / 2+\delta / 2}} \sum_{t=1}^T z_{t-1}^* u_t
$$
and
$$
\begin{aligned}
{B_T}=- & \frac{1}{2} \left[\frac{1}{T^{1+\delta }} \sum_{t=1}^T z_{t-1}^* (z_{t-1}^*)^\top u_t^2- \operatorname{I_K} \right]
\left(\frac{1}{T^{1 / 2+\delta / 2}} \sum_{t=1}^T z_{t-1}^* u_t\right)
\end{aligned}
$$
By the online appendix of \cite{Kostakisetal2015} and the continuous mapping theorem, it is known that ${Z_T} \xrightarrow{d} \operatorname{N}(0_K,\operatorname{I_K})$.

Since $\operatorname{E}
\left(\frac{1}{T^{1 / 2+\delta / 2}} \sum_{t=1}^T z_{t-1}^* u_t\right)=0$, then
\begin{align}\label{a7dgwpe}
&T^{(1-\delta)/2}\operatorname{E}\left({B_T} \right) \\
&=- \frac{1}{2} T^{(1-\delta)/2}\operatorname{E}\left[\left(\frac{1}{T^{1+\delta }} \sum_{t=1}^T z_{t-1}^* (z_{t-1}^*)^\top u_t^2- \operatorname{I_K} \right)
\left(\frac{1}{T^{1 / 2+\delta / 2}} \sum_{t=1}^T z_{t-1}^* u_t\right)\right]\nonumber\\
&=- \frac{1}{2} T^{(1-\delta)/2} \operatorname{E}\left[\left(\frac{1}{T^{1+\delta }} \sum_{t=1}^T z_{t-1}^* (z_{t-1}^*)^\top u_t^2 \right)
\left(\frac{1}{T^{1 / 2+\delta / 2}} \sum_{t=1}^T z_{t-1}^* u_t\right)\right]\nonumber
\end{align}
And
\begin{align}\label{apdu6pe}
&T^{(1-\delta)/2}\operatorname{E}\left[\left(\frac{1}{T^{1+\delta }} \sum_{t=1}^T z_{t-1}^* (z_{t-1}^*)^\top u_t^2 \right)
\left(\frac{1}{T^{1 / 2+\delta / 2}} \sum_{t=1}^T z_{t-1}^* u_t\right)\right]\\
& =\frac{1}{T^{1+2\delta} }\mathrm{E}\left(\sum_{t=1}^T z_{t-1}^* (z_{t-1}^*)^\top z_{t-1}^* u_t^3\right)
+ \frac{1}{T^{1+2\delta} } \mathrm{E}\left(\sum_{t=1}^T \sum_{s=1}^{t-1} z_{s-1}^*(z_{s-1}^*)^\top z_{t-1}^* u_s^2 u_t \right) \nonumber\\
& +\frac{1}{T^{1+2\delta} } \mathrm{E}\left(\sum_{t=1}^{T-1} \sum_{s=t+1}^T z_{s-1}^*(z_{s-1}^*)^\top z_{t-1}^* u_t u_s^2\right) \nonumber\\
& = \frac{1}{T^{1+2\delta} }\mathrm{E}\left(\sum_{t=1}^T z_{t-1}^* (z_{t-1}^*)^\top z_{t-1}^* u_t^3\right)
+\frac{1}{T^{1+2\delta} }\mathrm{E}\left(\sum_{t=1}^{T-1} \sum_{s=t+1}^T z_{s-1}^*(z_{s-1}^*)^\top z_{t-1}^* u_t u_s^2\right). \nonumber
\end{align}
Let $S_0=S_{0,1}+S_{0,2}$, with $S_{0,1}=\mathrm{E}\left(\sum_{t=1}^T z_{t-1}^* (z_{t-1}^*)^\top z_{t-1}^* u_t^3\right)$ and
$$
S_{0,2}=\mathrm{E}\left(\sum_{t=1}^{T-1} \sum_{s=t+1}^T z_{s-1}^*(z_{s-1}^*)^\top z_{t-1}^* u_t u_s^2\right).
$$ We work out $S_{0,2}$ first. By the proof of \cite{HosseinkouchackDemetrescu2021},
$$
z_{t-1}^* =\sum_{j=1}^{t-1} c_{j, t-1}  v_j^*  \text { with } c_{j, t-1}=\frac{\rho_z^{t-1-j}(1-\rho_z)-\rho_j^{t-1-j}(1-\rho_j)}{\rho_j-\rho_z} .
$$
Then for $s\geq t+1$, it follows that
\begin{small}
\begin{align}\label{ap35kep}
&\frac{1}{T^{1+2\delta} }\mathrm{E}\left[ z_{s-1}^*(z_{s-1}^*)^\top z_{t-1}^* u_t u_s^2\right]\\
&=\mathrm{E}(u_s^2)\frac{1}{T^{1+2\delta} } \mathrm{E}\left[\left(\sum_{j=1}^{t-1} c_{j, s-1} (v_j^*)+\sum_{j=t}^{s-1} c_{j, s-1} (v_j^*)\right) \left(\sum_{j=1}^{t-1} c_{j, s-1} (v_j^*)+\sum_{j=t}^{s-1} c_{j, s-1} (v_j^*)\right)^\top \left(\sum_{j=1}^{t-1} c_{j, t-1} (v_j^*)\right) u_t\right] \nonumber\\
& =\Sigma_{uu}\frac{1}{T^{1+2\delta} } \mathrm{E}\left[\left(\sum_{j=1}^{t-1} c_{j, s-1} (v_j^*)\right)\left(\sum_{j=t}^{s-1} c_{j, s-1} (v_j^*)\right)^\top \left(\sum_{j=1}^{t-1} c_{j, t-1} (v_j^*)\right) u_t \right] \nonumber\\
&+\Sigma_{uu}\frac{1}{T^{1+2\delta} }\mathrm{E}\left[\left(\sum_{j=t}^{s-1} c_{j, s-1} (v_j^*)\right) \left(\sum_{j=1}^{t-1} c_{j, s-1} (v_j^*)\right)^\top \left(\sum_{j=1}^{t-1} c_{j, t-1} (v_j^*)\right) u_t \right]. \nonumber
\end{align}
\end{small}
Thus by the independence of innovations and equation (\ref{mulhj7che1}), it follows that
\begin{align}\label{appe73y}
&\frac{1}{T^{1+2\delta} } \mathrm{E}\left[\left(\sum_{j=1}^{t-1} c_{j, s-1} (v_j^*)\right)\left( \sum_{j=t}^{s-1} c_{j, s-1} (v_j^*)\right)^\top \left(\sum_{j=1}^{t-1} c_{j, t-1} (v_j^*)\right) u_t \right] \\
&=\frac{1}{T^{1+2\delta} } \mathrm{E}\left[\left(\sum_{j=1}^{t-1} c_{j, s-1} (v_j^*)\right)\left(c_{t, s-1} (v_t^*)+\sum_{j=t+1}^{s-1} c_{j, s-1} (v_j^*)\right)^\top \left(\sum_{j=1}^{t-1} c_{j, t-1} (v_j^*)\right) u_t \right] \nonumber\\
&= \frac{1}{T^{1+2\delta} } \mathrm{E}\left[\left(\sum_{j=1}^{t-1} c_{j, s-1} (v_j^*)\right) c_{t, s-1} (v_t^*)^\top \left(\sum_{j=1}^{t-1} c_{j, t-1} (v_j^*)\right) u_t \right] \nonumber\\
&= \frac{1}{T^{1+2\delta} } \mathrm{E}\left( \sum_{j=1}^{t-1} c_{j, s-1} v_j^* c_{t, s-1} (v_t^*)^\top c_{j, t-1} v_j^* u_t \right) \nonumber\\
&= \frac{1}{T^{1+2\delta} } \mathrm{E}\left( \sum_{j=1}^{t-1} c_{j, s-1} c_{t, s-1} c_{j, t-1} v_j^* (v_t^*)^\top v_j^* u_t \right) \nonumber \\
& = \frac{1}{T^{1+2\delta} } \sum_{j=1}^{t-1} c_{j, s-1} c_{t, s-1} c_{j, t-1} \mathrm{E}\left( v_j^* (v_t^*)^\top v_j^* u_t \right). \nonumber\\
& = \frac{1}{T^{1+2\delta} } \sum_{j=1}^{t-1} c_{j, s-1} c_{t, s-1} c_{j, t-1} \mathrm{E}\left( v_j^* v_t^* (v_j^*)^\top u_t \right) \nonumber\\
& = \frac{1}{T^{1+2\delta} } \sum_{j=1}^{t-1} c_{j, s-1} c_{t, s-1} c_{j, t-1} \mathrm{E}\left(u_t v_t^* (v_j^*)^\top v_j^* \right) \nonumber\\
& = \frac{1}{T^{1+2\delta} } \sum_{j=1}^{t-1} c_{j, s-1} c_{t, s-1} c_{j, t-1} \mathrm{E}\left(u_t v_t^* \right) \mathrm{E}\left[ (v_j^*)^\top v_j^* \right] \nonumber\\
&= \Omega_{zz}^{-1/2} \Sigma_{vu} \left( - 2c_z/\Sigma_{uu}\right) \frac{1}{T^{1+2\delta} } \sum_{j=1}^{t-1} c_{j, s-1} c_{t, s-1} c_{j, t-1}. \nonumber
\end{align}
The last four lines of equation (\ref{appe73y}) hold since $(v_t^*)^\top  v_j^* = v_t^* (v_j^*)^\top $ is a scalar and $j<t$.

And by equations (S.2)-(S.5) of the appendix of \cite{HosseinkouchackDemetrescu2021}, we have
\begin{align}\label{appe75y}
\frac{1}{T^{1+2\delta} }\sum_{j=1}^{t-1} c_{j, s-1} c_{t, s-1} c_{j, t-1} \rightarrow \frac{1}{4c_z^2}
\end{align}
Then by equations (\ref{appe73y}) and (\ref{appe75y}), it follows that
\begin{align}\label{ap36kep}
&\frac{1}{T^{1+2\delta} } \mathrm{E}\left[\left(\sum_{j=1}^{t-1} c_{j, s-1} (v_j^*)\right)\left( \sum_{j=t}^{s-1} c_{j, s-1} (v_j^*)\right)^\top \left(\sum_{j=1}^{t-1} c_{j, t-1} (v_j^*)\right) u_t \right] \\
&\rightarrow (4c_z^2)^{-1}\Omega_{zz}^{-1/2} \Sigma_{vu} \left( - 2c_z/\Sigma_{uu}\right)= - \Omega_{zz}^{-1/2} \Sigma_{vu} /(2c_z\Sigma_{uu})
\nonumber
\end{align}

On the other hand,
\begin{align}\label{ap6fl3y}
&\frac{1}{T^{1+2\delta} } \mathrm{E}\left[\left( \sum_{j=t}^{s-1} c_{j, s-1} (v_j^*)\right) \left(\sum_{j=1}^{t-1} c_{j, s-1} (v_j^*)\right)^\top \left(\sum_{j=1}^{t-1} c_{j, t-1} (v_j^*)\right) u_t \right] \\
&=\frac{1}{T^{1+2\delta} } \mathrm{E}\left[\left(c_{t, s-1} (v_t^*)+\sum_{j=t+1}^{s-1} c_{j, s-1} (v_j^*)\right) \left(\sum_{j=1}^{t-1} c_{j, s-1} (v_j^*)\right)^\top \left(\sum_{j=1}^{t-1} c_{j, t-1} (v_j^*)\right) u_t \right] \nonumber\\
&= \frac{1}{T^{1+2\delta} } \mathrm{E}\left[c_{t, s-1} (v_t^*) \left(\sum_{j=1}^{t-1} c_{j, s-1} (v_j^*)^\top\right) \left(\sum_{j=1}^{t-1} c_{j, t-1} (v_j^*)\right) u_t \right] \nonumber\\
&= \frac{1}{T^{1+2\delta} } \mathrm{E}\left( \sum_{j=1}^{t-1} c_{t, s-1} (v_t^*) c_{j, s-1} (v_j^*)^\top c_{j, t-1} (v_j^*) u_t \right) \nonumber\\
&= \frac{1}{T^{1+2\delta} } \mathrm{E}\left( \sum_{j=1}^{t-1} c_{j, s-1} c_{t, s-1} c_{j, t-1} (v_t^*) u_t (v_j^*)^\top (v_j^*) \right)
= \frac{1}{T^{1+2\delta} } \sum_{j=1}^{t-1} c_{j, s-1} c_{t, s-1} c_{j, t-1} \mathrm{E}\left( (v_t^*) u_t (v_j^*)^\top (v_j^*) \right). \nonumber\\
&= \mathrm{E}\left[  v_t^*  u_t \right] \mathrm{E}\left[(v_j^*)^\top  v_j^*  \right] \frac{1}{T^{1+2\delta} } \sum_{j=1}^{t-1} c_{j, s-1} c_{t, s-1} c_{j, t-1} \rightarrow - K\Omega_{zz}^{-1/2} \Sigma_{vu} /(2c_z\Sigma_{uu}) . \nonumber
\end{align}

By equations (\ref{ap35kep}), (\ref{ap36kep}) and (\ref{ap6fl3y}), it follows that
\begin{align}\label{a9kdqwp}
 \frac{1}{T^{1+2\delta} } \mathrm{E}\left( z_{s-1}^*(z_{s-1}^*)^\top z_{t-1}^* u_t u_s^2\right) \rightarrow - \Omega_{zz}^{-1/2} \Sigma_{vu} /c_z ,\quad s \geq t + 1,
 \end{align}
By the same procedure of the appendix of \cite{HosseinkouchackDemetrescu2021}, we have
\begin{align}\label{a3kd7qdwp}
\frac{1}{T^{1+2\delta} }\mathrm{E}\left(\sum_{t=1}^T z_{t-1}^* (z_{t-1}^*)^\top z_{t-1}^* u_t^3\right) \rightarrow 0.
\end{align}
By equations (\ref{a7dgwpe}), (\ref{apdu6pe}), (\ref{a9kdqwp}) and (\ref{a3kd7qdwp})
\begin{align}\label{dgm3ul7jsd}
 T^{(1-\delta)/2}\operatorname{E}\left({B_T} \right) &\rightarrow - \frac{K+1}{2} \Omega_{zz}^{-1/2} \Sigma_{vu} /(-2 c_z).
\end{align}
Since  $\Omega_{zz}= \Sigma_{uu} \Omega_{vv}/(-2c_z)$, equation (\ref{dgm3ul7jsd}) is the same as
\begin{align}\label{gdp6pjk}
 T^{(1-\delta)/2}\operatorname{E}\left({B_T} \right) &\rightarrow - \frac{K+1}{2}{\rho_{uv^*}}/ \sqrt{-2 c_z}.
\end{align}
Thus by equations (\ref{khgpl7hi1}) and (\ref{gdp6pjk}), Proposition \ref{mulpropfdie3} holds.
\end{proof}

\begin{proof}[Proof of Theorem \ref{thmnew1}]
By the definitions of $Q_{ivx}$ and $\tilde{Q}_{ivx}$, it follows that
\begin{align}\label{ok784mh1}
&Q_{ivx}-\tilde{Q}_{ivx}=
\mathbf{t_{ivx}^\top} H_{ivx}^\top \left(H_{ivx}  H_{ivx}^\top \right)^{-1} H_{ivx} \mathbf{t_{ivx}}
-\mathbf{t_{ivx}^\top} H_{ivx}^\top \left[H_{ivx} (\operatorname{I_K} +\hat{\varpi}_b  \hat{\varpi}_b^\top) H_{ivx}^\top \right]^{-1} H_{ivx} \mathbf{t_{ivx}} \nonumber\\
&=
\mathbf{t_{ivx}^\top} H_{ivx}^\top \left(H_{ivx}  H_{ivx}^\top \right)^{-1} \left[H_{ivx} (\operatorname{I_K} +\hat{\varpi}_b  \hat{\varpi}_b^\top) H_{ivx}^\top \right]
\left[H_{ivx} (\operatorname{I_K} +\hat{\varpi}_b  \hat{\varpi}_b^\top) H_{ivx}^\top \right]^{-1}H_{ivx} \mathbf{t_{ivx}} \nonumber\\
&-\mathbf{t_{ivx}^\top} H_{ivx}^\top \left(H_{ivx}  H_{ivx}^\top \right)^{-1} \left(H_{ivx}  H_{ivx}^\top \right)
\left[H_{ivx} (\operatorname{I_K} +\hat{\varpi}_b  \hat{\varpi}_b^\top) H_{ivx}^\top \right]^{-1} H_{ivx} \mathbf{t_{ivx}} \nonumber\\
&=
\mathbf{t_{ivx}^\top} H_{ivx}^\top \left(H_{ivx}  H_{ivx}^\top \right)^{-1} \left[H_{ivx} (\operatorname{I_K} +\hat{\varpi}_b  \hat{\varpi}_b^\top) H_{ivx}^\top - H_{ivx} H_{ivx}^\top\right]
\left[H_{ivx} (\operatorname{I_K} +\hat{\varpi}_b  \hat{\varpi}_b^\top) H_{ivx}^\top \right]^{-1}H_{ivx} \mathbf{t_{ivx}} \nonumber\\
&=
\mathbf{t_{ivx}^\top} H_{ivx}^\top \left(H_{ivx}  H_{ivx}^\top \right)^{-1} \left(H_{ivx}  \hat{\varpi}_b  \hat{\varpi}_b^\top  H_{ivx}^\top  \right)
\left[H_{ivx} (\operatorname{I_K} +\hat{\varpi}_b  \hat{\varpi}_b^\top) H_{ivx}^\top \right]^{-1}H_{ivx} \mathbf{t_{ivx}}>0.
\end{align}
The last line of equation (\ref{ok784mh1}) holds since $\left(H_{ivx}  H_{ivx}^\top \right)^{-1}  $,
$H_{ivx}  \hat{\varpi}_b  \hat{\varpi}_b^\top  H_{ivx}^\top$ and
$\left[H_{ivx} (\operatorname{I_K} +\hat{\varpi}_b  \hat{\varpi}_b^\top) H_{ivx}^\top \right]^{-1}$ are positive definition.
And
\begin{align}\label{ok784mh2}
&\mathbf{t_{ivx}^\top} H_{ivx}^\top \left(H_{ivx}   H_{ivx}^\top \right)^{-1}
 H_{ivx}  \hat{\varpi}_b  \hat{\varpi}_b^\top  H_{ivx}^\top
\left[H_{ivx} (\operatorname{I_K} +\hat{\varpi}_b  \hat{\varpi}_b^\top) H_{ivx}^\top \right]^{-1} H_{ivx} \mathbf{t_{ivx}}  \\
& -  \mathbf{t_{ivx}^\top} H_{ivx}^\top \left(H_{ivx}   H_{ivx}^\top \right)^{-1}
 H_{ivx}  \hat{\varpi}_b  \hat{\varpi}_b^\top  H_{ivx}^\top
\left(H_{ivx}  H_{ivx}^\top \right)^{-1} H_{ivx} \mathbf{t_{ivx}} \nonumber\\
&= \mathbf{t_{ivx}^\top} H_{ivx}^\top \left(H_{ivx}   H_{ivx}^\top \right)^{-1}
 H_{ivx}  \hat{\varpi}_b  \hat{\varpi}_b^\top  H_{ivx}^\top \left(H_{ivx}  H_{ivx}^\top \right)^{-1} \left(H_{ivx}  H_{ivx}^\top \right)
\left[H_{ivx} (\operatorname{I_K} +\hat{\varpi}_b  \hat{\varpi}_b^\top) H_{ivx}^\top \right]^{-1} H_{ivx} \mathbf{t_{ivx}} \nonumber \\
& -  \mathbf{t_{ivx}^\top} H_{ivx}^\top \left(H_{ivx}   H_{ivx}^\top \right)^{-1}
 H_{ivx}  \hat{\varpi}_b  \hat{\varpi}_b^\top  H_{ivx}^\top
\left(H_{ivx}  H_{ivx}^\top \right)^{-1} \left[H_{ivx} (\operatorname{I_K} +\hat{\varpi}_b  \hat{\varpi}_b^\top) H_{ivx}^\top \right] \nonumber\\
 & \left[H_{ivx} (\operatorname{I_K} +\hat{\varpi}_b  \hat{\varpi}_b^\top) H_{ivx}^\top \right]^{-1} H_{ivx} \mathbf{t_{ivx}} \nonumber\\
 &= -  \mathbf{t_{ivx}^\top} H_{ivx}^\top \left(H_{ivx}   H_{ivx}^\top \right)^{-1}
 H_{ivx}  \hat{\varpi}_b  \hat{\varpi}_b^\top  H_{ivx}^\top
\left(H_{ivx}  H_{ivx}^\top \right)^{-1} \left[H_{ivx} (\operatorname{I_K} +\hat{\varpi}_b  \hat{\varpi}_b^\top) H_{ivx}^\top
- H_{ivx} H_{ivx}^\top
\right] \nonumber\\
 &\left[H_{ivx} (\operatorname{I_K} +\hat{\varpi}_b  \hat{\varpi}_b^\top) H_{ivx}^\top \right]^{-1} H_{ivx} \mathbf{t_{ivx}} \nonumber\\
  &= -  \mathbf{t_{ivx}^\top} H_{ivx}^\top \left(H_{ivx}   H_{ivx}^\top \right)^{-1}
 H_{ivx}  \hat{\varpi}_b  \hat{\varpi}_b^\top  H_{ivx}^\top
\left(H_{ivx}  H_{ivx}^\top \right)^{-1} \left[H_{ivx}  \hat{\varpi}_b  \hat{\varpi}_b^\top  H_{ivx}^\top\right] \nonumber\\
 &\left[H_{ivx} (\operatorname{I_K} +\hat{\varpi}_b  \hat{\varpi}_b^\top) H_{ivx}^\top \right]^{-1} H_{ivx} \mathbf{t_{ivx}} \nonumber\\
 & = O_p( \hat{\varpi}_b  \hat{\varpi}_b^\top ) = o_p\left[T^{-(1-\delta)/2}\right]\nonumber.
\end{align}
 The last line holds of equation (\ref{ok784mh2}) since $ \hat{\varpi}_b=O_p\left[T^{-(1-\delta)/2}\right]$. By equations (\ref{ok784mh1}) and (\ref{ok784mh2}), Theorem \ref{thmnew1} holds.

\end{proof}

\begin{proof}[Proof of equation (\ref{sasb69})]
We first consider the SD case for $S_a$. By the definition of $S_a$ and $S_b$ and part (i) of Lemma B1 of Online Appendix for \cite{Kostakisetal2015} , it follows that
\begin{align}
&S_a = \frac{1}{T}\sum\limits_{t=1}^{T} z_{t-1}\left(\frac{1}{T_0}\sum\limits_{t=1}^{T_0} z_{t-1}^\top \right) \left(\frac{1}{T_0}\sum\limits_{t=1}^{T_0} z_{t-1}^\top \frac{1}{T_0}\sum\limits_{t=1}^{T_0} z_{t-1}\right)^{-1}\\
&= \lambda \sum\limits_{t=1}^{T} z_{t-1}\left( \sum\limits_{t=1}^{T_0} z_{t-1}^\top \right) \left( \sum\limits_{t=1}^{T_0} z_{t-1}^\top  \sum\limits_{t=1}^{T_0} z_{t-1}\right)^{-1}\nonumber\\
&= \lambda \frac{1}{T^{1/2+\delta}}\sum\limits_{t=1}^{T} z_{t-1}\left( \frac{1}{T^{1/2+\delta}}\sum\limits_{t=1}^{T_0} z_{t-1}^\top \right) \left( \frac{1}{T^{1/2+\delta}}\sum\limits_{t=1}^{T_0} z_{t-1}^\top  \frac{1}{T^{1/2+\delta}}\sum\limits_{t=1}^{T_0} z_{t-1}\right)^{-1}\nonumber\\
&= \lambda \frac{1}{T^{1/2+\delta}}\sum\limits_{t=1}^{T} z_{t-1}\left( \frac{1}{T^{1/2+\delta}}\sum\limits_{t=1}^{T_0} z_{t-1}^\top \right) \left( \frac{1}{T^{1/2+\delta}}\sum\limits_{t=1}^{T_0} z_{t-1}^\top  \frac{1}{T^{1/2+\delta}}\sum\limits_{t=1}^{T_0} z_{t-1}\right)^{-1}\nonumber\\
&= \lambda \frac{1}{T^{1/2}}x_{T}\frac{1}{T^{1/2}}x_{T_0}^\top \left(  \frac{1}{T^{1/2}}x_{T_0}^\top \frac{1}{T^{1/2}}x_{T_0}  \right)^{-1}+o_p(1) \nonumber\\
&\Rightarrow \tilde{S}_a= \lambda  J_x^c(1) J_x^c(\lambda)^\top \left[ J_x^c(\lambda)^\top J_x^c(\lambda) \right]^{-1}.\nonumber
\end{align}
Similarly,
\begin{align}
&S_b = \frac{1}{T}\sum\limits_{t=1}^{T} z_{t-1}\left(\frac{1}{T-T_0}\sum\limits_{t=T_0+1}^{T} z_{t-1}^\top \right) \left(\frac{1}{T-T_0}\sum\limits_{t=T_0+1}^{T} z_{t-1}^\top \frac{1}{T-T_0}\sum\limits_{t=T_0+1}^{T} z_{t-1}\right)^{-1} \nonumber\\
&= \lambda \sum\limits_{t=1}^{T} z_{t-1}\left( \sum\limits_{t=T_0+1}^{T} z_{t-1}^\top \right) \left( \sum\limits_{t=T_0+1}^{T} z_{t-1}^\top  \sum\limits_{t=T_0+1}^{T} z_{t-1}\right)^{-1}\nonumber\\
&= \lambda \frac{1}{T^{1/2+\delta}}\sum\limits_{t=1}^{T} z_{t-1}\left( \frac{1}{T^{1/2+\delta}}\sum\limits_{t=T_0+1}^{T} z_{t-1}^\top \right) \left( \frac{1}{T^{1/2+\delta}}\sum\limits_{t=T_0+1}^{T} z_{t-1}^\top  \frac{1}{T^{1/2+\delta}}\sum\limits_{t=T_0+1}^{T} z_{t-1}\right)^{-1}\nonumber\\
&= \lambda \frac{1}{T^{1/2+\delta}}\sum\limits_{t=1}^{T} z_{t-1}\left( \frac{1}{T^{1/2+\delta}}\sum\limits_{t=T_0+1}^{T} z_{t-1}^\top \right) \left( \frac{1}{T^{1/2+\delta}}\sum\limits_{t=T_0+1}^{T} z_{t-1}^\top  \frac{1}{T^{1/2+\delta}}\sum\limits_{t=T_0+1}^{T} z_{t-1}\right)^{-1}\nonumber\\
&= \lambda \frac{1}{T^{1/2}}x_{T}\frac{1}{T^{1/2}}(x_{T}-x_{T_0})^\top \left[  \frac{1}{T^{1/2}}(x_{T}-x_{T_0})^\top \frac{1}{T^{1/2}}(x_{T}-x_{T_0})  \right]^{-1} + o_p(1)\nonumber\\
&\Rightarrow \tilde{S}_b= \lambda  J_x^c(1) \left[ J_x^c(1) - J_x^c(\lambda) \right]^\top \left\{ \left[ J_x^c(1) - J_x^c(\lambda) \right]^\top \left[ J_x^c(1) - J_x^c(\lambda) \right] \right\}^{-1}.\nonumber
\end{align}

Second, by equation (7) of the supplementary material for \cite{Kostakisetal2015}, for WD predictors, we have
\begin{align}\label{mulap2dsa}
\frac{1}{\sqrt{T}}\sum\limits_{t=1}^{T} z_{i,t-1} = \frac{1}{\sqrt{T}}\sum\limits_{t=1}^{T}  x_{i,t-1} +o_p(1),\\ \frac{1}{\sqrt{T}}\sum\limits_{t=1}^{T_0} z_{i,t-1} = \frac{1}{\sqrt{T}}\sum\limits_{t=1}^{T_0}  x_{i,t-1}  +o_p(1)\nonumber
\end{align}
By equation (\ref{mulgtuA1}), it follows that
\begin{align}\label{mulap3dsa}
\frac{1}{\sqrt{T}}\sum\limits_{t=1}^{T}  x_{i,t-1} = \rho_i \frac{1}{\sqrt{T}}\sum\limits_{t=2}^{T}  x_{i,t-2}  +\frac{1}{\sqrt{T}}\sum\limits_{t=1}^{T} v_{i,t}.
\end{align}
So
\begin{align}\label{mulap4dsa}
\frac{1}{\sqrt{T}}\sum\limits_{t=1}^{T}  x_{i,t-1} = \frac{1}{1-\rho_i}\frac{1}{\sqrt{T}}\sum\limits_{t=1}^{T} v_{i,t} + o_p(1)\Rightarrow B_{v_i}(1)/(1-\rho_i).\\
\frac{1}{\sqrt{T}}\sum\limits_{t=1}^{T_0}  x_{i,t-1} = \frac{1}{1-\rho_i}\frac{1}{\sqrt{T}}\sum\limits_{t=1}^{T_0} v_{i,t}+ o_p(1)\Rightarrow B_{v_i}(\lambda)/(1-\rho_i), \nonumber
\end{align}
where $B_v(\lambda)=\left[B_{v_1}(\lambda),B_{v_2}(\lambda),\cdots,B_{v_K}(\lambda)\right]^\top$.
By equations (\ref{7hj4e45d}), (\ref{mulap2dsa}), (\ref{mulap3dsa}) and (\ref{mulap4dsa}), for WD predictors,  we have
\begin{align*}
S_a \Rightarrow \tilde{S}_a =
  B_v(1) B_v(\lambda)^\top \left[ B_v(\lambda)^\top B_v(\lambda) \right]^{-1}.
\end{align*}
Similarly, for WD predictors, by the continuous mapping theorem and the definition of $S_a$ and $S_b$ , it follows that
\begin{align}\label{dkhw4uyy}
S_b  \Rightarrow \tilde{S}_b= B_v(1) \left[B_v(1) - B_v(\lambda) \right]^\top \left\{ \left[B_v(1) - B_v(\lambda) \right]^\top \left[B_v(1) - B_v(\lambda) \right] \right\}^{-1}.\nonumber
\end{align}

\end{proof}

\begin{lemma}\label{leea4pp}
Under Assumption \ref{Assumption A.1}, for SD and WD predictors, it follows that
\begin{align}
D_T^{-2} \sum\limits_{t=1}^{T} \tilde{z}_{t-1} x_{t-1}^\top \Rightarrow  \Sigma_{zx}
\end{align}
\end{lemma}
\begin{proof}[Proof of Lemma \ref{leea4pp}]
 By the definition of $\tilde{z}_{t-1}$ in equation (\ref{7hj4e45d}), it follows that
 \begin{align}\label{dmul9jas1}
D_T^{-2} \sum\limits_{t=1}^{T} \tilde{z}_{t-1} x_{t-1}^\top = \left(\operatorname{I_K}-S_a \right) D_T^{-2}\sum_{t=1}^{T_0} z_{t-1} x_{t-1}^\top+ \left(\operatorname{I_K}-S_b \right) D_T^{-2}\sum_{t=T_0+1}^{T} z_{t-1} x_{t-1}^\top.
\end{align}
Also
\begin{align}\label{op8pt1}
x_{t-1} = x_{t-2} + \Delta x_{t-1}.
\end{align}
The result of equation (\ref{op8pt1}) times equation (\ref{mulivz}) is as follows.
\begin{align*}
 &\frac{1}{T} \sum_{t=1}^{T_0} z_{t-1}x_{t-1}^\top \\
 &= \rho_z \frac{1}{T} \sum_{t=1}^{T_0}z_{t-2} x_{t-2}^\top+ \rho_z \frac{1}{T} \sum_{t=1}^{T_0}z_{t-2} \Delta x_{t-1}^\top + \frac{1}{T} \sum_{t=2}^{T_0} \Delta x_{t-1}  x_{t-2}^\top + \frac{1}{T}\sum_{t=1}^{T_0} \Delta x_{t-1} \Delta x_{t-1}^\top \nonumber
\end{align*}
Since $\frac{1}{T^{(1+\delta)/2}} \sum_{t=1}^{T_0}z_{t-2} \Delta x_{t-1}^\top =O_p(1)$, then
\begin{align}
(1-\rho_z)\frac{1}{T} \sum_{t=1}^{T_0} z_{t-1}x_{t-1}^\top
 &= \frac{1}{T} \sum_{t=1}^{T_0}\Delta x_{t-1}  x_{t-2}^\top + \frac{1}{T}\sum_{t=2}^{T_0} \Delta x_{t-1}  \Delta x_{t-1}^\top + o_p(1). \nonumber
\end{align}
As a result,
\begin{align}\label{hg3jj6e}
&(\operatorname{I_K}-S_a)(1-\rho_z)\frac{1}{T} \sum_{t=1}^{T_0} z_{t-1}x_{t-1}^\top \\
 &= (\operatorname{I_K}-S_a)\frac{1}{T} \sum_{t=1}^{T_0} \Delta x_{t-1} x_{t-2}^\top + (\operatorname{I_K}-S_a) \frac{1}{T}\sum_{t=1}^{T_0} \Delta x_{t-1} \Delta x_{t-1}^\top + o_p(1). \nonumber\\
 & \Rightarrow (\operatorname{I_K}-\tilde{S}_a)\int_0^{\lambda} d J_x^c(r)J_x^c(r)^\top + \lambda (\operatorname{I_K}-\tilde{S}_a) \operatorname{E}(v_tv_t^\top). \nonumber
\end{align}
Similarly,
\begin{align}\label{hg4jj6e}
& (\operatorname{I_K}-S_b)(1-\rho_z)\frac{1}{T} \sum_{t=T_0+1}^{T} z_{t-1}x_{t-1}^\top \\
 &= (\operatorname{I_K}-S_b)\frac{1}{T} \sum_{t=T_0+1}^{T} \Delta x_{t-1} x_{t-2}^\top + (\operatorname{I_K}-S_b) \frac{1}{T}\sum_{t=T_0+1}^{T} \Delta x_{t-1} \Delta x_{t-1}^\top + o_p(1) \nonumber\\
 & \Rightarrow (\operatorname{I_K}-\tilde{S}_a)\int_0^{\lambda} d J_x^c(r) J_x^c(r)^\top  + (1 -\lambda) (\operatorname{I_K}-\tilde{S}_b) \nonumber \operatorname{E}(v_tv_t^\top).
\end{align}
By equations (\ref{dmul9jas1}), (\ref{hg3jj6e}) and (\ref{hg4jj6e}), for SD predictors, we have
\begin{align}\label{cl9ihe}
D_T^{-2} \sum\limits_{t=1}^{T} \tilde{z}_{t-1} x_{t-1}^\top \xrightarrow{P} \Sigma_{zx}
\end{align}

By the similar procedure, we have the following equation for WD predictors.
\begin{align}\label{hg5jj6e}
\frac{1}{T} \sum_{t=1}^{T_0} \tilde{z}_{t-1}x_{t-1}^\top & =(\operatorname{I_K}-S_a) \frac{1}{T} \sum_{t=1}^{T_0} z_{t-1}x_{t-1}^\top
 = (\operatorname{I_K}-S_a)\frac{1}{T} \sum_{t=1}^{T_0} x_{t-1} x_{t-1}^\top + o_p(1)  \\
 &\Rightarrow \lambda(\operatorname{I_K}-\tilde{S}_a) \operatorname{E}\left( x_{t-1}x_{t-1}^\top\right) \nonumber
\end{align}
and
\begin{align}\label{hg6jj6e}
\frac{1}{T} \sum_{t=T_0+1}^{T} \tilde{z}_{t-1}x_{t-1}
 &= (\operatorname{I_K}-S_b) \frac{1}{T} \sum_{t=T_0+1}^{T} z_{t-1}x_{t-1}
  = (\operatorname{I_K}-S_b)\frac{1}{T} \sum_{t=T_0+1}^{T} x_{t-1} x_{t-1}^\top + o_p(1)  \\
 &\Rightarrow \lambda(1-\tilde{S}_b) \operatorname{E}\left( x_{t-1}x_{t-1}^\top\right) \nonumber
\end{align}
By equations (\ref{dmul9jas1}), (\ref{hg5jj6e}) and (\ref{hg6jj6e}), for WD predictors, we also have
\begin{align}\label{cl0ihe}
D_T^{-2} \sum\limits_{t=1}^{T} \tilde{z}_{t-1} x_{t-1}^\top \xrightarrow{P} \Sigma_{zx}
\end{align}
By equations (\ref{cl9ihe}) and (\ref{cl0ihe}), Lemma \ref{leea4pp} holds.
\end{proof}

\begin{proof}[Proof of Theorem \ref{multh1m}]
The illustration of the joint convergence  between $D_T^{-2} \sum\limits_{t=1}^{T} \tilde{z}_{t-1} x_{t-1} $ and $D_T^{-1} \sum\limits_{t=1}^{T} \tilde{z}_{t-1} u_t$ is the same as the last part of the proof of Lemma \ref{mulleea3pp}. Therefore, Theorem \ref{multh1m} holds by Lemma \ref{mulleea3pp} and Lemma \ref{leea4pp}.
\end{proof}

The proof of equations (\ref{mulkee3y1}) and (\ref{mulkee3y2}) is very similar to the proof of Proposition \ref{mulpropfdie3}, so it is omitted here.

\begin{lemma}\label{opv7as}
Under Assumption \ref{Assumption A.1}, for SD and WD predictors, we have
$$D_T^{-2}\sum_{t=1}^T \tilde{z}_{t-1} \tilde{z}_{t-1}^\top \hat{u}_t^2\Rightarrow \Sigma_{zz}.$$
\end{lemma}
\begin{proof}[Proof of Lemma \ref{opv7as}]
Following the similar procedure for the stability condition shown in equation (\ref{mullindb10}) for $\{ \frac{1}{T^{(1+\delta)/2}} \breve{z}_{t-1} u_t\}_{t=1}^T$ of proof of Lemma \ref{mulleea3pp}, Lemma \ref{opv7as} is proved.
\end{proof}

\begin{proof}[Proof of Proposition \ref{mulpropp2}]
This proof is quite similar to the proof of Proposition 1 of \cite{HosseinkouchackDemetrescu2021}.

By Lemma \ref{porlem2}, it follows that
\begin{align}\label{gk66jjh}
\mathbf{t_l} &\equiv \left( \sum_{t=1}^T \tilde{z}_{t-1} \tilde{z}_{t-1}^\top \hat{u}_t^2 \right)^{-1/2} \sum_{t=1}^T \tilde{z}_{t-1} u_t \\
&=\left( \frac{1}{T^{1+\delta}} \sum_{t=1}^T \tilde{z}_{t-1} \tilde{z}_{t-1}^\top \hat{u}_t^2 \right)^{-1/2}\frac{1}{T^{1 / 2+\delta / 2}} \sum_{t=1}^T \tilde{z}_{t-1} u_t \nonumber\\
&=\left[ \frac{1}{T^{1+\delta}} \sum_{t=1}^T \tilde{z}_{t-1}^* (\tilde{z}_{t-1}^*)^\top \hat{u}_t^2 \right]^{-1/2}\frac{1}{T^{1 / 2+\delta / 2}} \sum_{t=1}^T \tilde{z}_{t-1}^* u_t \nonumber\\
&= \left\{\operatorname{I_K} -\frac{1}{2} \left(\frac{1}{T^{1+\delta }} \sum_{t=1}^T \tilde{z}_{t-1}^*(\tilde{z}_{t-1}^*)^\top u_t^2 -\operatorname{I_K}\right)+o_p\left[ T^{-(1-\delta)/2} \right] \right\}
\frac{1}{T^{1 / 2+\delta / 2}} \sum_{t=1}^T \tilde{z}_{t-1}^* u_t  \nonumber \\
&=  \frac{1}{T^{1 / 2+\delta / 2}} \sum_{t=1}^T \tilde{z}_{t-1}^* u_t   -\frac{1}{2} \left(\frac{1}{T^{1+\delta }} \sum_{t=1}^T \tilde{z}_{t-1}^*(\tilde{z}_{t-1}^*)^\top u_t^2 -\operatorname{I_K}\right)\frac{1}{T^{1 / 2+\delta / 2}} \sum_{t=1}^T \tilde{z}_{t-1}^* u_t  +o_p\left[ T^{-(1-\delta)/2} \right]\nonumber\\
&=Z_T^l+B_T^l+o_p\left[ T^{-(1-\delta)/2} \right],
\nonumber
\end{align}
where
$Z_T^l = \left[ \Sigma_{zz} \right]^{-1/2} \frac{1}{T^{(1+\delta)/2}}\sum\nolimits_{t=1}^T \tilde{z}_{t-1} u_t$, $B_T^l  = \varpi_l Z_T^l$ and $\varpi_l = -\frac{1}{2}\left[\sum\nolimits_{t=1}^T \tilde{z}_{t-1}^* (\tilde{z}_{t-1}^*)^\top \hat{u}_t^2 -1\right]$.

By Lemma \ref{mulleea3pp} and Lemma \ref{opv7as}, we have
\begin{align}\label{ahj3j4gd}
 Z_T^l\xrightarrow{d}\operatorname{N}(0_K,\operatorname{I_K}).
\end{align}

By equations (\ref{mulkee3y1}), (\ref{mulkee3y2}) and (\ref{mulcom4r}), it follows that
\begin{small}
\begin{align}\label{kee3y4yd}
&Z_T^l - W_a Z_T^{p_a} - W_b Z_T^{p_b} = \Sigma_{zz}^{-1/2}\frac{1}{T^{(1+\delta)/2}} \sum\nolimits_{t=1}^T \tilde{z}_{t-1} u_t \\
& - \left( \frac{\sum_{t=1}^{T} \tilde{z}_{t-1} \tilde{z}_{t-1}^\top \hat{u}_t^2}{T^{1+\delta}}  \right)^{-1/2}
(\operatorname{I_K}-S_a)
\left( \frac{\sum_{t=1}^{T_0} z_{t-1} z_{t-1}^\top \hat{u}_t^2}{T^{1+\delta}}  \right)^{1/2}
\left( \frac{\sum_{t=1}^{T_0} z_{t-1} z_{t-1}^\top \hat{u}_t^2 }{T^{1+\delta}} \right)^{-1/2}
\frac{ \sum\nolimits_{t=1}^{T_0} z_{t-1} u_t }{T^{(1+\delta)/2}}
 \nonumber\\
&-\left( \frac{\sum_{t=1}^{T} \tilde{z}_{t-1} \tilde{z}_{t-1}^\top \hat{u}_t^2}{T^{1+\delta}}  \right)^{-1/2}
(\operatorname{I_K}-S_b)
\left( \frac{\sum_{t=T_0+1}^{T} z_{t-1} z_{t-1}^\top \hat{u}_t^2}{T^{1+\delta}}  \right)^{1/2}  \nonumber\\
&\cdot \left( \frac{ \sum_{t=T_0+1}^{T} z_{t-1} z_{t-1}^\top \hat{u}_t^2}{T^{1+\delta}} \right)^{-1/2}
\frac{ \sum\nolimits_{t=T_0+1}^T z_{t-1} u_t }{T^{(1+\delta)/2}}
\nonumber\\
&=   \Sigma_{zz}^{-1/2}\frac{1}{T^{(1+\delta)/2}} \sum\nolimits_{t=1}^T \tilde{z}_{t-1} u_t   - \left( \frac{\sum_{t=1}^{T} \tilde{z}_{t-1} \tilde{z}_{t-1}^\top \hat{u}_t^2}{T^{1+\delta}}  \right)^{-1/2}
(\operatorname{I_K}-S_a)
\frac{ \sum\nolimits_{t=1}^{T_0} z_{t-1} u_t }{T^{(1+\delta)/2}}
 \nonumber\\
&-\left( \frac{\sum_{t=1}^{T} \tilde{z}_{t-1} \tilde{z}_{t-1}^\top \hat{u}_t^2}{T^{1+\delta}}  \right)^{-1/2}
(\operatorname{I_K}-S_b)
\frac{ \sum\nolimits_{t=T_0+1}^T z_{t-1} u_t }{T^{(1+\delta)/2}}
\nonumber
\end{align}
\end{small}
\vspace{-0.9cm}
\begin{small}
\begin{align}
&=   \Sigma_{zz}^{-1/2}\frac{1}{T^{(1+\delta)/2}} \sum\nolimits_{t=1}^T \tilde{z}_{t-1} u_t   - \left( \frac{\sum_{t=1}^{T} \tilde{z}_{t-1} \tilde{z}_{t-1}^\top \hat{u}_t^2}{T^{1+\delta}}  \right)^{-1/2}
\frac{ \sum\nolimits_{t=1}^{T_0} \tilde{z}_{t-1} u_t }{T^{(1+\delta)/2}}
 \nonumber\\
&-\left( \frac{\sum_{t=1}^{T} \tilde{z}_{t-1} \tilde{z}_{t-1}^\top \hat{u}_t^2}{T^{1+\delta}}  \right)^{-1/2}
\frac{ \sum\nolimits_{t=T_0+1}^T z_{t-1} u_t }{T^{(1+\delta)/2}}
\nonumber\\
&=  \left[ \Sigma_{zz}^{-1/2}   - \left( \frac{\sum_{t=1}^{T} \tilde{z}_{t-1} \tilde{z}_{t-1}^\top \hat{u}_t^2}{T^{1+\delta}}  \right)^{-1/2} \right]
\frac{ \sum\nolimits_{t=1}^{T_0} \tilde{z}_{t-1} u_t }{T^{(1+\delta)/2}}   -  \left[ \Sigma_{zz}^{-1/2} -\left( \frac{\sum_{t=1}^{T} \tilde{z}_{t-1} \tilde{z}_{t-1}^\top \hat{u}_t^2}{T^{1+\delta}}  \right)^{-1/2}\right]
\frac{ \sum\nolimits_{t=T_0+1}^T z_{t-1} u_t }{T^{(1+\delta)/2}}
\nonumber\\
 &= o_p\left[T^{(1-\delta)/2}\right]O_p(1)+ o_p\left[T^{(1-\delta)/2}\right]O_p(1) = o_p\left[T^{(1-\delta)/2}\right]\nonumber
\end{align}
\end{small}
The last step holds since   $\frac{1}{T^{1+\delta}}\sum_{t=1}^{T_0} \tilde{z}_{t-1} \tilde{z}_{t-1}^\top \hat{u}_t^2-\Sigma_{zz}=o_p\left[T^{(1-\delta)/2}\right]$ and $\frac{1}{T^{1+\delta}}\sum_{t=T_0+1}^{T} \tilde{z}_{t-1} \tilde{z}_{t-1}^\top \hat{u}_t^2-\Sigma_{zz}=o_p\left[T^{(1-\delta)/2}\right]$ by the proof of Proposition 1 of \cite{HosseinkouchackDemetrescu2021}.

Additionally, by equations (\ref{mulkee3y1}), (\ref{mulkee3y2}), (\ref{mulcom4r}) and (\ref{gk66jjh}), it follows that
\begin{align}\label{oop7hg}
B_T^l &= t_l - Z_T^l +o_p\left[ T^{-(1-\delta)/2} \right]
=  W_a t_a^p + W_b t_b^p - Z_T^l +o_p\left[ T^{-(1-\delta)/2} \right] \\
& = (W_a Z_T^{p_a} + W_b Z_T^{p_b} -Z_T^l) + W_a B_T^{p_a}+ W_b B_T^{p_b} + o_p\left[T^{-(1-\delta)/2}\right] \nonumber\\
& = W_a B_T^{p_a}+ W_b B_T^{p_b} + o_p\left[T^{-(1-\delta)/2}\right]. \nonumber
\end{align}
The last step holds by equation (\ref{kee3y4yd}). By equation (\ref{oop7hg})
 \begin{align}\label{app2conl1}
 R_T^l&= T^{(1 -\delta) / 2} B_T^l - W_l \operatorname{plim}_{T\rightarrow \infty} \; T^{(1 -\delta) / 2} \operatorname{E}\left(B_T\right) \\
& = W_a B_T^{p_a}+ W_b B_T^{p_b}- \left[W_a/\sqrt{\lambda} + W_b/\sqrt{1-\lambda} \right]\operatorname{plim}_{T\rightarrow \infty} \; T^{(1 -\delta) / 2} \operatorname{E}\left(B_T\right) + o_p\left[T^{-(1-\delta)/2}\right] \nonumber\\
& = W_a B_T^{p_a}+ W_b B_T^{p_b}- \left[W_a/\sqrt{\lambda} + W_b/\sqrt{1-\lambda} \right] \left[- \frac{K+1}{2}{\rho_{u v^*}}/ \sqrt{-2 c_z} \right] + o_p\left[T^{-(1-\delta)/2}\right] \nonumber\\
&=W_a R_{1,T}^l+W_b R_{2,T}^l \nonumber
 \end{align}
where $ R_{1,T}^l=T^{(1 -\delta) / 2}\left[ B_T^{p_a}- \operatorname{E}(B_T^{p_a}) \right]$ and $ R_{2,T}^l=T^{(1 -\delta) / 2}\left[ B_T^{p_b}- \operatorname{E}(B_T^{p_b}) \right]$.

By equations (\ref{gk66jjh}), (\ref{ahj3j4gd}), (\ref{oop7hg}) and (\ref{app2conl1}), Proposition \ref{mulpropp2} holds.
\end{proof}

\begin{proof}[Proof of Theorem \ref{mulkeythe2}]
When $J=K=1$, by equations (\ref{mulpredModel}) and (\ref{mulkjeyi4ns}), we have
\begin{align}\label{hh4su948}
Q_l^t&= \frac{\sum_{t=1}^T \tilde{z}_{t-1}y_t}{\sqrt{\sum_{t=1}^T \tilde{z}_{t-1}\tilde{z}_{t-1}^\top \hat{u}_t^2}} = \frac{\sum_{t=1}^T \tilde{z}_{t-1}(\mu + x_{t-1}^\top \beta + u_t)}{\sqrt{\sum_{t=1}^T \tilde{z}_{t-1}\tilde{z}_{t-1}^\top \hat{u}_t^2}} \\
&= \frac{\sum_{t=1}^T \tilde{z}_{t-1}  x_{t-1}^\top \beta  }{\sqrt{\sum_{t=1}^T \tilde{z}_{t-1}\tilde{z}_{t-1}^\top \hat{u}_t^2}}
+ \frac{\sum_{t=1}^T \tilde{z}_{t-1} u_t }{\sqrt{\sum_{t=1}^T \tilde{z}_{t-1}\tilde{z}_{t-1}^\top \hat{u}_t^2}} \nonumber \\
&=b \frac{D_T^{-2}\sum_{t=1}^T \tilde{z}_{t-1}  x_{t-1}^\top    }{\sqrt{D_T^{-2}\sum_{t=1}^T \tilde{z}_{t-1}\tilde{z}_{t-1}^\top \hat{u}_t^2}}
+ \frac{D_T^{-1}\sum_{t=1}^T \tilde{z}_{t-1} u_t }{\sqrt{D_T^{-2}\sum_{t=1}^T \tilde{z}_{t-1}\tilde{z}_{t-1}^\top \hat{u}_t^2}} \nonumber
\end{align}
By equation (\ref{hh4su948}) and Theorem \ref{multh1m} and  Lemma \ref{mulopv7as} and \ref{leea4pp}, Theorem \ref{mulkeythe2} holds.
\end{proof}

\begin{proof}[Proof of Theorem \ref{multh1m23}]
By equation (\ref{dsjkih6}) and Lemma \ref{mulopv7as} and \ref{leea4pp}, it follows that
\begin{align}\label{appgh4e}
D_T(\tilde{\beta}_m - \hat{\beta}_{l} ) = O_p\left[ T^{-(1-\delta)/2} \right].
\end{align}
By equation (\ref{appgh4e}) and Theorem \ref{multh1m}, Theorem \ref{multh1m23} holds.
\end{proof}

\begin{proof}[Proof of Proposition \ref{thmpropnew1}]
By the similar procedure of the proof of Theorem \ref{thmnew1} and the definitions of $ Q_l$ and ${\tilde{Q}_m}$, Proposition \ref{thmpropnew1} holds.
\end{proof}

\begin{proof}[Proof of Theorem \ref{thm8f1}]
By  Theorem \ref{multh1m23} and Lemma \ref{mulopv7as} and \ref{leea4pp}, Theorem \ref{thm8f1} holds.
\end{proof}

\begin{proof}[Proof of Proposition \ref{thmpropnew2}]
For SD predictors,
by the definitions of $\tilde{\beta}_m$ and $\hat{\beta}_m$ in equations (\ref{dsjkih6}) and (\ref{uusjkih6}), $$\tilde{\beta}_m -\hat{\beta}_m = O_p({T^{-1}}). $$
Therefore, by the definitions of $Q_m$ and $\tilde{Q}_m$ , ${Q_m}={\tilde{Q}_m}+O_p(T^{-1})$. So
$$Q_l-Q_m= Q_l-\tilde{Q}_m + \tilde{Q}_m-Q_m= Q_l-\tilde{Q}_m+O_p(T^{-1}).$$
Similarly,  for WD predictors, by $ W_z =  o_p({T^{-1}})$,
$$Q_l-Q_m=  O_p(T^{-1}).$$
 By these two results and Proposition \ref{thmpropnew1},  Proposition \ref{thmpropnew2} holds.
\end{proof}

\begin{proof}[Proof of Theorem \ref{thm4}]
By equation (\ref{appgh4e}), for SD and WD predictors, it follows that
\begin{align}\label{hd4fsj}
 D_T(\tilde{\beta}_m- \beta)=D_T(\hat{\beta}_l- \beta)+o_p(1)\xrightarrow{d}
 \operatorname{MN}\left[0,\Sigma_{zx}^{-1}\Sigma_{zz}\left(\Sigma_{zx}^{-1}\right)^\top \right], \quad \text{SD}
\end{align}
for SD and WD predictors.
By equation (\ref{hd4fsj}) and Lemma \ref{mulopv7as} and \ref{leea4pp}, Theorem \ref{thm4} holds.
\end{proof}

\section{Additional Simulation Results}\label{app:b}

\subsection{Example 1 (univariate models)}
\noindent    In this example for the univariate model, the DGP is set up by equations (\ref{mulpredModel}), (\ref{mulgtuA1}) and (\ref{garchset}), in which $K=1$, $\mu=1$ and $\varphi_0=1$. In the univariate model, we define the notation $Q_m^l$ and $Q_m^t$ as the t-test statistics $t_l$ and $t_m$, respectively.  To create the embedded endogeneity among innovations, the innovation processes are generated as $(\eta_t,\varepsilon_t)^\top \;\sim \; i.i.d. \, \operatorname{N}(0_{2\times 1},\Sigma_{2\times 2})$, where $\Sigma=\left(\begin{array}{ccc}1 & \phi  \\ \phi & 1 \end{array}\right)$. To save space, we only report the simulation results for GARCH model with the sample $T=750$. \footnote{To save space, we did not report the simulation results for GARCH model with the sample size $T=500$ and $T=250$ and  ARCH and i.i.d. model with the sample size $T=750$, $T=500$ and $T=250$  in the paper.  Their results are similar to the results of GARCH model with the sample size $T=750$. The codes and results are available upon request.}

\begin{figure}[H]
\centering
\subfigure[Case 1 ($\alpha=1$, $c=0$)]{
\includegraphics[width=5.5cm]{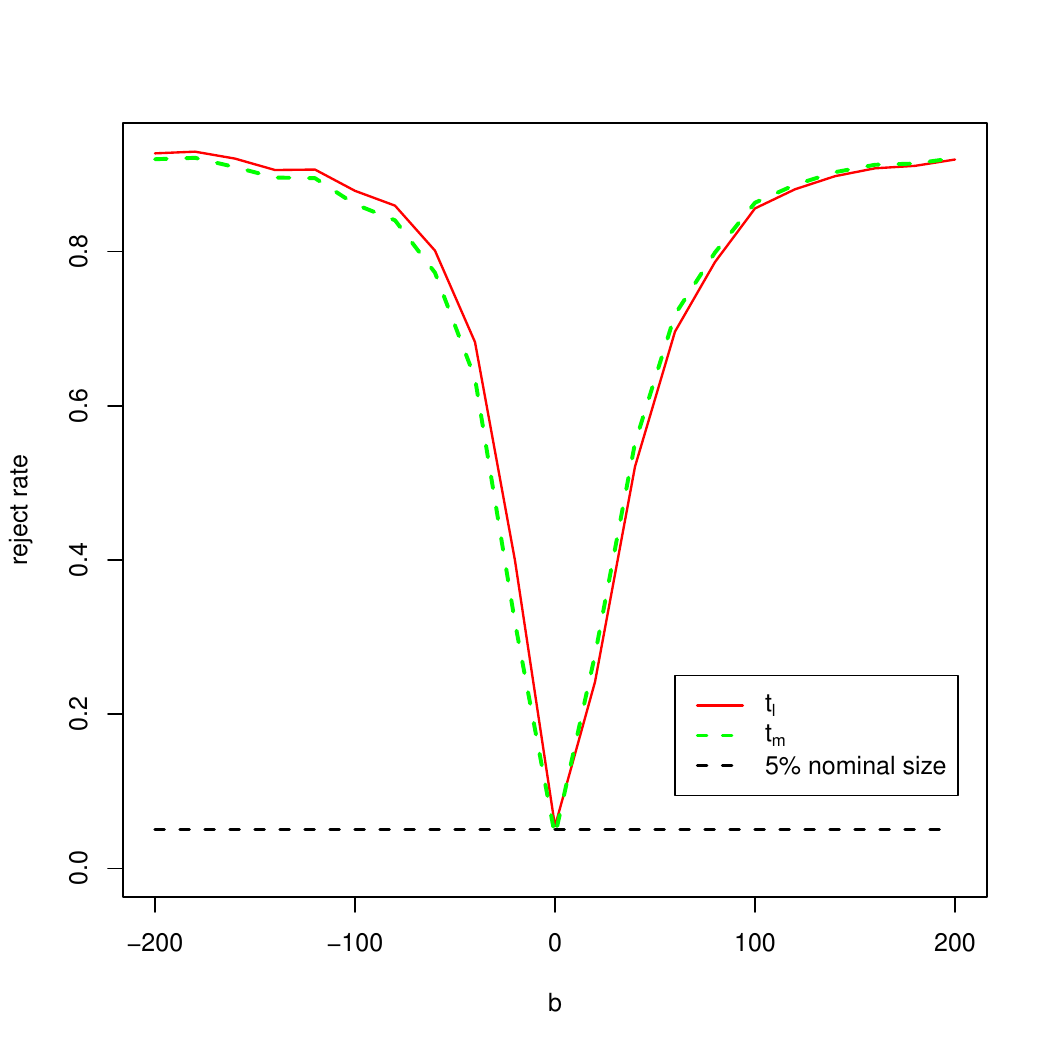}
%\caption{fig1}
}
\quad
\subfigure[Case 2 ($\alpha=1$, $c=-5$)]{
\includegraphics[width=5.5cm]{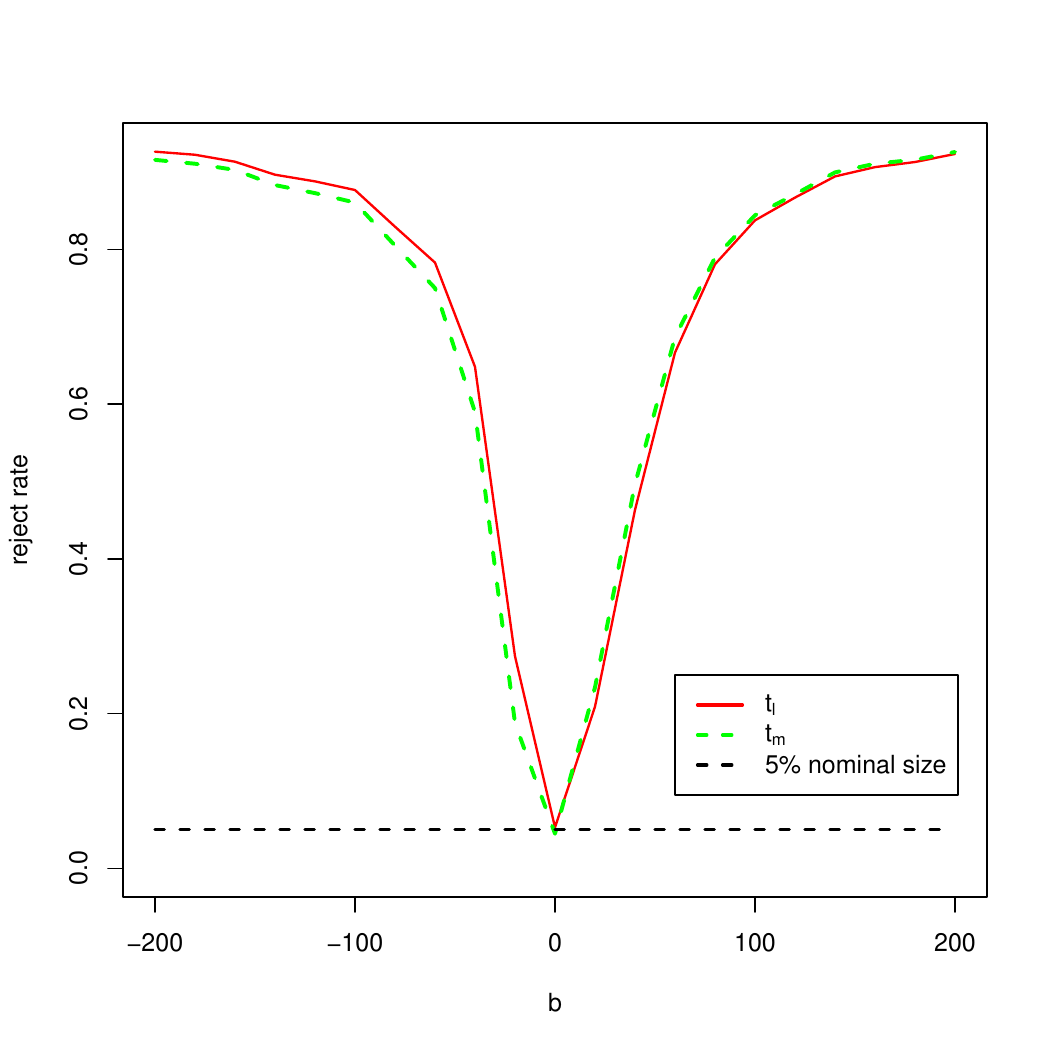}
}
\quad
\subfigure[Case 3 ($\alpha=1$, $c=-10$)]{
\includegraphics[width=5.5cm]{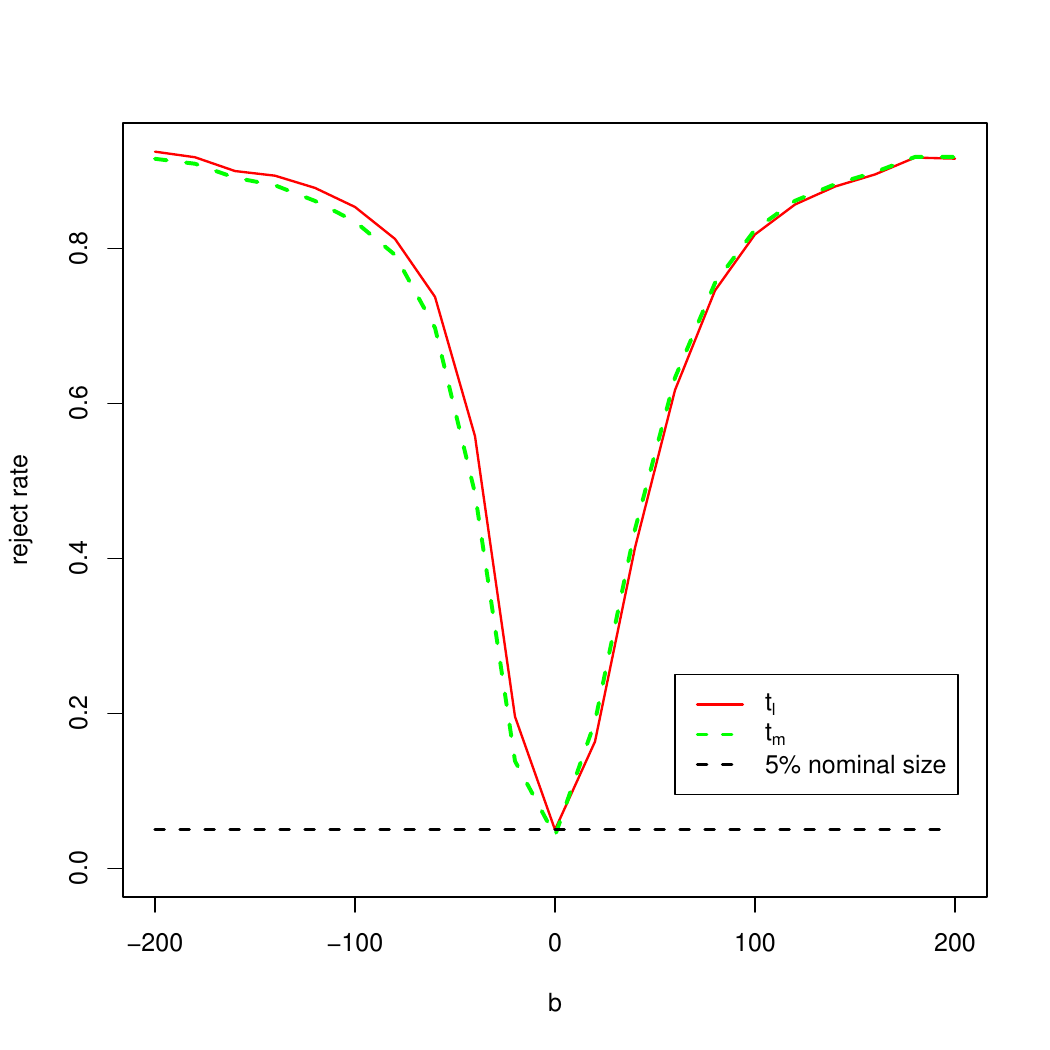}
%\caption{fig1}
}
\quad
\subfigure[Case 4 ($\alpha=1$, $c=-15$)]{
\includegraphics[width=5.5cm]{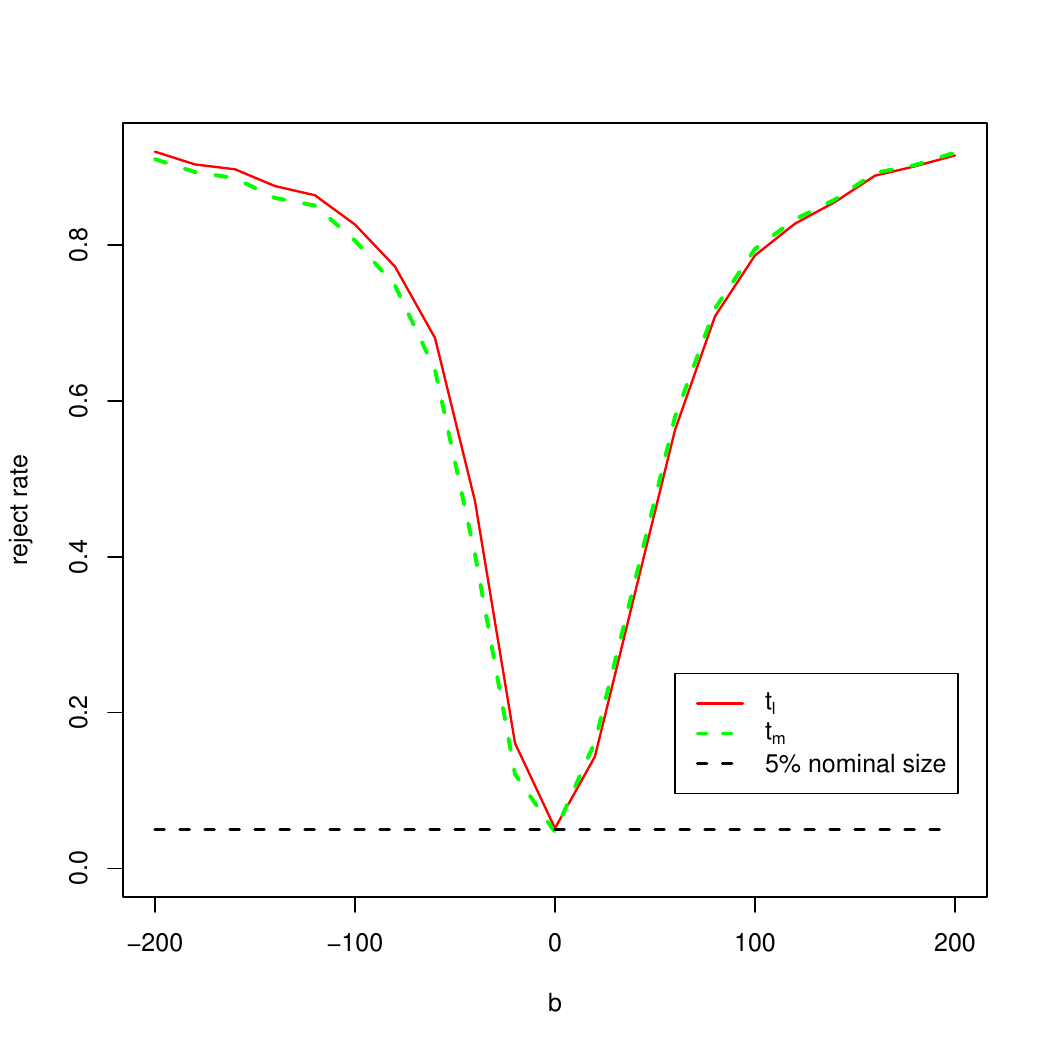}
}
\quad
\subfigure[Case 5 ($\alpha=0$, $c=-0.05$)]{
\includegraphics[width=5.5cm]{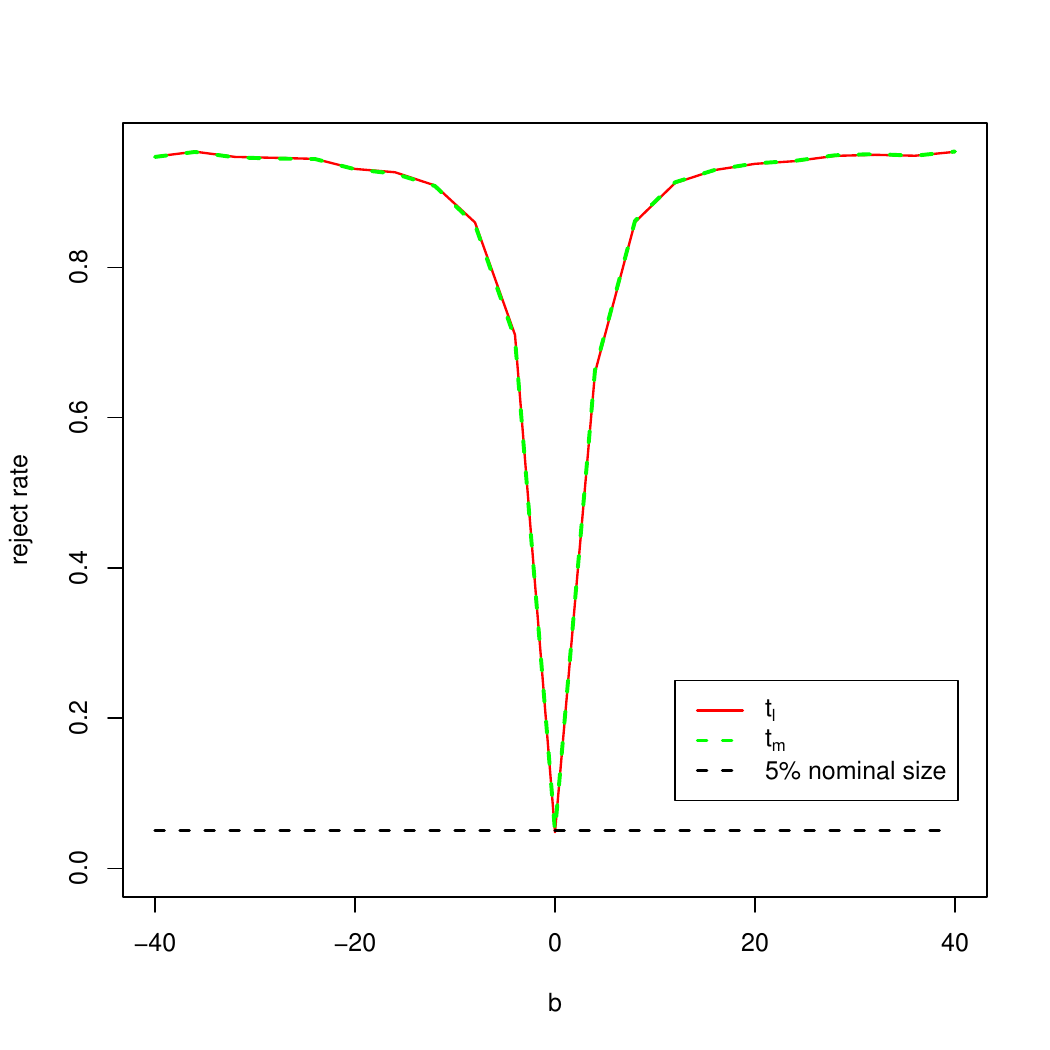}
%\caption{fig1}
}
\quad
\subfigure[Case 6 ($\alpha=0$, $c=-0.1$)]{
\includegraphics[width=5.5cm]{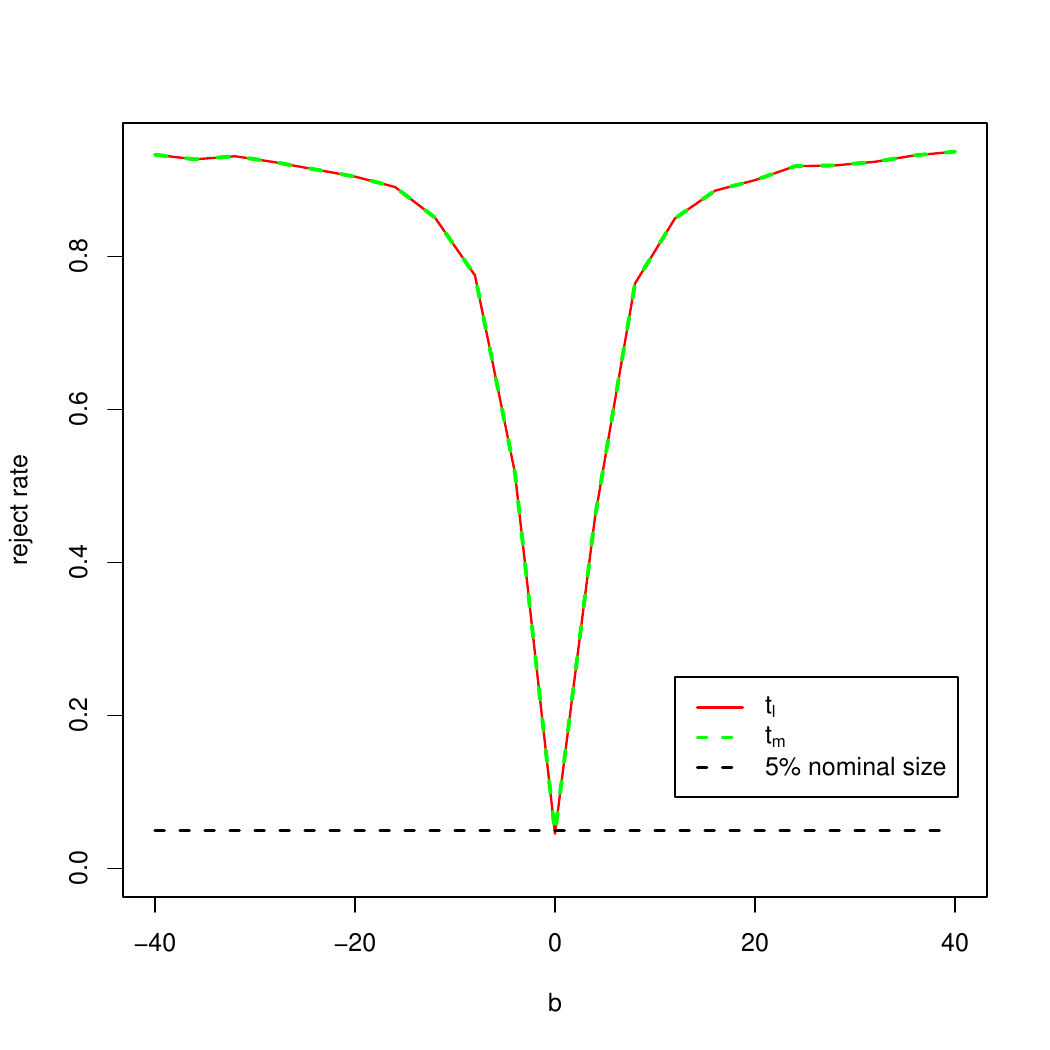}
}
\caption{Power of Two-sided Test $H_0:\beta = 0$ vs $H_a:\beta \neq 0$ with $\phi=0.95$ and $\lambda=0.5$}
\label{power1}
\end{figure}

\begin{figure}[H]
\centering
\subfigure[Case 1 ($\alpha=1$, $c=0$)]{
\includegraphics[width=5.5cm]{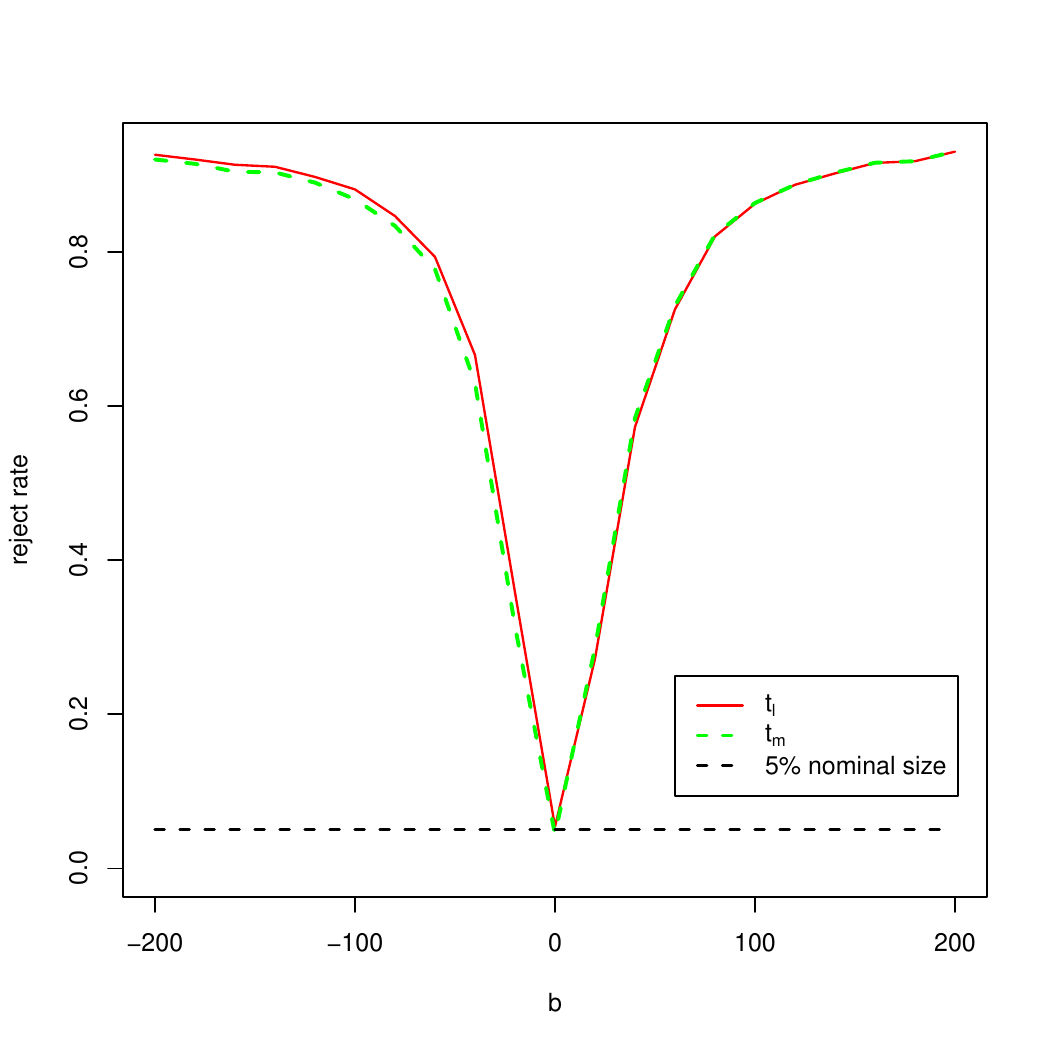}
%\caption{fig1}
}
\quad
\subfigure[Case 2 ($\alpha=1$, $c=-5$)]{
\includegraphics[width=5.5cm]{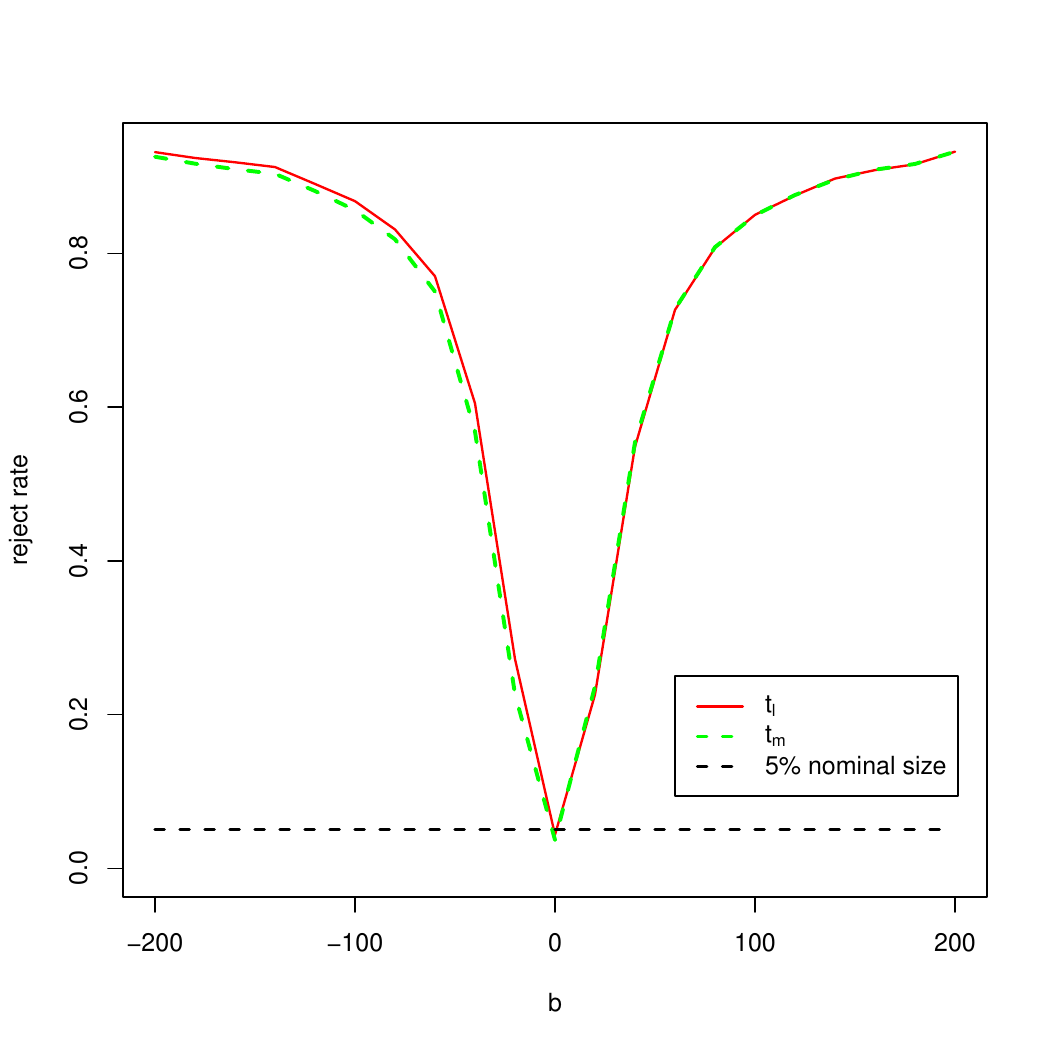}
}
\quad
\subfigure[Case 3 ($\alpha=1$, $c=-10$)]{
\includegraphics[width=5.5cm]{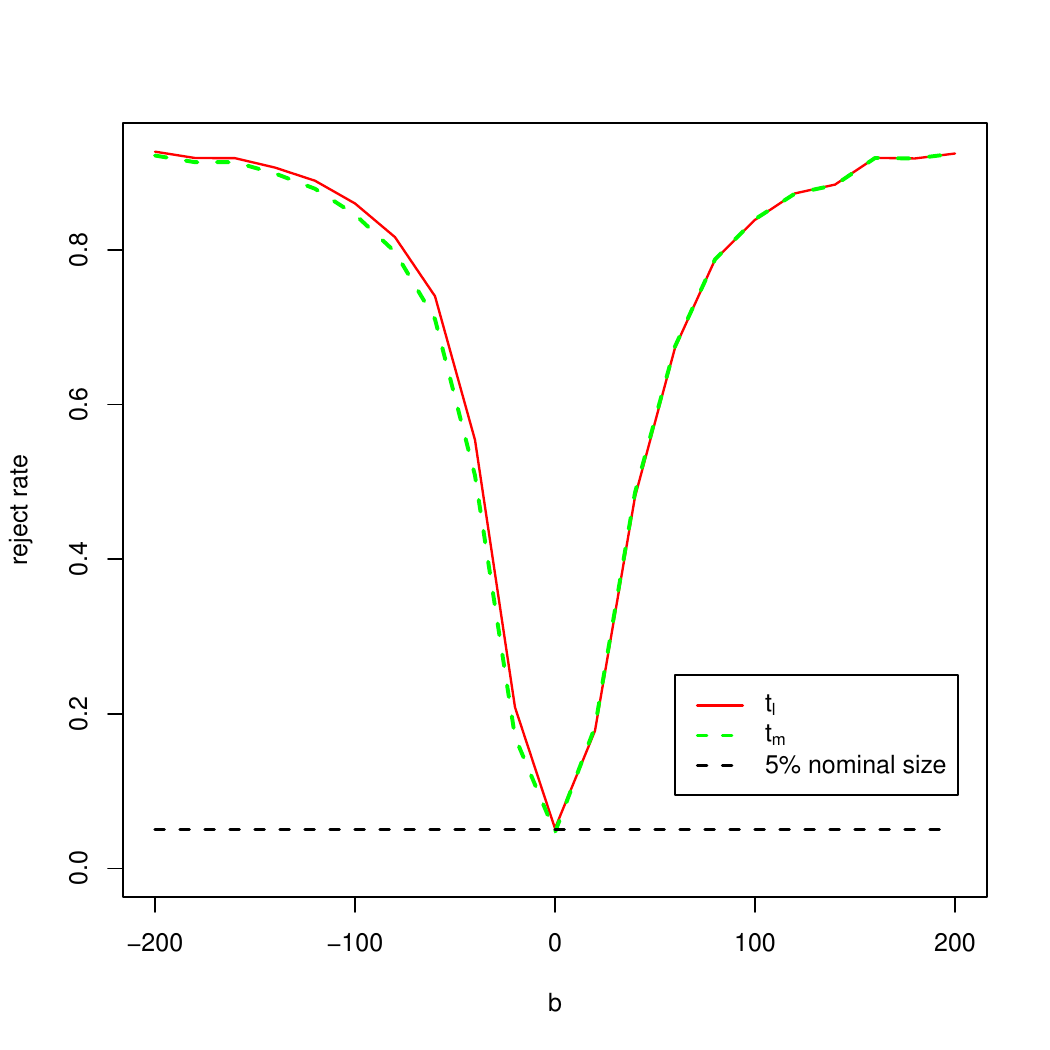}
%\caption{fig1}
}
\quad
\subfigure[Case 4 ($\alpha=1$, $c=-15$)]{
\includegraphics[width=5.5cm]{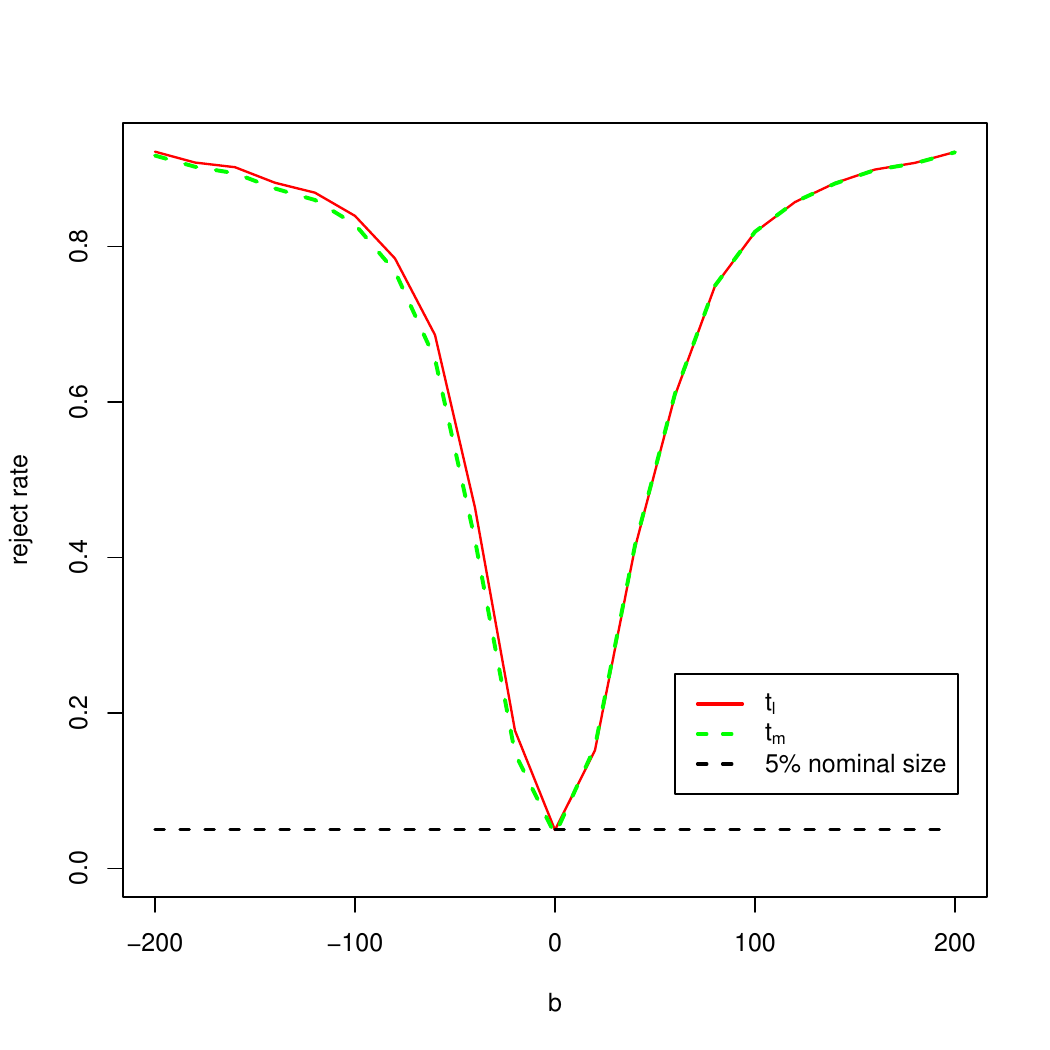}
}
\quad
\subfigure[Case 5 ($\alpha=0$, $c=-0.05$)]{
\includegraphics[width=5.5cm]{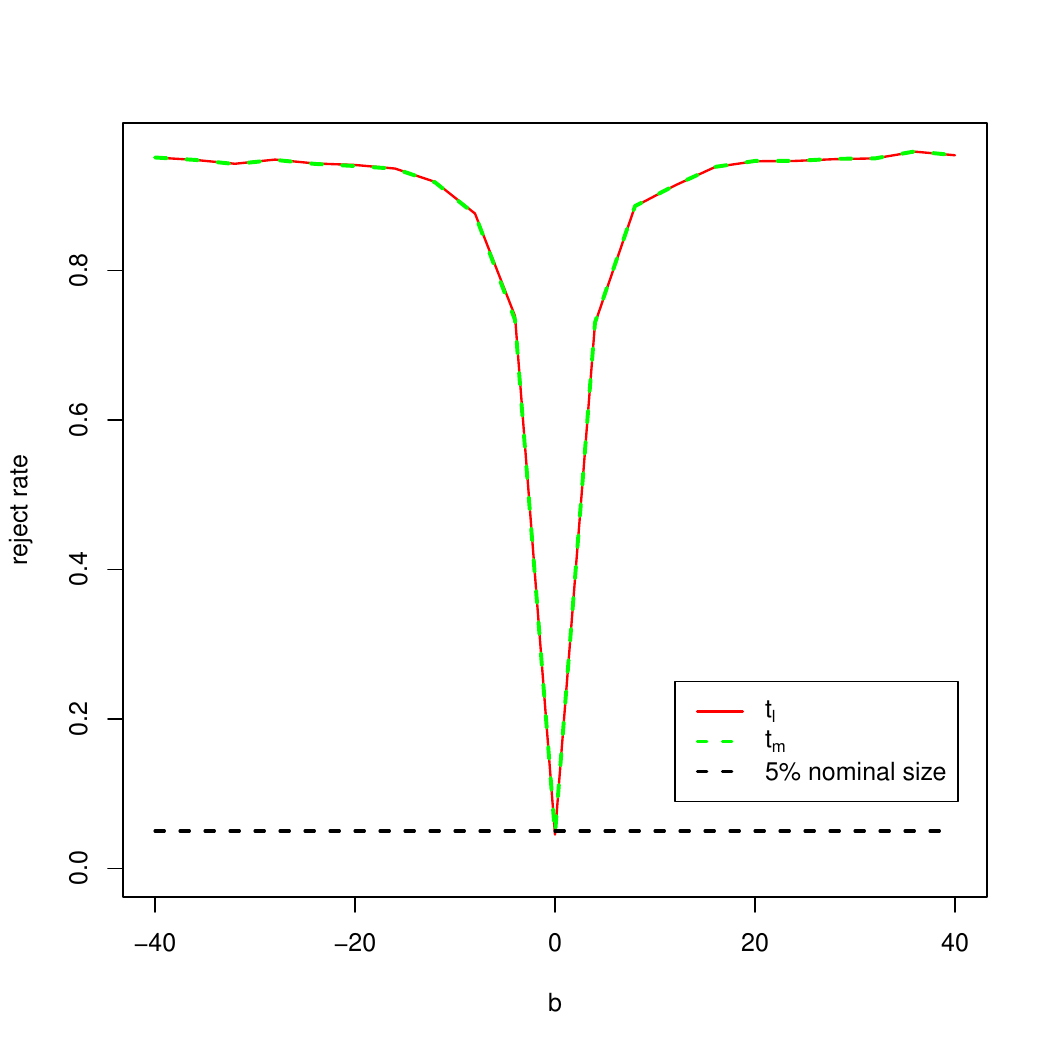}
%\caption{fig1}
}
\quad
\subfigure[Case 6 ($\alpha=0$, $c=-0.1$)]{
\includegraphics[width=5.5cm]{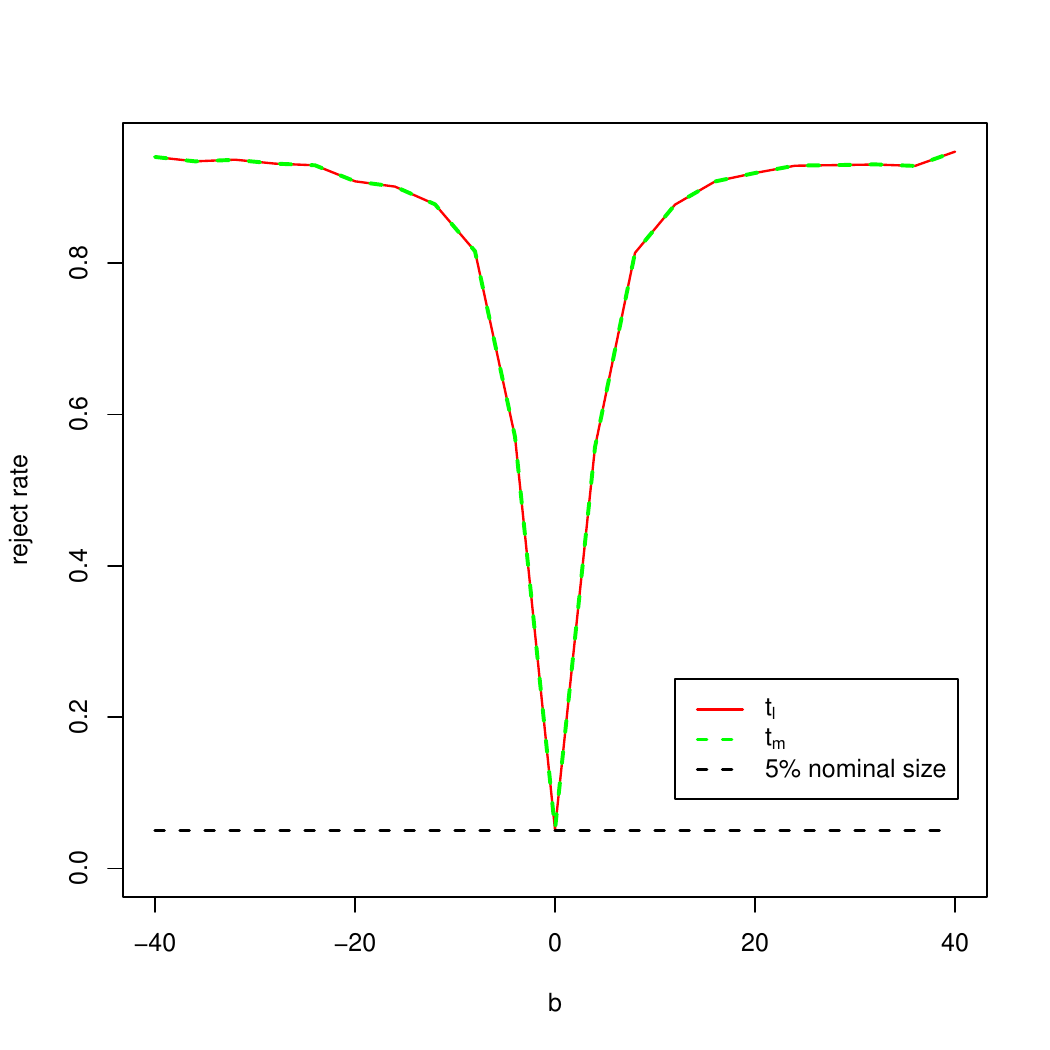}
}
\caption{Power of Two-sided Test $H_0:\beta = 0$ vs $H_a:\beta \neq 0$ with $\phi=0.5$ and $\lambda=0.5$}
\label{power2}
\end{figure}

\begin{figure}[H]
\centering
\subfigure[Case 1 ($\alpha=1$, $c=0$)]{
\includegraphics[width=5.5cm]{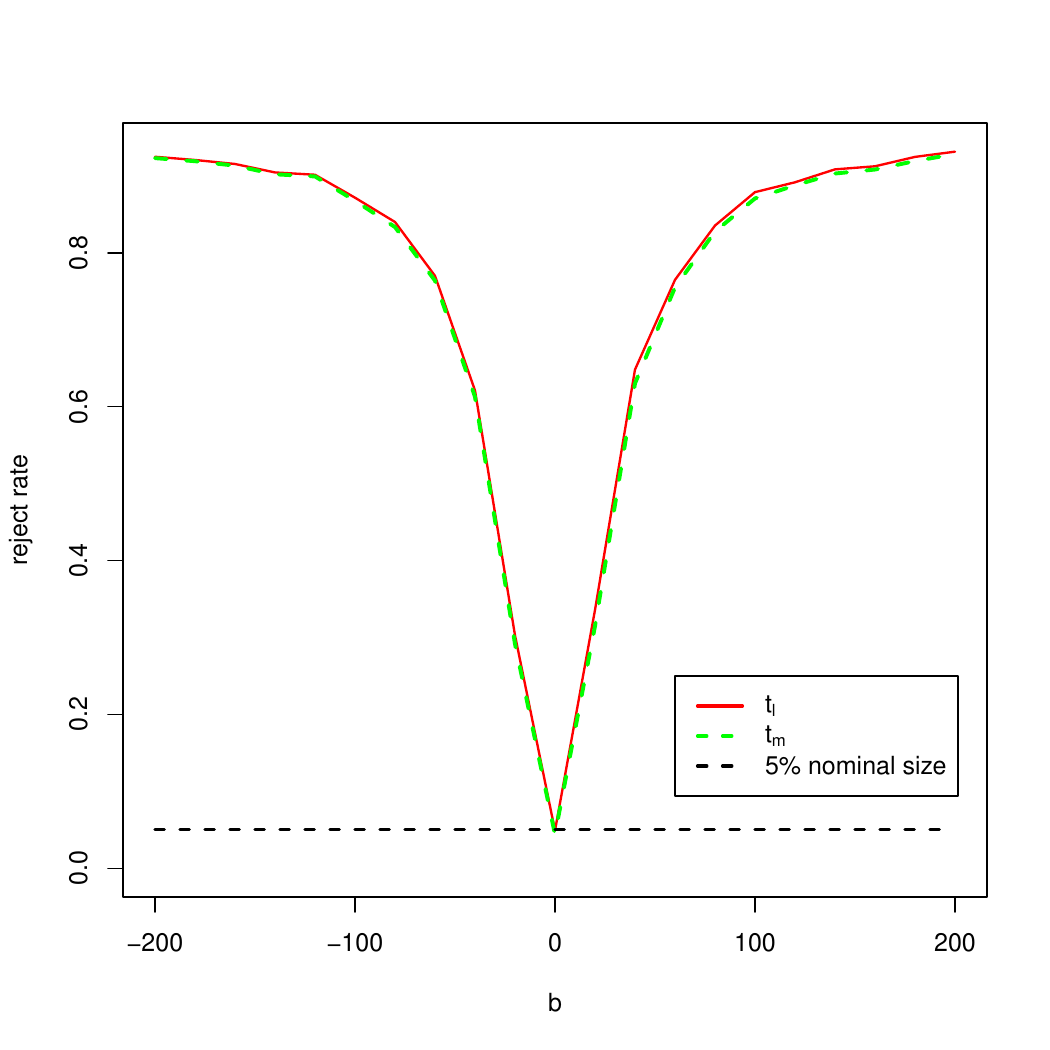}
%\caption{fig1}
}
\quad
\subfigure[Case 2 ($\alpha=1$, $c=-5$)]{
\includegraphics[width=5.5cm]{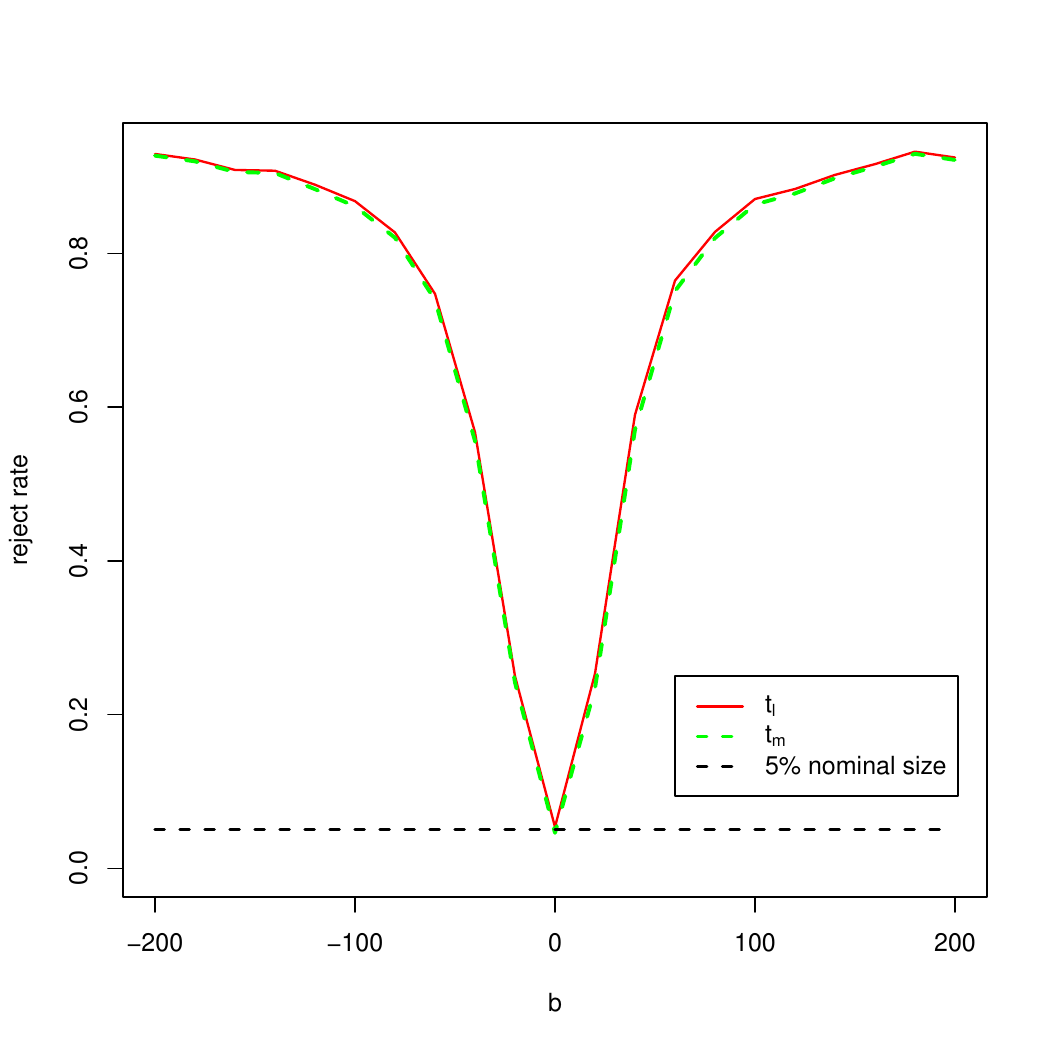}
}
\quad
\subfigure[Case 3 ($\alpha=1$, $c=-10$)]{
\includegraphics[width=5.5cm]{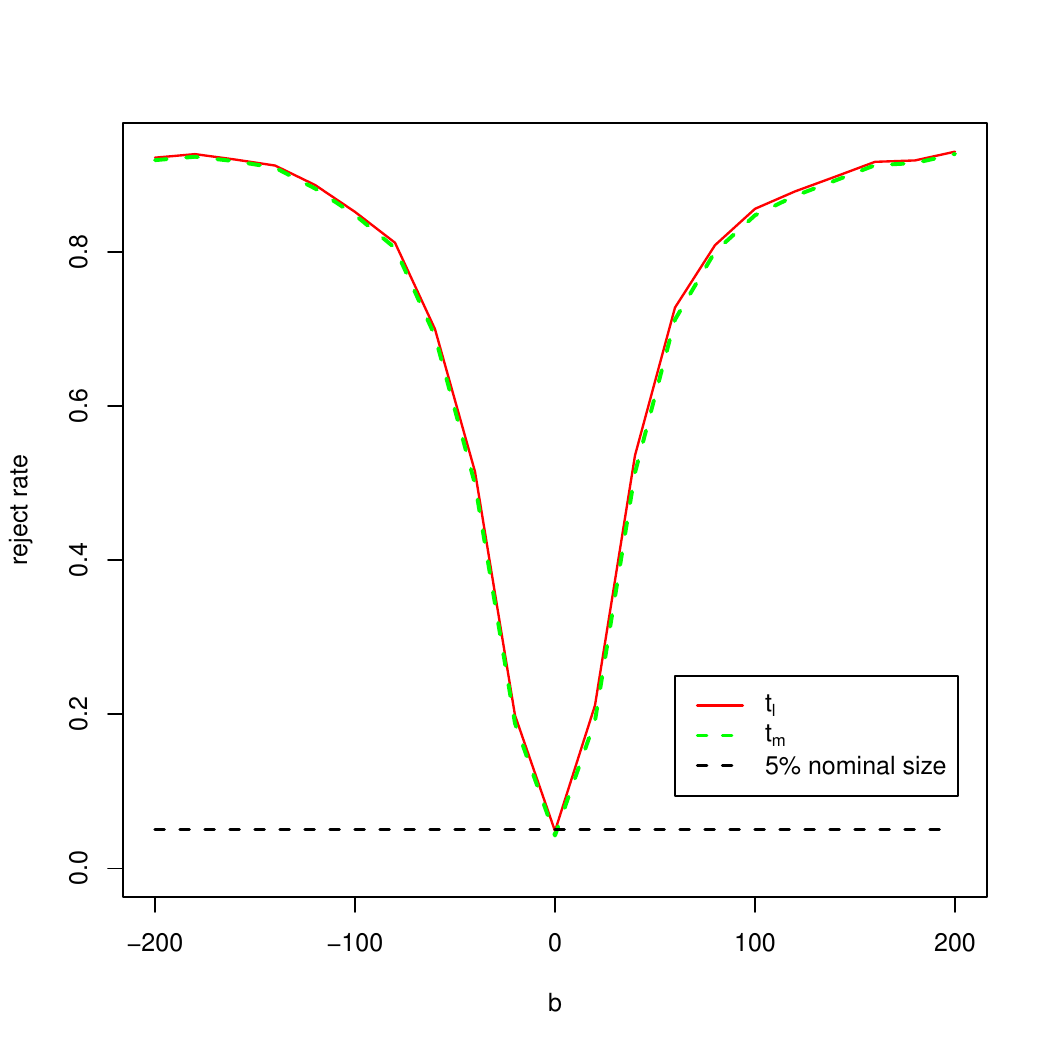}
%\caption{fig1}
}
\quad
\subfigure[Case 4 ($\alpha=1$, $c=-15$)]{
\includegraphics[width=5.5cm]{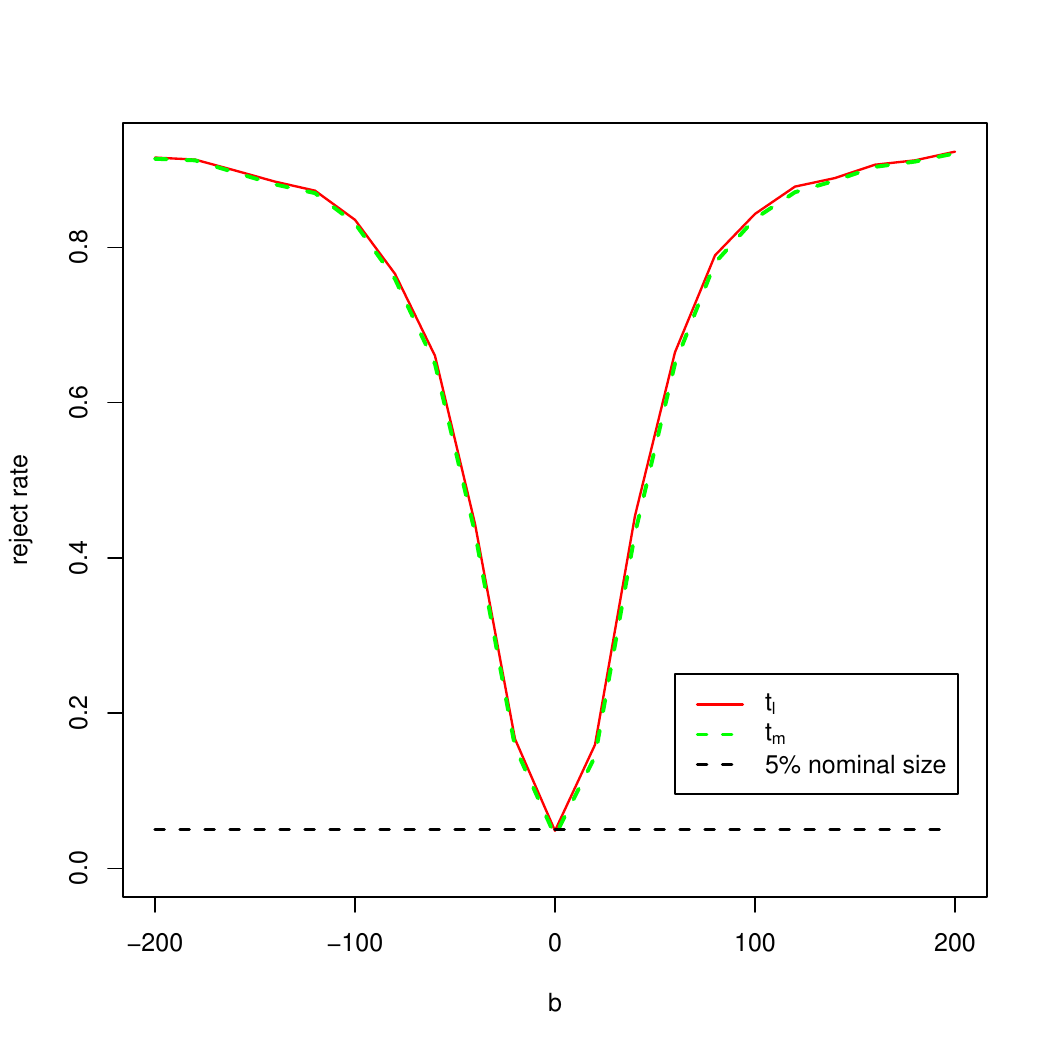}
}
\quad
\subfigure[Case 5 ($\alpha=0$, $c=-0.05$)]{
\includegraphics[width=5.5cm]{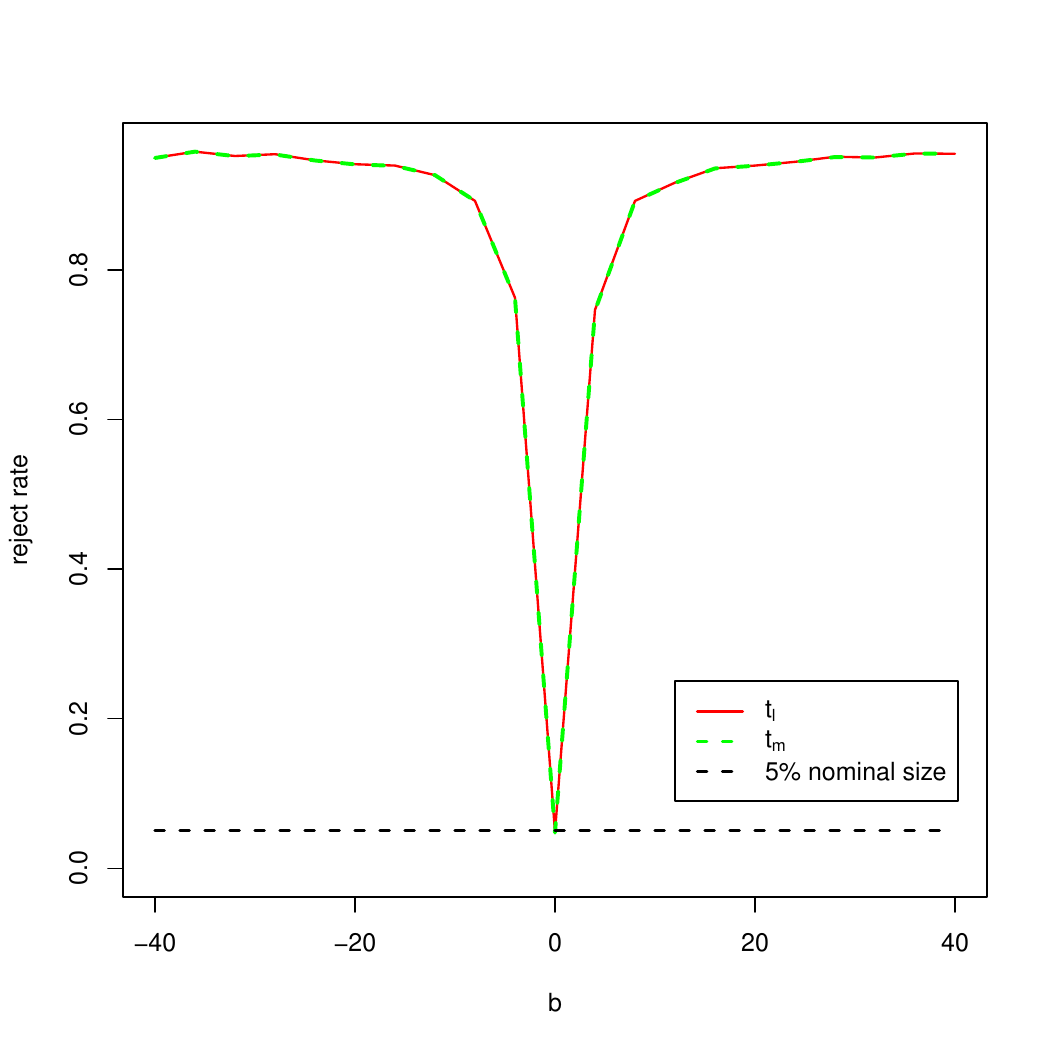}
%\caption{fig1}
}
\quad
\subfigure[Case 6 ($\alpha=0$, $c=-0.1$)]{
\includegraphics[width=5.5cm]{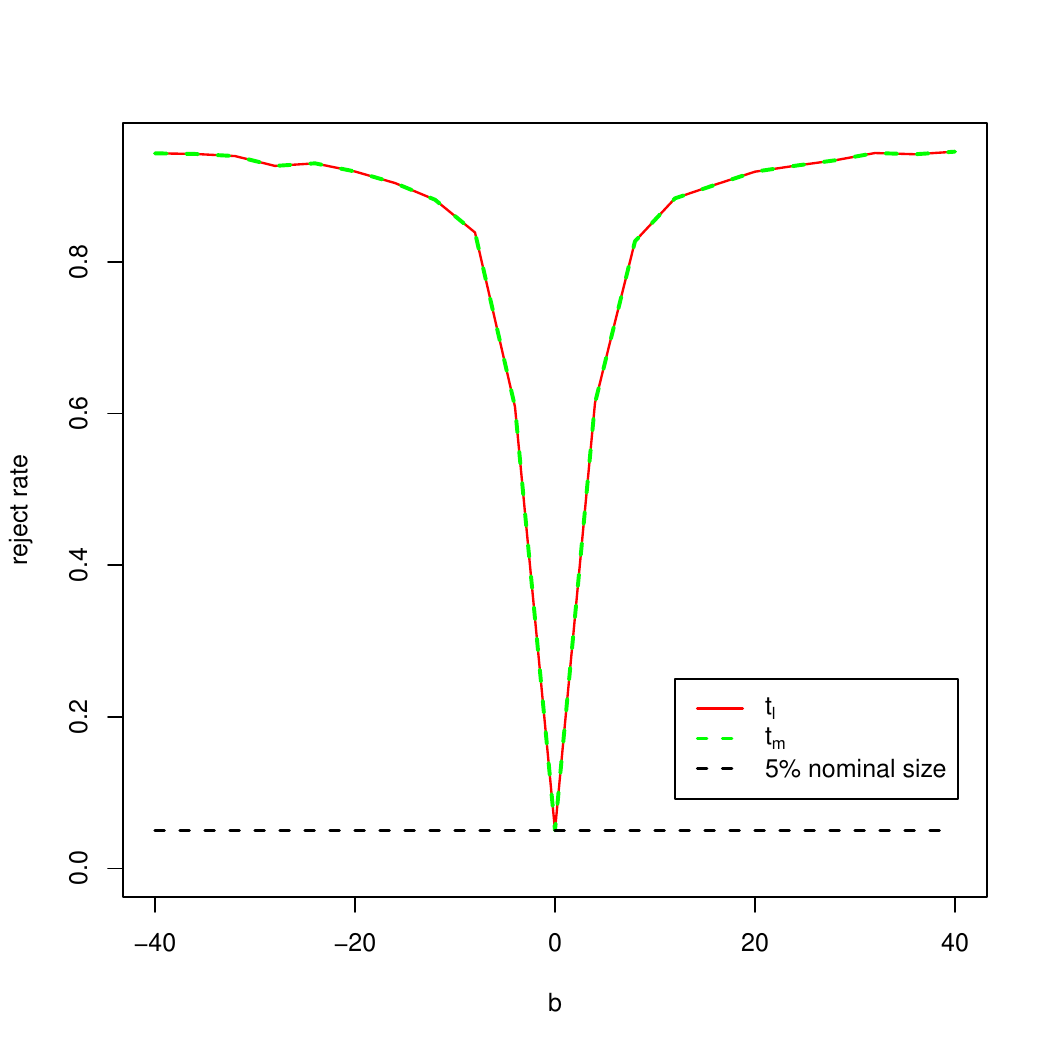}
}
\caption{Power of Two-sided Test $H_0:\beta = 0$ vs $H_a:\beta \neq 0$ with $\phi=-0.1$ and $\lambda=0.5$}
\label{power3}
\end{figure}

\begin{figure}[H]
\centering
\subfigure[Case 1 ($\alpha=1$, $c=0$)]{
\includegraphics[width=5.5cm]{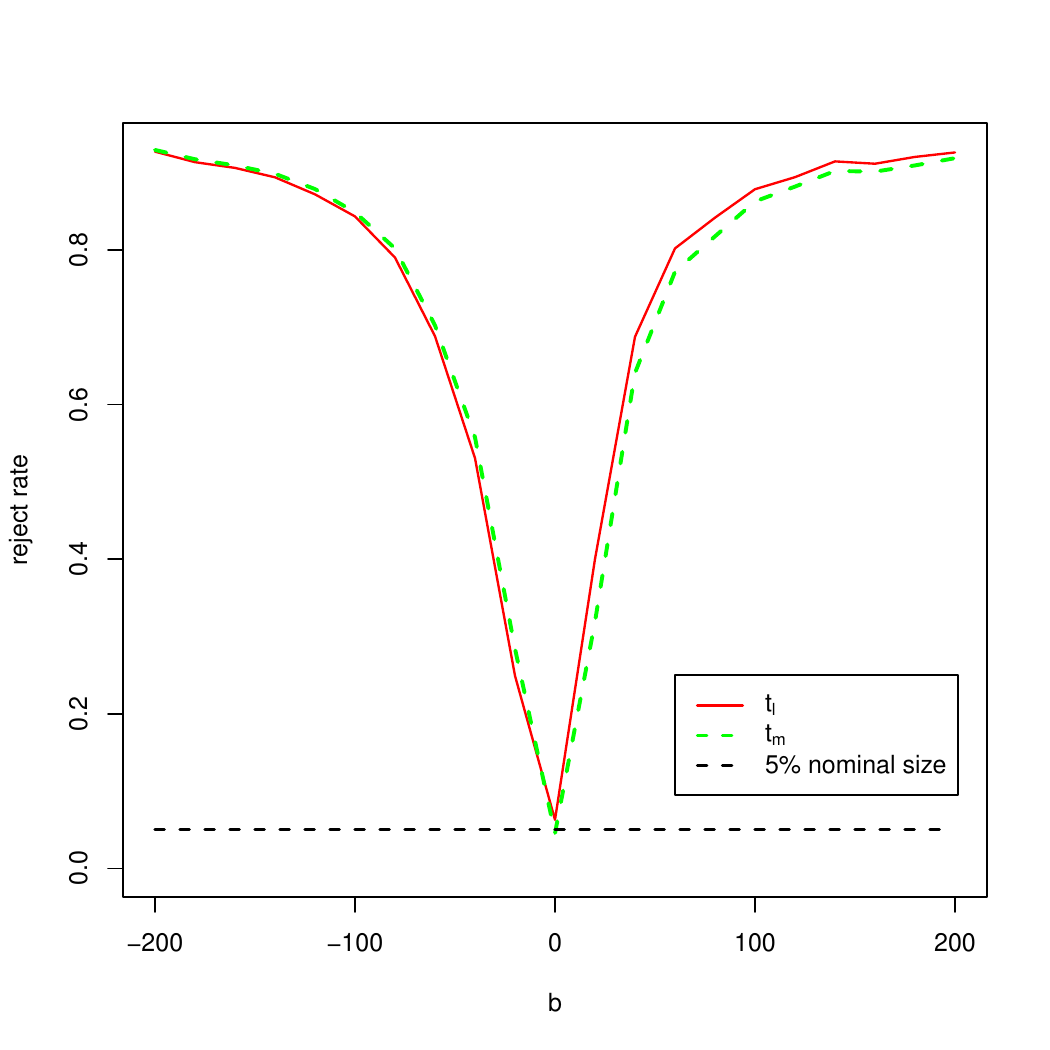}
%\caption{fig1}
}
\quad
\subfigure[Case 2 ($\alpha=1$, $c=-5$)]{
\includegraphics[width=5.5cm]{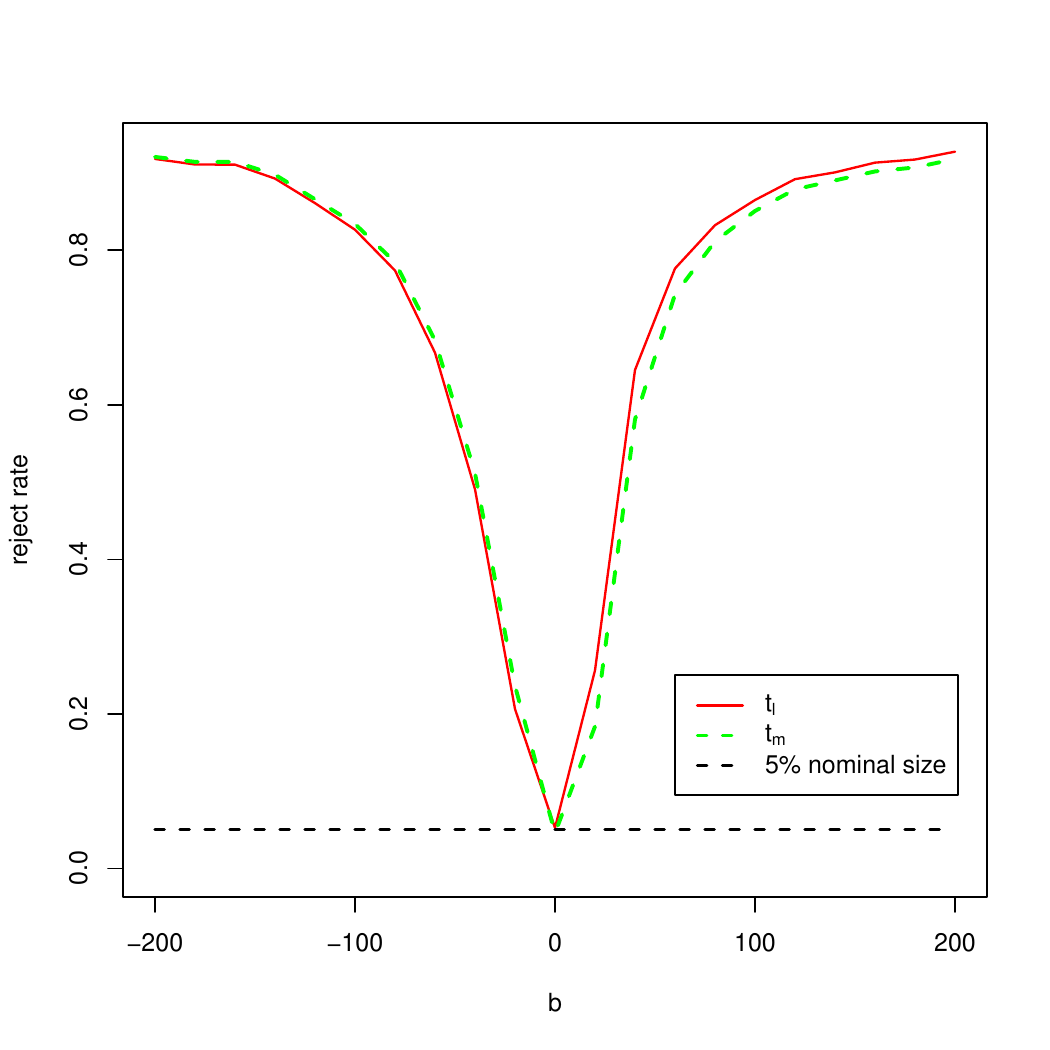}
}
\quad
\subfigure[Case 3 ($\alpha=1$, $c=-10$)]{
\includegraphics[width=5.5cm]{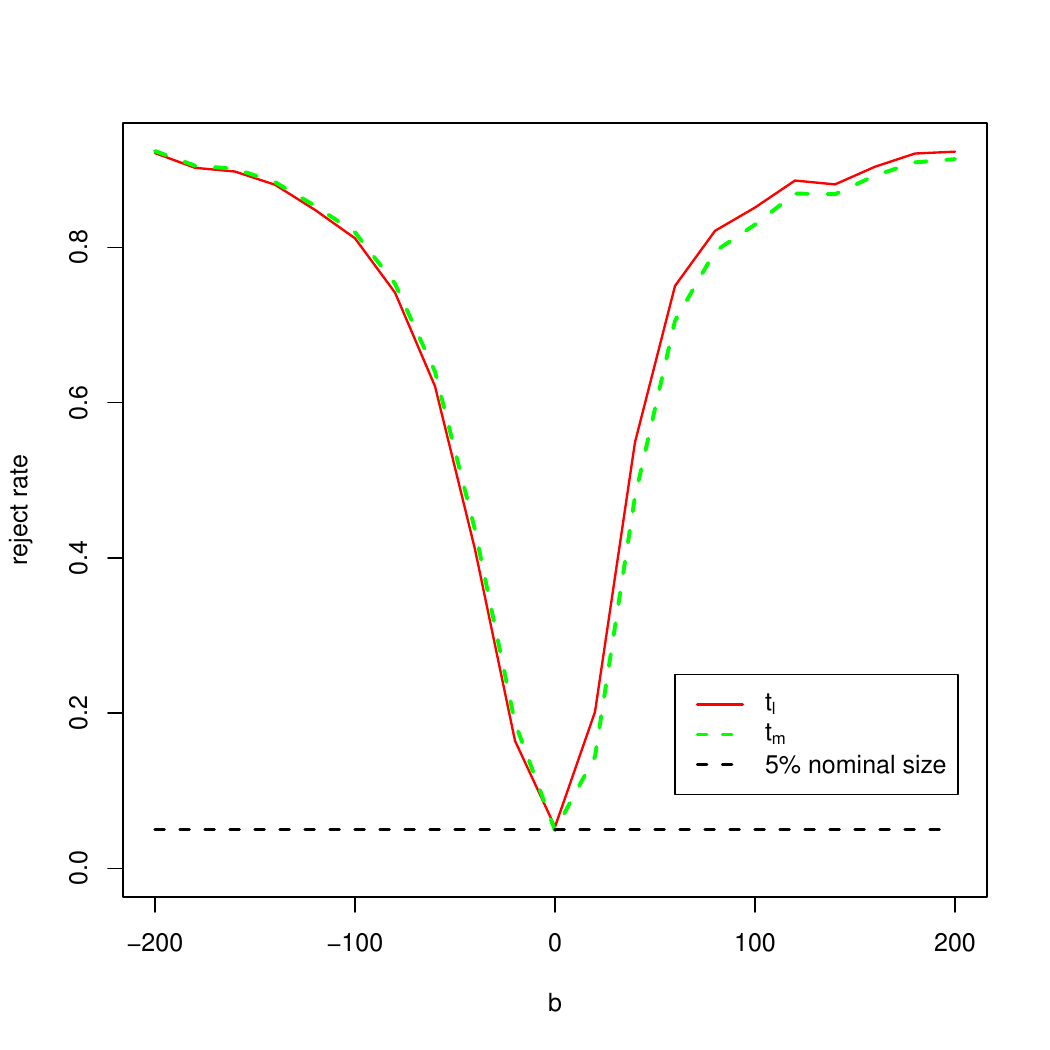}
%\caption{fig1}
}
\quad
\subfigure[Case 4 ($\alpha=1$, $c=-15$)]{
\includegraphics[width=5.5cm]{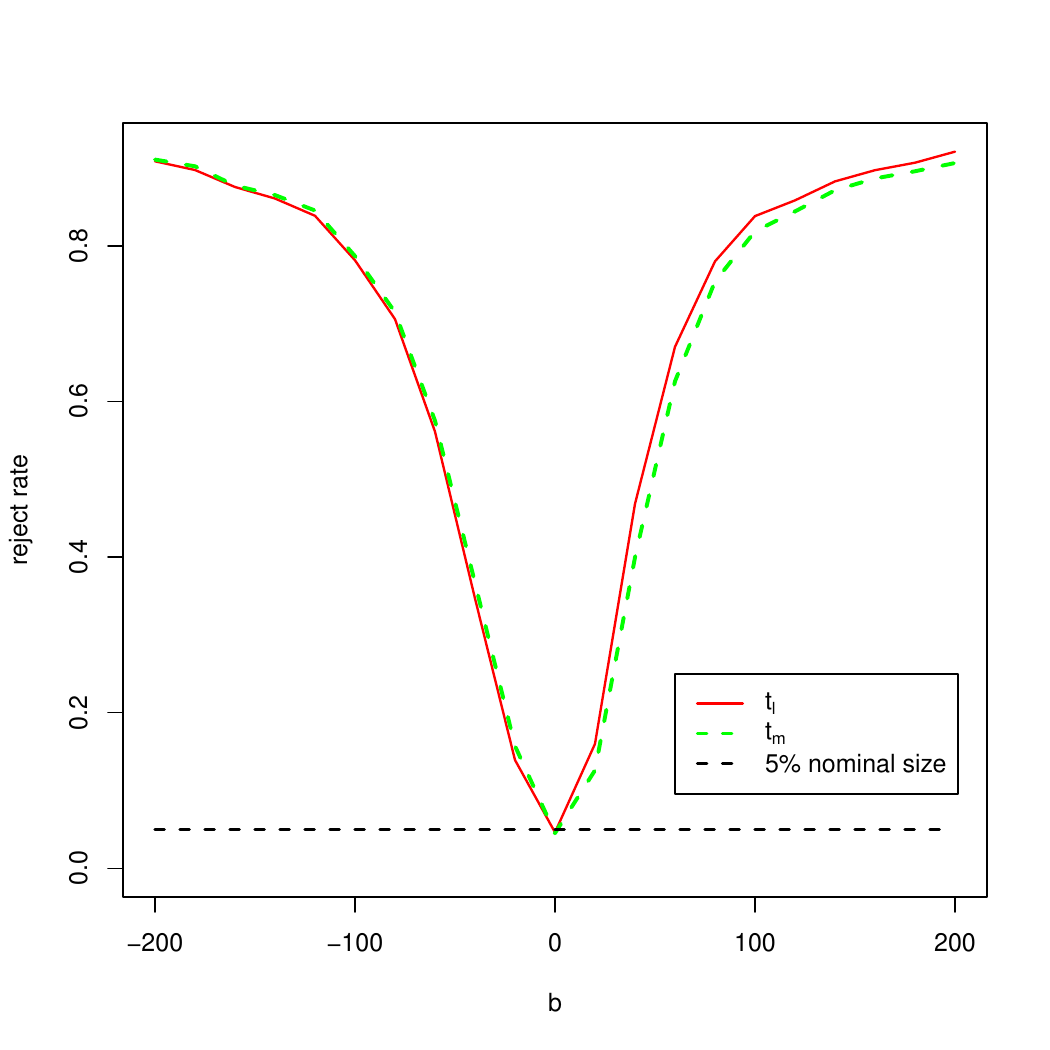}
}
\quad
\subfigure[Case 5 ($\alpha=0$, $c=-0.05$)]{
\includegraphics[width=5.5cm]{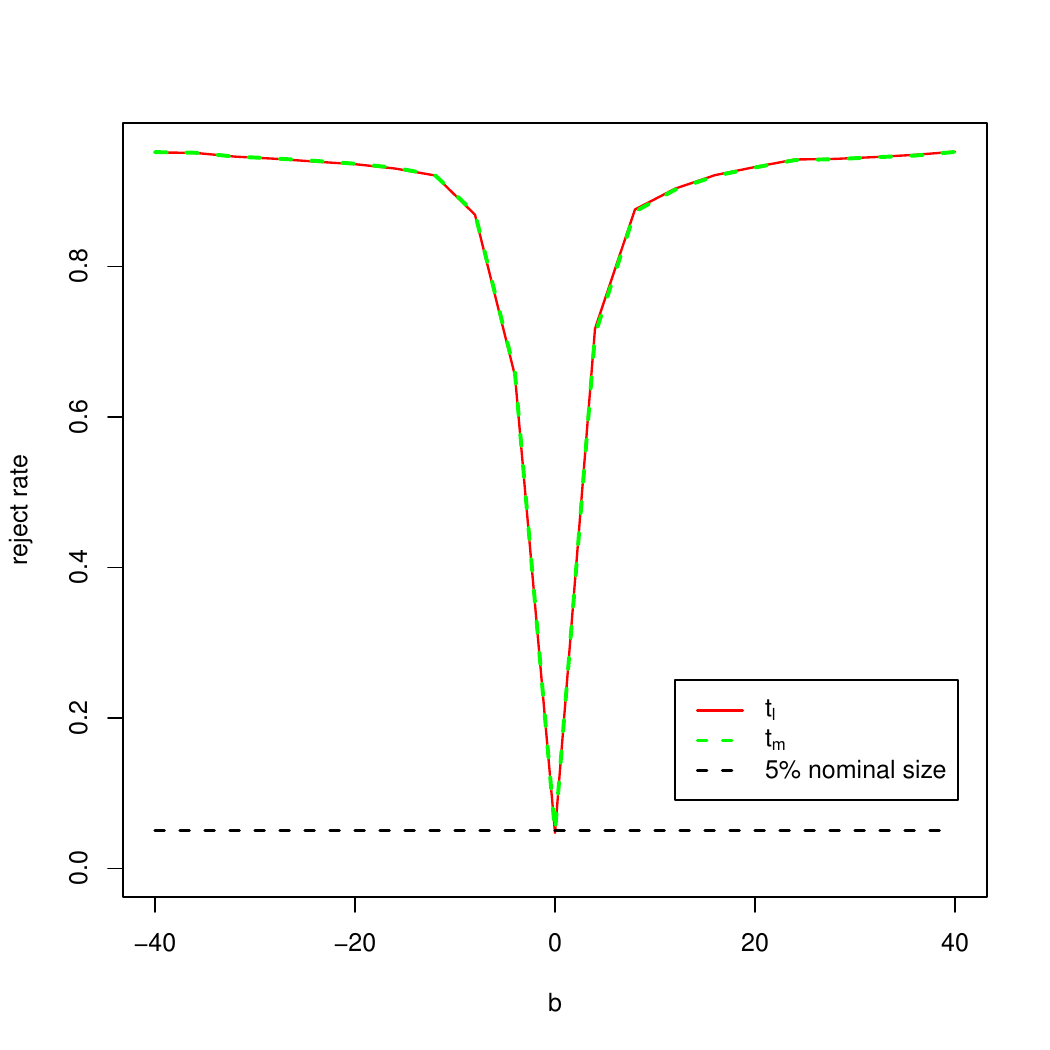}
%\caption{fig1}
}
\quad
\subfigure[Case 6 ($\alpha=0$, $c=-0.1$)]{
\includegraphics[width=5.5cm]{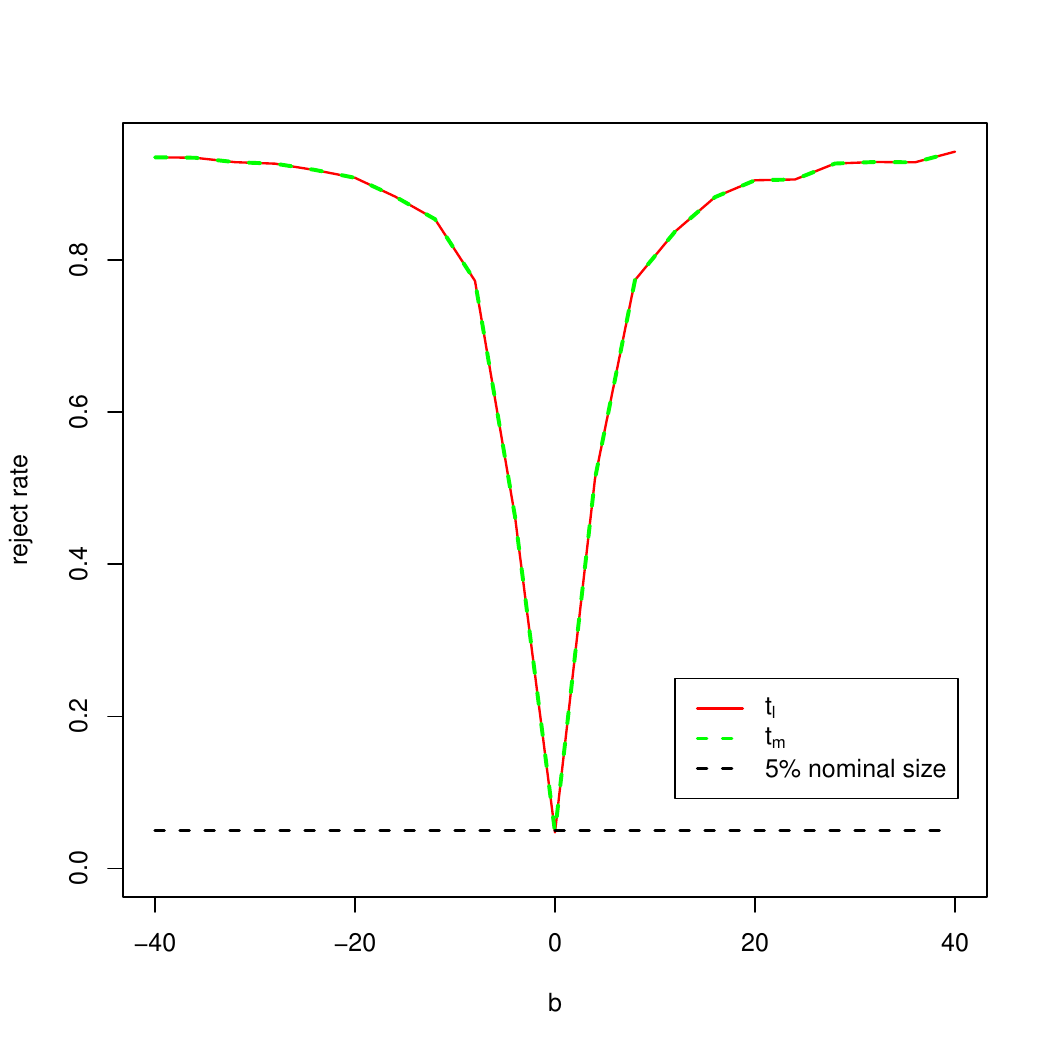}
}
\caption{Power of Two-sided Test $H_0:\beta = 0$ vs $H_a:\beta \neq 0$ with $\phi=-0.95$ and $\lambda=0.5$}
\label{power4}
\end{figure}

\begin{figure}[H]
\centering
\subfigure[Case 1 ($\alpha=1$, $c=0$)]{
\includegraphics[width=5.5cm]{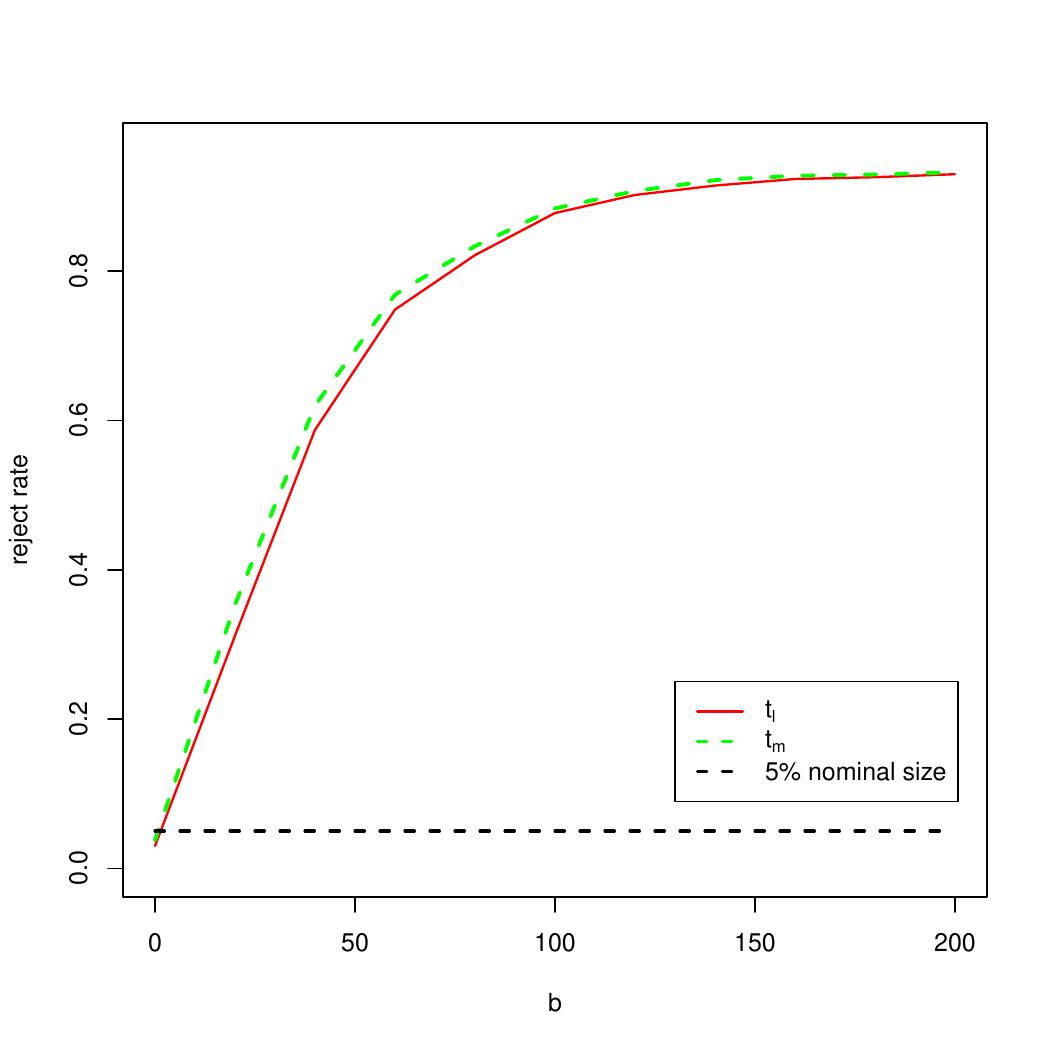}
%\caption{fig1}
}
\quad
\subfigure[Case 2 ($\alpha=1$, $c=-5$)]{
\includegraphics[width=5.5cm]{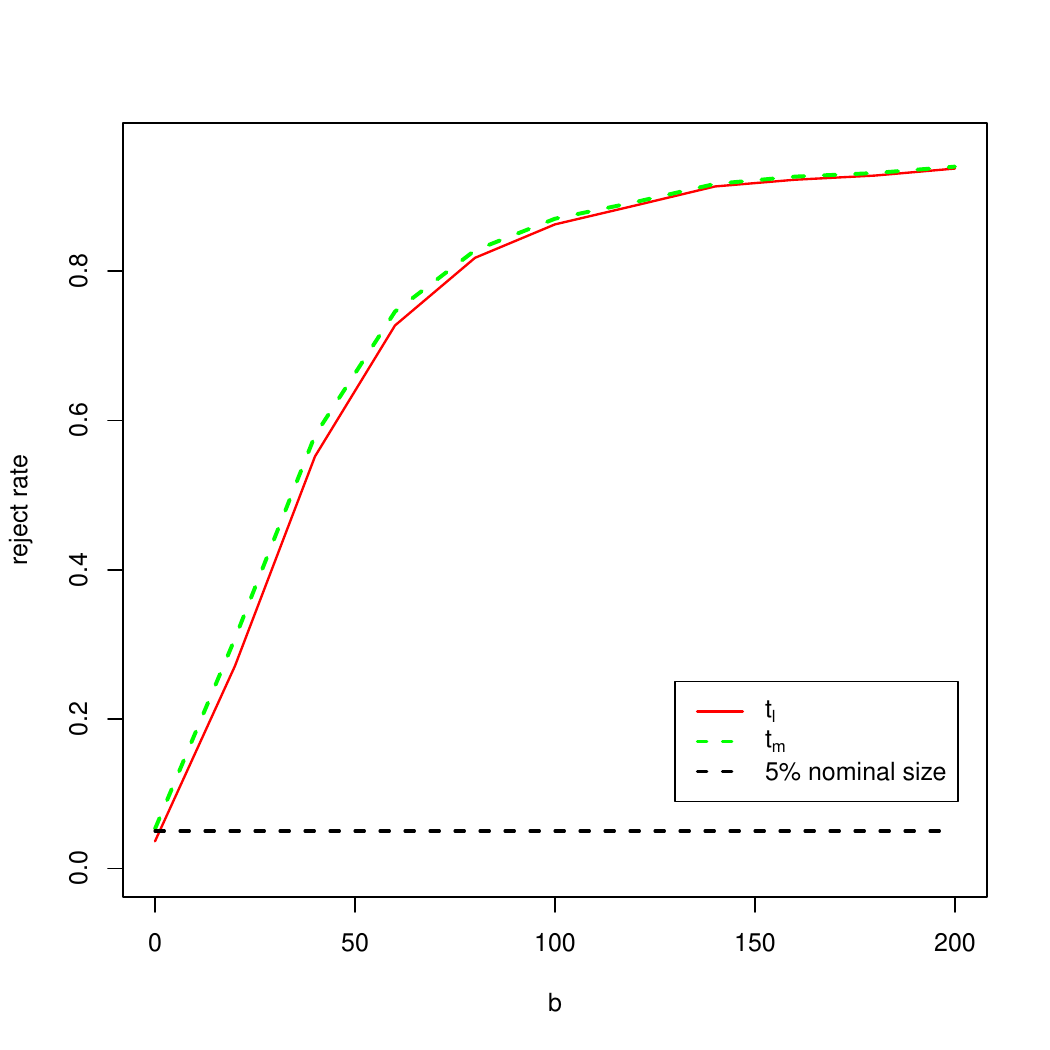}
}
\quad
\subfigure[Case 3 ($\alpha=1$, $c=-10$)]{
\includegraphics[width=5.5cm]{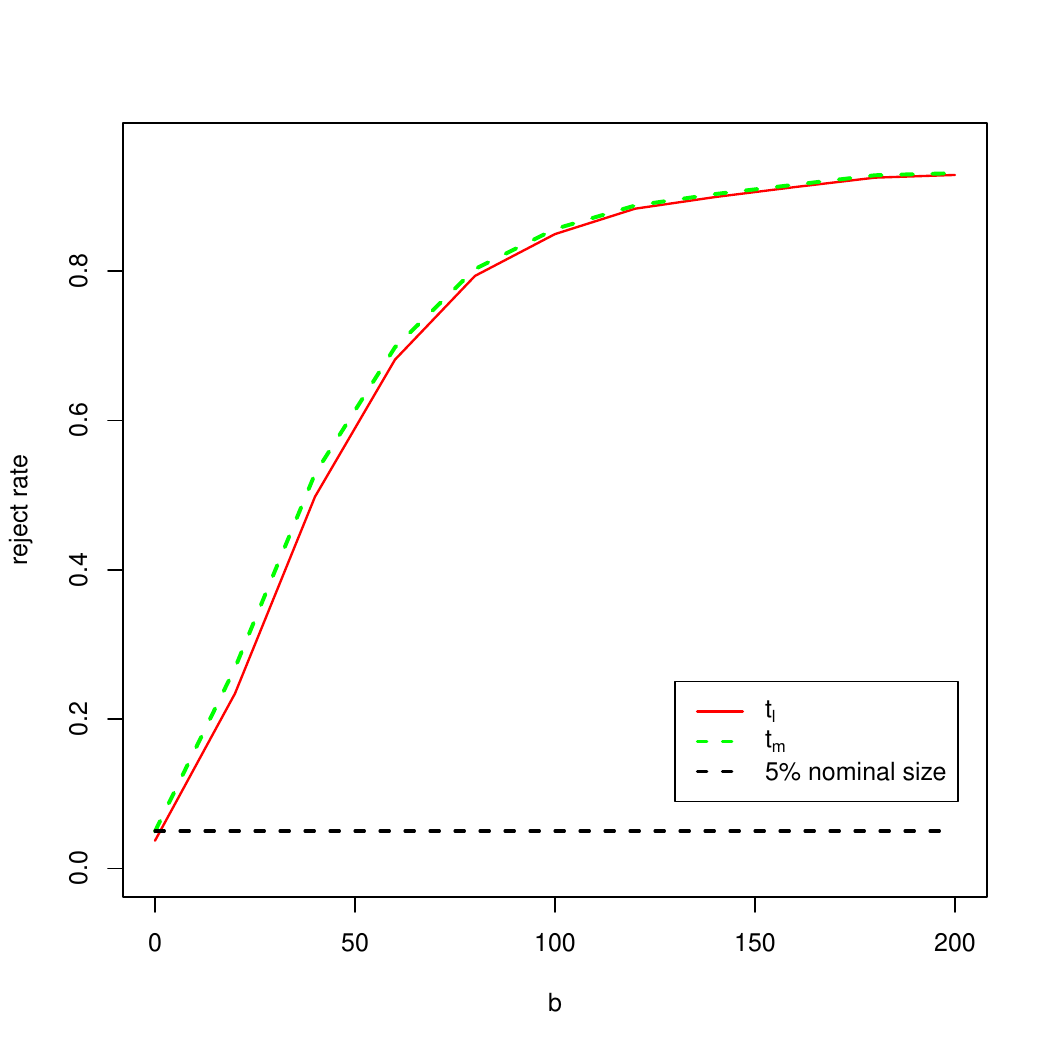}
%\caption{fig1}
}
\quad
\subfigure[Case 4 ($\alpha=1$, $c=-15$)]{
\includegraphics[width=5.5cm]{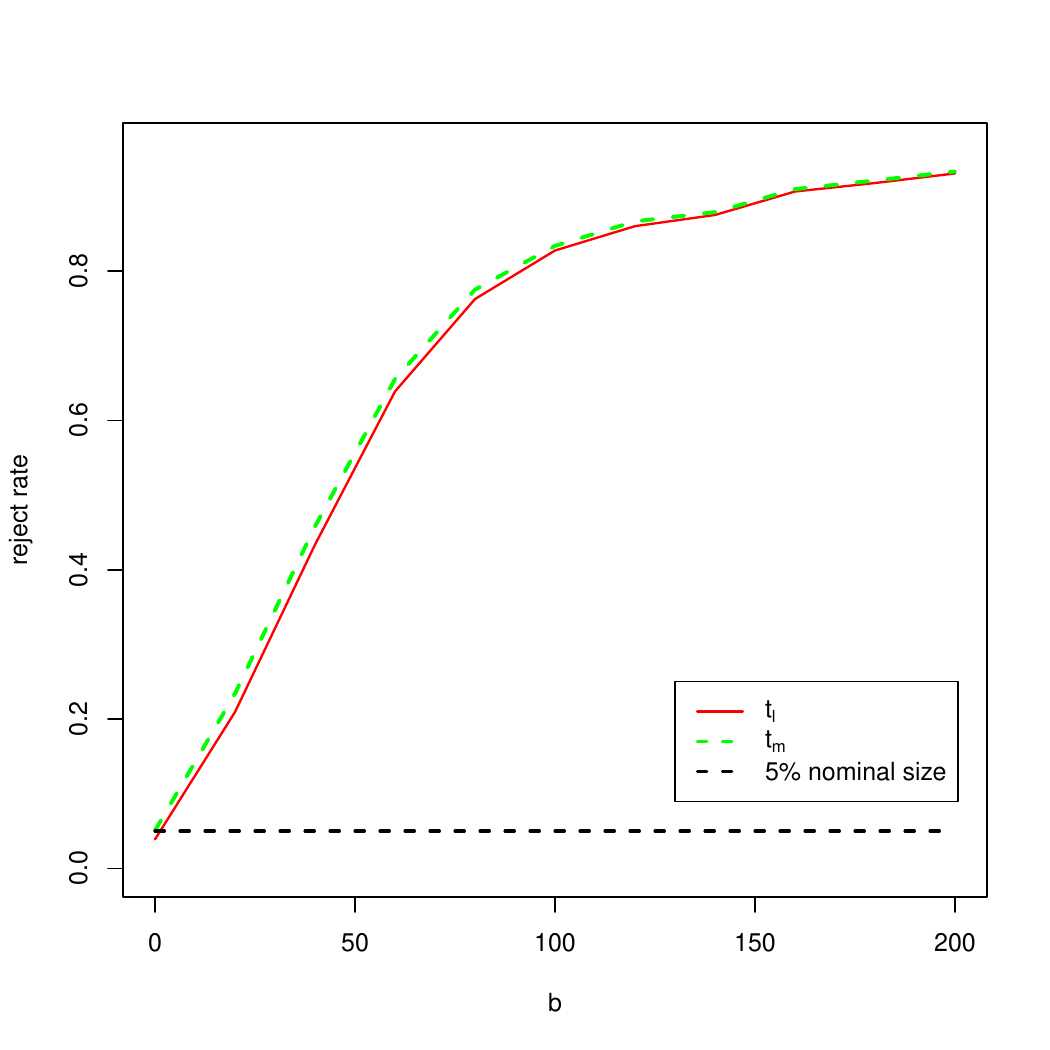}
}
\quad
\subfigure[Case 5 ($\alpha=0$, $c=-0.05$)]{
\includegraphics[width=5.5cm]{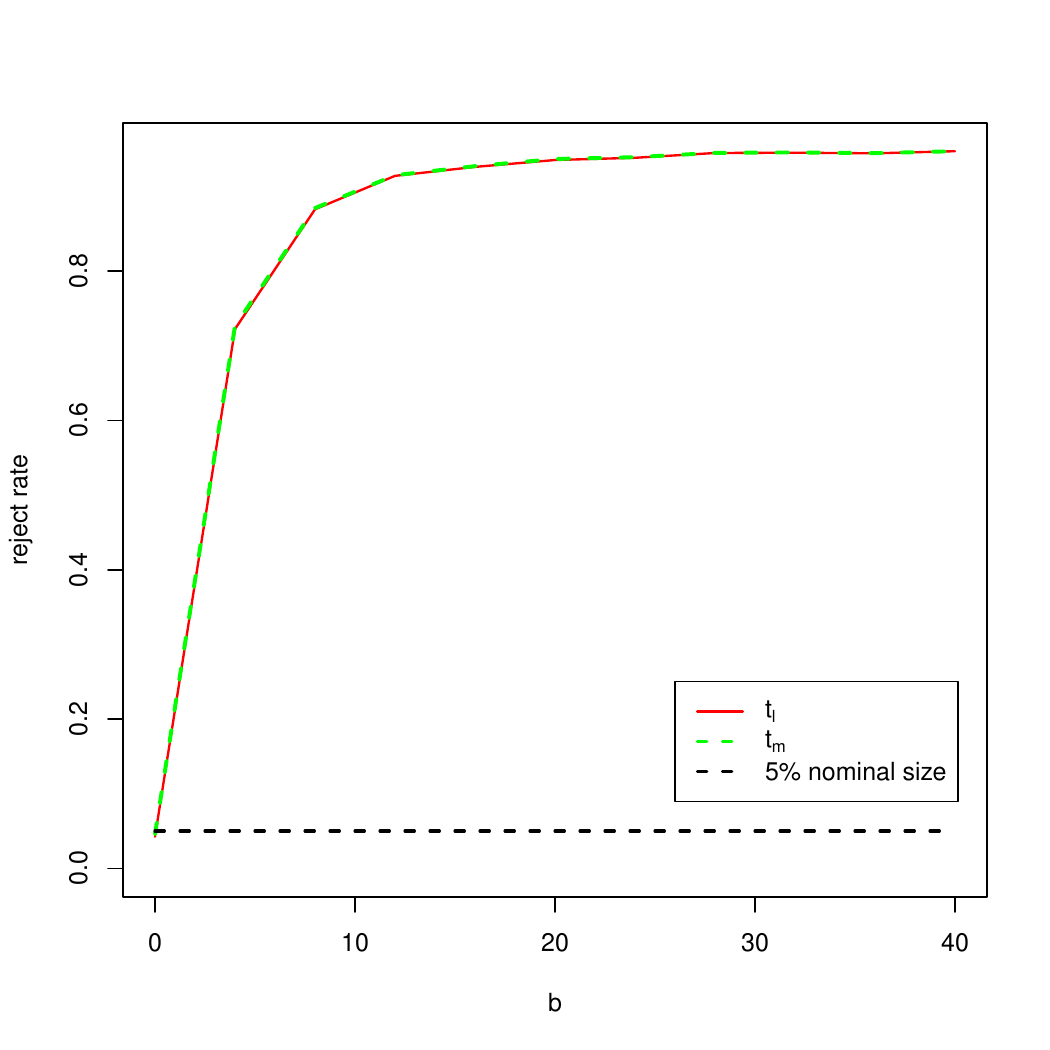}
%\caption{fig1}
}
\quad
\subfigure[Case 6 ($\alpha=0$, $c=-0.1$)]{
\includegraphics[width=5.5cm]{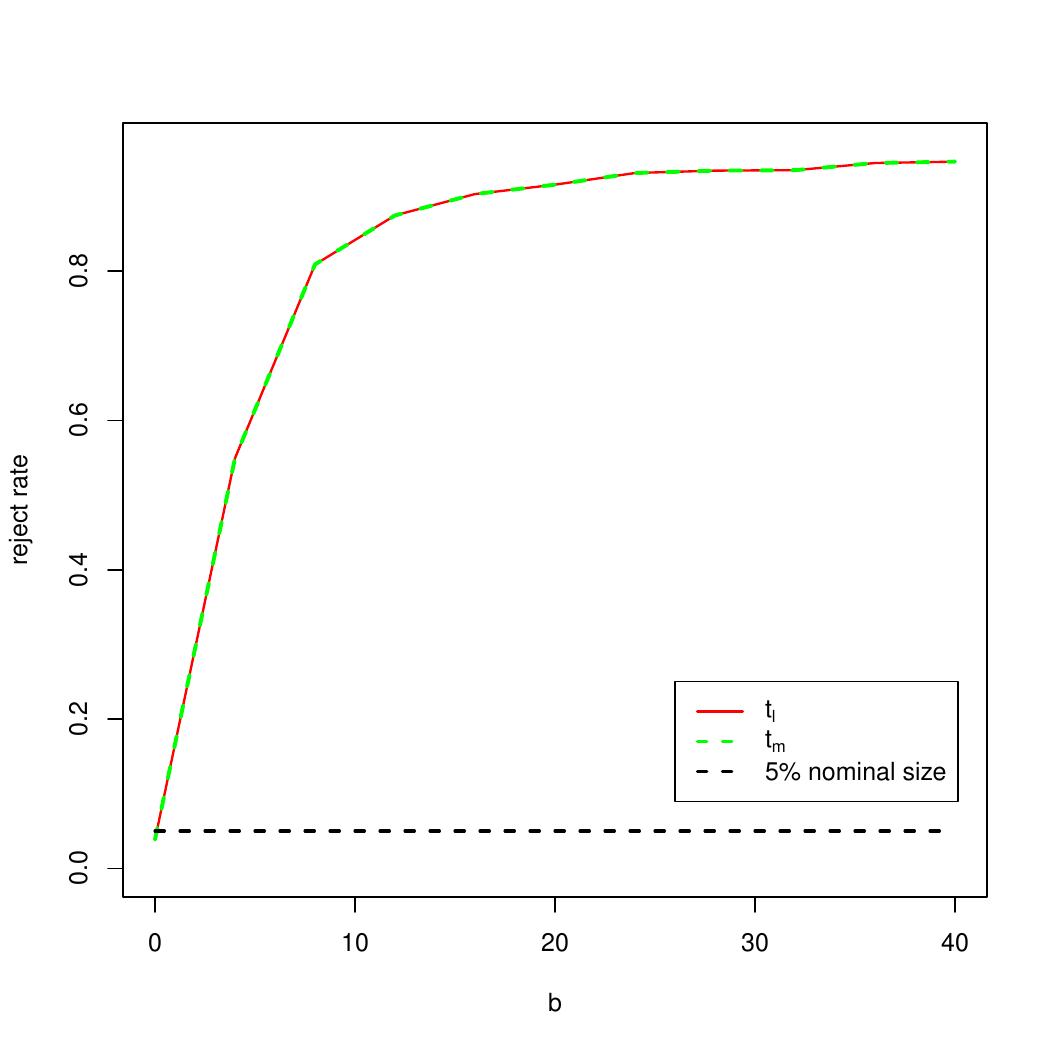}
}
\caption{Power of Right Side Test $H_0:\beta = 0$ vs $H_a:\beta > 0$ with $\phi=0.95$ and $\lambda=0.5$}
\label{power5}
\end{figure}

\begin{figure}[H]
\centering
\subfigure[Case 1 ($\alpha=1$, $c=0$)]{
\includegraphics[width=5.5cm]{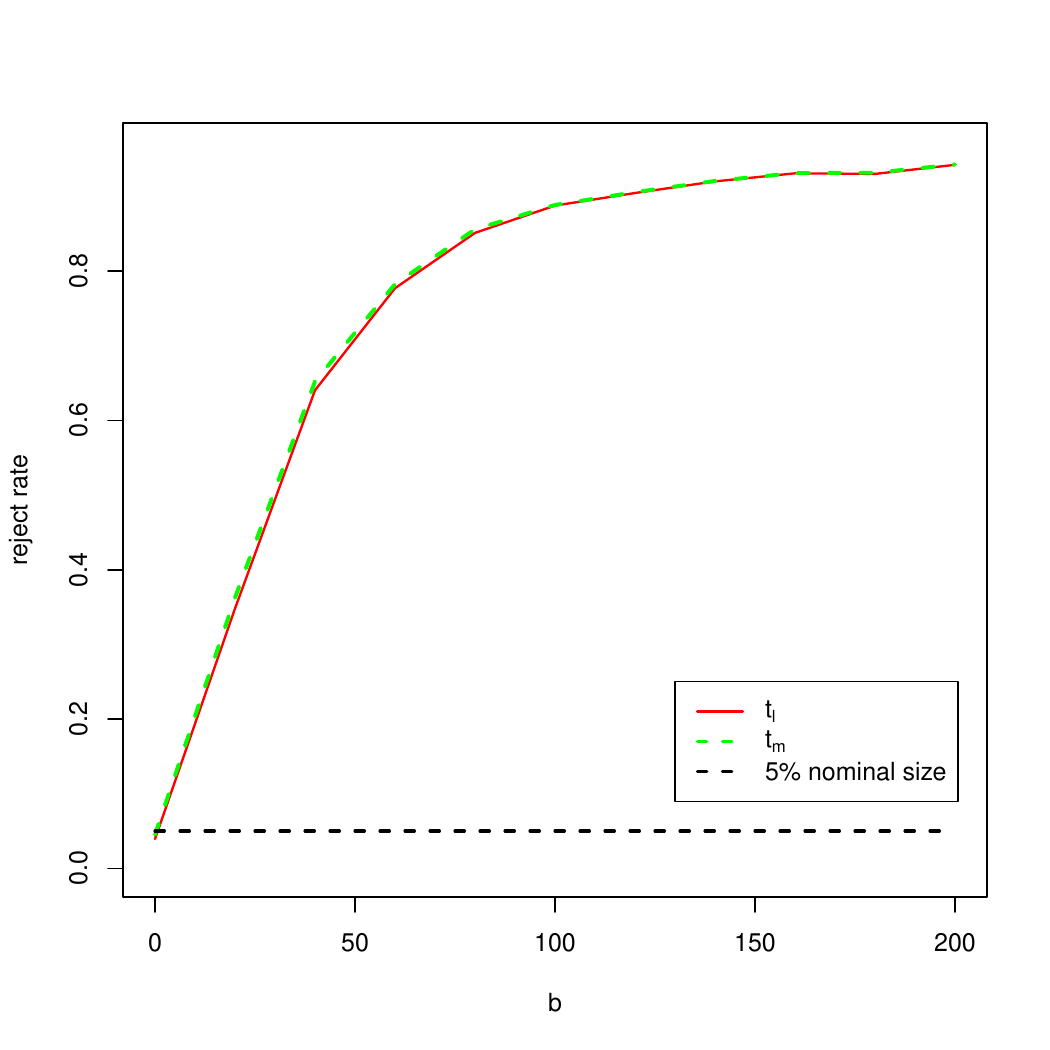}
%\caption{fig1}
}
\quad
\subfigure[Case 2 ($\alpha=1$, $c=-5$)]{
\includegraphics[width=5.5cm]{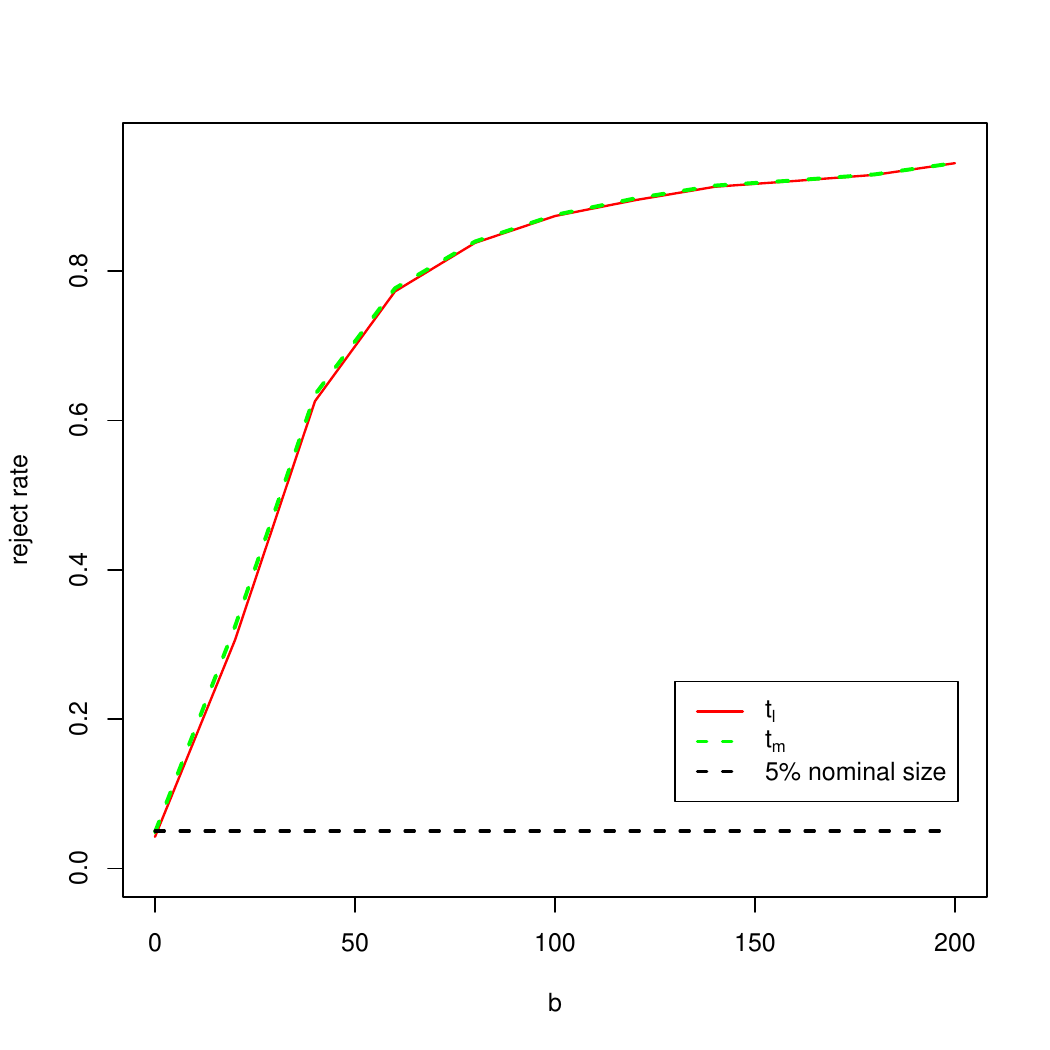}
}
\quad
\subfigure[Case 3 ($\alpha=1$, $c=-10$)]{
\includegraphics[width=5.5cm]{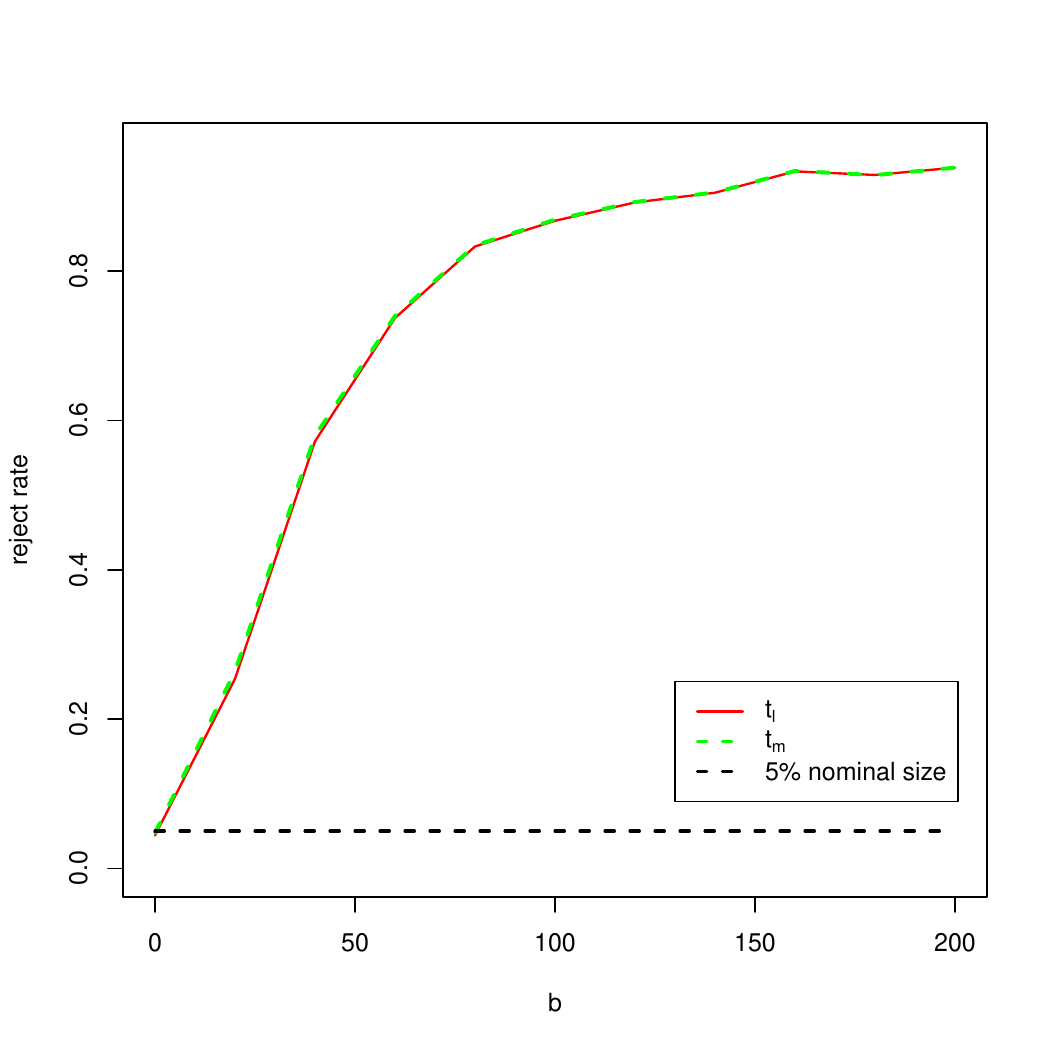}
%\caption{fig1}
}
\quad
\subfigure[Case 4 ($\alpha=1$, $c=-15$)]{
\includegraphics[width=5.5cm]{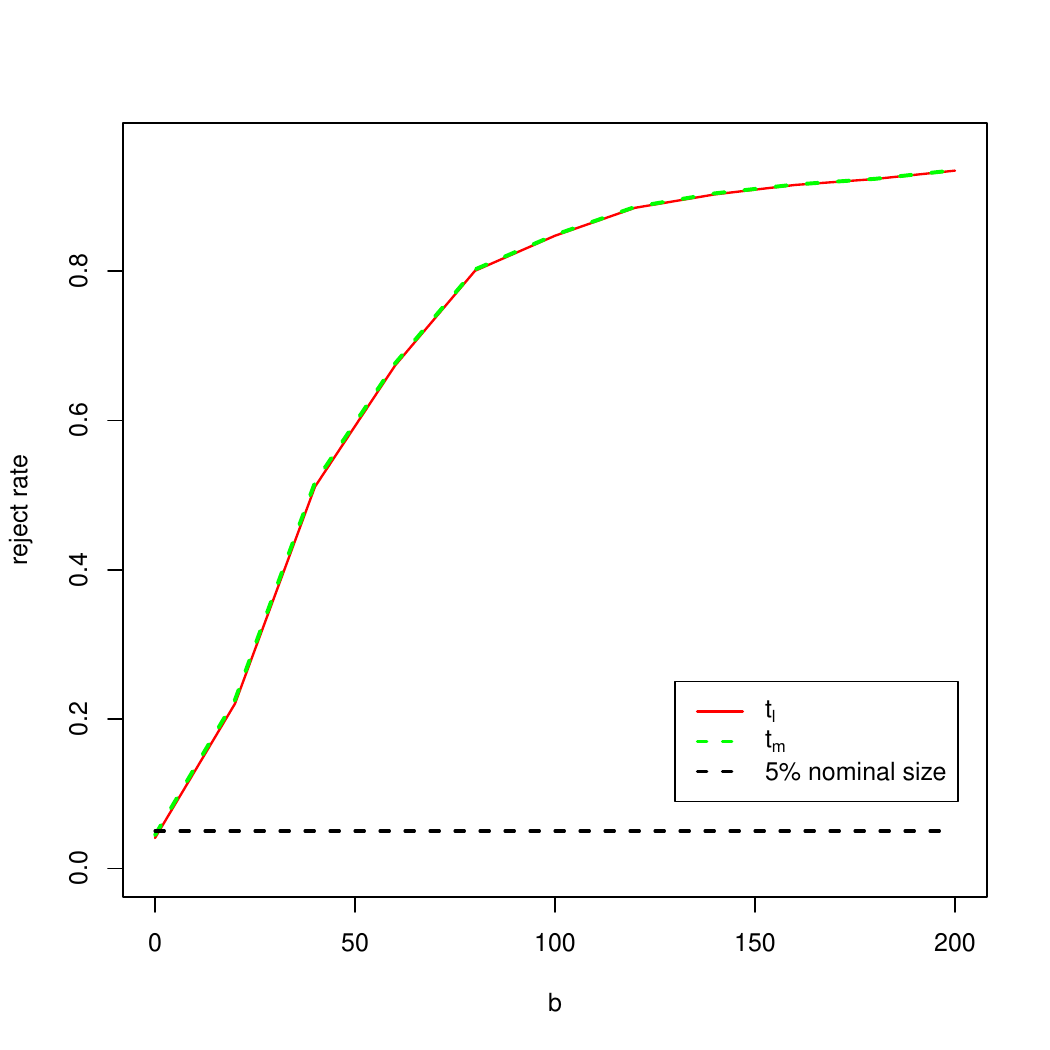}
}
\quad
\subfigure[Case 5 ($\alpha=0$, $c=-0.05$)]{
\includegraphics[width=5.5cm]{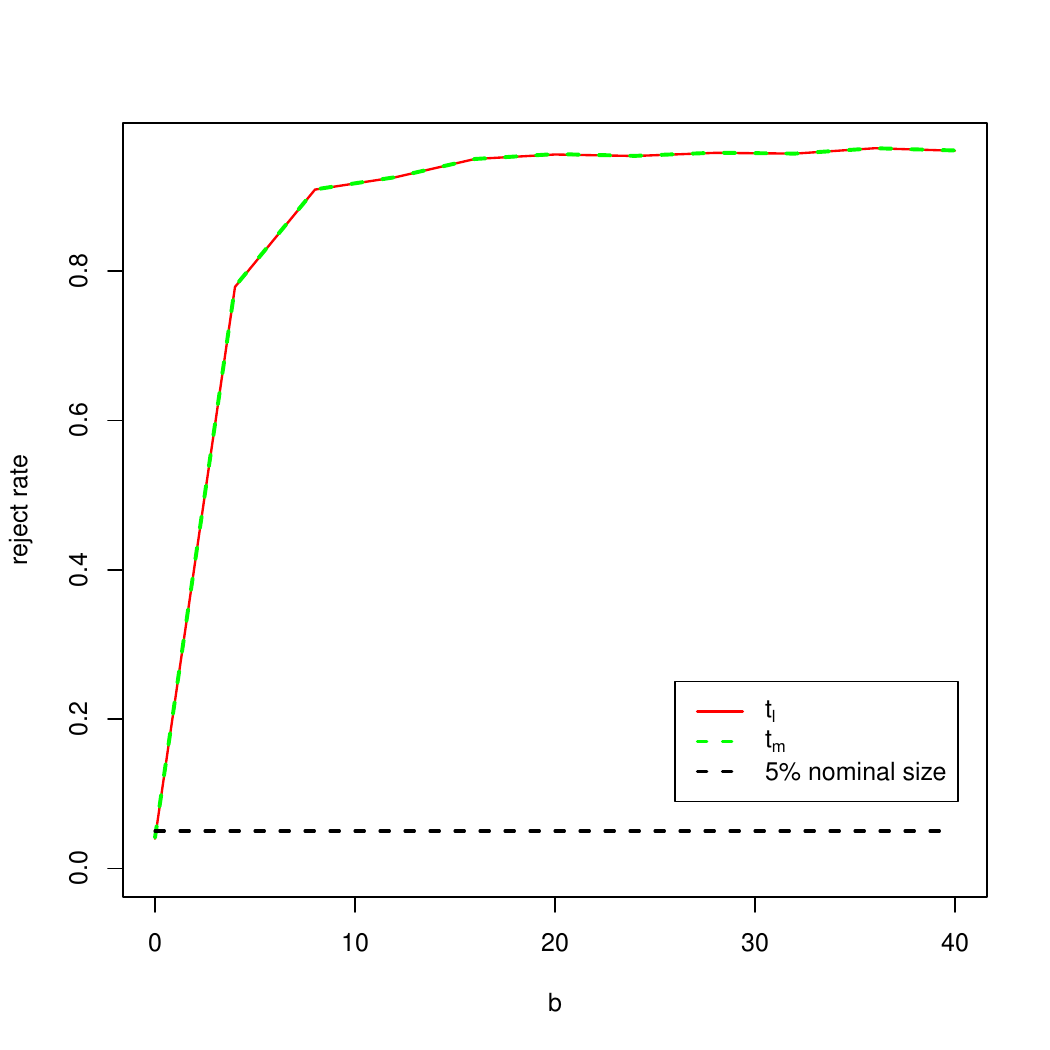}
%\caption{fig1}
}
\quad
\subfigure[Case 6 ($\alpha=0$, $c=-0.1$)]{
\includegraphics[width=5.5cm]{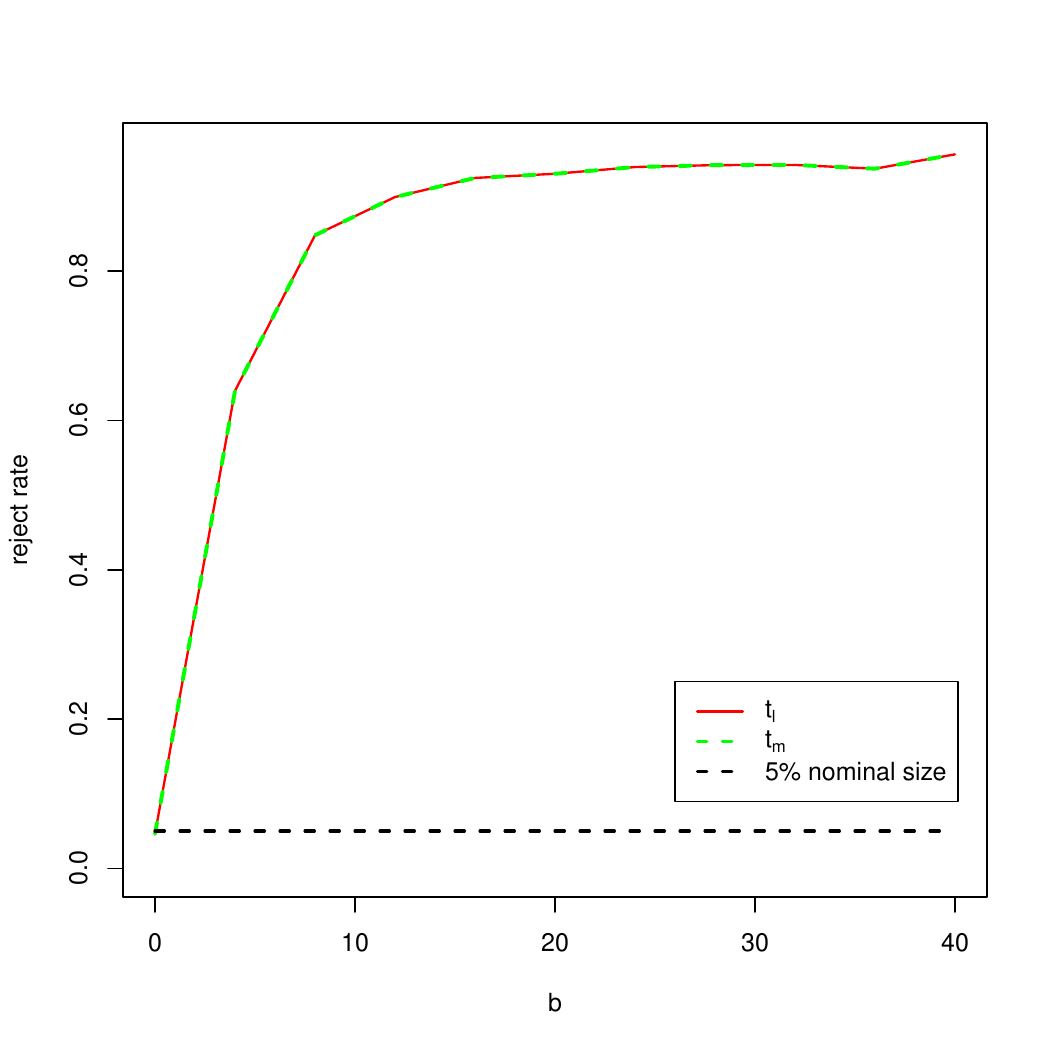}
}
\caption{Power of Right Side Test $H_0:\beta = 0$ vs $H_a:\beta > 0$ with $\phi=0.5$ and $\lambda=0.5$}
\label{power6}
\end{figure}

\begin{figure}[H]
\centering
\subfigure[Case 1 ($\alpha=1$, $c=0$)]{
\includegraphics[width=5.5cm]{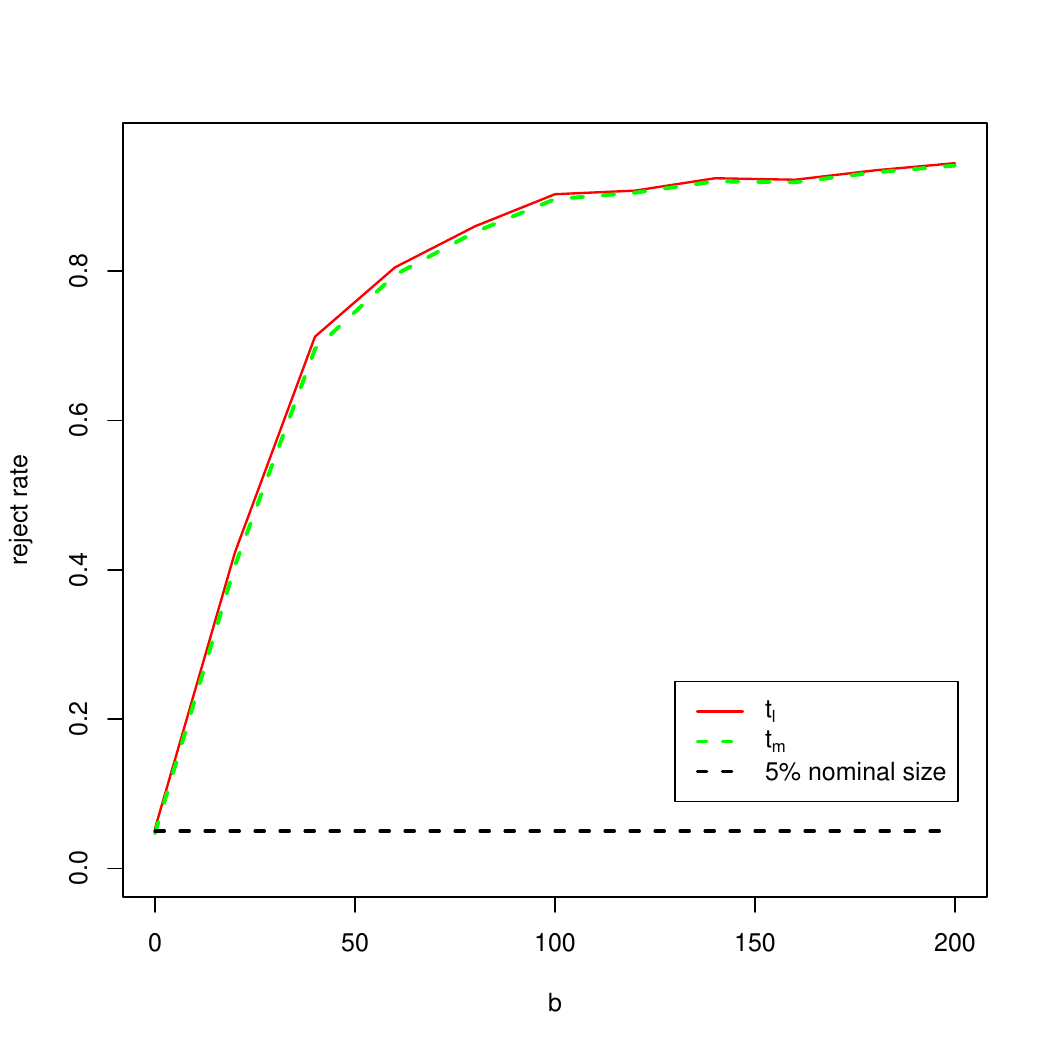}
%\caption{fig1}
}
\quad
\subfigure[Case 2 ($\alpha=1$, $c=-5$)]{
\includegraphics[width=5.5cm]{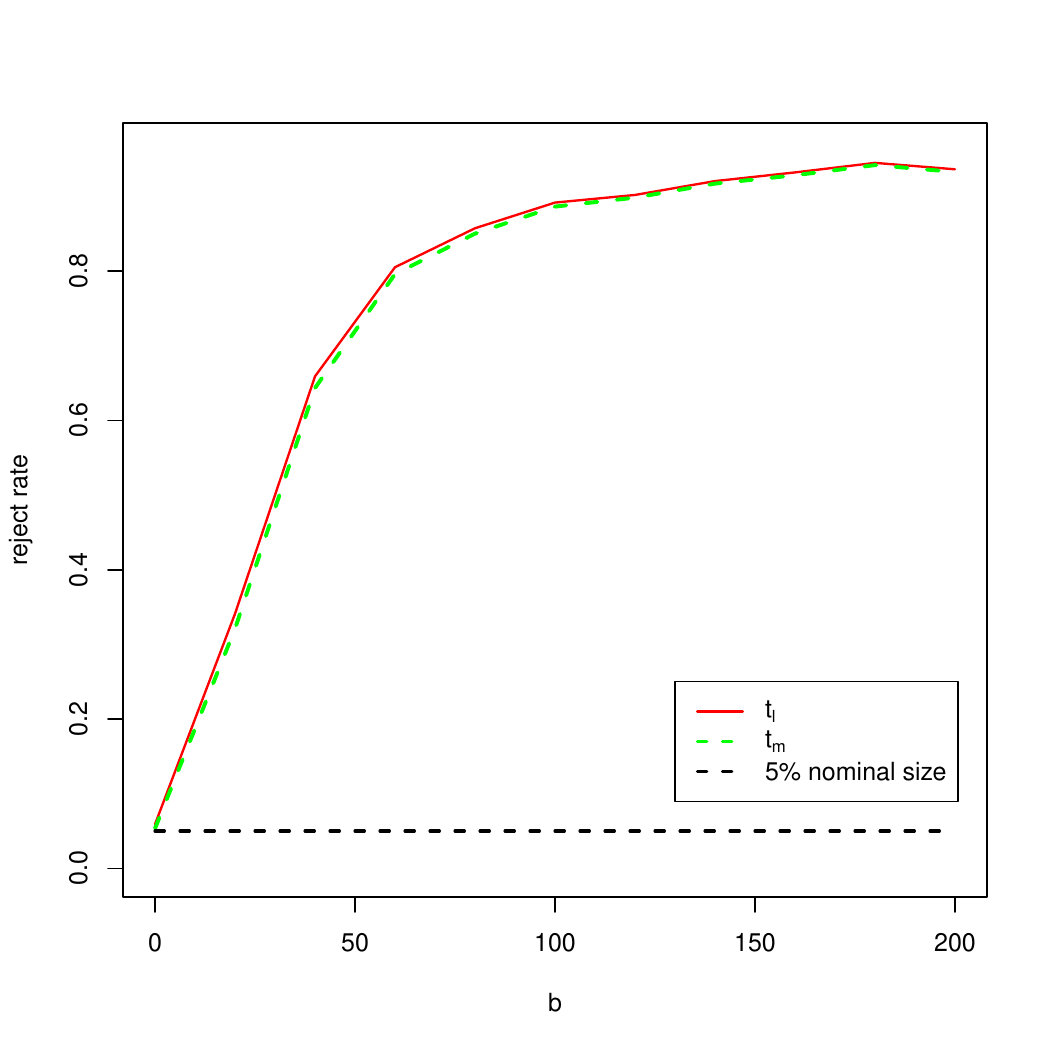}
}
\quad
\subfigure[Case 3 ($\alpha=1$, $c=-10$)]{
\includegraphics[width=5.5cm]{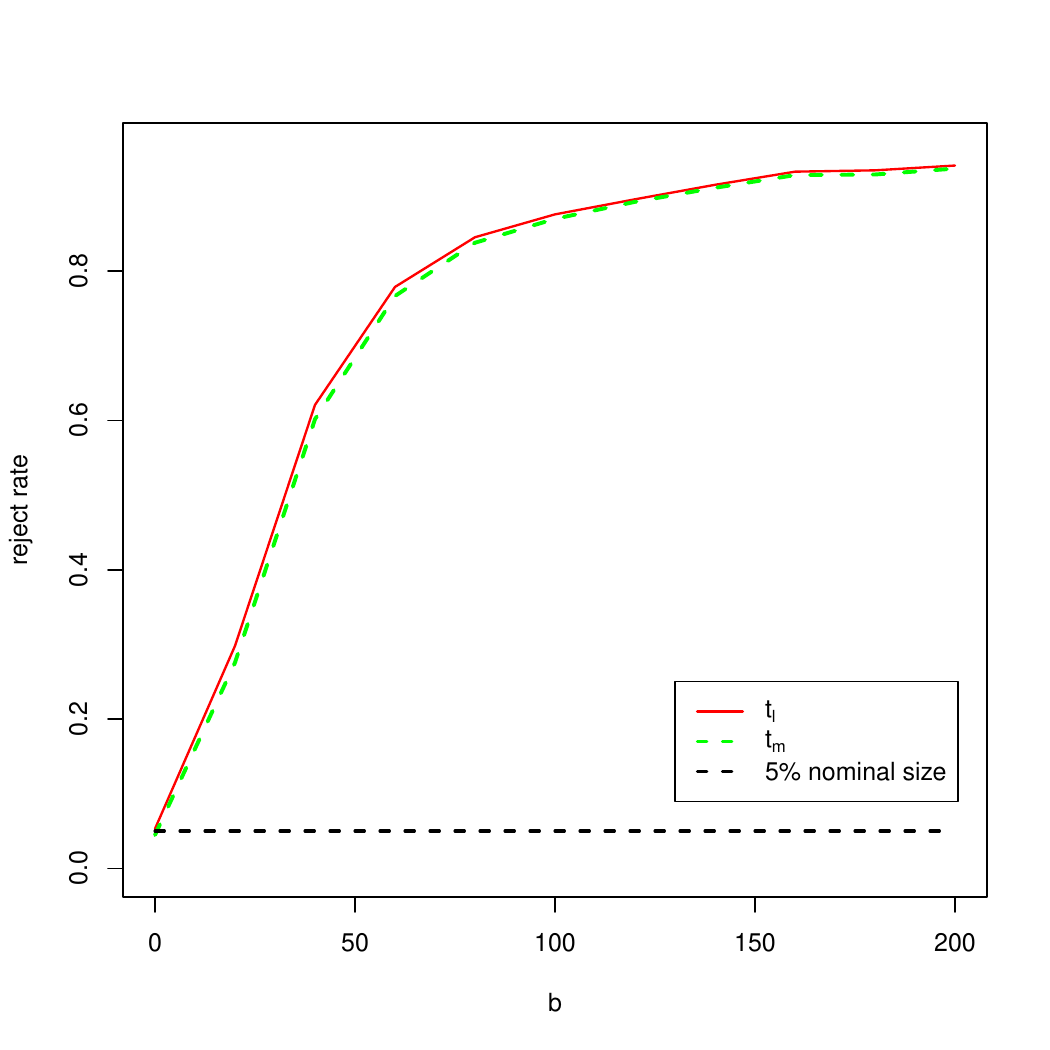}
%\caption{fig1}
}
\quad
\subfigure[Case 4 ($\alpha=1$, $c=-15$)]{
\includegraphics[width=5.5cm]{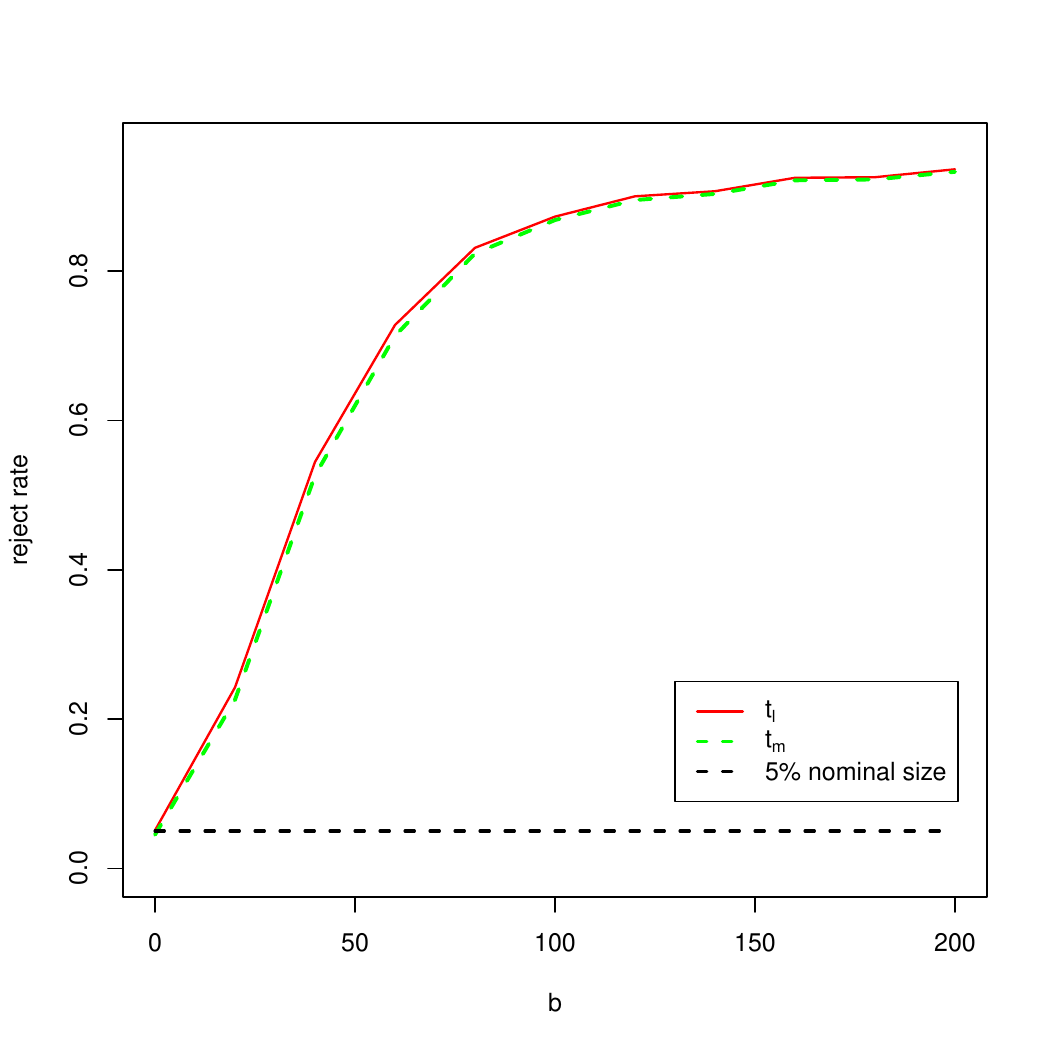}
}
\quad
\subfigure[Case 5 ($\alpha=0$, $c=-0.05$)]{
\includegraphics[width=5.5cm]{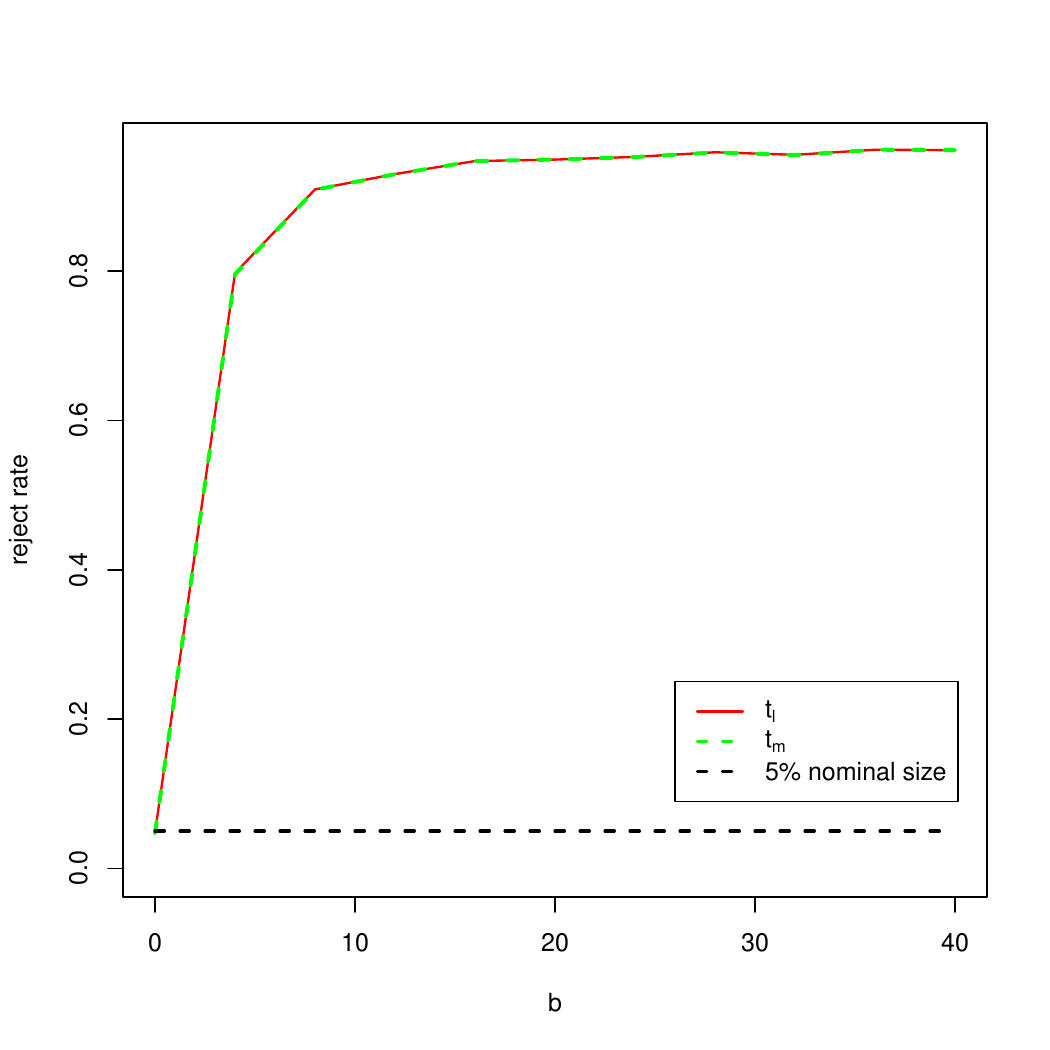}
%\caption{fig1}
}
\quad
\subfigure[Case 6 ($\alpha=0$, $c=-0.1$)]{
\includegraphics[width=5.5cm]{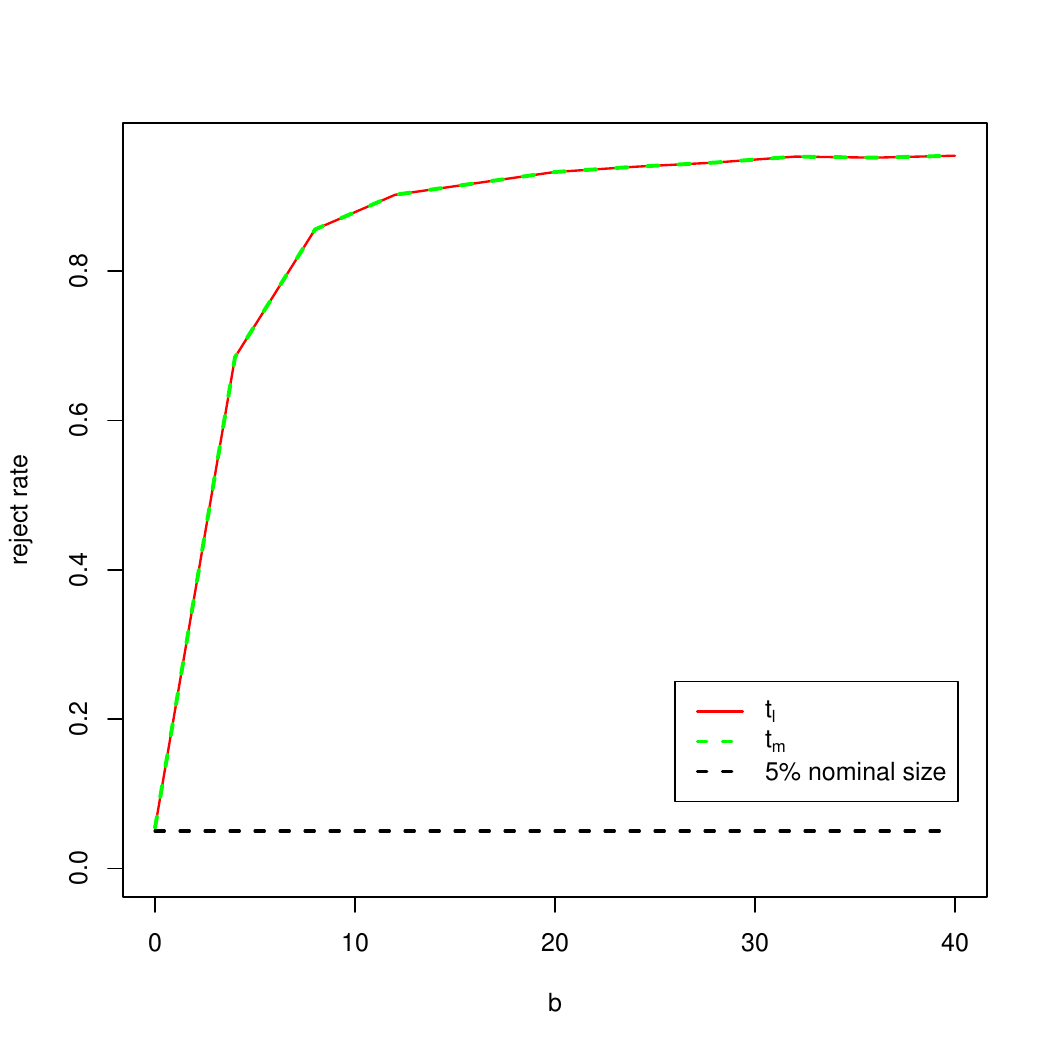}
}
\caption{Power of Right Side Test $H_0:\beta = 0$ vs $H_a:\beta > 0$ with $\phi=-0.1$ and $\lambda=0.5$}
\label{power7}
\end{figure}

\begin{figure}[H]
\centering
\subfigure[Case 1 ($\alpha=1$, $c=0$)]{
\includegraphics[width=5.5cm]{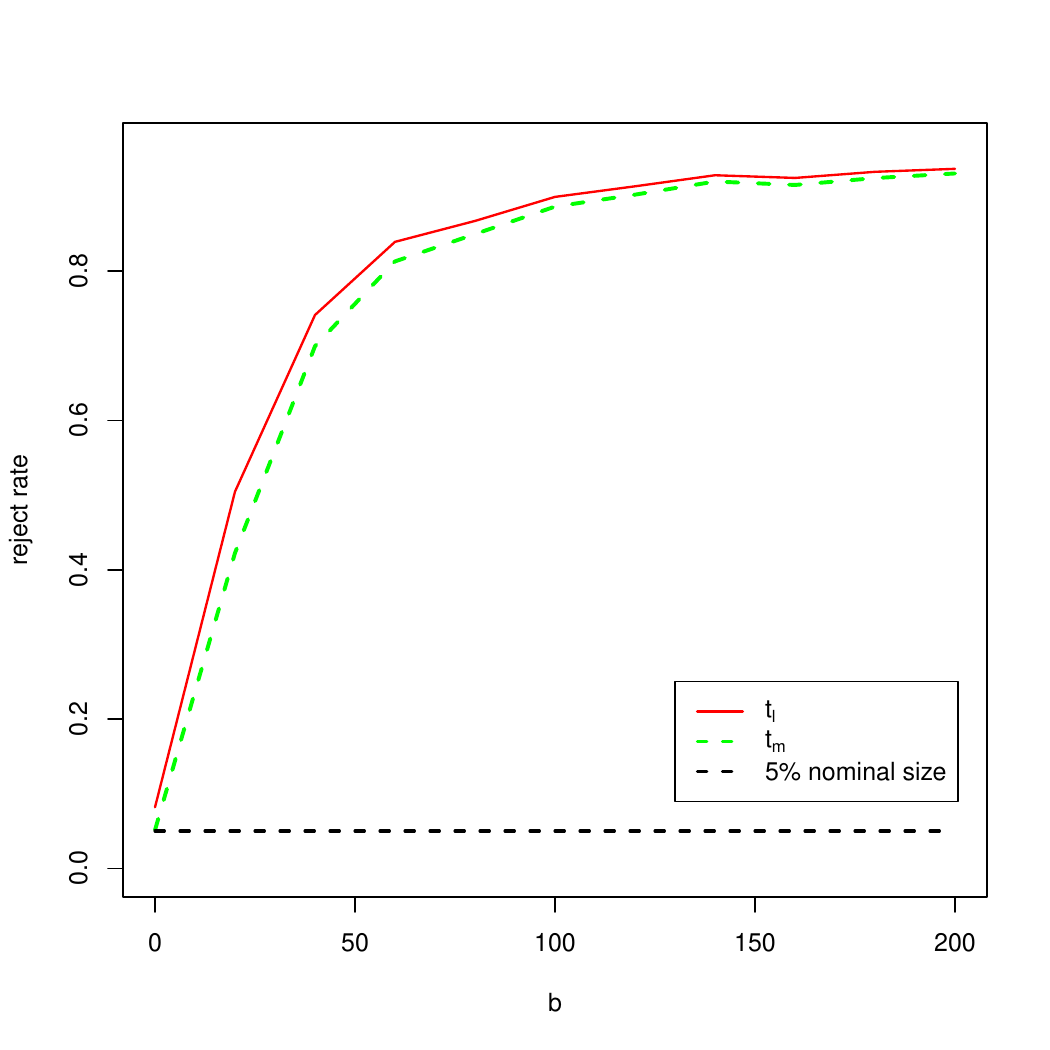}
%\caption{fig1}
}
\quad
\subfigure[Case 2 ($\alpha=1$, $c=-5$)]{
\includegraphics[width=5.5cm]{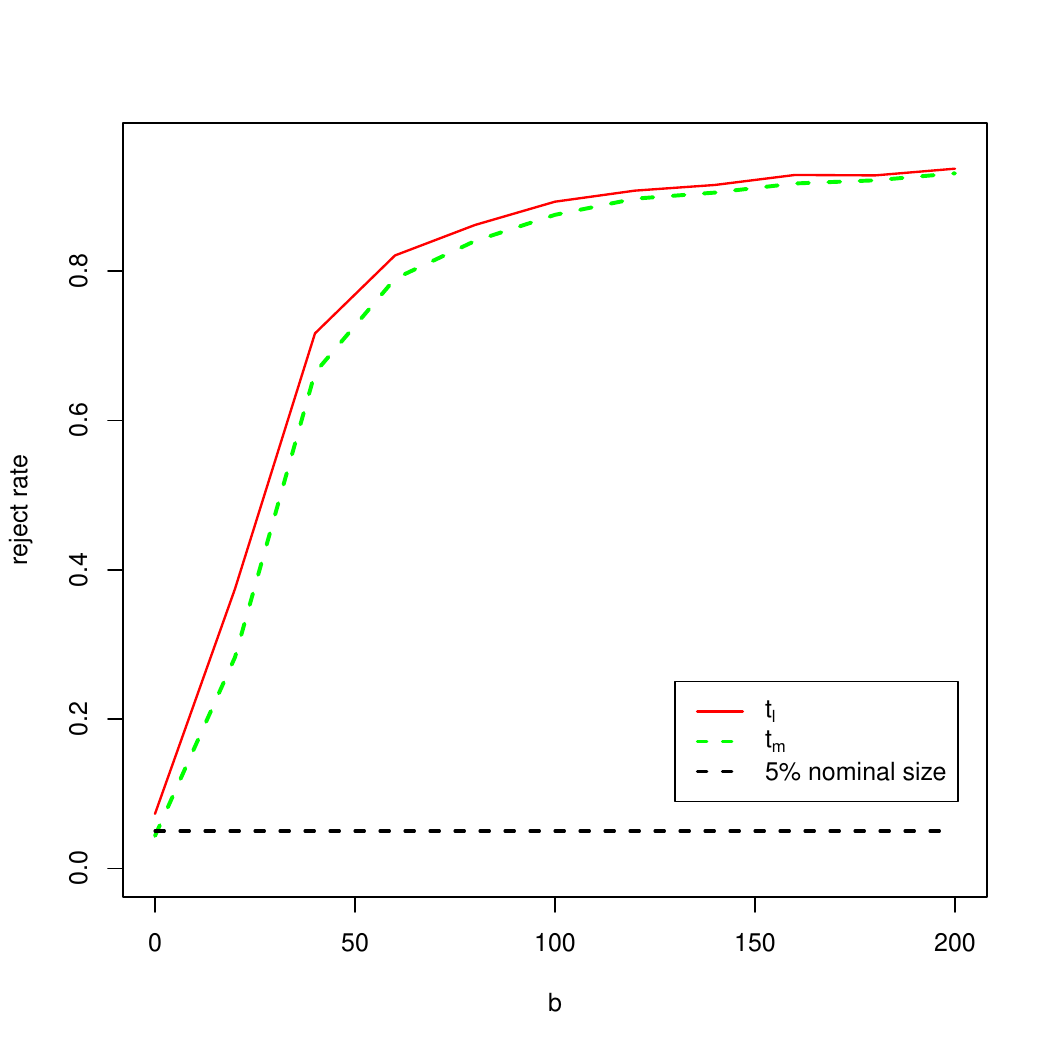}
}
\quad
\subfigure[Case 3 ($\alpha=1$, $c=-10$)]{
\includegraphics[width=5.5cm]{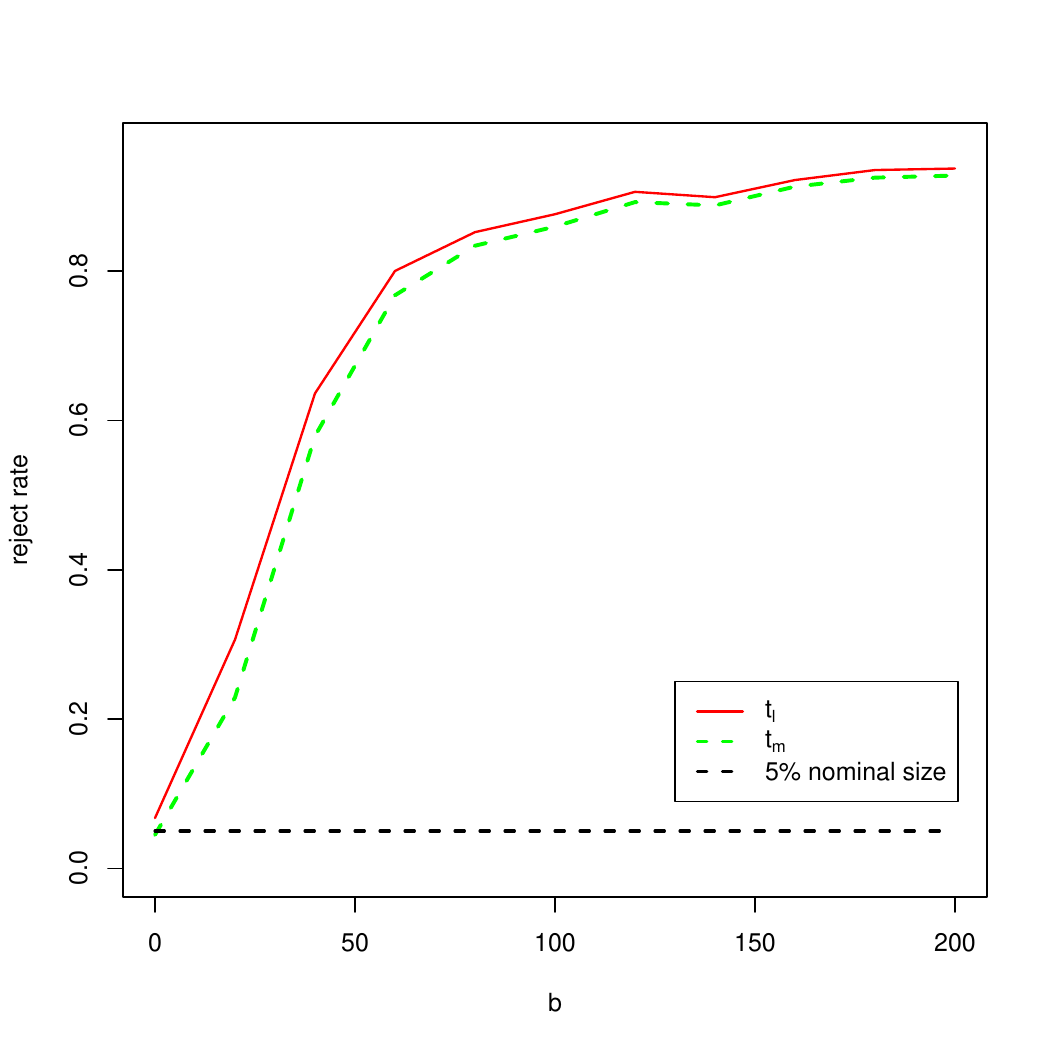}
%\caption{fig1}
}
\quad
\subfigure[Case 4 ($\alpha=1$, $c=-15$)]{
\includegraphics[width=5.5cm]{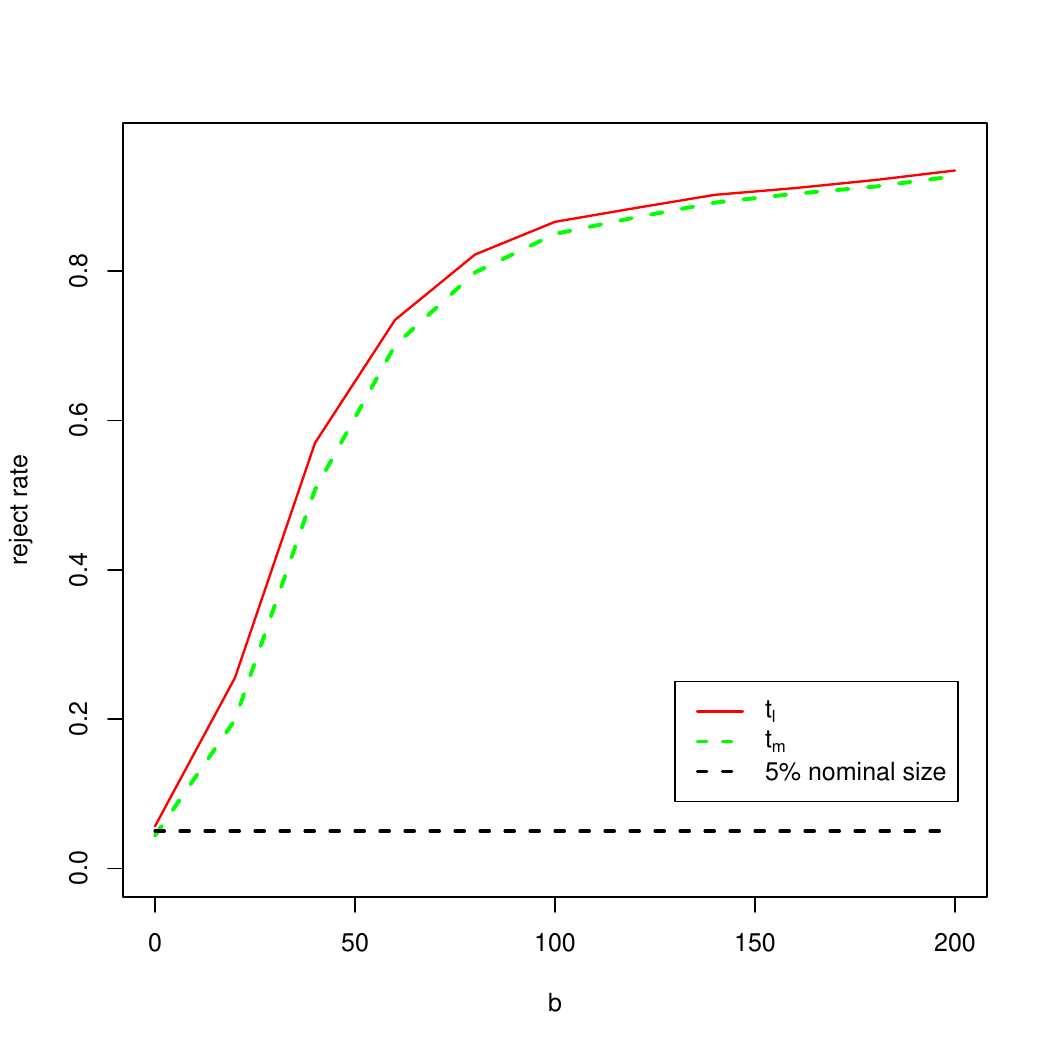}
}
\quad
\subfigure[Case 5 ($\alpha=0$, $c=-0.05$)]{
\includegraphics[width=5.5cm]{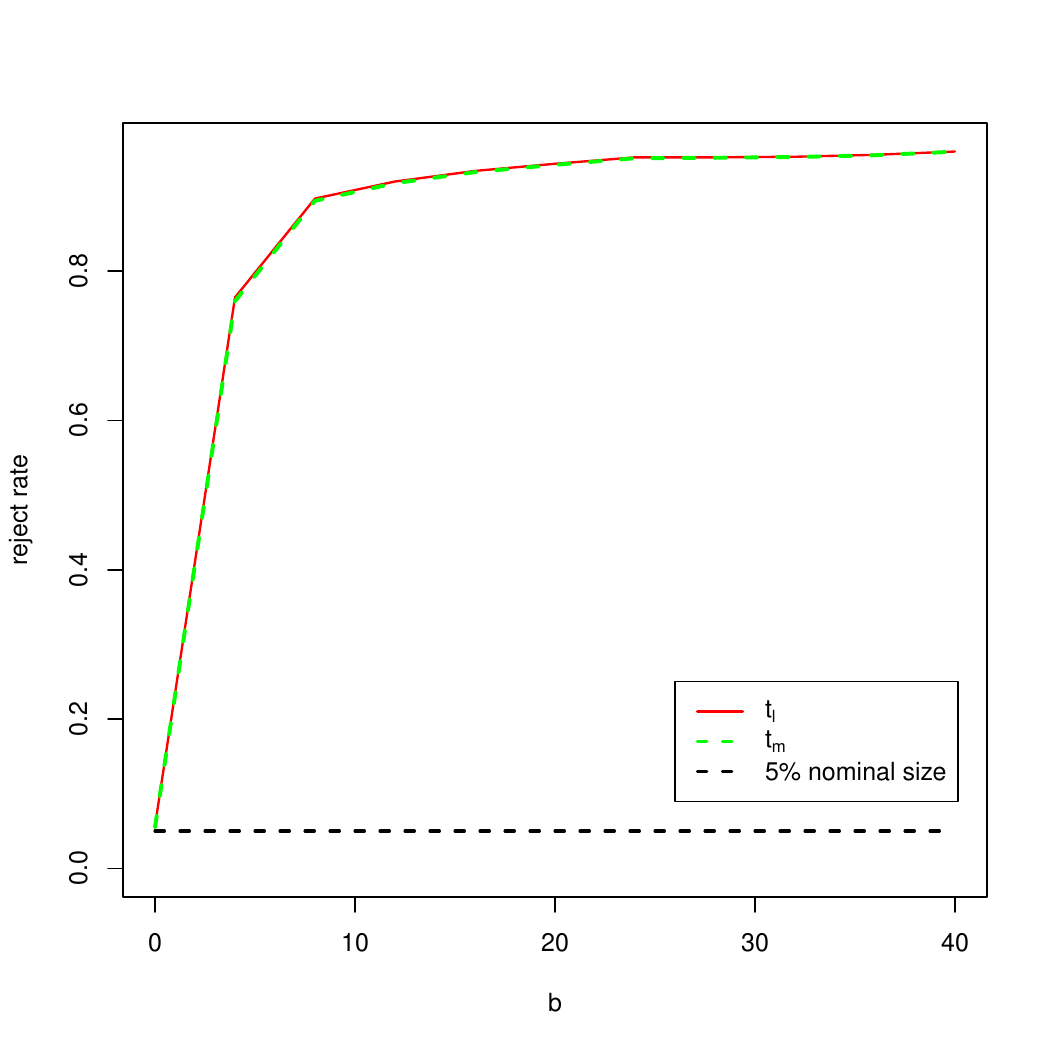}
%\caption{fig1}
}
\quad
\subfigure[Case 6 ($\alpha=0$, $c=-0.1$)]{
\includegraphics[width=5.5cm]{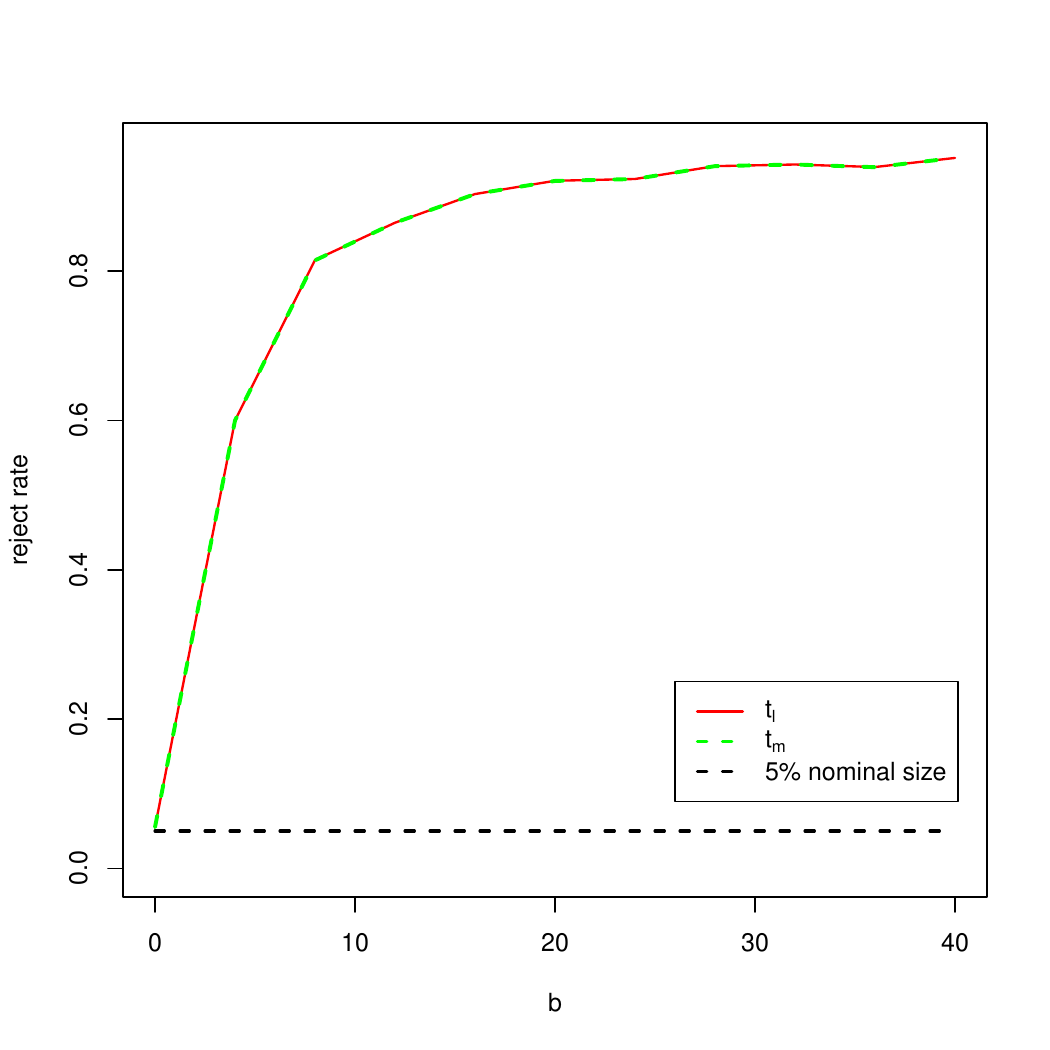}
}
\caption{Power of Right Side Test $H_0:\beta = 0$ vs $H_a:\beta > 0$ with $\phi=-0.95$ and $\lambda=0.5$}
\label{power8}
\end{figure}

\begin{figure}[H]
\centering
\subfigure[Case 1 ($\alpha=1$, $c=0$)]{
\includegraphics[width=5.5cm]{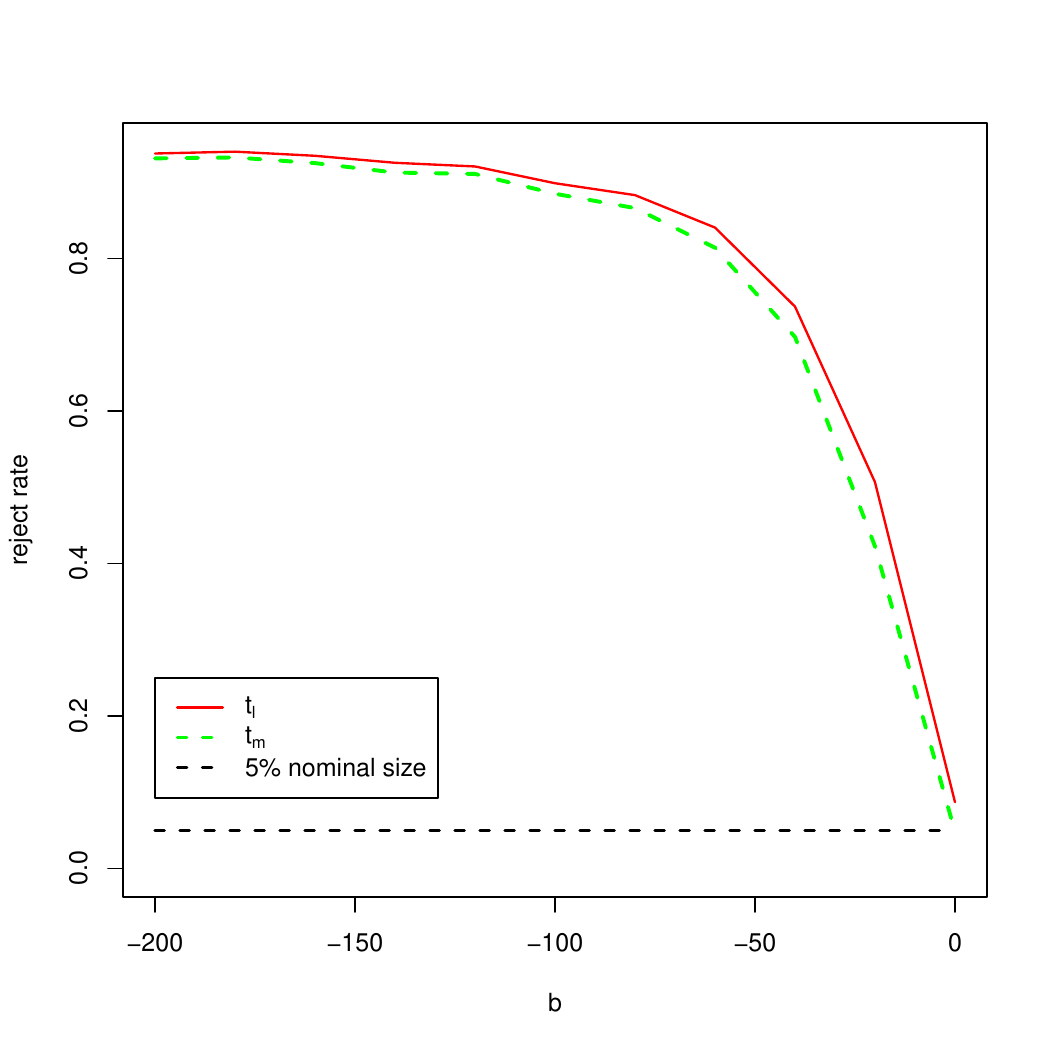}
%\caption{fig1}
}
\quad
\subfigure[Case 2 ($\alpha=1$, $c=-5$)]{
\includegraphics[width=5.5cm]{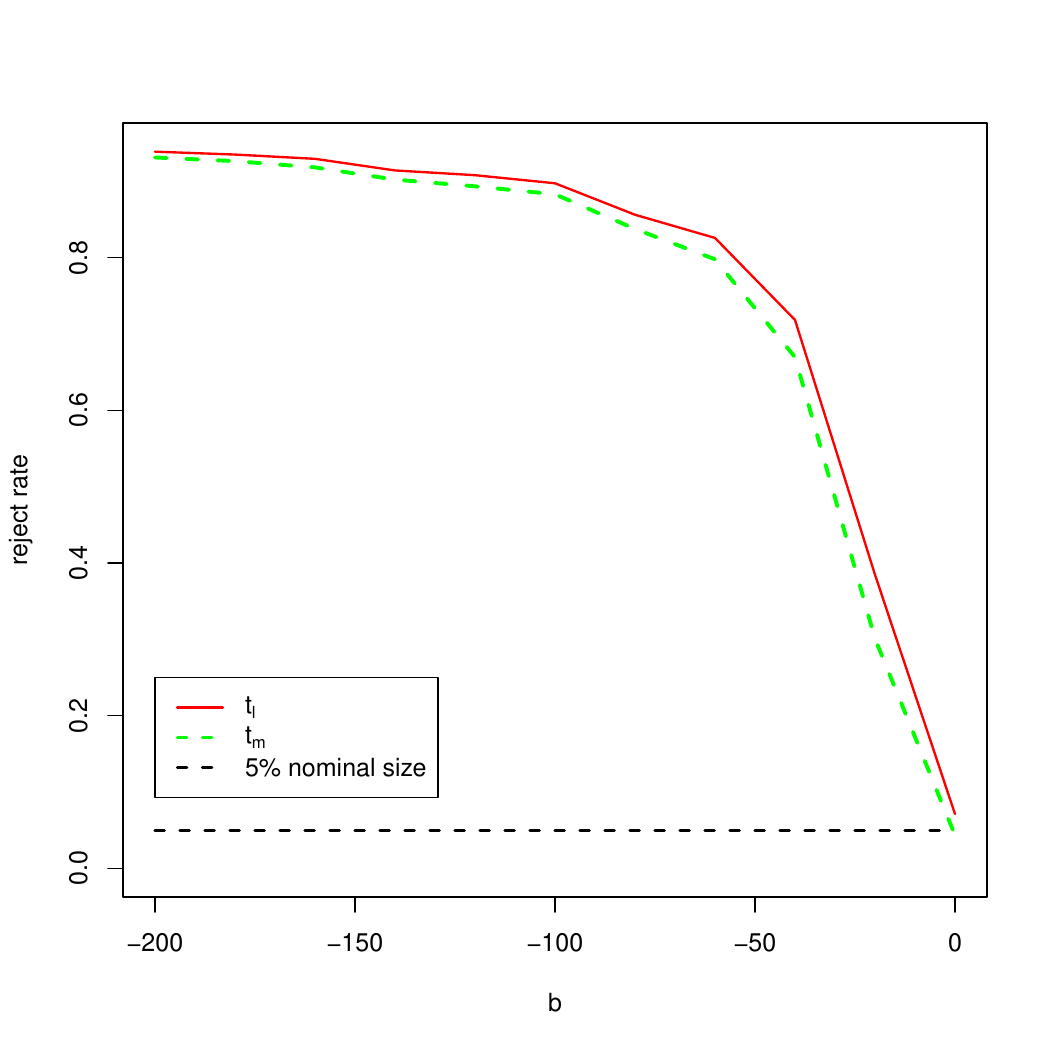}
}
\quad
\subfigure[Case 3 ($\alpha=1$, $c=-10$)]{
\includegraphics[width=5.5cm]{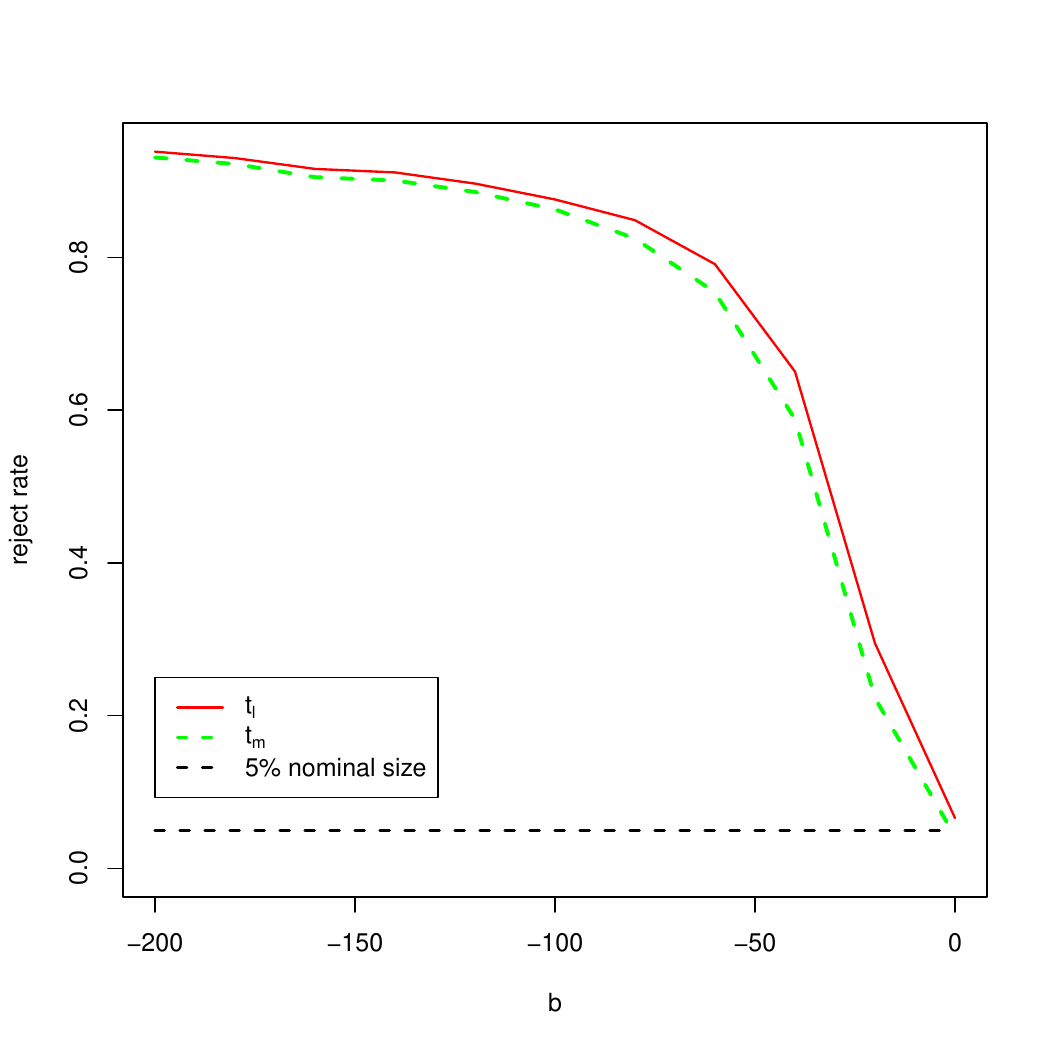}
%\caption{fig1}
}
\quad
\subfigure[Case 4 ($\alpha=1$, $c=-15$)]{
\includegraphics[width=5.5cm]{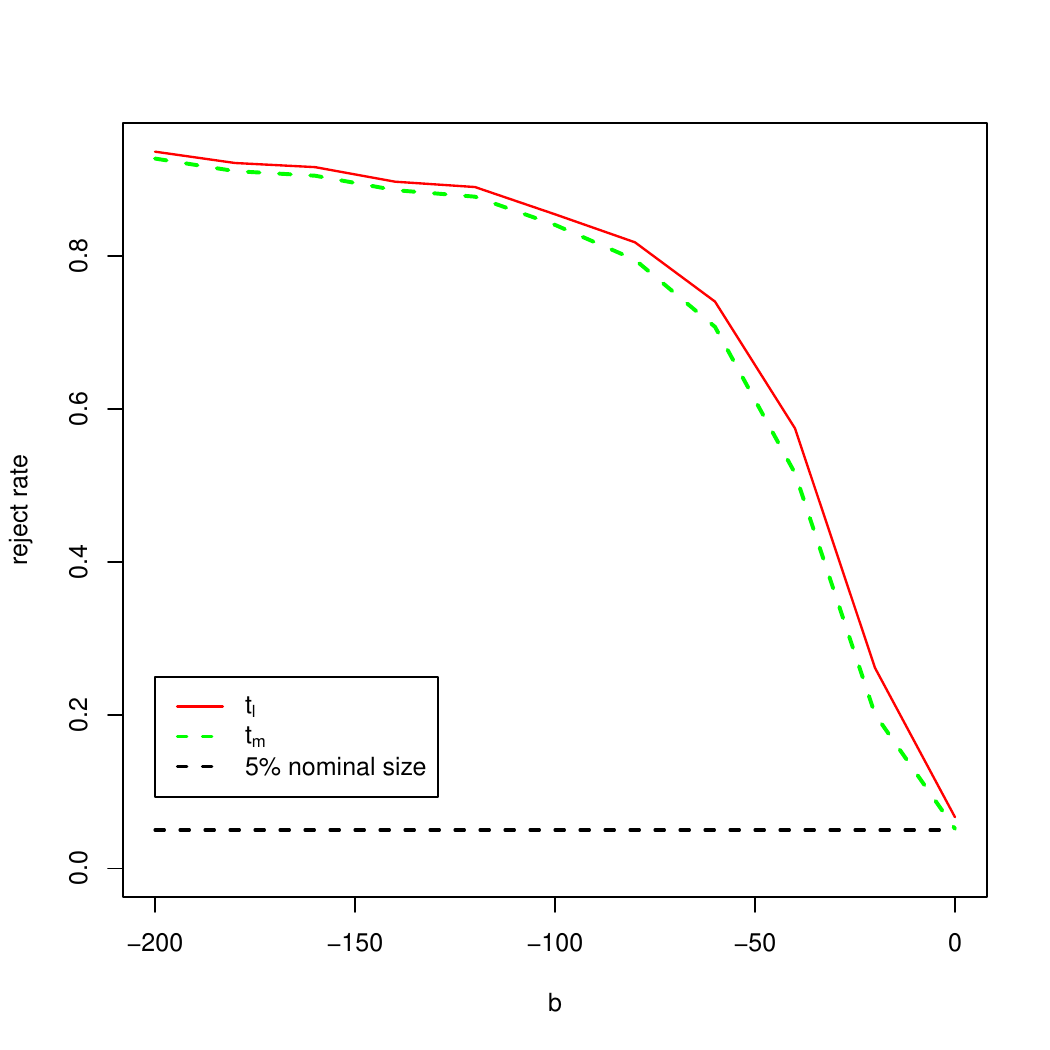}
}
\quad
\subfigure[Case 5 ($\alpha=0$, $c=-0.05$)]{
\includegraphics[width=5.5cm]{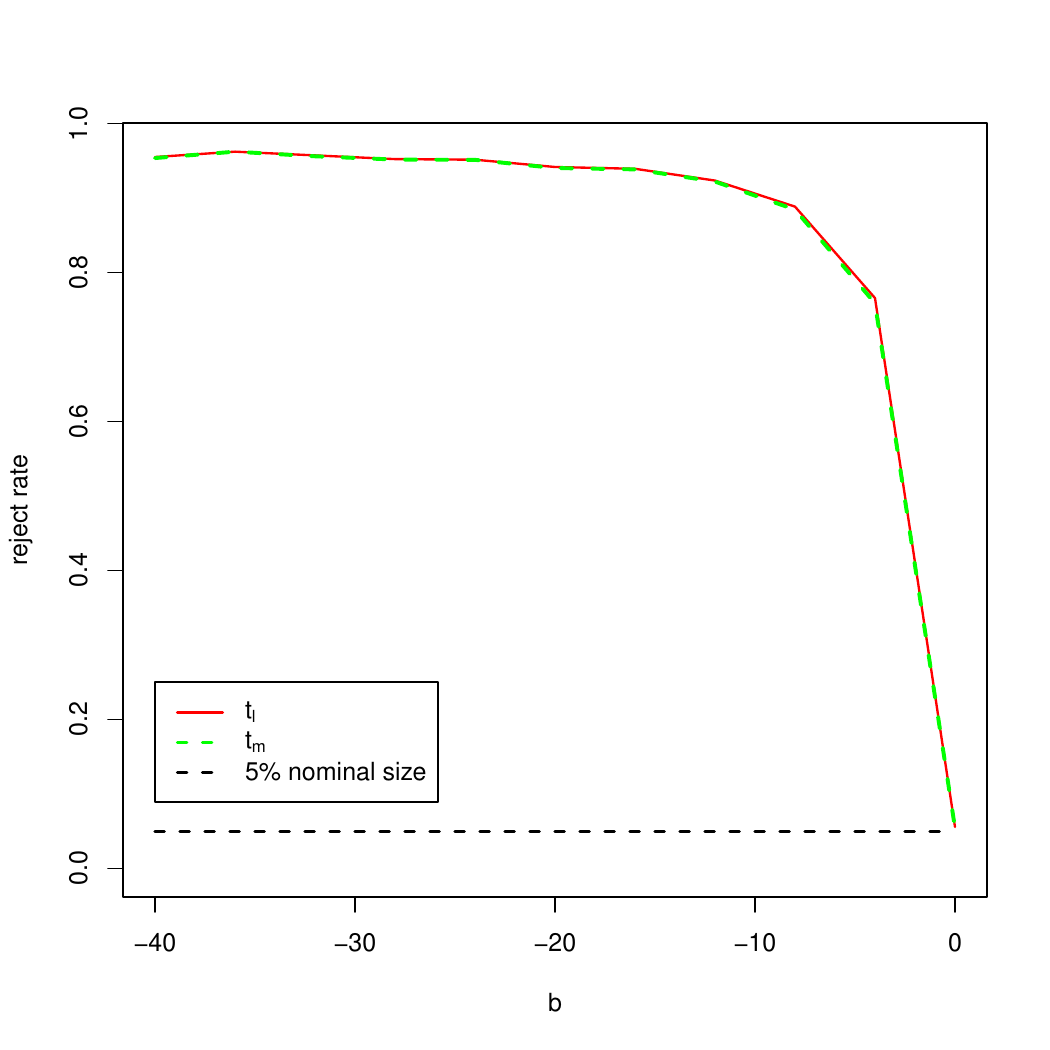}
%\caption{fig1}
}
\quad
\subfigure[Case 6 ($\alpha=0$, $c=-0.1$)]{
\includegraphics[width=5.5cm]{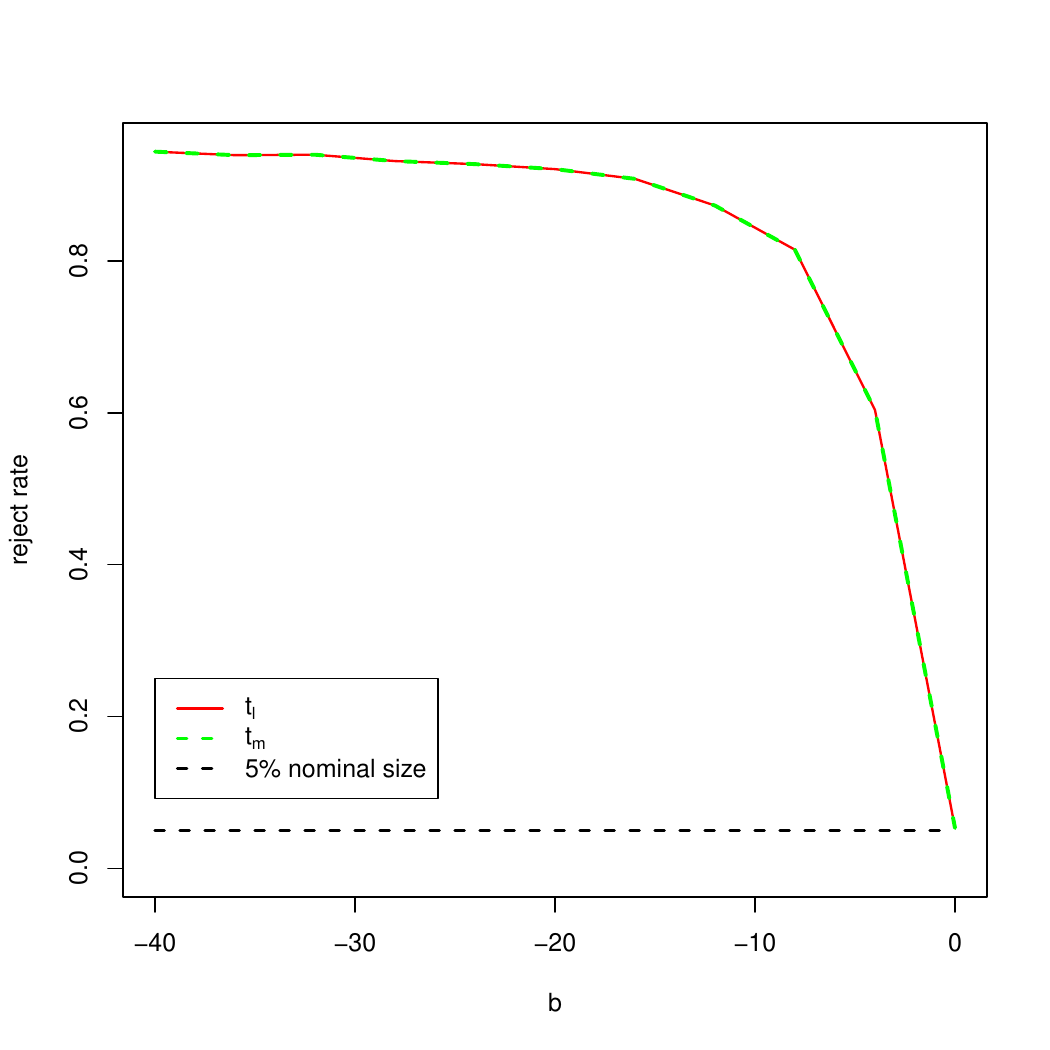}
}
\caption{Power of Left Side Test $H_0:\beta = 0$ vs $H_a:\beta < 0$ with $\phi=0.95$ and $\lambda=0.95$}
\label{power9}
\end{figure}

\begin{figure}[H]
\centering
\subfigure[Case 1 ($\alpha=1$, $c=0$)]{
\includegraphics[width=5.5cm]{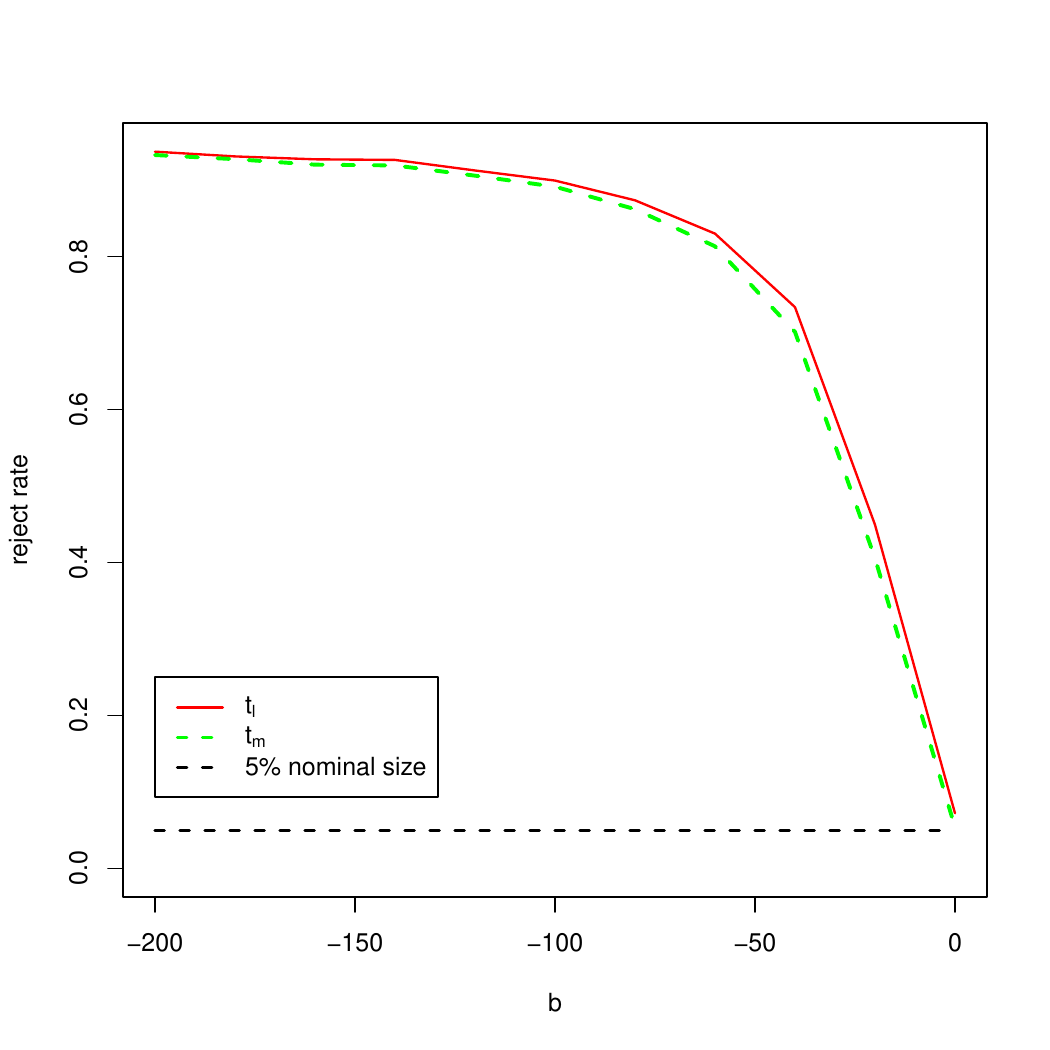}
%\caption{fig1}
}
\quad
\subfigure[Case 2 ($\alpha=1$, $c=-5$)]{
\includegraphics[width=5.5cm]{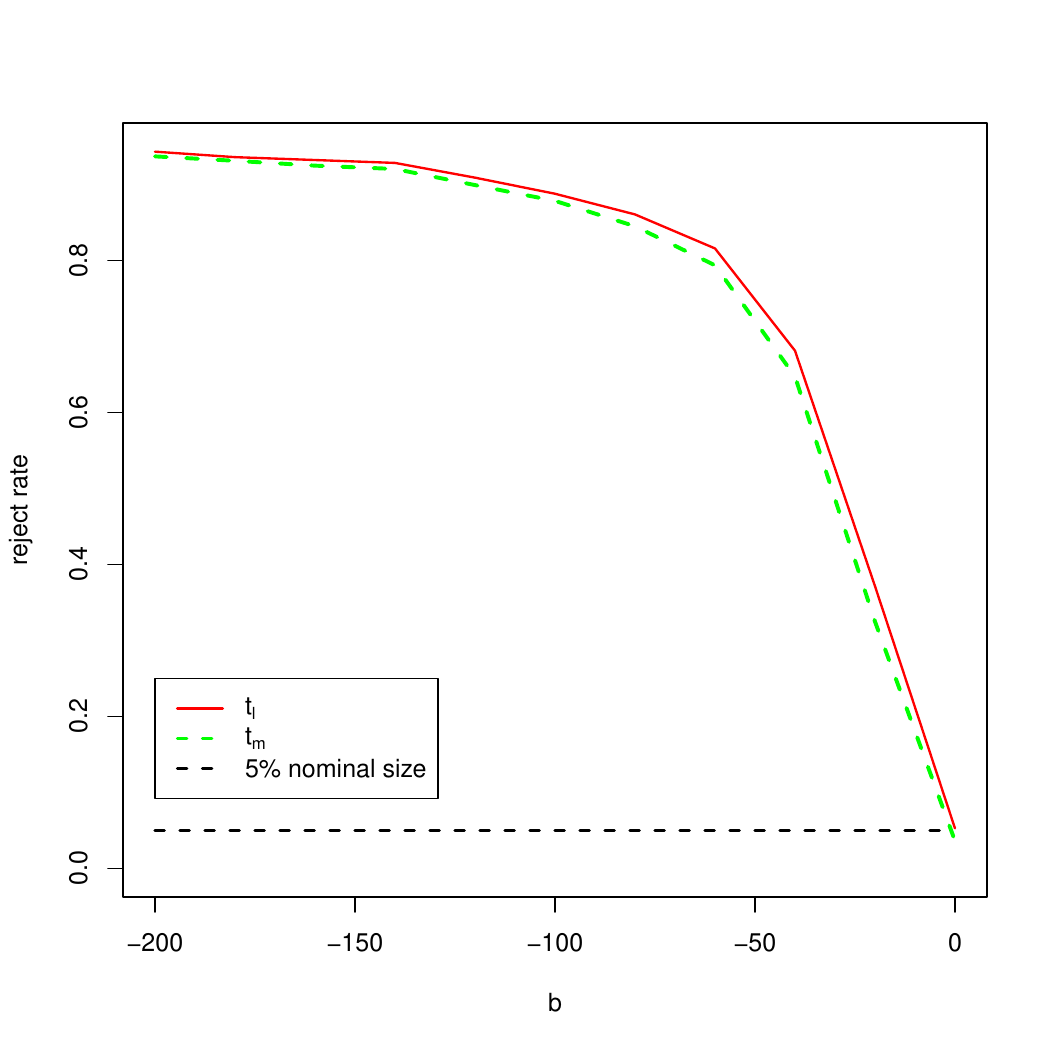}
}
\quad
\subfigure[Case 3 ($\alpha=1$, $c=-10$)]{
\includegraphics[width=5.5cm]{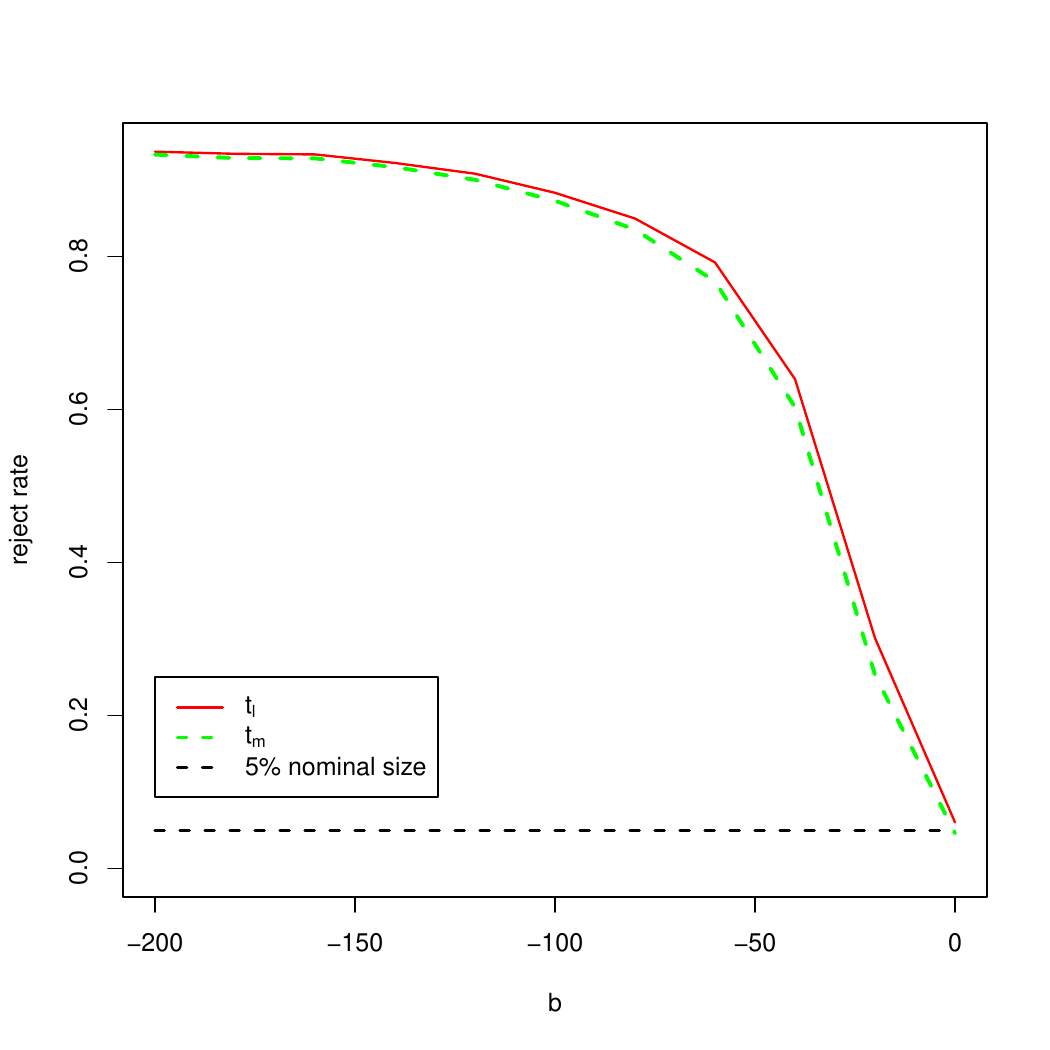}
%\caption{fig1}
}
\quad
\subfigure[Case 4 ($\alpha=1$, $c=-15$)]{
\includegraphics[width=5.5cm]{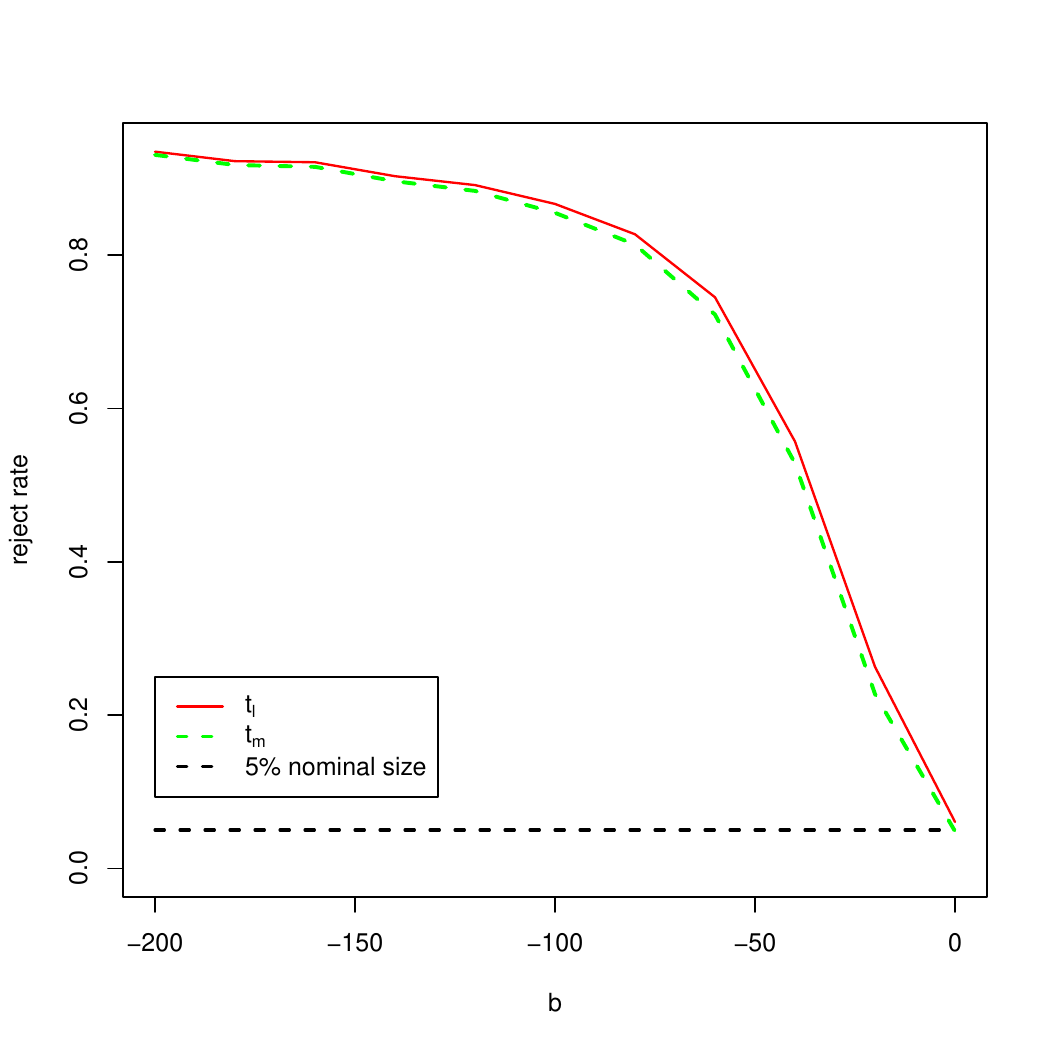}
}
\quad
\subfigure[Case 5 ($\alpha=0$, $c=-0.05$)]{
\includegraphics[width=5.5cm]{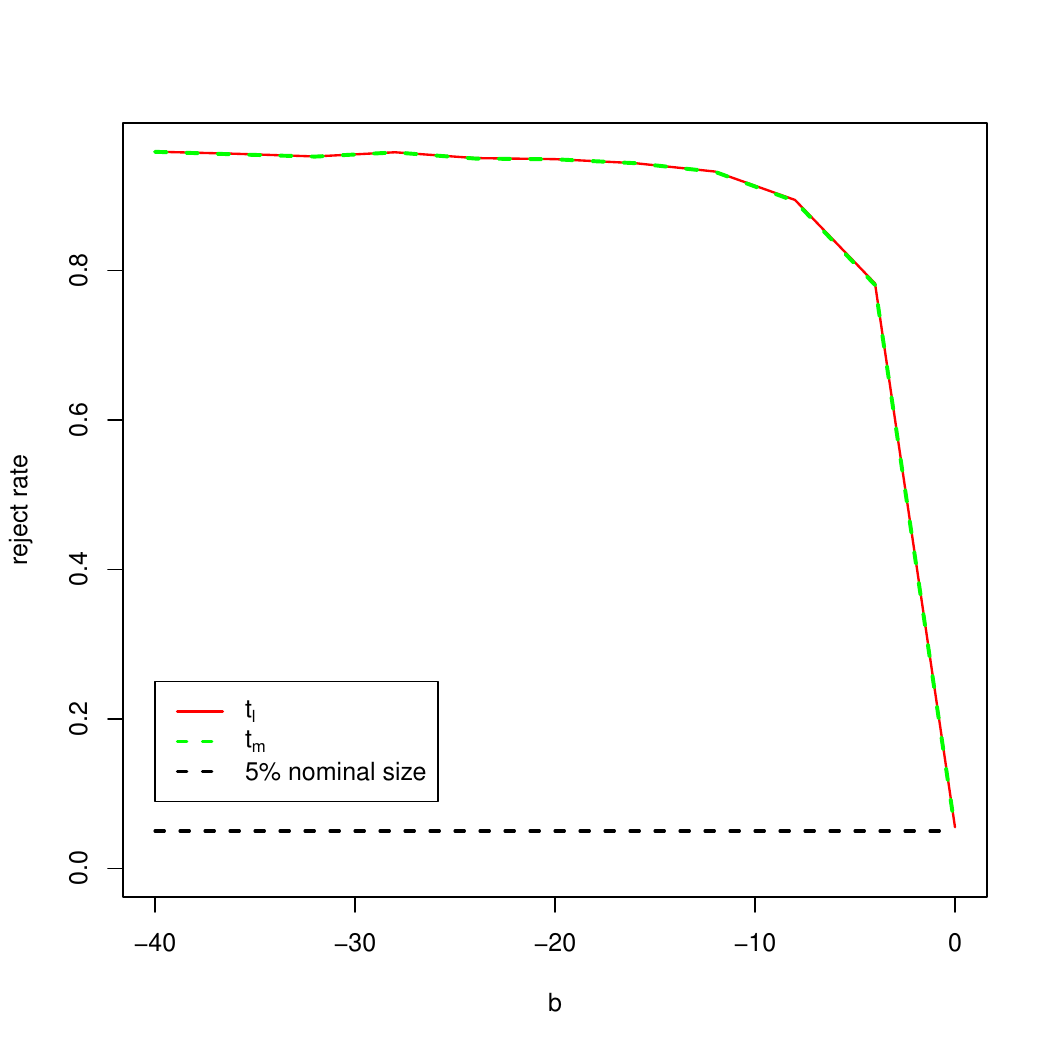}
%\caption{fig1}
}
\quad
\subfigure[Case 6 ($\alpha=0$, $c=-0.1$)]{
\includegraphics[width=5.5cm]{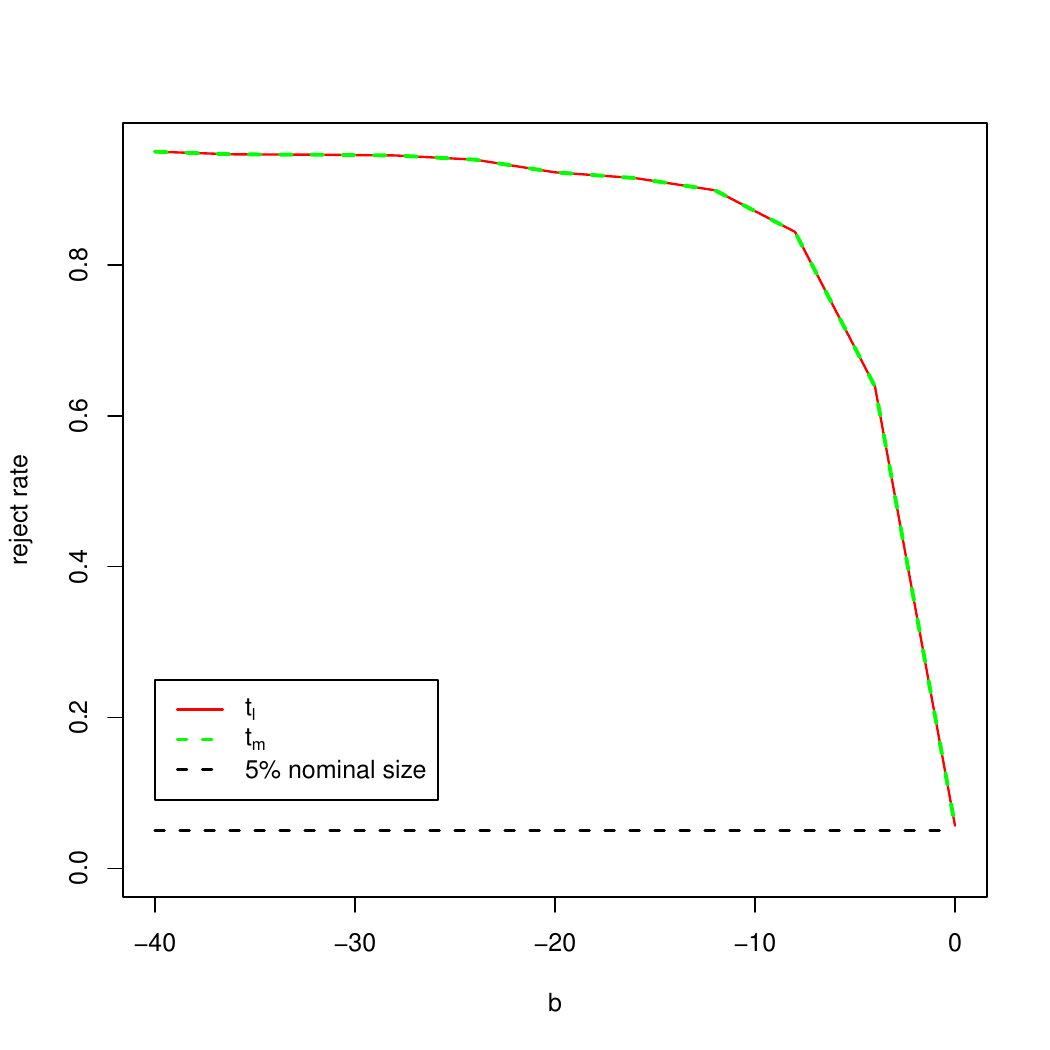}
}
\caption{Power of Left Side Test $H_0:\beta = 0$ vs $H_a:\beta < 0$ with $\phi=0.5$ and $\lambda=0.5$}
\label{power10}
\end{figure}

\begin{figure}[H]
\centering
\subfigure[Case 1 ($\alpha=1$, $c=0$)]{
\includegraphics[width=5.5cm]{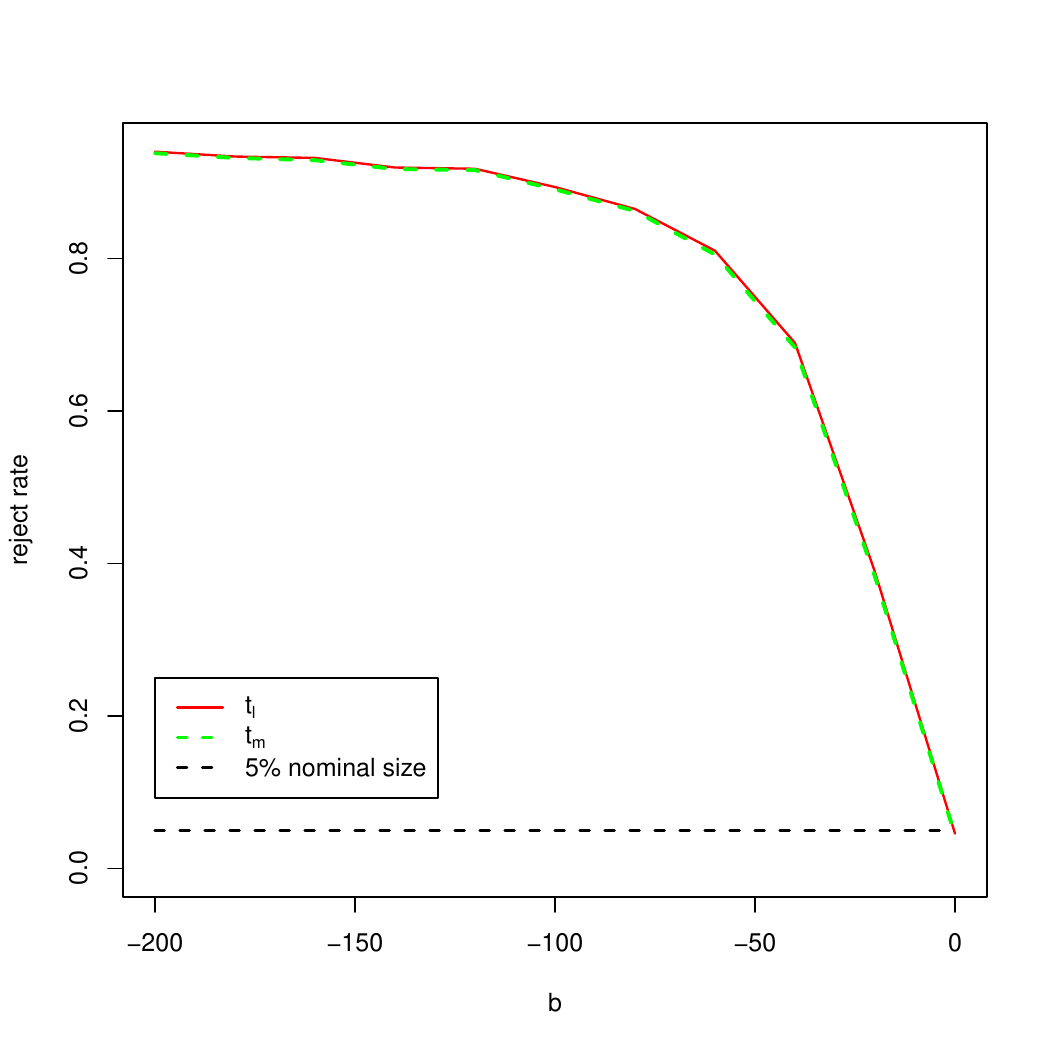}
%\caption{fig1}
}
\quad
\subfigure[Case 2 ($\alpha=1$, $c=-5$)]{
\includegraphics[width=5.5cm]{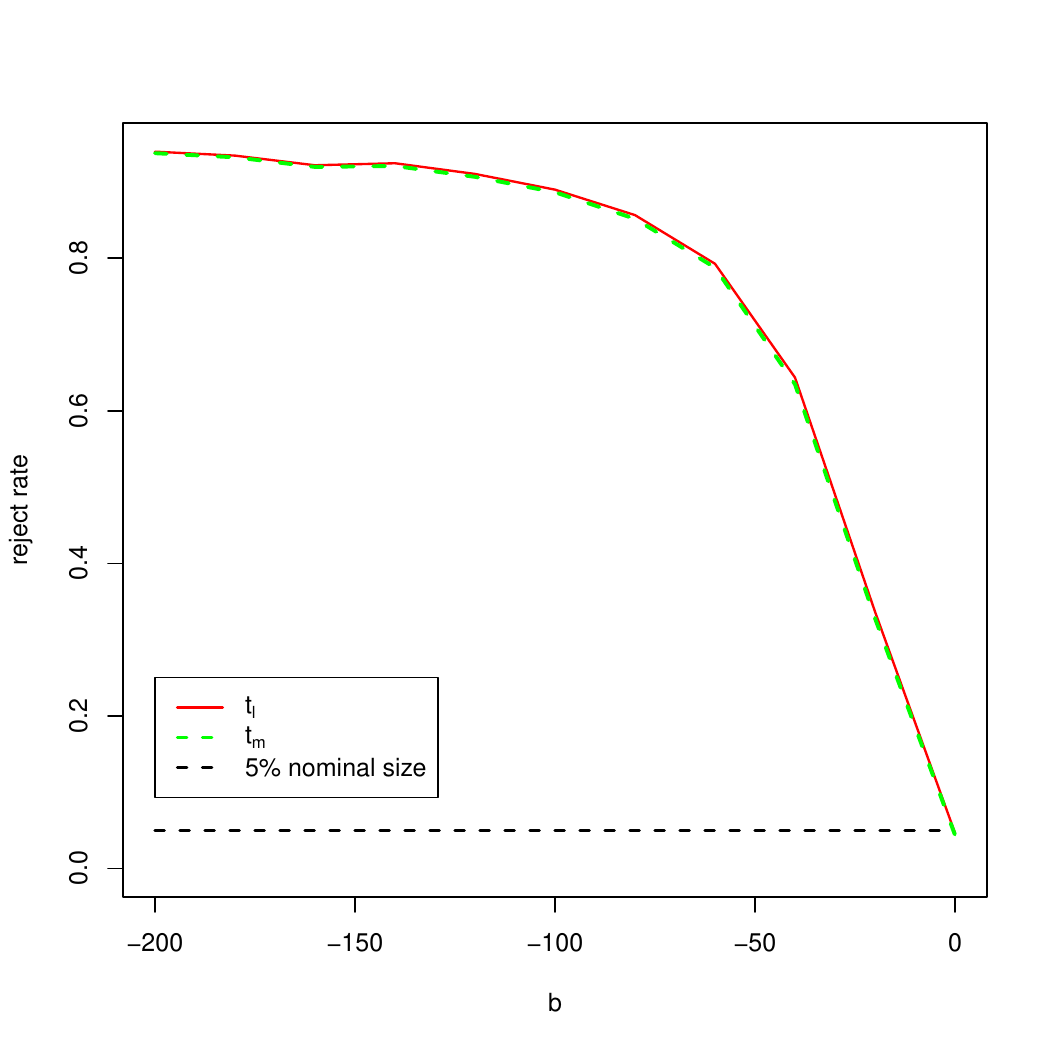}
}
\quad
\subfigure[Case 3 ($\alpha=1$, $c=-10$)]{
\includegraphics[width=5.5cm]{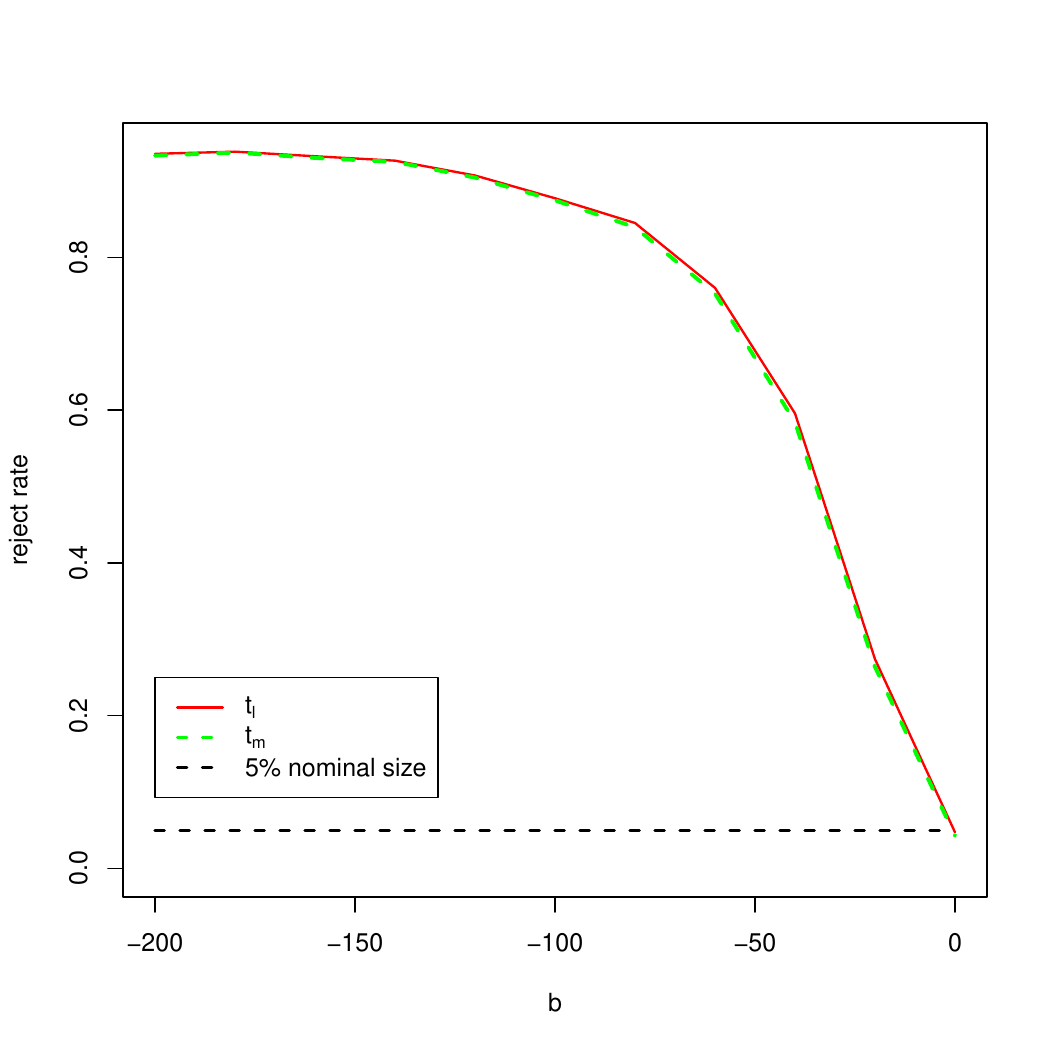}
%\caption{fig1}
}
\quad
\subfigure[Case 4 ($\alpha=1$, $c=-15$)]{
\includegraphics[width=5.5cm]{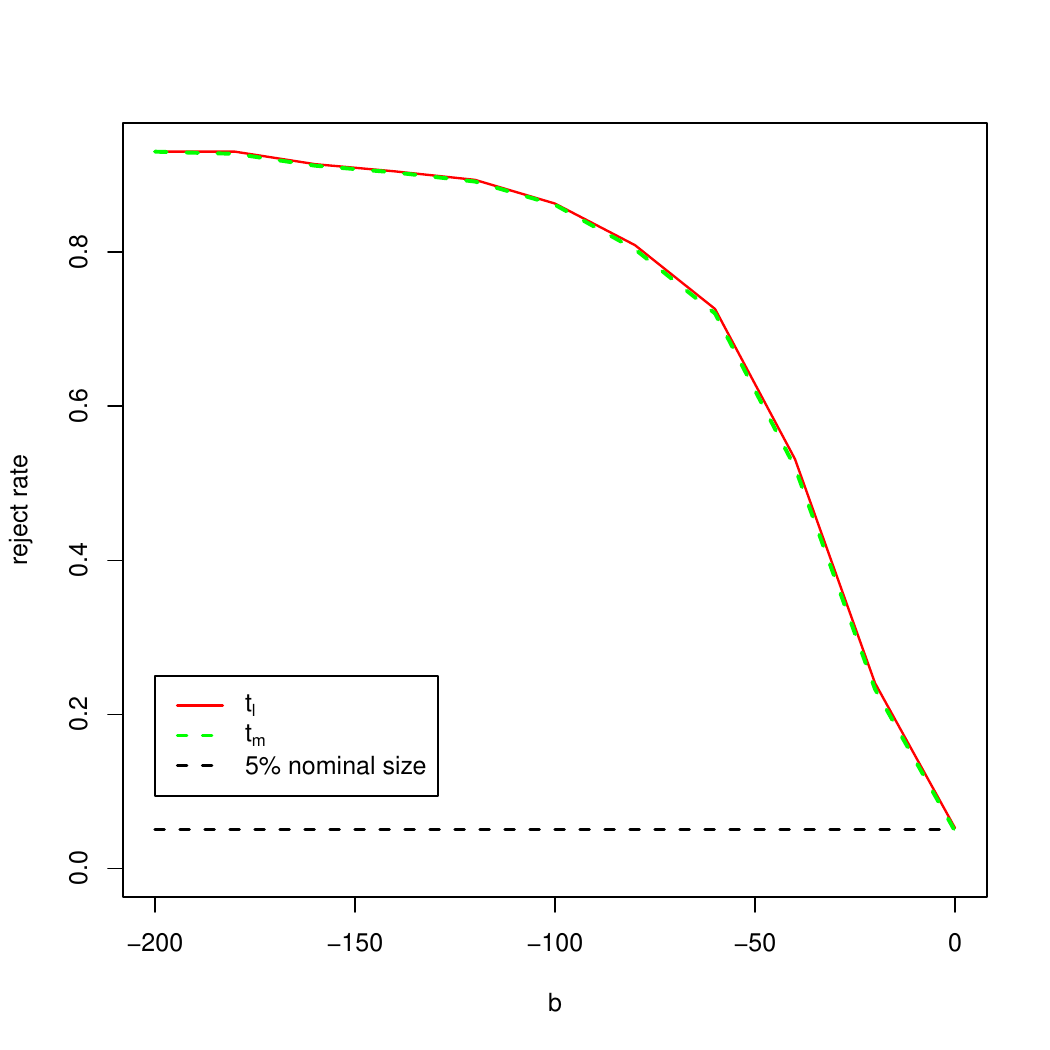}
}
\quad
\subfigure[Case 5 ($\alpha=0$, $c=-0.05$)]{
\includegraphics[width=5.5cm]{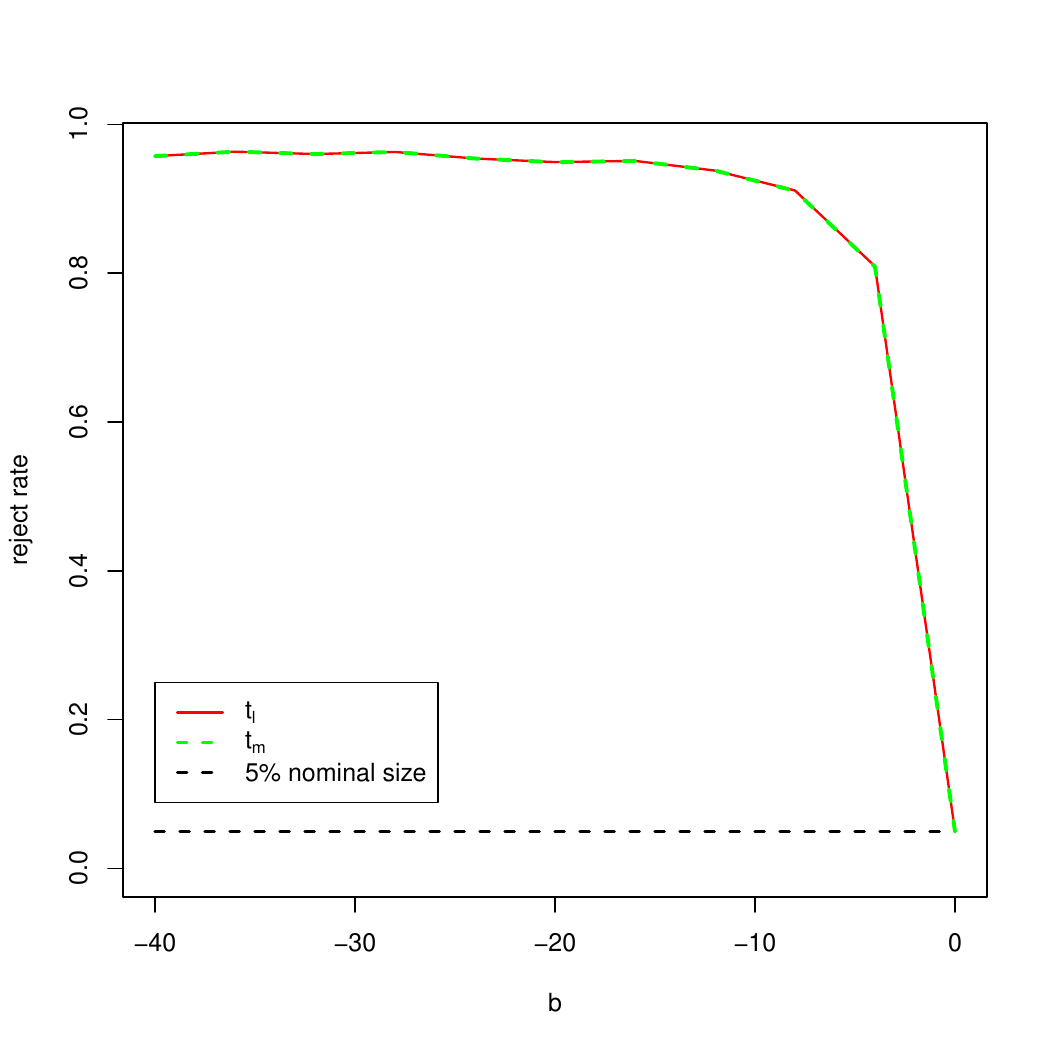}
%\caption{fig1}
}
\quad
\subfigure[Case 6 ($\alpha=0$, $c=-0.1$)]{
\includegraphics[width=5.5cm]{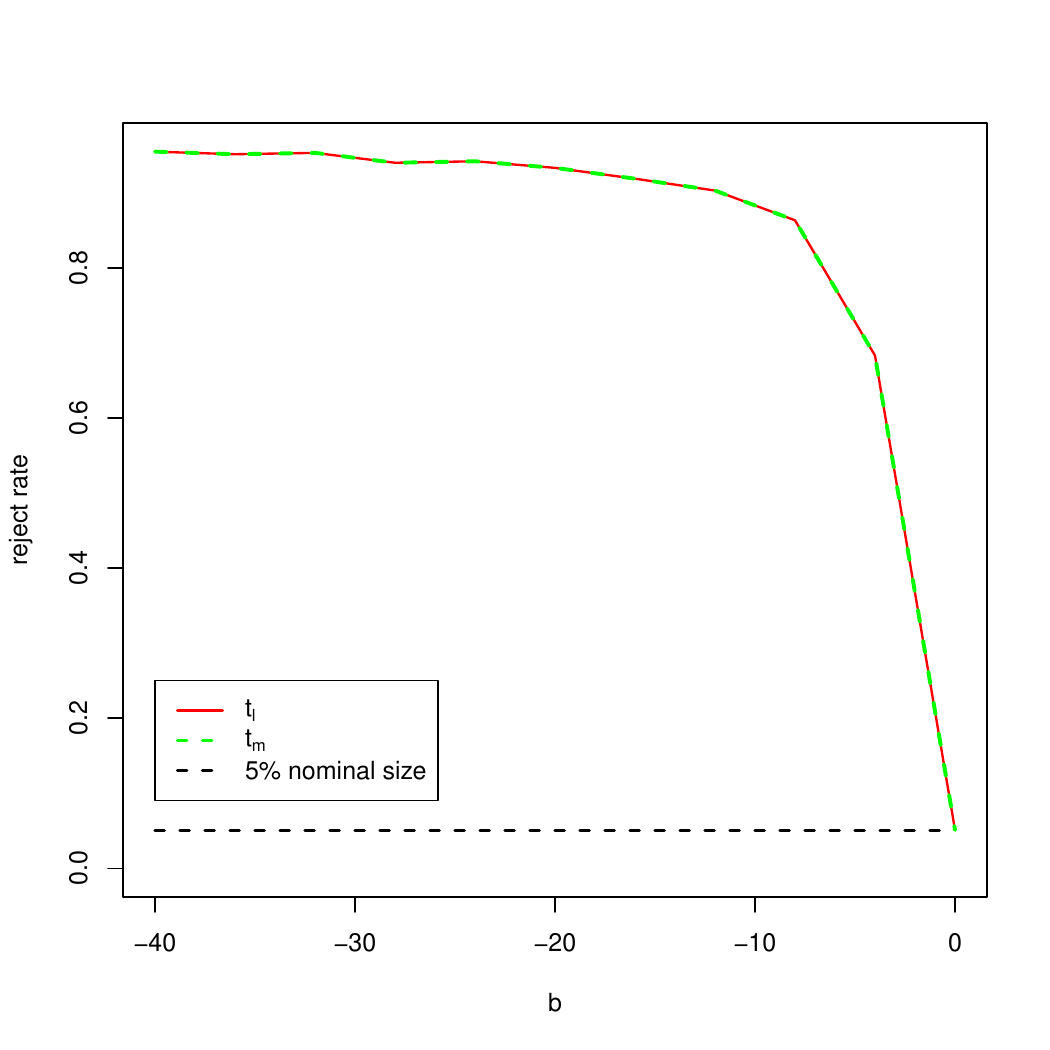}
}
\caption{Power of Left Side Test $H_0:\beta = 0$ vs $H_a:\beta < 0$ with $\phi=-0.1$ and $\lambda=0.5$}
\label{power11}
\end{figure}

\begin{figure}[H]
\centering
\subfigure[Case 1 ($\alpha=1$, $c=0$)]{
\includegraphics[width=5.5cm]{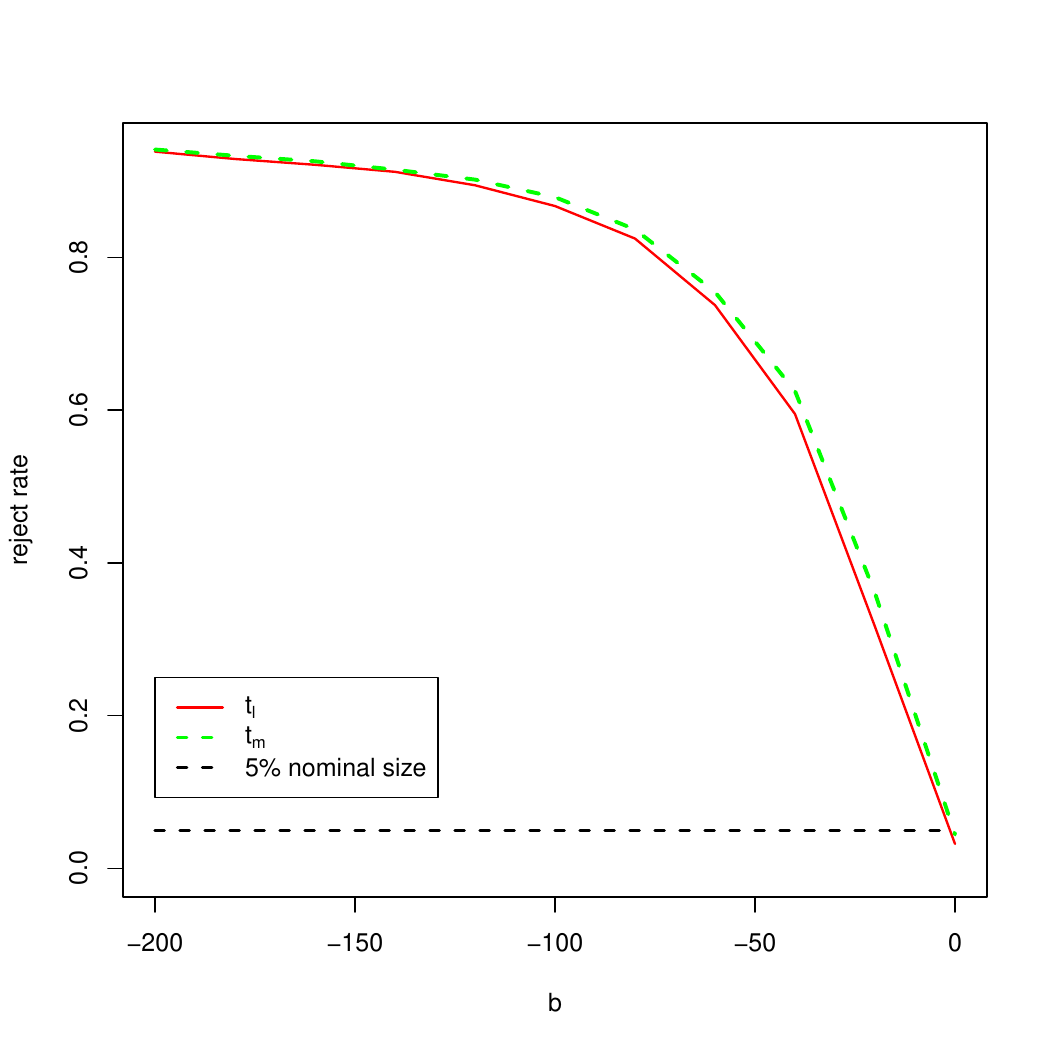}
%\caption{fig1}
}
\quad
\subfigure[Case 2 ($\alpha=1$, $c=-5$)]{
\includegraphics[width=5.5cm]{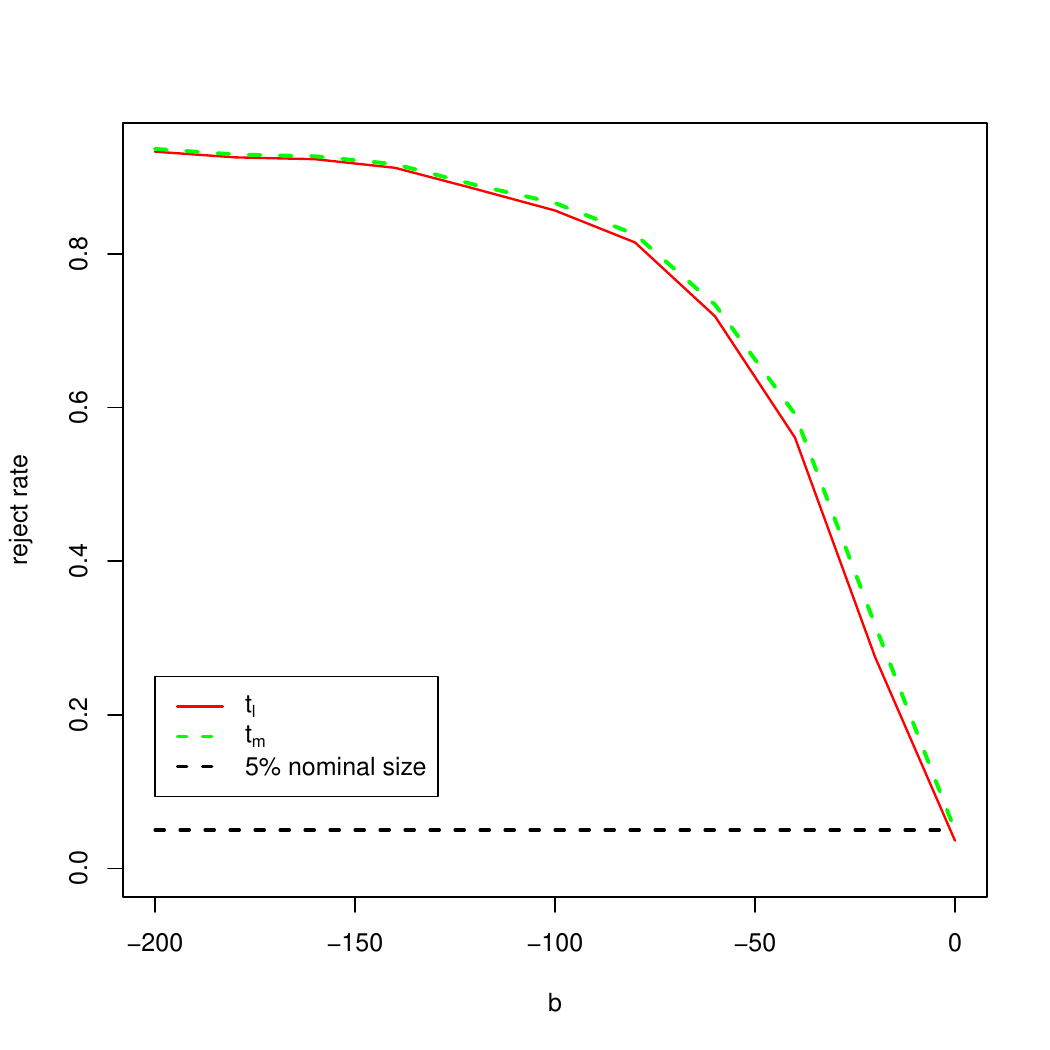}
}
\quad
\subfigure[Case 3 ($\alpha=1$, $c=-10$)]{
\includegraphics[width=5.5cm]{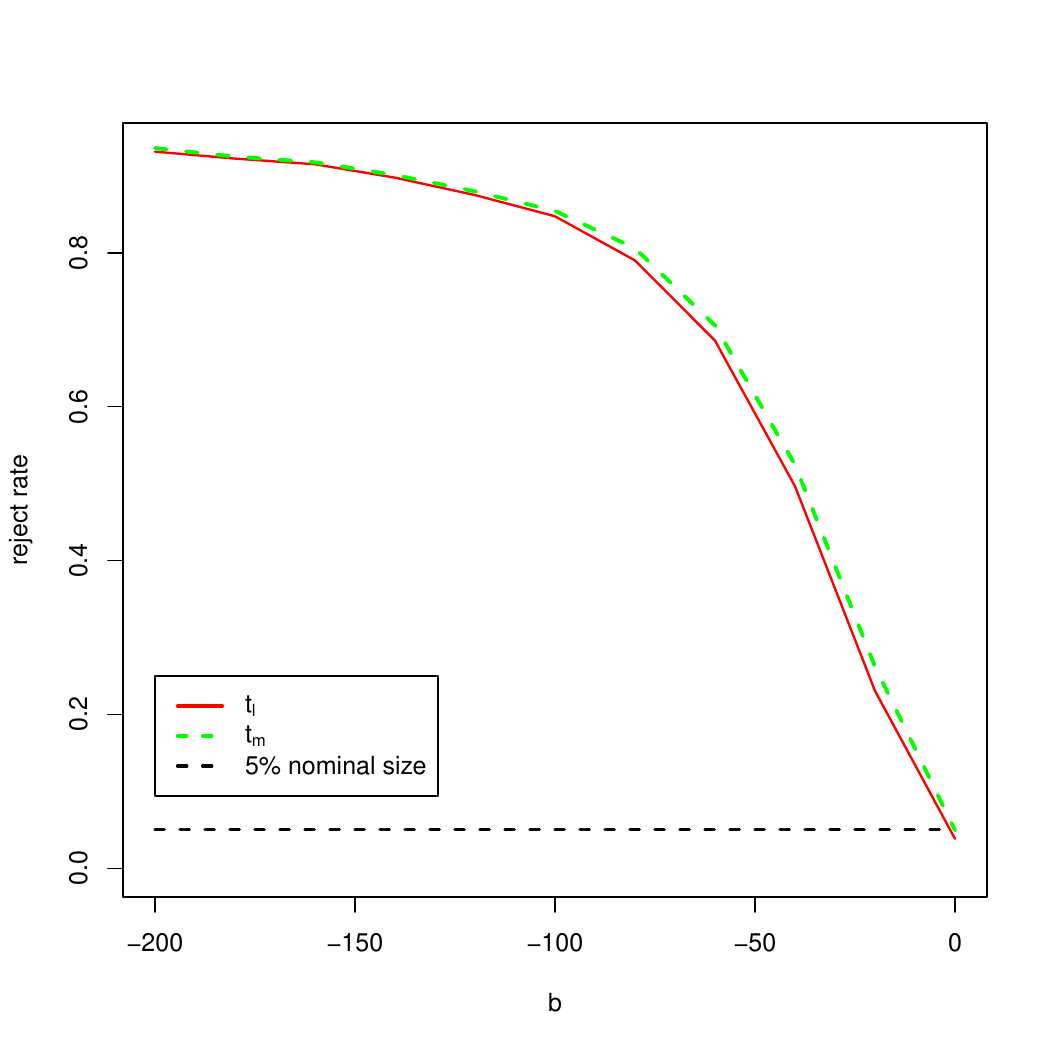}
%\caption{fig1}
}
\quad
\subfigure[Case 4 ($\alpha=1$, $c=-15$)]{
\includegraphics[width=5.5cm]{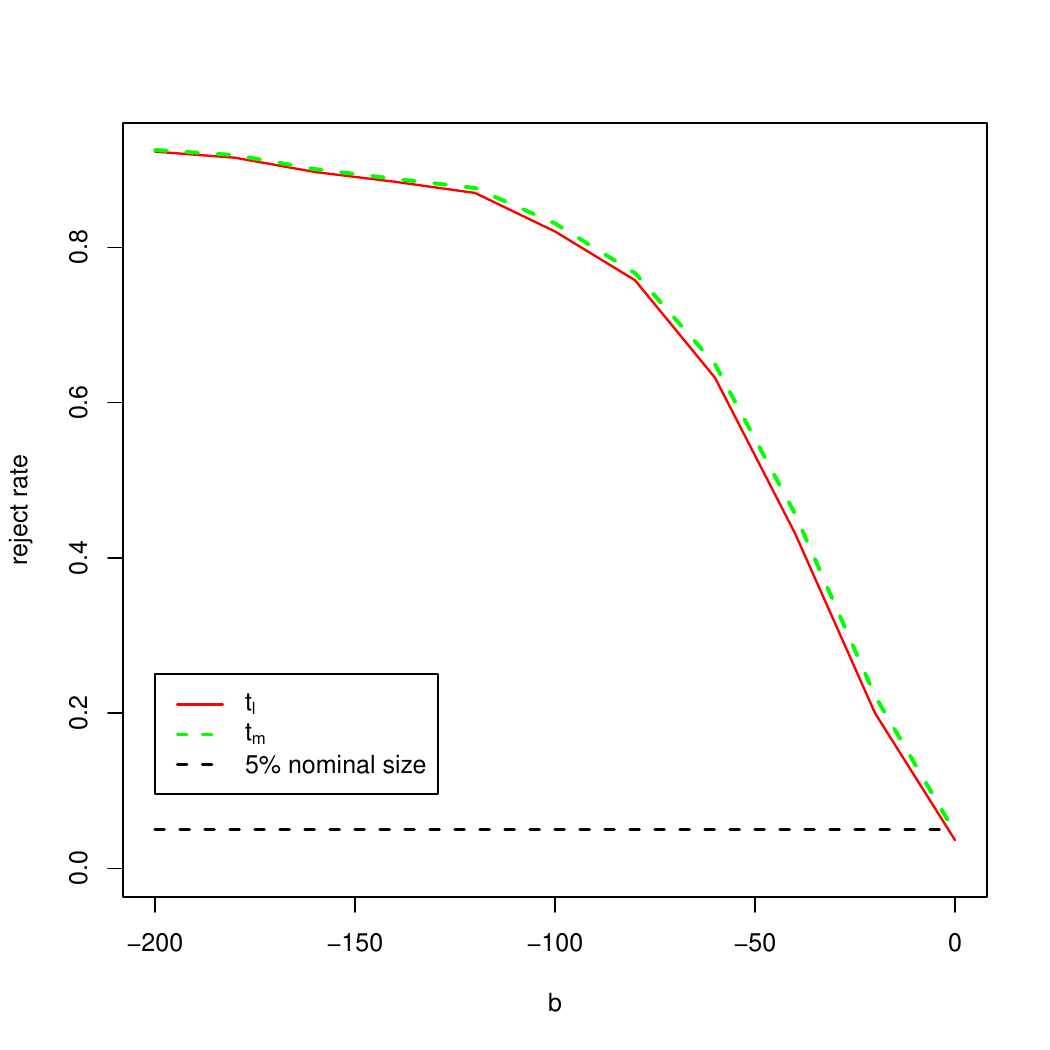}
}
\quad
\subfigure[Case 5 ($\alpha=0$, $c=-0.05$)]{
\includegraphics[width=5.5cm]{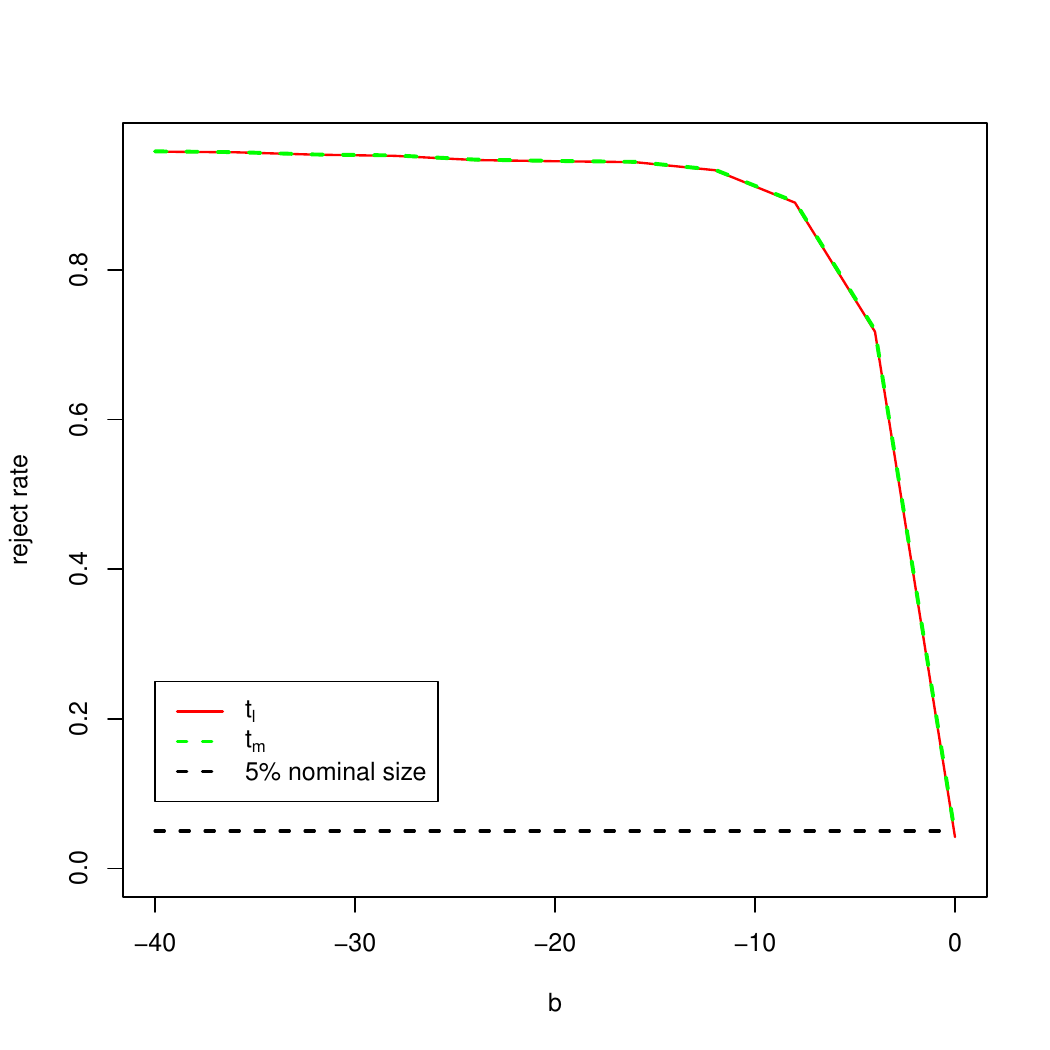}
%\caption{fig1}
}
\quad
\subfigure[Case 6 ($\alpha=0$, $c=-0.1$)]{
\includegraphics[width=5.5cm]{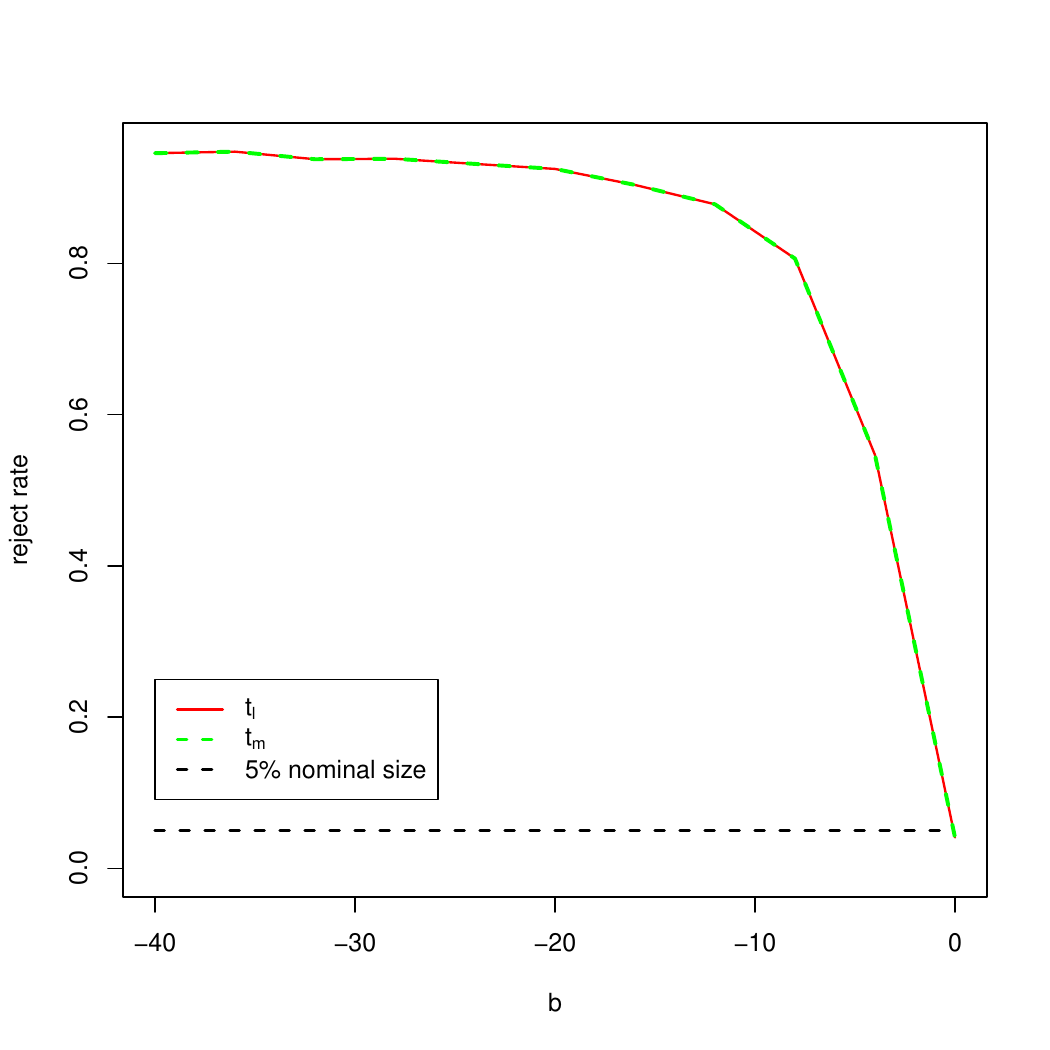}
}
\caption{Power of Left Side Test $H_0:\beta = 0$ vs $H_a:\beta < 0$ with $\phi=-0.95$ and $\lambda=0.5$}
\label{power12}
\end{figure}

First, the results for the comparison of the size performances of the proposed test statistics $t_l$ and $t_m$ for seven cases  in GARCH(1,1) model are shown in Tables \ref{size1}-\ref{size3} while the power performances are shown in Figures \ref{power1}-\ref{power12}. The size performance of two-sided test $H_0:\beta=0$ vs $H_a:\beta\neq 0$ and the right side test $H_0:\beta=0$ vs $H_a:\beta> 0$ and the left side test $H_0:\beta=0$ vs $H_a:\beta < 0$ are shown in Tables \ref{size1}-\ref{size3}, respectively. Meanwhile, the  power performances of two-sided test $H_0:\beta=0$ vs $H_a:\beta\neq 0$ and the right side test $H_0:\beta=0$ vs $H_a:\beta> 0$ and left side test $H_0:\beta=0$ vs $H_a:\beta < 0$ are shown in Figures \ref{power1}-\ref{power4}, \ref{power5}-\ref{power8} and \ref{power9}-\ref{power12}  respectively. In each hypothesis, we show the simulation results for $\phi=0.95$, $0.5$, $-0.1$ and $-0.95$. Additionally, six persistence  including two categories  are  considered in each hypothesis. The first category including case 1 - case 4 is SD with $\alpha=1$, while $c_z=0,-5,-10,-15$ respectively. The second category including case 5 - case 6 is WD with $\alpha=0$, while $c_z=-0.05,-0.1,-0.15$, respectively. And we set $\lambda=0.1,0.25,0.5,0.75,0.9$ in Tables \ref{size1}-\ref{size3} to compare the size performance with different $\lambda$,  $\beta= b/T^{(1+\alpha)/2}$ to see the local power  and $\lambda=0.5$ in Figures \ref{power1}-\ref{power12}, \footnote{Since the following part states that the proposed test statistics perform the best with $\lambda=0.5$, we only show the power performance  with $\lambda=0.5$.} and $\varphi_1=0.1$, $\varphi_1=0.1$, $\bar{ \varphi}_1=0.85$ in GARCH(1,1) model and the nominal size to be $5\%$.   Simulations are repeated $10,000$ times for each setting.

% Table generated by Excel2LaTeX from sheet 'table2'

\begin{sidewaystable}
  \centering
  \caption{Size Performance ($\%$) of $t_l$ and $t_m$ for  $H_0:\beta=0$ vs $H_a:\beta\neq 0$}
  \resizebox{\textwidth}{!}{
    \begin{tabular}{c|c|cccc|cccc|cccc|cccc|cccc}
    \hline
          & $\lambda$ & \multicolumn{4}{c|}{0.1}       & \multicolumn{4}{c|}{0.25}      & \multicolumn{4}{c|}{0.5}       & \multicolumn{4}{c|}{0.75}      & \multicolumn{4}{c}{0.9} \\
    \hline
          & $\phi$ & 0.95  & 0.5   & -0.1  & -0.95 & 0.95  & 0.5   & -0.1  & -0.95 & 0.95  & 0.5   & -0.1  & -0.95 & 0.95  & 0.5   & -0.1  & -0.95 & 0.95  & 0.5   & -0.1  & -0.95 \\
    \hline
    \multirow{7}[0]{*}{$t_l$} & Case 1 & 5.8   & 5.6   & 5.1   & 5.7   & 5.9   & 5.2   & 5.0   & 6.2   & 5.8   & 5.3   & 5.1   & 5.9   & 5.7   & 5.4   & 4.5   & 5.6   & 5.7   & 5.6   & 4.7   & 5.5  \\
          & Case 2 & 5.5   & 5.4   & 4.4   & 5.5   & 5.3   & 5.1   & 4.8   & 5.4   & 5.7   & 4.9   & 5.0   & 5.4   & 5.6   & 5.3   & 5.2   & 5.5   & 5.3   & 4.9   & 4.9   & 5.6  \\
          & Case 3 & 4.8   & 5.1   & 5.0   & 5.4   & 5.3   & 4.9   & 5.0   & 5.0   & 4.5   & 4.9   & 4.6   & 5.0   & 5.1   & 5.3   & 4.8   & 5.1   & 5.6   & 4.5   & 5.0   & 5.1  \\
          & Case 4 & 5.0   & 5.0   & 4.5   & 5.4   & 5.1   & 4.9   & 5.1   & 5.0   & 5.1   & 4.8   & 5.0   & 5.2   & 4.8   & 4.7   & 4.9   & 5.0   & 5.4   & 4.8   & 4.7   & 5.7  \\
          & Case 5 & 5.0   & 5.0   & 4.8   & 4.6   & 4.8   & 4.7   & 5.2   & 4.5   & 4.6   & 4.7   & 4.8   & 5.1   & 4.8   & 5.0   & 5.3   & 4.8   & 5.0   & 4.9   & 4.8   & 5.1  \\
          & Case 6 & 4.5   & 4.5   & 4.9   & 4.5   & 4.3   & 5.2   & 4.9   & 4.8   & 4.8   & 4.9   & 4.6   & 4.5   & 4.7   & 5.0   & 4.6   & 4.6   & 5.2   & 4.7   & 5.0   & 4.3  \\
          & Case 7 & 4.6   & 4.6   & 4.5   & 4.5   & 4.6   & 4.9   & 4.7   & 4.9   & 5.0   & 4.8   & 5.0   & 4.6   & 4.9   & 4.8   & 4.7   & 5.1   & 4.3   & 4.5   & 4.3   & 4.4  \\
    \hline
    \multirow{7}[0]{*}{$t_m$} & Case 1 & 3.8   & 4.3   & 4.0   & 3.6   & 4.3   & 4.2   & 4.3   & 4.3   & 4.2   & 4.3   & 4.4   & 4.2   & 3.9   & 4.4   & 3.9   & 3.6   & 3.6   & 4.4   & 3.8   & 3.8  \\
          & Case 2 & 4.6   & 4.2   & 3.6   & 4.3   & 4.5   & 4.3   & 4.2   & 4.6   & 4.8   & 4.3   & 4.2   & 4.5   & 4.5   & 4.7   & 4.6   & 4.5   & 4.1   & 3.8   & 4.1   & 4.5  \\
          & Case 3 & 4.3   & 4.3   & 4.2   & 4.7   & 4.8   & 4.2   & 4.3   & 4.4   & 4.2   & 4.4   & 4.0   & 4.3   & 4.6   & 4.5   & 4.1   & 4.3   & 4.6   & 3.9   & 4.3   & 4.3  \\
          & Case 4 & 4.9   & 4.3   & 3.9   & 5.1   & 4.8   & 4.4   & 4.4   & 4.7   & 4.7   & 4.4   & 4.3   & 4.9   & 4.3   & 4.1   & 4.3   & 4.5   & 4.9   & 4.2   & 4.2   & 5.1  \\
          & Case 5 & 5.2   & 4.9   & 4.7   & 4.8   & 5.0   & 4.7   & 5.1   & 4.7   & 4.8   & 4.6   & 4.7   & 5.2   & 4.8   & 5.0   & 5.1   & 4.9   & 5.1   & 4.8   & 4.7   & 5.2  \\
          & Case 6 & 4.5   & 4.5   & 4.9   & 4.5   & 4.3   & 5.2   & 4.9   & 4.8   & 4.8   & 4.9   & 4.6   & 4.5   & 4.7   & 5.0   & 4.6   & 4.6   & 5.2   & 4.7   & 5.0   & 4.3  \\
          & Case 7 & 4.6   & 4.6   & 4.5   & 4.5   & 4.6   & 4.9   & 4.7   & 4.9   & 5.0   & 4.8   & 5.0   & 4.6   & 4.9   & 4.8   & 4.7   & 5.1   & 4.3   & 4.5   & 4.3   & 4.4  \\
    \hline
    \end{tabular}%
    }
  \label{size1}%
\end{sidewaystable}

\begin{sidewaystable}
  \centering
  \caption{Size Performance  ($\%$)  of $t_l$ and $t_m$ for  $H_0:\beta=0$ vs $H_a:\beta> 0$}
  \resizebox{\textwidth}{!}{
    \begin{tabular}{c|c|cccc|cccc|cccc|cccc|cccc}
    \hline
          & $\lambda$ & \multicolumn{4}{c|}{0.1}       & \multicolumn{4}{c|}{0.25}      & \multicolumn{4}{c|}{0.5}       & \multicolumn{4}{c|}{0.75}      & \multicolumn{4}{c}{0.9} \\
    \hline
          & $\phi$ & 0.95  & 0.5   & -0.1  & -0.95 & 0.95  & 0.5   & -0.1  & -0.95 & 0.95  & 0.5   & -0.1  & -0.95 & 0.95  & 0.5   & -0.1  & -0.95 & 0.95  & 0.5   & -0.1  & -0.95 \\
    \hline
    \multirow{7}[0]{*}{$t_l$} & Case 1 & 3.6   & 4.1   & 5.7   & 8.4   & 3.1   & 3.8   & 5.7   & 9.1   & 3.1   & 3.6   & 5.6   & 8.7   & 2.9   & 3.6   & 4.9   & 9.0   & 2.8   & 4.2   & 5.5   & 8.4  \\
          & Case 2 & 4.0   & 4.4   & 4.6   & 7.1   & 4.0   & 4.1   & 5.0   & 6.8   & 3.6   & 4.1   & 5.2   & 7.2   & 3.4   & 4.3   & 5.3   & 7.6   & 3.6   & 3.9   & 5.2   & 7.5  \\
          & Case 3 & 4.3   & 4.4   & 5.5   & 6.5   & 4.0   & 4.1   & 4.9   & 6.1   & 3.9   & 4.2   & 5.3   & 7.1   & 3.5   & 4.0   & 5.2   & 6.8   & 3.4   & 4.0   & 5.3   & 7.2  \\
          & Case 4 & 4.3   & 4.5   & 5.3   & 6.3   & 3.9   & 4.3   & 5.3   & 6.3   & 3.8   & 4.2   & 5.3   & 6.7   & 3.5   & 4.1   & 5.2   & 6.8   & 3.6   & 3.9   & 5.0   & 7.1  \\
          & Case 5 & 4.3   & 4.2   & 5.4   & 5.8   & 4.5   & 4.4   & 5.5   & 5.7   & 4.4   & 4.5   & 5.1   & 6.1   & 3.9   & 4.6   & 5.2   & 6.1   & 3.7   & 4.3   & 5.2   & 6.5  \\
          & Case 6 & 4.6   & 4.3   & 4.9   & 5.4   & 4.5   & 4.9   & 5.4   & 5.4   & 4.2   & 4.7   & 4.9   & 5.3   & 4.5   & 4.6   & 5.0   & 5.5   & 4.2   & 4.3   & 5.2   & 5.6  \\
          & Case 7 & 4.3   & 4.8   & 5.0   & 5.0   & 4.3   & 4.7   & 5.1   & 5.4   & 4.6   & 4.6   & 5.1   & 5.6   & 4.3   & 4.8   & 5.1   & 5.7   & 4.4   & 4.5   & 4.7   & 5.4  \\
    \hline
    \multirow{7}[0]{*}{$t_m$} & Case 1 & 4.3   & 4.3   & 4.6   & 4.6   & 4.1   & 4.3   & 4.9   & 5.1   & 4.2   & 4.0   & 4.8   & 5.1   & 4.0   & 4.3   & 4.1   & 5.0   & 3.6   & 4.5   & 4.6   & 4.4  \\
          & Case 2 & 5.3   & 4.7   & 3.9   & 4.3   & 5.4   & 4.7   & 4.4   & 4.1   & 4.9   & 4.6   & 4.4   & 4.4   & 4.8   & 4.9   & 4.5   & 4.6   & 5.0   & 4.4   & 4.4   & 4.5  \\
          & Case 3 & 5.5   & 4.8   & 4.7   & 4.4   & 5.4   & 4.4   & 4.2   & 4.0   & 5.0   & 4.6   & 4.6   & 4.6   & 4.6   & 4.4   & 4.5   & 4.6   & 4.5   & 4.3   & 4.6   & 4.8  \\
          & Case 4 & 5.5   & 4.7   & 4.6   & 4.8   & 5.1   & 4.5   & 4.7   & 4.7   & 4.9   & 4.5   & 4.6   & 4.9   & 4.6   & 4.4   & 4.6   & 5.1   & 4.6   & 4.1   & 4.4   & 5.4  \\
          & Case 5 & 4.7   & 4.4   & 5.3   & 5.6   & 4.9   & 4.5   & 5.5   & 5.5   & 4.7   & 4.7   & 5.0   & 5.9   & 4.3   & 4.8   & 5.1   & 6.0   & 4.0   & 4.5   & 5.1   & 6.3  \\
          & Case 6 & 4.6   & 4.4   & 4.9   & 5.4   & 4.5   & 4.9   & 5.4   & 5.4   & 4.2   & 4.7   & 4.9   & 5.3   & 4.5   & 4.6   & 5.0   & 5.5   & 4.2   & 4.3   & 5.2   & 5.6  \\
          & Case 7 & 4.3   & 4.8   & 5.0   & 5.0   & 4.3   & 4.7   & 5.1   & 5.4   & 4.6   & 4.6   & 5.1   & 5.6   & 4.3   & 4.8   & 5.1   & 5.7   & 4.4   & 4.5   & 4.7   & 5.4  \\
    \hline
    \end{tabular}%
    }
  \label{size2}%
\end{sidewaystable}

Clearly, the following findings can be observed from Tables \ref{size1}-\ref{size3}. First, the proposed test statistics $t_l$ and $t_m$ generally perform well in terms of size for the joint test $H_0:\beta=0$ vs $H_a:\beta\neq 0$. $t_l$ performs well in terms of size for different $\lambda$ while $t_m$ suffers a little bit of size distortion with $\lambda=0.1,0.75,0.9$ for SD predictors. Second, the proposed test statistics $t_l$ suffers size distortion  for the right side test $H_0:\beta=0$ vs $H_a:\beta> 0$ and left side test $H_0:\beta=0$ vs $H_a:\beta < 0$ with SD predictors and $\phi=0.95,0.5,-0.95$, while $t_m$  performs well in terms of size, especially with $\lambda=0.5$. In one word, $t_m$ with $\lambda=0.5$ performs well in terms of size for all cases while $t_l$ only performs well in terms of size for the joint test, which is according to the results in Section \ref{section4} in the main text.

Next, it is clearly observed from Figures \ref{power1}-\ref{power12} that the power performance of the proposed test statistics $t_l$ and $t_m$ are quite well and comparable.  To sum up, $t_m$ with $\lambda=0.5$ performs well in terms of size and power, which is according to the statements in Section \ref{section4} that we reduce DiE and VEE while keeping its power performance.

%%%%%%%%%%%%%%rightside test power

%%%%%%%%%%%%%%%%%%%%leftside test

%

\subsection{Example 2}

%\textcolor{green}{The second experiment}
We observe the following findings from Tables \ref{joint1}-\ref{joint5}, \ref{mar_t1}-\ref{mar_t5}, \ref{mar_r1}-\ref{mar_r5} and \ref{mar_l1}-\ref{mar_l5}. First, for the joint test $H_0:\beta=0$, the proposed test statistic ${Q_l}$ still suffers from multiple predictors ($K\geq 3$) and the size distortion grows bigger as the number of predictors K grows bigger. Meanwhile, the proposed test statistic ${Q_m}$ is free of size distortion for different settings with $\lambda=0.1,0.25,0.5,0.75,0.9$. With $\lambda=0.1$ and $K=6$, ${Q_m}$ tends to slightly under-reject the null hypothesis $H_0:\beta=0$. It is not surprising since ${Q_l}$ still suffers the DiE and the VEE while ${Q_m}$ does not. Second, the size performance of ${Q_m^t}$ is better than that of ${Q_l^t}$ for both two-sided and one-sided marginal tests although ${Q_m^t}$ suffers a little size distortion. In particular, the size performance of ${Q_l^t}$ and ${Q_m^t}$ are comparable for right side marginal test $H_0:\beta_{i}=0$. And the size performance of ${Q_m^t}$ is much better than that of ${Q_l^t}$ for two-sided marginal test $H_0:\beta_{i}=0$ vs $H_a:\beta_{i}\neq 0$ and left side marginal test $H_0:\beta_{i}=0$ vs $H_a:\beta_{i}< 0$. Third, for two-sided marginal test $H_0:\beta_{i}=0$ vs $H_a:\beta_{i}\neq 0$, right side marginal test $H_0:\beta_{i}=0$ vs $H_a:\beta_{i}> 0$ and left side marginal test $H_0:\beta_{i}=0$ vs $H_a:\beta_{i}< 0$, the size performances of ${Q_l^t}$ and ${Q_m^t}$ become slightly better as $\lambda$ grows bigger while their power performances become worse. Therefore, we recommend the empirical researchers to set $\lambda=0.5$. Fourth, the power performances of ${Q_l^t}$ and ${Q_m^t}$ are comparable and quite well.

\begin{sidewaystable}
 \centering
 \caption{Size Performance ($\%$) of $t_l$ and $t_m$ for $H_0:\beta=0$ vs $H_a:\beta<0$}
 \resizebox{\textwidth}{!}{
 \begin{tabular}{c|c|cccc|cccc|cccc|cccc|cccc}
 \hline
   & $\lambda$ & \multicolumn{4}{c|}{0.1}  & \multicolumn{4}{c|}{0.25}  & \multicolumn{4}{c|}{0.5}  & \multicolumn{4}{c|}{0.75}  & \multicolumn{4}{c}{0.9} \\
 \hline
   & $\phi$ & 0.95 & 0.5 & -0.1 & -0.95 & 0.95 & 0.5 & -0.1 & -0.95 & 0.95 & 0.5 & -0.1 & -0.95 & 0.95 & 0.5 & -0.1 & -0.95 & 0.95 & 0.5 & -0.1 & -0.95 \\
 \hline
 \multirow{7}[0]{*}{$t_l$} & Case 1 & 8.6 & 6.9 & 4.8 & 3.6 & 8.9 & 6.8 & 4.6 & 3.4 & 8.4 & 7.1 & 4.6 & 2.9 & 8.6 & 7.0 & 4.6 & 2.9 & 9.0 & 6.9 & 4.6 & 3.2 \\
   & Case 2 & 7.1 & 6.4 & 4.5 & 3.9 & 6.7 & 6.3 & 4.7 & 4.0 & 7.8 & 5.9 & 5.0 & 3.5 & 7.9 & 6.3 & 4.8 & 3.7 & 7.6 & 6.1 & 4.8 & 4.0 \\
   & Case 3 & 5.8 & 6.2 & 4.8 & 4.1 & 6.6 & 6.0 & 4.9 & 4.1 & 6.3 & 6.2 & 4.5 & 3.6 & 7.0 & 6.5 & 4.8 & 3.6 & 7.6 & 5.8 & 4.5 & 3.5 \\
   & Case 4 & 6.1 & 5.7 & 4.5 & 4.4 & 6.4 & 5.9 & 5.1 & 4.1 & 6.8 & 6.0 & 5.0 & 3.9 & 6.1 & 5.5 & 4.9 & 4.0 & 7.3 & 6.2 & 5.1 & 3.8 \\
   & Case 5 & 5.9 & 5.6 & 4.8 & 4.3 & 5.3 & 5.2 & 5.0 & 4.2 & 5.7 & 5.7 & 5.0 & 4.5 & 6.4 & 5.8 & 5.0 & 3.9 & 6.9 & 5.9 & 5.0 & 4.1 \\
   & Case 6 & 5.4 & 5.4 & 5.7 & 4.6 & 5.0 & 5.5 & 4.7 & 4.9 & 5.3 & 5.2 & 4.9 & 4.1 & 5.5 & 5.7 & 4.8 & 4.3 & 6.2 & 5.9 & 5.1 & 3.8 \\
   & Case 7 & 5.2 & 4.8 & 5.0 & 4.7 & 5.1 & 5.5 & 4.8 & 4.8 & 5.1 & 5.3 & 4.8 & 4.3 & 5.6 & 5.4 & 4.8 & 4.7 & 5.6 & 5.2 & 4.9 & 4.2 \\
 \hline
 \multirow{7}[0]{*}{$t_m$} & Case 1 & 4.6 & 4.9 & 4.2 & 4.6 & 5.0 & 4.6 & 4.3 & 4.3 & 4.9 & 4.9 & 4.3 & 4.0 & 4.8 & 5.2 & 4.3 & 3.9 & 4.7 & 4.9 & 4.2 & 4.0 \\
   & Case 2 & 4.2 & 4.7 & 4.1 & 5.2 & 4.0 & 4.6 & 4.2 & 5.4 & 4.7 & 4.3 & 4.9 & 4.9 & 4.8 & 4.5 & 4.4 & 5.0 & 4.5 & 4.3 & 4.3 & 5.1 \\
   & Case 3 & 3.9 & 4.8 & 4.4 & 5.3 & 4.3 & 4.4 & 4.5 & 5.5 & 4.0 & 4.8 & 4.2 & 4.8 & 4.7 & 5.0 & 4.4 & 4.9 & 5.1 & 4.1 & 4.0 & 4.7 \\
   & Case 4 & 4.4 & 4.8 & 4.0 & 5.4 & 4.7 & 4.8 & 4.8 & 5.1 & 5.2 & 4.8 & 4.7 & 5.0 & 4.6 & 4.4 & 4.5 & 5.0 & 5.3 & 4.9 & 4.7 & 4.8 \\
   & Case 5 & 5.8 & 5.5 & 4.8 & 4.7 & 5.1 & 5.0 & 5.0 & 4.7 & 5.4 & 5.5 & 5.0 & 4.9 & 6.1 & 5.6 & 5.0 & 4.3 & 6.6 & 5.8 & 4.9 & 4.4 \\
   & Case 6 & 5.4 & 5.4 & 5.7 & 4.6 & 5.0 & 5.5 & 4.7 & 4.9 & 5.3 & 5.2 & 4.9 & 4.1 & 5.5 & 5.7 & 4.8 & 4.3 & 6.2 & 5.9 & 5.1 & 3.8 \\
   & Case 7 & 5.2 & 4.8 & 5.0 & 4.7 & 5.1 & 5.5 & 4.8 & 4.8 & 5.1 & 5.3 & 4.8 & 4.3 & 5.6 & 5.4 & 4.8 & 4.7 & 5.6 & 5.2 & 4.9 & 4.2 \\
 \hline
 \end{tabular}%
 }
 \label{size3}%
\end{sidewaystable}

% Table generated by Excel2LaTeX from sheet 'lam0.1'
\begin{table}[H]
 \centering
 \caption{Result (\%) for Joint Test $H_0:\beta=0$ with $\lambda=0.1$}
 \begin{tabular}{c|c|ccccccccc}
 \hline
 \multicolumn{2}{c|}{b} & 0  & 0.02 & 0.04 & 0.06 & 0.08 & 0.1 & 0.12 & 0.14 & 0.16 \\
 \hline
 \multirow{9}[0]{*}{${Q_l}$} & K=2 & 5.8 & 24.0 & 63.2 & 83.5 & 91.3 & 94.3 & 96.2 & 97.0 & 98.0 \\
   & K=3 & 6.7 & 15.9 & 43.9 & 70.0 & 83.5 & 91.3 & 95.3 & 97.0 & 98.0 \\
   & K=4 & 6.4 & 18.3 & 50.6 & 77.4 & 91.2 & 95.3 & 98.0 & 98.7 & 99.5 \\
   & K=5 & 6.7 & 16.4 & 47.2 & 74.5 & 88.6 & 95.2 & 98.4 & 99.1 & 99.6 \\
   & K=6 & 6.1 & 14.3 & 40.7 & 68.6 & 86.3 & 93.6 & 97.3 & 98.6 & 99.2 \\
   & K=7 & 6.8 & 11.8 & 33.7 & 60.8 & 79.7 & 90.9 & 95.7 & 97.6 & 98.9 \\
   & K=8 & 7.2 & 12.0 & 32.3 & 57.9 & 77.6 & 88.9 & 94.9 & 97.5 & 99.0 \\
   & K=9 & 7.4 & 10.3 & 26.2 & 50.3 & 71.4 & 85.2 & 91.9 & 96.5 & 98.3 \\
   & K=10 & 7.2 & 8.5 & 21.8 & 42.5 & 64.2 & 79.3 & 89.2 & 93.7 & 97.2 \\
 \hline
 \multirow{9}[0]{*}{${Q_m}$} & K=2 & 4.3 & 19.0 & 58.3 & 81.1 & 90.2 & 93.5 & 95.5 & 96.6 & 97.7 \\
   & K=3 & 4.7 & 12.6 & 39.4 & 65.8 & 80.8 & 89.8 & 94.6 & 96.6 & 97.8 \\
   & K=4 & 4.5 & 19.1 & 53.4 & 79.0 & 92.0 & 95.6 & 98.1 & 98.8 & 99.5 \\
   & K=5 & 4.4 & 17.6 & 51.0 & 76.8 & 89.7 & 95.8 & 98.5 & 99.1 & 99.7 \\
   & K=6 & 3.7 & 15.1 & 44.4 & 72.1 & 87.9 & 94.6 & 97.6 & 98.7 & 99.2 \\
   & K=7 & 4.5 & 12.4 & 37.2 & 65.8 & 82.9 & 92.2 & 96.5 & 97.8 & 99.0 \\
   & K=8 & 4.8 & 11.9 & 35.2 & 62.5 & 81.3 & 90.8 & 95.5 & 97.9 & 99.2 \\
   & K=9 & 4.8 & 9.7 & 30.0 & 55.7 & 76.0 & 88.1 & 93.8 & 97.1 & 98.5 \\
   & K=10 & 4.4 & 8.1 & 24.9 & 47.6 & 70.3 & 83.4 & 91.8 & 95.0 & 97.8 \\
 \hline
 \end{tabular}%
 \label{joint1}%
 \begin{tablenotes}
  % \scriptsize
  \item[1] $\beta = \frac{b}{(1+K)/2}$.
 \end{tablenotes}
\end{table}%

% Table generated by Excel2LaTeX from sheet 'lam0.1'
\begin{table}[H]
 \centering
 \caption{Result (\%) for Joint Test $H_0:\beta=0$ with $\lambda=0.25$}
 \begin{tabular}{c|c|ccccccccc}
 \hline
 \multicolumn{2}{c|}{b} & 0  & 0.02 & 0.04 & 0.06 & 0.08 & 0.1 & 0.12 & 0.14 & 0.16 \\
 \hline
  \multirow{9}[0]{*}{${Q_l}$} & K=2 & 5.7 & 24.9 & 64.8 & 83.7 & 91.6 & 94.4 & 96.0 & 96.7 & 97.5 \\
   & K=3 & 7.7 & 17.4 & 44.9 & 69.9 & 84.6 & 91.8 & 95.2 & 96.8 & 97.9 \\
   & K=4 & 7.4 & 18.0 & 52.1 & 78.2 & 90.5 & 95.7 & 98.1 & 99.0 & 99.5 \\
   & K=5 & 7.7 & 16.0 & 47.4 & 74.2 & 89.0 & 95.3 & 98.0 & 99.0 & 99.5 \\
   & K=6 & 7.2 & 14.9 & 41.2 & 68.9 & 86.4 & 94.0 & 97.7 & 98.7 & 99.3 \\
   & K=7 & 6.6 & 12.5 & 34.6 & 61.4 & 80.4 & 91.1 & 96.0 & 98.2 & 98.7 \\
   & K=8 & 7.2 & 11.8 & 31.5 & 58.6 & 77.0 & 89.3 & 95.3 & 97.7 & 98.9 \\
   & K=9 & 6.9 & 10.1 & 26.9 & 51.0 & 71.8 & 85.0 & 93.0 & 96.3 & 98.0 \\
   & K=10 & 7.2 & 9.9 & 22.6 & 44.3 & 65.0 & 80.7 & 89.2 & 94.4 & 97.4 \\
 \hline
 \multirow{9}[0]{*}{${Q_m}$} & K=2 & 4.3 & 19.9 & 59.9 & 81.5 & 90.3 & 93.7 & 95.7 & 96.5 & 97.3 \\
   & K=3 & 5.1 & 13.7 & 40.6 & 66.2 & 82.5 & 90.4 & 94.5 & 96.4 & 97.6 \\
   & K=4 & 4.8 & 20.3 & 55.7 & 80.4 & 91.4 & 96.0 & 98.2 & 99.0 & 99.4 \\
   & K=5 & 5.3 & 17.4 & 51.4 & 77.2 & 90.1 & 95.9 & 98.4 & 99.1 & 99.6 \\
   & K=6 & 4.8 & 16.3 & 45.9 & 73.1 & 89.2 & 95.2 & 98.2 & 98.9 & 99.4 \\
   & K=7 & 4.7 & 13.7 & 39.1 & 66.8 & 83.9 & 92.5 & 96.8 & 98.5 & 98.9 \\
   & K=8 & 4.9 & 12.4 & 36.6 & 64.0 & 81.4 & 91.4 & 96.4 & 97.9 & 99.0 \\
   & K=9 & 4.5 & 10.4 & 31.0 & 57.4 & 77.0 & 88.5 & 94.6 & 97.2 & 98.7 \\
   & K=10 & 5.0 & 9.8 & 26.6 & 50.9 & 70.7 & 84.8 & 91.8 & 95.9 & 98.1 \\
 \hline
 \end{tabular}%
 \label{joint2}%
 \begin{tablenotes}
  % \scriptsize
  \item[1] $\beta = \frac{b}{(1+K)/2}$.
 \end{tablenotes}
\end{table}%

% Table generated by Excel2LaTeX from sheet 'lam0.1'
\begin{table}[H]
 \centering
 \caption{Result (\%) for Joint Test $H_0:\beta=0$ with $\lambda=0.75$}
 \begin{tabular}{c|c|ccccccccc}
 \hline
 \multicolumn{2}{c|}{b} & 0  & 0.02 & 0.04 & 0.06 & 0.08 & 0.1 & 0.12 & 0.14 & 0.16 \\
 \hline
  \multirow{9}[0]{*}{${Q_l}$} & K=2 & 5.8 & 21.7 & 54.7 & 76.1 & 85.2 & 89.5 & 91.8 & 94.1 & 95.2 \\
   & K=3 & 6.7 & 15.9 & 40.8 & 65.0 & 79.2 & 88.1 & 91.7 & 94.2 & 96.1 \\
   & K=4 & 7.5 & 19.2 & 51.3 & 76.5 & 88.5 & 94.3 & 97.4 & 98.4 & 99.4 \\
   & K=5 & 7.2 & 16.7 & 45.3 & 72.8 & 86.9 & 94.6 & 97.5 & 98.8 & 99.2 \\
   & K=6 & 7.6 & 14.6 & 41.3 & 65.7 & 85.0 & 93.1 & 97.4 & 98.8 & 99.4 \\
   & K=7 & 6.4 & 11.8 & 33.1 & 59.1 & 79.3 & 89.6 & 95.4 & 98.1 & 99.0 \\
   & K=8 & 7.7 & 11.9 & 31.0 & 57.1 & 77.1 & 89.2 & 94.7 & 97.6 & 98.6 \\
   & K=9 & 6.9 & 11.3 & 25.6 & 48.9 & 69.7 & 85.1 & 91.8 & 96.2 & 97.9 \\
   & K=10 & 7.4 & 10.1 & 22.0 & 42.7 & 62.6 & 78.8 & 88.6 & 94.2 & 96.5 \\
 \hline
 \multirow{9}[0]{*}{${Q_m}$} & K=2 & 4.7 & 17.4 & 49.7 & 73.5 & 83.4 & 88.4 & 90.6 & 93.5 & 94.9 \\
   & K=3 & 4.4 & 12.2 & 36.2 & 60.9 & 76.9 & 86.2 & 90.4 & 93.5 & 95.7 \\
   & K=4 & 4.9 & 20.1 & 54.0 & 78.2 & 90.0 & 94.9 & 97.5 & 98.4 & 99.4 \\
   & K=5 & 5.0 & 18.0 & 49.6 & 76.0 & 88.6 & 94.9 & 97.7 & 98.8 & 99.3 \\
   & K=6 & 5.1 & 15.2 & 45.9 & 70.0 & 87.5 & 94.0 & 98.0 & 98.9 & 99.4 \\
   & K=7 & 4.0 & 12.3 & 37.9 & 64.5 & 82.6 & 91.5 & 96.1 & 98.1 & 99.1 \\
   & K=8 & 5.2 & 12.4 & 34.9 & 62.6 & 81.3 & 91.2 & 95.5 & 98.0 & 98.9 \\
   & K=9 & 4.7 & 11.7 & 29.1 & 54.0 & 74.9 & 88.1 & 93.7 & 96.9 & 98.3 \\
   & K=10 & 4.9 & 9.6 & 25.4 & 49.5 & 68.3 & 83.2 & 90.9 & 95.5 & 97.0 \\
 \hline
 \end{tabular}%
 \label{joint4}%
 \begin{tablenotes}
  % \scriptsize
  \item[1] $\beta = \frac{b}{(1+K)/2}$.
 \end{tablenotes}
\end{table}%

% Table generated by Excel2LaTeX from sheet 'lam0.1'
\begin{table}[H]
 \centering
 \caption{Result (\%) for Joint Test $H_0:\beta=0$ with $\lambda=0.9$}
 \begin{tabular}{c|c|ccccccccc}
 \hline
 \multicolumn{2}{c|}{b} & 0  & 0.02 & 0.04 & 0.06 & 0.08 & 0.1 & 0.12 & 0.14 & 0.16 \\
 \hline
  \multirow{9}[0]{*}{${Q_l}$} & K=2 & 5.6 & 20.6 & 50.9 & 72.3 & 81.6 & 87.3 & 90.8 & 91.8 & 93.3 \\
   & K=3 & 6.5 & 14.8 & 39.7 & 61.3 & 78.1 & 86.3 & 91.0 & 93.1 & 95.2 \\
   & K=4 & 6.9 & 17.9 & 47.1 & 73.7 & 87.8 & 94.4 & 96.9 & 98.5 & 99.1 \\
   & K=5 & 6.9 & 16.3 & 43.5 & 70.0 & 86.7 & 93.5 & 97.4 & 98.4 & 99.2 \\
   & K=6 & 7.2 & 13.6 & 39.1 & 65.2 & 83.0 & 92.5 & 96.7 & 98.1 & 99.0 \\
   & K=7 & 6.8 & 11.4 & 31.4 & 58.5 & 78.5 & 89.3 & 94.7 & 97.6 & 98.6 \\
   & K=8 & 7.5 & 10.9 & 30.2 & 55.5 & 75.2 & 88.1 & 93.9 & 97.1 & 98.7 \\
   & K=9 & 7.3 & 11.0 & 25.9 & 47.3 & 68.4 & 82.6 & 91.6 & 95.7 & 97.8 \\
   & K=10 & 6.9 & 10.3 & 21.6 & 40.9 & 62.4 & 78.1 & 88.0 & 93.5 & 96.4 \\
 \hline
 \multirow{9}[0]{*}{${Q_m}$} & K=2 & 4.2 & 16.4 & 46.2 & 68.8 & 79.5 & 85.6 & 89.6 & 91.0 & 92.3 \\
   & K=3 & 4.1 & 10.9 & 35.5 & 57.6 & 75.2 & 84.4 & 89.7 & 92.5 & 94.6 \\
   & K=4 & 4.6 & 18.1 & 50.4 & 75.2 & 88.5 & 94.7 & 97.2 & 98.4 & 99.1 \\
   & K=5 & 4.9 & 16.3 & 46.7 & 72.3 & 87.5 & 94.0 & 97.4 & 98.5 & 99.3 \\
   & K=6 & 4.4 & 14.0 & 42.0 & 68.9 & 85.3 & 93.3 & 97.0 & 98.2 & 99.0 \\
   & K=7 & 4.6 & 11.1 & 34.3 & 62.5 & 81.4 & 91.0 & 95.3 & 97.9 & 98.8 \\
   & K=8 & 4.5 & 10.7 & 33.0 & 59.3 & 78.6 & 90.5 & 94.9 & 97.3 & 98.8 \\
   & K=9 & 4.9 & 10.3 & 28.8 & 52.3 & 72.4 & 85.6 & 93.0 & 96.6 & 98.2 \\
   & K=10 & 4.1 & 9.5 & 23.0 & 45.6 & 67.1 & 81.7 & 90.4 & 94.7 & 96.9 \\
 \hline
 \end{tabular}%
 \label{joint5}%
 \begin{tablenotes}
  % \scriptsize
  \item[1] $\beta = \frac{b}{(1+K)/2}$.
 \end{tablenotes}
\end{table}%

% Table generated by Excel2LaTeX from sheet 'H0x1=0'
\begin{table}[H]
 \centering
 \caption{Result (\%) for Two-sided Marginal Test $H_0:\beta_i=0$ vs $H_0:\beta_i \neq 0$ with $\lambda=0.1$}
% \resizebox{\textwidth}{!}{
  \begin{tabular}{c|c|ccccccccc}
  \hline
  \multicolumn{2}{c|}{$\beta$} & 0  & 0.05 & 0.1 & 0.15 & 0.2 & 0.25 & 0.3 & 0.35 & 0.4 \\
  \hline
  \multirow{10}[0]{*}{${Q_l^t}$} & i=1 & 7.0 & 70.0 & 90.7 & 95.4 & 96.8 & 97.3 & 97.5 & 98.0 & 98.2 \\
   & i=2 & 6.9 & 45.8 & 82.7 & 93.0 & 95.5 & 96.6 & 97.6 & 97.9 & 98.2 \\
   & i=3 & 6.9 & 53.1 & 83.1 & 90.5 & 93.3 & 94.8 & 95.3 & 96.1 & 95.9 \\
   & i=4 & 6.8 & 54.7 & 88.5 & 94.8 & 97.1 & 97.5 & 98.4 & 98.3 & 98.6 \\
   & i=5 & 5.9 & 29.5 & 71.2 & 89.7 & 94.6 & 96.4 & 97.3 & 98.1 & 98.2 \\
   & i=6 & 5.6 & 24.7 & 64.4 & 85.6 & 93.7 & 95.7 & 97.4 & 97.6 & 98.1 \\
   & i=7 & 5.3 & 16.9 & 49.1 & 76.5 & 88.8 & 93.6 & 95.6 & 96.7 & 97.7 \\
   & i=8 & 6.2 & 33.6 & 74.4 & 89.8 & 94.4 & 96.7 & 97.5 & 97.7 & 98.3 \\
   & i=9 & 5.3 & 19.3 & 52.5 & 78.7 & 89.9 & 94.3 & 96.2 & 96.9 & 97.9 \\
   & i=10 & 5.3 & 23.0 & 58.5 & 82.2 & 91.6 & 95.0 & 96.8 & 97.5 & 97.8 \\
  \hline
 \multirow{10}[0]{*}{${Q_m^t}$} & i=1 & 5.8 & 69.2 & 90.2 & 95.2 & 96.4 & 97.1 & 97.4 & 97.8 & 98.1 \\
   & i=2 & 5.8 & 43.4 & 81.6 & 92.5 & 95.1 & 96.4 & 97.4 & 97.8 & 98.1 \\
   & i=3 & 6.1 & 50.1 & 81.6 & 89.8 & 92.8 & 94.4 & 95.0 & 95.6 & 95.6 \\
   & i=4 & 5.7 & 56.6 & 88.9 & 95.0 & 97.1 & 97.4 & 98.2 & 98.2 & 98.6 \\
   & i=5 & 5.1 & 30.7 & 71.8 & 89.8 & 94.4 & 96.2 & 97.3 & 98.0 & 98.0 \\
   & i=6 & 5.1 & 25.9 & 65.1 & 85.7 & 93.6 & 95.7 & 97.4 & 97.6 & 98.1 \\
   & i=7 & 5.1 & 18.5 & 51.2 & 77.5 & 89.0 & 93.7 & 95.7 & 96.8 & 97.7 \\
   & i=8 & 5.5 & 32.3 & 73.0 & 89.0 & 94.0 & 96.4 & 97.4 & 97.6 & 98.2 \\
   & i=9 & 5.1 & 19.4 & 52.3 & 78.3 & 89.6 & 94.1 & 96.0 & 96.8 & 97.7 \\
   & i=10 & 4.8 & 21.7 & 56.5 & 81.2 & 91.0 & 94.6 & 96.6 & 97.2 & 97.7 \\
   \hline
 \end{tabular}%
% }
 \label{mar_t1}%
%  \begin{tablenotes}
%  \scriptsize
%  \item[1] $\beta_1 = 10^{-\alpha}b/T^{(1+\alpha)/2}$ and $\beta_2 = 10^{-\alpha}/T^{(1+\alpha)/2}$. T=750.
% \end{tablenotes}
\end{table}%

\begin{table}[H]
 \centering
 \caption{Result (\%) for Two-sided Marginal Test $H_0:\beta_i=0$ vs $H_0:\beta_i \neq 0$ with $\lambda=0.25$}
% \resizebox{\textwidth}{!}{
  \begin{tabular}{c|c|ccccccccc}
  \hline
  \multicolumn{2}{c|}{$\beta$} & 0  & 0.05 & 0.1 & 0.15 & 0.2 & 0.25 & 0.3 & 0.35 & 0.4 \\
  \hline
    & i=2 & 6.9 & 45.6 & 80.4 & 90.9 & 94.4 & 96.1 & 97.0 & 97.2 & 97.9 \\
   & i=3 & 6.8 & 49.3 & 78.7 & 88.2 & 91.6 & 93.8 & 94.3 & 95.0 & 95.4 \\
   & i=4 & 6.6 & 53.2 & 86.8 & 94.1 & 96.4 & 97.3 & 97.5 & 98.2 & 98.5 \\
   & i=5 & 5.8 & 29.2 & 69.5 & 87.9 & 93.9 & 95.7 & 97.0 & 97.7 & 98.0 \\
   & i=6 & 6.1 & 24.4 & 63.5 & 84.9 & 92.5 & 95.4 & 96.4 & 97.4 & 97.6 \\
   & i=7 & 5.5 & 16.5 & 48.2 & 74.7 & 87.9 & 93.1 & 95.3 & 97.1 & 97.5 \\
   & i=8 & 6.7 & 32.9 & 73.0 & 88.9 & 93.9 & 95.9 & 96.4 & 97.5 & 97.7 \\
   & i=9 & 5.4 & 18.9 & 52.1 & 77.5 & 88.9 & 93.5 & 95.8 & 96.8 & 97.7 \\
   & i=10 & 5.9 & 21.6 & 58.1 & 82.1 & 91.0 & 94.4 & 96.2 & 97.1 & 97.6 \\
  \hline
 \multirow{10}[0]{*}{${Q_m^t}$} & i=1 & 6.4 & 66.4 & 89.1 & 93.6 & 95.6 & 96.5 & 97.2 & 97.6 & 97.6 \\
   & i=2 & 6.0 & 43.4 & 79.2 & 90.4 & 94.1 & 96.0 & 96.9 & 97.0 & 97.7 \\
   & i=3 & 6.1 & 46.7 & 77.1 & 87.4 & 91.2 & 93.4 & 93.9 & 94.7 & 95.1 \\
   & i=4 & 5.9 & 55.8 & 87.0 & 94.2 & 96.3 & 97.3 & 97.3 & 98.1 & 98.5 \\
   & i=5 & 5.4 & 31.0 & 70.3 & 88.0 & 93.8 & 95.7 & 97.0 & 97.6 & 98.0 \\
   & i=6 & 5.5 & 26.2 & 64.7 & 85.1 & 92.5 & 95.4 & 96.4 & 97.3 & 97.6 \\
   & i=7 & 5.5 & 18.7 & 50.6 & 76.1 & 88.5 & 93.2 & 95.4 & 97.1 & 97.4 \\
   & i=8 & 6.0 & 32.0 & 72.0 & 88.3 & 93.7 & 95.6 & 96.2 & 97.3 & 97.7 \\
   & i=9 & 5.3 & 18.9 & 51.9 & 76.8 & 88.8 & 93.1 & 95.6 & 96.6 & 97.5 \\
   & i=10 & 5.7 & 21.0 & 56.8 & 80.7 & 90.3 & 94.0 & 95.9 & 97.0 & 97.4 \\
   \hline
 \end{tabular}%
% }
 \label{mar_t2}%
%  \begin{tablenotes}
%  \scriptsize
%  \item[1] $\beta_1 = 10^{-\alpha}b/T^{(1+\alpha)/2}$ and $\beta_2 = 10^{-\alpha}/T^{(1+\alpha)/2}$. T=750.
% \end{tablenotes}
\end{table}%

\begin{table}[H]
 \centering
 \caption{Result (\%) for Two-sided Marginal Test $H_0:\beta_i=0$ vs $H_0:\beta_i \neq 0$ with $\lambda=0.75$}
  \begin{tabular}{c|c|ccccccccc}
  \hline
  \multicolumn{2}{c|}{$\beta$} & 0  & 0.05 & 0.1 & 0.15 & 0.2 & 0.25 & 0.3 & 0.35 & 0.4 \\
  \hline
  \multirow{10}[0]{*}{${Q_l^t}$} & i=1 & 7.3 & 51.1 & 74.9 & 83.0 & 86.8 & 89.4 & 90.9 & 90.8 & 92.0 \\
   & i=2 & 7.4 & 34.6 & 66.1 & 78.9 & 84.1 & 87.2 & 89.6 & 91.2 & 91.7 \\
   & i=3 & 7.1 & 33.5 & 58.4 & 69.7 & 75.8 & 79.1 & 81.4 & 82.8 & 84.3 \\
   & i=4 & 6.7 & 42.5 & 75.2 & 85.0 & 89.4 & 90.7 & 92.6 & 93.2 & 94.1 \\
   & i=5 & 6.3 & 25.2 & 59.3 & 78.0 & 85.7 & 89.3 & 91.9 & 93.0 & 93.0 \\
   & i=6 & 6.0 & 21.3 & 53.9 & 76.1 & 85.1 & 89.0 & 91.9 & 92.7 & 93.7 \\
   & i=7 & 5.4 & 15.0 & 42.3 & 66.2 & 80.4 & 85.9 & 89.8 & 91.6 & 93.0 \\
   & i=8 & 6.6 & 26.8 & 61.9 & 78.2 & 84.7 & 89.4 & 90.6 & 92.2 & 93.1 \\
   & i=9 & 5.7 & 16.6 & 44.6 & 68.3 & 80.5 & 86.1 & 89.6 & 91.8 & 93.0 \\
   & i=10 & 6.0 & 19.5 & 50.5 & 71.4 & 82.8 & 87.6 & 90.3 & 92.3 & 93.2 \\
 \hline
 \multirow{10}[0]{*}{${Q_m^t}$} & i=1 & 6.3 & 49.5 & 73.4 & 82.2 & 86.1 & 88.9 & 90.4 & 90.3 & 91.3 \\
   & i=2 & 6.2 & 32.4 & 64.4 & 78.0 & 83.3 & 86.9 & 89.1 & 90.7 & 91.3 \\
   & i=3 & 6.3 & 31.9 & 57.2 & 68.7 & 74.8 & 78.4 & 80.7 & 82.2 & 83.6 \\
   & i=4 & 6.0 & 44.0 & 75.1 & 84.5 & 89.0 & 90.6 & 92.4 & 92.7 & 93.7 \\
   & i=5 & 5.7 & 25.9 & 59.2 & 77.9 & 85.3 & 88.8 & 91.4 & 92.8 & 93.0 \\
   & i=6 & 5.5 & 22.3 & 54.3 & 76.0 & 85.0 & 89.0 & 91.7 & 92.5 & 93.6 \\
   & i=7 & 5.1 & 16.5 & 44.2 & 67.2 & 80.7 & 86.0 & 89.8 & 91.4 & 92.8 \\
   & i=8 & 5.8 & 25.7 & 60.9 & 77.3 & 83.7 & 88.9 & 90.2 & 92.1 & 92.8 \\
   & i=9 & 5.4 & 16.8 & 43.9 & 67.3 & 79.8 & 85.8 & 89.4 & 91.7 & 92.8 \\
   & i=10 & 5.6 & 18.6 & 49.2 & 70.0 & 82.1 & 87.0 & 89.8 & 91.6 & 92.9 \\
   \hline
 \end{tabular}%
 \label{mar_t4}%
%  \begin{tablenotes}
%  \scriptsize
%  \item[1] $\beta_1 = 10^{-\alpha}b/T^{(1+\alpha)/2}$ and $\beta_2 = 10^{-\alpha}/T^{(1+\alpha)/2}$. T=750.
% \end{tablenotes}
\end{table}%

\begin{table}[H]
 \centering
 \caption{Result (\%) for Two-sided Marginal Test $H_0:\beta_i=0$ vs $H_0:\beta_i \neq 0$ with $\lambda=0.9$}
  \begin{tabular}{c|c|ccccccccc}
  \hline
  \multicolumn{2}{c|}{$\beta$} & 0  & 0.05 & 0.1 & 0.15 & 0.2 & 0.25 & 0.3 & 0.35 & 0.4 \\
  \hline
  \multirow{10}[0]{*}{${Q_l^t}$} & i=1 & 6.8 & 44.2 & 69.0 & 77.2 & 81.8 & 84.2 & 85.1 & 87.0 & 87.3 \\
   & i=2 & 6.7 & 30.3 & 60.5 & 72.3 & 78.7 & 82.4 & 84.5 & 86.2 & 88.4 \\
   & i=3 & 6.2 & 27.9 & 50.0 & 61.8 & 67.1 & 71.3 & 73.7 & 75.8 & 76.8 \\
   & i=4 & 6.8 & 38.3 & 69.2 & 80.2 & 85.3 & 87.5 & 89.4 & 90.2 & 90.9 \\
   & i=5 & 6.1 & 22.8 & 54.5 & 73.5 & 81.6 & 85.9 & 87.8 & 90.3 & 91.3 \\
   & i=6 & 6.1 & 19.7 & 49.8 & 70.8 & 79.9 & 85.0 & 88.6 & 90.1 & 90.9 \\
   & i=7 & 5.8 & 14.5 & 38.2 & 62.8 & 75.1 & 82.1 & 86.1 & 88.7 & 90.6 \\
   & i=8 & 6.2 & 24.4 & 56.1 & 73.1 & 80.2 & 84.1 & 86.6 & 89.3 & 90.5 \\
   & i=9 & 5.5 & 15.3 & 41.6 & 63.8 & 76.8 & 82.6 & 86.5 & 88.1 & 89.7 \\
   & i=10 & 5.8 & 18.2 & 45.9 & 66.8 & 77.8 & 83.3 & 86.9 & 89.4 & 91.0 \\
 \hline
 \multirow{10}[0]{*}{${Q_m^t}$} & i=1 & 5.7 & 42.1 & 66.6 & 76.2 & 80.8 & 83.3 & 84.1 & 86.3 & 86.6 \\
   & i=2 & 5.7 & 28.1 & 58.9 & 71.0 & 77.9 & 81.6 & 83.8 & 85.7 & 87.8 \\
   & i=3 & 5.5 & 26.2 & 48.4 & 60.3 & 65.8 & 70.0 & 72.7 & 74.8 & 75.8 \\
   & i=4 & 5.5 & 38.5 & 68.3 & 79.4 & 84.7 & 86.8 & 88.9 & 89.8 & 90.6 \\
   & i=5 & 5.4 & 22.8 & 54.0 & 72.7 & 80.9 & 85.6 & 87.4 & 89.7 & 90.9 \\
   & i=6 & 5.3 & 20.1 & 50.0 & 70.8 & 79.5 & 84.6 & 88.2 & 89.8 & 90.4 \\
   & i=7 & 5.1 & 14.6 & 39.2 & 63.7 & 75.3 & 82.1 & 85.6 & 88.5 & 90.4 \\
   & i=8 & 5.3 & 22.7 & 54.2 & 71.8 & 79.4 & 83.8 & 85.9 & 88.7 & 89.9 \\
   & i=9 & 5.1 & 14.9 & 40.9 & 63.0 & 76.0 & 82.2 & 85.9 & 87.5 & 89.2 \\
   & i=10 & 5.2 & 17.1 & 43.9 & 65.2 & 76.8 & 82.6 & 86.4 & 88.8 & 90.6 \\
   \hline
 \end{tabular}%
 \label{mar_t5}%
%  \begin{tablenotes}
%  \scriptsize
%  \item[1] $\beta_1 = 10^{-\alpha}b/T^{(1+\alpha)/2}$ and $\beta_2 = 10^{-\alpha}/T^{(1+\alpha)/2}$. T=750.
% \end{tablenotes}
\end{table}%

% Table generated by Excel2LaTeX from sheet 'H0x1=0'
\begin{table}[H]
 \centering
 \caption{Result (\%) for Right Side Marginal Test $H_0:\beta_i=0$ vs $H_0:\beta_i>0$ with $\lambda=0.1$}
% \resizebox{\textwidth}{!}{
  \begin{tabular}{c|c|ccccccccc}
  \hline
  \multicolumn{2}{c|}{$\beta$} & 0  & 0.05 & 0.1 & 0.15 & 0.2 & 0.25 & 0.3 & 0.35 & 0.4 \\
  \hline
 \multirow{10}[0]{*}{${Q_l^t}$} & i=1 & 7.7 & 76.5 & 92.6 & 96.4 & 97.4 & 97.8 & 98.0 & 98.4 & 98.5 \\
   & i=2 & 6.0 & 55.1 & 87.2 & 94.7 & 96.3 & 97.2 & 98.0 & 98.4 & 98.5 \\
   & i=3 & 5.1 & 60.9 & 86.3 & 92.1 & 94.4 & 95.5 & 96.1 & 96.7 & 96.6 \\
   & i=4 & 5.3 & 63.3 & 91.4 & 96.0 & 97.6 & 98.0 & 98.6 & 98.6 & 98.9 \\
   & i=5 & 5.3 & 39.4 & 78.4 & 92.6 & 95.7 & 97.2 & 97.9 & 98.4 & 98.5 \\
   & i=6 & 5.1 & 34.7 & 72.3 & 89.6 & 95.5 & 96.8 & 97.9 & 98.1 & 98.4 \\
   & i=7 & 4.7 & 25.4 & 59.7 & 82.8 & 91.7 & 95.2 & 96.6 & 97.5 & 98.3 \\
   & i=8 & 5.7 & 44.1 & 81.1 & 92.4 & 95.9 & 97.3 & 98.0 & 98.1 & 98.6 \\
   & i=9 & 5.1 & 28.0 & 62.7 & 84.5 & 92.9 & 95.8 & 97.2 & 97.6 & 98.4 \\
   & i=10 & 5.2 & 31.6 & 68.2 & 86.8 & 93.7 & 96.2 & 97.4 & 98.0 & 98.3 \\
 \hline
 \multirow{10}[0]{*}{${Q_m^t}$} & i=1 & 7.6 & 76.2 & 92.3 & 96.2 & 97.2 & 97.7 & 97.9 & 98.3 & 98.4 \\
   & i=2 & 5.3 & 53.3 & 86.0 & 94.2 & 96.2 & 97.1 & 98.0 & 98.2 & 98.5 \\
   & i=3 & 4.2 & 58.3 & 85.6 & 91.6 & 94.2 & 95.4 & 95.9 & 96.4 & 96.3 \\
   & i=4 & 6.1 & 65.4 & 91.8 & 96.0 & 97.6 & 97.9 & 98.5 & 98.5 & 98.9 \\
   & i=5 & 6.0 & 41.0 & 78.3 & 92.5 & 95.8 & 97.1 & 97.8 & 98.4 & 98.4 \\
   & i=6 & 5.9 & 36.3 & 73.4 & 89.8 & 95.4 & 96.7 & 97.9 & 98.1 & 98.4 \\
   & i=7 & 5.8 & 27.6 & 61.7 & 83.5 & 92.0 & 95.3 & 96.7 & 97.6 & 98.2 \\
   & i=8 & 5.4 & 43.1 & 80.1 & 91.9 & 95.5 & 97.2 & 97.9 & 98.0 & 98.6 \\
   & i=9 & 5.4 & 28.4 & 62.6 & 84.2 & 92.6 & 95.7 & 97.1 & 97.5 & 98.2 \\
   & i=10 & 5.1 & 30.9 & 66.6 & 86.4 & 93.4 & 96.0 & 97.3 & 97.8 & 98.1 \\
   \hline
 \end{tabular}%
% }
 \label{mar_r1}%
%  \begin{tablenotes}
%  \scriptsize
%  \item[1] $\beta_1 = 10^{-\alpha}b/T^{(1+\alpha)/2}$ and $\beta_2 = 10^{-\alpha}/T^{(1+\alpha)/2}$. T=750.
% \end{tablenotes}
\end{table}%

\begin{table}[H]
 \centering
 \caption{Result (\%) for Right Side Marginal Test $H_0:\beta_i=0$ vs $H_0:\beta_i>0$ with $\lambda=0.25$}
% \resizebox{\textwidth}{!}{
  \begin{tabular}{c|c|ccccccccc}
  \hline
  \multicolumn{2}{c|}{$\beta$} & 0  & 0.05 & 0.1 & 0.15 & 0.2 & 0.25 & 0.3 & 0.35 & 0.4 \\
  \hline
 \multirow{10}[0]{*}{${Q_l^t}$} & i=1 & 8.3 & 74.1 & 91.9 & 95.3 & 96.5 & 97.3 & 97.7 & 98.0 & 98.1 \\
   & i=2 & 5.7 & 55.0 & 84.4 & 92.7 & 95.6 & 96.8 & 97.6 & 97.7 & 98.3 \\
   & i=3 & 5.1 & 57.3 & 83.0 & 90.5 & 93.0 & 94.9 & 95.2 & 96.0 & 96.2 \\
   & i=4 & 5.4 & 61.9 & 89.8 & 95.4 & 97.2 & 97.8 & 97.9 & 98.5 & 98.8 \\
   & i=5 & 5.7 & 39.4 & 76.9 & 91.0 & 95.3 & 96.5 & 97.7 & 98.2 & 98.4 \\
   & i=6 & 5.4 & 33.7 & 72.1 & 88.7 & 94.5 & 96.3 & 97.1 & 97.8 & 98.1 \\
   & i=7 & 5.1 & 24.7 & 58.9 & 81.3 & 91.0 & 94.8 & 96.4 & 97.8 & 98.0 \\
   & i=8 & 6.4 & 43.2 & 79.6 & 92.0 & 95.3 & 96.8 & 97.2 & 98.0 & 98.2 \\
   & i=9 & 5.5 & 26.8 & 62.1 & 83.2 & 92.1 & 95.2 & 96.6 & 97.6 & 98.1 \\
   & i=10 & 5.7 & 31.0 & 67.5 & 87.2 & 93.3 & 95.7 & 97.0 & 97.7 & 97.9 \\
 \hline
 \multirow{10}[0]{*}{${Q_m^t}$} & i=1 & 8.1 & 73.5 & 91.4 & 95.0 & 96.3 & 97.1 & 97.6 & 97.9 & 98.0 \\
   & i=2 & 5.3 & 53.4 & 84.0 & 92.4 & 95.3 & 96.8 & 97.5 & 97.6 & 98.2 \\
   & i=3 & 4.3 & 55.0 & 81.5 & 89.7 & 92.6 & 94.5 & 95.0 & 95.7 & 96.0 \\
   & i=4 & 6.7 & 64.6 & 89.9 & 95.6 & 97.0 & 97.7 & 97.8 & 98.4 & 98.8 \\
   & i=5 & 6.5 & 41.2 & 77.5 & 91.1 & 95.1 & 96.5 & 97.6 & 98.1 & 98.5 \\
   & i=6 & 6.5 & 36.1 & 72.9 & 88.8 & 94.6 & 96.4 & 97.0 & 97.7 & 98.0 \\
   & i=7 & 6.3 & 27.6 & 61.5 & 82.2 & 91.3 & 94.9 & 96.5 & 97.8 & 97.9 \\
   & i=8 & 6.1 & 42.1 & 78.8 & 91.2 & 95.1 & 96.5 & 97.0 & 97.9 & 98.0 \\
   & i=9 & 5.8 & 27.5 & 62.4 & 82.8 & 91.8 & 94.9 & 96.5 & 97.4 & 98.0 \\
   & i=10 & 5.7 & 30.2 & 66.6 & 86.3 & 92.9 & 95.6 & 96.8 & 97.6 & 97.8 \\
   \hline
 \end{tabular}%
% }
 \label{mar_r2}%
%  \begin{tablenotes}
%  \scriptsize
%  \item[1] $\beta_1 = 10^{-\alpha}b/T^{(1+\alpha)/2}$ and $\beta_2 = 10^{-\alpha}/T^{(1+\alpha)/2}$. T=750.
% \end{tablenotes}
\end{table}%

\begin{table}[H]
 \centering
 \caption{Result (\%) for Right Side Marginal Test $H_0:\beta_i=0$ vs $H_0:\beta_i>0$ with $\lambda=0.75$}
  \begin{tabular}{c|c|ccccccccc}
  \hline
  \multicolumn{2}{c|}{$\beta$} & 0  & 0.05 & 0.1 & 0.15 & 0.2 & 0.25 & 0.3 & 0.35 & 0.4 \\
  \hline
 \multirow{10}[0]{*}{${Q_l^t}$} & i=1 & 8.3 & 58.9 & 79.1 & 85.8 & 88.9 & 90.9 & 92.3 & 92.2 & 93.4 \\
   & i=2 & 6.2 & 42.8 & 71.6 & 82.5 & 86.5 & 89.1 & 91.5 & 92.5 & 93.1 \\
   & i=3 & 6.0 & 41.2 & 64.2 & 73.9 & 79.3 & 82.4 & 84.2 & 85.5 & 86.8 \\
   & i=4 & 5.3 & 51.0 & 79.8 & 87.5 & 91.0 & 92.2 & 93.8 & 94.3 & 94.9 \\
   & i=5 & 5.3 & 34.1 & 67.1 & 82.1 & 88.3 & 91.0 & 93.2 & 94.1 & 94.1 \\
   & i=6 & 5.3 & 29.6 & 62.1 & 80.9 & 88.2 & 91.2 & 93.2 & 93.9 & 94.7 \\
   & i=7 & 4.8 & 22.7 & 52.1 & 73.2 & 84.9 & 88.9 & 91.9 & 92.9 & 94.2 \\
   & i=8 & 5.9 & 35.5 & 68.9 & 82.1 & 86.9 & 91.1 & 92.0 & 93.6 & 94.3 \\
   & i=9 & 5.4 & 24.5 & 54.1 & 74.3 & 84.1 & 88.8 & 91.4 & 93.3 & 94.3 \\
   & i=10 & 5.8 & 28.1 & 59.7 & 77.0 & 86.3 & 90.0 & 91.8 & 93.4 & 94.5 \\
 \hline
 \multirow{10}[0]{*}{${Q_m^t}$} & i=1 & 7.7 & 57.1 & 77.7 & 85.0 & 88.2 & 90.5 & 91.8 & 91.9 & 93.0 \\
   & i=2 & 5.6 & 40.9 & 70.5 & 81.8 & 86.0 & 88.9 & 91.0 & 92.3 & 92.6 \\
   & i=3 & 5.6 & 39.7 & 63.3 & 73.3 & 78.4 & 81.7 & 83.8 & 85.1 & 86.3 \\
   & i=4 & 6.1 & 52.4 & 79.6 & 87.3 & 90.7 & 92.2 & 93.6 & 94.0 & 94.9 \\
   & i=5 & 5.8 & 34.8 & 67.1 & 82.1 & 88.2 & 90.6 & 92.9 & 93.9 & 94.1 \\
   & i=6 & 6.1 & 31.3 & 62.9 & 80.9 & 88.0 & 91.0 & 93.2 & 93.9 & 94.6 \\
   & i=7 & 5.7 & 24.7 & 53.9 & 73.8 & 84.9 & 88.6 & 91.9 & 92.8 & 94.1 \\
   & i=8 & 5.8 & 34.6 & 68.2 & 81.4 & 86.5 & 90.6 & 91.7 & 93.3 & 94.1 \\
   & i=9 & 5.7 & 24.6 & 54.4 & 74.1 & 83.7 & 88.6 & 91.2 & 93.1 & 94.2 \\
   & i=10 & 5.7 & 27.1 & 58.2 & 76.0 & 85.6 & 89.6 & 91.5 & 93.1 & 94.2 \\
   \hline
 \end{tabular}%
 \label{mar_r4}%
%  \begin{tablenotes}
%  \scriptsize
%  \item[1] $\beta_1 = 10^{-\alpha}b/T^{(1+\alpha)/2}$ and $\beta_2 = 10^{-\alpha}/T^{(1+\alpha)/2}$. T=750.
% \end{tablenotes}
\end{table}%

\begin{table}[H]
 \centering
 \caption{Result (\%) for Right Side Marginal Test $H_0:\beta_i=0$ vs $H_0:\beta_i>0$ with $\lambda=0.9$}
  \begin{tabular}{c|c|ccccccccc}
  \hline
  \multicolumn{2}{c|}{$\beta$} & 0  & 0.05 & 0.1 & 0.15 & 0.2 & 0.25 & 0.3 & 0.35 & 0.4 \\
  \hline
 \multirow{10}[0]{*}{${Q_l^t}$} & i=1 & 7.5 & 51.3 & 73.3 & 81.0 & 84.7 & 86.6 & 87.5 & 88.8 & 89.3 \\
   & i=2 & 6.2 & 38.0 & 66.3 & 76.4 & 82.0 & 85.3 & 87.0 & 88.2 & 90.1 \\
   & i=3 & 5.4 & 34.9 & 56.8 & 66.9 & 71.6 & 75.5 & 77.2 & 79.3 & 80.5 \\
   & i=4 & 5.7 & 46.3 & 74.2 & 83.5 & 87.6 & 89.2 & 91.2 & 91.5 & 92.2 \\
   & i=5 & 5.7 & 30.8 & 62.2 & 78.1 & 84.9 & 88.1 & 90.0 & 91.8 & 92.7 \\
   & i=6 & 5.6 & 27.9 & 58.7 & 77.0 & 83.4 & 87.4 & 90.5 & 91.8 & 92.3 \\
   & i=7 & 5.3 & 20.7 & 48.2 & 70.8 & 79.8 & 85.7 & 88.5 & 90.6 & 92.2 \\
   & i=8 & 6.1 & 33.1 & 63.2 & 77.6 & 83.3 & 86.7 & 88.9 & 91.1 & 91.8 \\
   & i=9 & 5.4 & 22.4 & 51.1 & 71.1 & 81.3 & 86.0 & 88.8 & 89.9 & 91.2 \\
   & i=10 & 5.8 & 25.7 & 55.5 & 73.0 & 81.9 & 86.3 & 89.3 & 91.2 & 92.6 \\
 \hline
 \multirow{10}[0]{*}{${Q_m^t}$} & i=1 & 6.7 & 49.6 & 71.6 & 79.7 & 83.7 & 86.0 & 86.7 & 88.4 & 88.5 \\
   & i=2 & 5.3 & 36.6 & 65.1 & 75.5 & 81.2 & 84.5 & 86.2 & 87.8 & 89.7 \\
   & i=3 & 4.6 & 32.9 & 55.1 & 65.5 & 70.8 & 74.2 & 76.4 & 78.5 & 79.6 \\
   & i=4 & 5.6 & 46.7 & 73.6 & 82.8 & 87.0 & 88.8 & 90.6 & 91.4 & 92.1 \\
   & i=5 & 5.9 & 30.8 & 61.8 & 77.7 & 84.0 & 87.7 & 89.6 & 91.4 & 92.5 \\
   & i=6 & 6.0 & 28.4 & 59.0 & 76.7 & 83.3 & 87.1 & 90.2 & 91.4 & 91.9 \\
   & i=7 & 5.7 & 21.8 & 49.0 & 71.1 & 79.7 & 85.7 & 88.1 & 90.2 & 92.0 \\
   & i=8 & 5.5 & 31.2 & 62.0 & 76.4 & 82.4 & 86.1 & 88.4 & 90.8 & 91.4 \\
   & i=9 & 5.3 & 22.1 & 50.5 & 70.2 & 80.9 & 85.5 & 88.5 & 89.6 & 91.1 \\
   & i=10 & 5.4 & 24.9 & 53.6 & 71.8 & 81.0 & 85.6 & 88.6 & 90.8 & 92.1 \\
   \hline
 \end{tabular}%
 \label{mar_r5}%
%  \begin{tablenotes}
%  \scriptsize
%  \item[1] $\beta_1 = 10^{-\alpha}b/T^{(1+\alpha)/2}$ and $\beta_2 = 10^{-\alpha}/T^{(1+\alpha)/2}$. T=750.
% \end{tablenotes}
\end{table}%

% Table generated by Excel2LaTeX from sheet 'H0x1=0'
\begin{table}[H]
 \centering
 \caption{Result (\%) for Left Side Marginal Test $H_0:\beta_i=0$ vs $H_0:\beta_i<0$ with $\lambda=0.1$}
% \resizebox{\textwidth}{!}{
  \begin{tabular}{c|c|ccccccccc}
  \hline
 \multicolumn{2}{c|}{$\beta$} & 0  & -0.05 & -0.1 & -0.15 & -0.2 & -0.25 & -0.3 & -0.35 & -0.4 \\
 \hline
 \multirow{10}[0]{*}{${Q_l^t}$} & i=1 & 5.0 & 66.6 & 92.6 & 95.7 & 96.5 & 97.4 & 98.0 & 98.2 & 98.5 \\
   & i=2 & 6.9 & 59.6 & 87.4 & 95.3 & 96.5 & 96.8 & 98.7 & 98.2 & 98.5 \\
   & i=3 & 7.9 & 68.6 & 89.5 & 93.9 & 95.3 & 95.4 & 96.5 & 96.0 & 96.8 \\
   & i=4 & 7.2 & 69.1 & 93.0 & 96.1 & 97.8 & 98.2 & 99.0 & 98.9 & 98.9 \\
   & i=5 & 6.1 & 43.8 & 81.2 & 93.4 & 95.4 & 97.4 & 97.6 & 98.4 & 98.7 \\
   & i=6 & 6.5 & 40.5 & 75.5 & 91.3 & 94.6 & 96.9 & 97.5 & 98.0 & 97.9 \\
   & i=7 & 6.0 & 28.8 & 63.9 & 82.8 & 92.4 & 95.2 & 96.9 & 97.6 & 97.8 \\
   & i=8 & 6.1 & 44.6 & 81.5 & 92.9 & 95.8 & 96.8 & 98.6 & 98.2 & 98.7 \\
   & i=9 & 5.6 & 30.5 & 63.5 & 84.3 & 93.0 & 96.0 & 96.4 & 97.8 & 98.5 \\
   & i=10 & 5.9 & 31.8 & 68.2 & 86.5 & 93.6 & 96.6 & 97.3 & 97.8 & 98.5 \\
   \hline
 \multirow{10}[0]{*}{${Q_m^t}$} & i=1 & 4.1 & 63.7 & 91.5 & 95.2 & 96.2 & 97.5 & 97.9 & 98.1 & 98.4 \\
   & i=2 & 6.2 & 57.3 & 87.3 & 95.0 & 96.4 & 96.7 & 98.5 & 98.1 & 98.5 \\
   & i=3 & 7.2 & 66.6 & 89.1 & 93.1 & 95.0 & 95.1 & 96.4 & 95.9 & 96.6 \\
   & i=4 & 4.8 & 63.8 & 90.9 & 95.6 & 97.7 & 98.0 & 98.7 & 98.8 & 98.8 \\
   & i=5 & 4.5 & 37.2 & 77.6 & 91.7 & 94.8 & 97.2 & 97.6 & 98.2 & 98.6 \\
   & i=6 & 4.8 & 33.5 & 69.8 & 89.1 & 94.0 & 96.3 & 97.2 & 97.9 & 97.6 \\
   & i=7 & 4.8 & 24.6 & 59.3 & 80.2 & 90.7 & 94.4 & 96.4 & 97.4 & 97.6 \\
   & i=8 & 5.2 & 41.3 & 79.9 & 92.0 & 95.4 & 96.7 & 98.4 & 97.8 & 98.5 \\
   & i=9 & 4.9 & 27.9 & 60.0 & 82.6 & 91.6 & 95.4 & 96.1 & 97.6 & 98.4 \\
   & i=10 & 5.3 & 28.9 & 65.5 & 84.6 & 92.8 & 96.2 & 97.1 & 97.8 & 98.2 \\
   \hline
 \end{tabular}%
% }
 \label{mar_l1}%
%  \begin{tablenotes}
%  \scriptsize
%  \item[1] $\beta_1 = 10^{-\alpha}b/T^{(1+\alpha)/2}$ and $\beta_2 = 10^{-\alpha}/T^{(1+\alpha)/2}$. T=750.
% \end{tablenotes}
\end{table}%

\begin{table}[H]
 \centering
 \caption{Result (\%) for Left Side Marginal Test $H_0:\beta_i=0$ vs $H_0:\beta_i<0$ with $\lambda=0.25$}
% \resizebox{\textwidth}{!}{
  \begin{tabular}{c|c|ccccccccc}
  \hline
 \multicolumn{2}{c|}{$\beta$} & 0  & -0.05 & -0.1 & -0.15 & -0.2 & -0.25 & -0.3 & -0.35 & -0.4 \\
 \hline
  \multirow{10}[0]{*}{${Q_m^t}$} & i=1 & 5.3 & 67.3 & 89.6 & 94.3 & 96.6 & 97.1 & 97.6 & 97.8 & 97.8 \\
   & i=2 & 7.1 & 58.3 & 87.7 & 93.2 & 95.3 & 96.5 & 97.1 & 98.1 & 98.4 \\
   & i=3 & 7.7 & 62.8 & 85.6 & 90.8 & 92.5 & 94.0 & 94.9 & 95.2 & 96.4 \\
   & i=4 & 7.0 & 68.0 & 90.3 & 94.6 & 97.1 & 97.9 & 97.9 & 98.3 & 98.8 \\
   & i=5 & 6.1 & 42.6 & 78.9 & 91.7 & 95.5 & 96.4 & 97.6 & 98.2 & 98.0 \\
   & i=6 & 6.3 & 38.0 & 74.0 & 89.8 & 94.3 & 96.4 & 97.1 & 98.1 & 98.2 \\
   & i=7 & 6.3 & 29.7 & 61.4 & 81.4 & 91.4 & 93.8 & 96.6 & 97.4 & 97.4 \\
   & i=8 & 6.1 & 45.1 & 80.3 & 92.2 & 94.9 & 96.5 & 97.4 & 98.0 & 98.2 \\
   & i=9 & 5.6 & 27.3 & 63.7 & 84.1 & 91.8 & 94.6 & 96.6 & 97.8 & 98.1 \\
   & i=10 & 6.0 & 33.8 & 68.6 & 87.3 & 93.8 & 95.5 & 96.8 & 97.7 & 98.4 \\
   \hline
 \multirow{10}[0]{*}{${Q_l^t}$} & i=1 & 4.2 & 64.8 & 89.2 & 94.0 & 96.6 & 96.8 & 97.6 & 97.6 & 97.8 \\
   & i=2 & 6.5 & 56.1 & 86.8 & 92.3 & 95.0 & 96.1 & 96.9 & 97.8 & 98.4 \\
   & i=3 & 7.2 & 61.9 & 84.7 & 90.5 & 92.4 & 94.2 & 94.5 & 94.9 & 95.8 \\
   & i=4 & 4.8 & 61.3 & 88.8 & 93.9 & 96.7 & 97.7 & 97.6 & 98.2 & 98.7 \\
   & i=5 & 4.7 & 36.4 & 74.8 & 90.4 & 94.7 & 95.8 & 97.4 & 97.9 & 98.1 \\
   & i=6 & 4.7 & 31.3 & 69.4 & 87.9 & 93.2 & 96.0 & 96.8 & 97.7 & 98.0 \\
   & i=7 & 5.0 & 24.8 & 56.3 & 78.5 & 89.7 & 93.0 & 95.5 & 97.3 & 96.8 \\
   & i=8 & 5.3 & 42.4 & 78.8 & 91.2 & 94.4 & 96.1 & 97.2 & 97.7 & 98.1 \\
   & i=9 & 4.8 & 24.4 & 61.0 & 82.2 & 90.3 & 94.0 & 96.3 & 97.2 & 97.8 \\
   & i=10 & 5.3 & 30.9 & 66.0 & 85.6 & 92.9 & 95.0 & 96.6 & 97.4 & 98.4 \\
   \hline
 \end{tabular}%
% }
 \label{mar_l2}%
%  \begin{tablenotes}
%  \scriptsize
%  \item[1] $\beta_1 = 10^{-\alpha}b/T^{(1+\alpha)/2}$ and $\beta_2 = 10^{-\alpha}/T^{(1+\alpha)/2}$. T=750.
% \end{tablenotes}
\end{table}%

\begin{table}[H]
 \centering
 \caption{Result (\%) for Left Side Marginal Test $H_0:\beta_i=0$ vs $H_0:\beta_i<0$ with $\lambda=0.5$}
% \resizebox{\textwidth}{!}{
  \begin{tabular}{c|c|ccccccccc}
  \hline
 \multicolumn{2}{c|}{$\beta$} & 0  & -0.05 & -0.1 & -0.15 & -0.2 & -0.25 & -0.3 & -0.35 & -0.4 \\
 \hline
 \multirow{10}[0]{*}{${Q_l^t}$} & i=1 & 5.0 & 59.6 & 84.4 & 91.1 & 93.6 & 95.2 & 95.7 & 96.4 & 95.9 \\
   & i=2 & 6.7 & 52.4 & 81.6 & 89.8 & 94.0 & 93.8 & 95.9 & 95.9 & 96.6 \\
   & i=3 & 7.8 & 54.2 & 75.0 & 83.8 & 87.0 & 89.7 & 91.3 & 91.2 & 92.5 \\
   & i=4 & 7.2 & 63.2 & 87.4 & 92.0 & 95.2 & 95.5 & 96.7 & 97.3 & 97.3 \\
   & i=5 & 6.6 & 41.2 & 74.1 & 87.7 & 93.2 & 94.9 & 96.4 & 96.4 & 96.8 \\
   & i=6 & 6.0 & 33.8 & 71.2 & 86.0 & 91.8 & 94.4 & 95.7 & 97.1 & 97.5 \\
   & i=7 & 6.3 & 29.3 & 59.3 & 80.9 & 89.7 & 92.7 & 95.4 & 96.2 & 96.8 \\
   & i=8 & 6.1 & 40.7 & 74.9 & 87.9 & 91.9 & 94.1 & 95.6 & 96.2 & 96.9 \\
   & i=9 & 5.8 & 28.0 & 60.9 & 81.0 & 90.0 & 93.7 & 94.3 & 95.8 & 95.9 \\
   & i=10 & 5.9 & 30.1 & 64.7 & 83.2 & 90.1 & 93.4 & 95.2 & 96.3 & 96.6 \\
 \hline
 \multirow{10}[0]{*}{${Q_m^t}$} & i=1 & 4.3 & 57.8 & 84.0 & 90.5 & 93.1 & 94.7 & 95.5 & 96.1 & 95.9 \\
   & i=2 & 6.2 & 50.8 & 80.2 & 89.0 & 93.6 & 93.7 & 95.8 & 95.8 & 96.6 \\
   & i=3 & 7.2 & 52.1 & 73.9 & 83.3 & 86.5 & 88.8 & 91.1 & 91.0 & 92.3 \\
   & i=4 & 5.1 & 57.6 & 86.0 & 91.4 & 94.7 & 95.2 & 96.4 & 97.2 & 97.1 \\
   & i=5 & 5.1 & 35.4 & 70.1 & 86.1 & 92.2 & 94.2 & 96.0 & 96.0 & 96.7 \\
   & i=6 & 4.4 & 28.4 & 65.6 & 83.6 & 90.6 & 93.8 & 95.3 & 96.7 & 97.1 \\
   & i=7 & 5.2 & 25.1 & 54.3 & 77.6 & 88.1 & 91.4 & 94.7 & 95.6 & 96.3 \\
   & i=8 & 5.3 & 38.0 & 72.8 & 86.9 & 91.3 & 94.2 & 95.2 & 96.1 & 97.0 \\
   & i=9 & 5.1 & 25.8 & 57.6 & 78.7 & 89.4 & 93.1 & 94.0 & 95.3 & 95.7 \\
   & i=10 & 5.4 & 28.4 & 62.3 & 81.7 & 89.0 & 92.9 & 95.0 & 95.9 & 96.5 \\
   \hline
 \end{tabular}%
% }
 \label{mar_l3}%
%  \begin{tablenotes}
%  \scriptsize
%  \item[1] $\beta_1 = 10^{-\alpha}b/T^{(1+\alpha)/2}$ and $\beta_2 = 10^{-\alpha}/T^{(1+\alpha)/2}$. T=750.
% \end{tablenotes}
\end{table}%

\begin{table}[H]
 \centering
 \caption{Result (\%) for Left Side Marginal Test $H_0:\beta_i=0$ vs $H_0:\beta_i<0$ with $\lambda=0.75$}
  \begin{tabular}{c|c|ccccccccc}
  \hline
 \multicolumn{2}{c|}{$\beta$} & 0  & -0.05 & -0.1 & -0.15 & -0.2 & -0.25 & -0.3 & -0.35 & -0.4 \\
 \hline
  \multirow{10}[0]{*}{${Q_l^t}$} & i=1 & 5.2 & 53.1 & 75.1 & 84.9 & 87.6 & 90.9 & 91.3 & 92.4 & 92.8 \\
   & i=2 & 6.7 & 46.6 & 73.9 & 83.3 & 87.5 & 89.1 & 91.7 & 92.6 & 93.3 \\
   & i=3 & 7.6 & 44.5 & 66.1 & 74.9 & 80.4 & 84.5 & 84.1 & 85.4 & 86.1 \\
   & i=4 & 7.3 & 56.3 & 80.9 & 88.2 & 91.2 & 92.8 & 93.2 & 95.0 & 94.6 \\
   & i=5 & 6.4 & 37.5 & 70.3 & 83.6 & 88.6 & 92.2 & 93.2 & 94.0 & 94.4 \\
   & i=6 & 6.7 & 33.7 & 64.4 & 80.9 & 88.5 & 91.2 & 92.8 & 93.7 & 94.6 \\
   & i=7 & 6.2 & 25.6 & 56.0 & 74.4 & 85.5 & 88.3 & 92.4 & 93.0 & 94.6 \\
   & i=8 & 6.6 & 36.4 & 69.1 & 83.2 & 87.6 & 90.7 & 91.7 & 93.1 & 94.6 \\
   & i=9 & 6.2 & 26.3 & 56.6 & 76.0 & 84.7 & 90.0 & 91.0 & 93.2 & 94.3 \\
   & i=10 & 5.7 & 27.0 & 59.7 & 77.3 & 87.3 & 89.8 & 91.6 & 93.6 & 94.4 \\
 \hline
 \multirow{10}[0]{*}{${Q_m^t}$} & i=1 & 4.9 & 51.1 & 74.5 & 84.3 & 87.1 & 90.5 & 90.7 & 92.3 & 92.7 \\
   & i=2 & 6.1 & 44.5 & 72.0 & 82.7 & 87.1 & 88.4 & 90.9 & 92.0 & 92.9 \\
   & i=3 & 6.6 & 42.0 & 63.8 & 73.9 & 79.4 & 82.9 & 83.6 & 84.6 & 85.4 \\
   & i=4 & 5.4 & 51.4 & 79.2 & 88.0 & 91.0 & 92.5 & 93.0 & 94.6 & 94.5 \\
   & i=5 & 5.0 & 32.0 & 66.4 & 81.2 & 87.5 & 91.3 & 92.8 & 93.4 & 93.8 \\
   & i=6 & 5.2 & 29.4 & 61.2 & 78.2 & 87.1 & 90.2 & 91.8 & 93.6 & 94.2 \\
   & i=7 & 5.1 & 21.8 & 51.0 & 71.5 & 83.4 & 87.3 & 91.7 & 92.3 & 93.8 \\
   & i=8 & 5.6 & 33.0 & 67.7 & 81.8 & 86.8 & 90.6 & 91.5 & 92.6 & 93.9 \\
   & i=9 & 5.3 & 23.4 & 54.0 & 74.1 & 83.2 & 88.9 & 90.8 & 92.8 & 94.1 \\
   & i=10 & 5.2 & 25.3 & 57.1 & 75.5 & 86.2 & 89.2 & 91.0 & 93.0 & 94.0 \\
   \hline
 \end{tabular}%
 \label{mar_l4}%
%  \begin{tablenotes}
%  \scriptsize
%  \item[1] $\beta_1 = 10^{-\alpha}b/T^{(1+\alpha)/2}$ and $\beta_2 = 10^{-\alpha}/T^{(1+\alpha)/2}$. T=750.
% \end{tablenotes}
\end{table}%

\begin{table}[H]
 \centering
 \caption{Result (\%) for Left Side Marginal Test $H_0:\beta_i=0$ vs $H_0:\beta_i<0$ with $\lambda=0.9$}
  \begin{tabular}{c|c|ccccccccc}
  \hline
 \multicolumn{2}{c|}{$\beta$} & 0  & -0.05 & -0.1 & -0.15 & -0.2 & -0.25 & -0.3 & -0.35 & -0.4 \\
 \hline
   \multirow{10}[0]{*}{${Q_l^t}$} & i=1 & 6.0 & 46.8 & 69.9 & 78.7 & 83.3 & 85.4 & 86.8 & 88.9 & 89.7 \\
   & i=2 & 6.8 & 40.9 & 67.1 & 78.4 & 82.3 & 85.2 & 86.5 & 88.0 & 90.2 \\
   & i=3 & 6.6 & 36.7 & 57.2 & 67.8 & 72.9 & 75.2 & 77.7 & 80.0 & 81.7 \\
   & i=4 & 7.3 & 51.1 & 74.7 & 83.4 & 88.8 & 89.2 & 90.2 & 92.6 & 92.5 \\
   & i=5 & 6.5 & 34.6 & 62.0 & 79.8 & 84.8 & 89.0 & 90.6 & 91.4 & 92.2 \\
   & i=6 & 7.6 & 31.3 & 61.8 & 77.0 & 83.8 & 87.6 & 89.7 & 91.8 & 93.6 \\
   & i=7 & 6.1 & 23.3 & 52.1 & 69.2 & 81.1 & 85.7 & 89.1 & 90.2 & 91.8 \\
   & i=8 & 6.0 & 34.6 & 63.3 & 78.5 & 84.3 & 87.0 & 89.2 & 90.1 & 90.6 \\
   & i=9 & 6.2 & 24.8 & 54.2 & 71.7 & 80.5 & 85.6 & 88.9 & 89.7 & 91.7 \\
   & i=10 & 6.3 & 27.0 & 54.1 & 74.1 & 82.9 & 86.4 & 88.9 & 91.0 & 92.8 \\
 \hline
 \multirow{10}[0]{*}{${Q_m^t}$} & i=1 & 5.2 & 44.6 & 69.0 & 78.2 & 82.8 & 84.8 & 86.5 & 88.5 & 89.7 \\
   & i=2 & 5.9 & 37.9 & 66.0 & 77.3 & 80.5 & 84.7 & 86.2 & 87.6 & 89.4 \\
   & i=3 & 5.2 & 33.5 & 55.4 & 65.0 & 71.2 & 73.0 & 76.5 & 79.2 & 80.2 \\
   & i=4 & 5.4 & 46.3 & 72.6 & 82.6 & 88.0 & 88.8 & 89.6 & 92.0 & 91.9 \\
   & i=5 & 5.4 & 30.0 & 57.9 & 78.0 & 83.4 & 88.2 & 90.3 & 91.0 & 91.6 \\
   & i=6 & 5.8 & 27.4 & 56.7 & 73.8 & 82.5 & 86.5 & 89.0 & 90.9 & 92.7 \\
   & i=7 & 5.0 & 19.9 & 48.0 & 66.3 & 78.8 & 84.9 & 88.0 & 89.6 & 91.2 \\
   & i=8 & 5.5 & 31.3 & 61.3 & 77.2 & 83.1 & 86.0 & 88.5 & 89.5 & 90.2 \\
   & i=9 & 5.1 & 21.9 & 50.9 & 69.4 & 79.4 & 84.4 & 88.1 & 88.9 & 91.1 \\
   & i=10 & 5.6 & 24.8 & 51.3 & 71.7 & 81.9 & 85.6 & 88.2 & 90.7 & 92.4 \\
   \hline
 \end{tabular}%
 \label{mar_l5}%
%  \begin{tablenotes}
%  \scriptsize
%  \item[1] $\beta_1 = 10^{-\alpha}b/T^{(1+\alpha)/2}$ and $\beta_2 = 10^{-\alpha}/T^{(1+\alpha)/2}$. T=750.
% \end{tablenotes}
\end{table}%

\setcounter{table}{0}
\renewcommand{\thetable}{C\arabic{table}}
\setcounter{figure}{0}
\renewcommand{\thefigure}{C\arabic{figure}}

\section{Additional Empirical Studies}\label{section7.3}
In this section, we report the results of the two extra empirical studies.  Here we give more details of the linear combination variables. The CP factor \citep{2005Bond}  is the linear combination of the forward rate with $\mathrm{CP}_{t} = \hat{\gamma}_{cp}^{\top} \overrightarrow{\mathrm{\textbf{F}}}_{t}$. Here, $\overrightarrow{\mathrm{\textbf{F}}}_{t} =  (\mathrm{F1}_{t},\mathrm{F2}_{t},\mathrm{F3}_{t},\mathrm{F4}_{t},\mathrm{F5}_{t})$ and $\hat{\gamma}_{cp}$ is the OLS estimator of the regression, $\frac{1}{4}\sum_{n=2}^{5}rx(n)_{t+1} = \gamma_{cp}^{\top} \overrightarrow{\mathrm{\textbf{F}}}_{t} + \varepsilon_{cp,t+1}$. The {LN1} factor and {LN2} factor \citep{ludvigson2009macro} are the linear combination of the macroeconomic factors of $\overrightarrow{\mathrm{\textbf{Ma}}}_{t} = (\mathrm{M1}_{t}, \mathrm{M1}^{3}_{t}, \mathrm{M3}_{t}, \mathrm{M4}_{t}, \mathrm{M8}_{t})^{\top}$ and $\overrightarrow{\mathrm{\textbf{Mb}}}_{t} = (\mathrm{M1}_{t}, \mathrm{M1}^{3}_{t}, \mathrm{M2}_{t}, \mathrm{M3}_{t}, \mathrm{M4}_{t}, \mathrm{M8}_{t})^{\top}$ respectively with $\mathrm{LN1}_{t} = \hat{\vartheta}_{ln1}^{\top}\overrightarrow{\mathrm{\textbf{Ma}}}_{t}$ where $\hat{\vartheta}_{ln1}^{\top}$ is the OLS estimator of the regression $\frac{1}{4}\sum_{n=2}^{5}rx(n)_{t+1} = \vartheta_{ln1}^{\top}\overrightarrow{\mathrm{\textbf{Ma}}}_{t} + \varepsilon_{ln1,t+1}$ and $ \mathrm{LN2}_{t} = \hat{\vartheta}_{ln2}^{\top}\overrightarrow{\mathrm{\textbf{Mb}}}_{t}$ where $\hat{\vartheta}_{ln2}^{\top}$ is the OLS estimator of the regression $\frac{1}{4}\sum_{n=2}^{5}rx(n)_{t+1} = \vartheta_{ln2}^{\top} \overrightarrow{\mathrm{\textbf{Mb}}}_{t} + \varepsilon_{ln2,t+1}$. First, we show the comparison between the proposed test and \cite{ludvigson2009macro}, whose data sample spanning the period 1964:01-2003:12 and regressors are the linear combinations CP and {LN1}, and CP and {LN2}, respectively. Second, we conduct the empirical study during 2020:01-2022:12 to see whether the predictability of bond risk premia varies  during the COVID-19 epidemic period with the aggressive monetary policy. The results are shown in Table \ref{tab:Example one table2}.

Different from the original IVX and \cite{ludvigson2009macro}, Table \ref{tab:Example one table2} shows that the predicting power of CP,  {LN1}  and {LN2} are found by our method during 1964:01-2003:12 while not found during the COVID-19 epidemic period (2020:01-2022:12).  The  disappearance of predicting power of CP, {LN1} and {LN2} are most likely caused by the extensively loose monetary policy. Such monetary policy leads to the rapid growth of the debt and deficit of the U.S.; thus, the pricing of the bond is driven by the aggressively eased monetary policy rather than CP,  {LN1}  and {LN2}. As shown in Table \ref{tab:Example one table2}, during the COVID-19 epidemic period, the predicting power of CP, {LN1}, and {LN2} is not found by our method. On the contrary, IVX finds the CP can be useful in predicting bond risk premia, and CP, {LN1} and {LN2}  are all found by \cite{ludvigson2009macro} to be useful predictors except for the capacity of CP to predict $rx(4)$ and $rx(5)$ in Panel B.

\begin{table}[H]
  \centering
  \caption{P Value (\%) of Inference for CP and ${LN1}$ and ${LN2}$}
  \makebox[\textwidth][c]{
  \scalebox{0.76}{
   \begin{threeparttable}
    \begin{tabular}{l|lll|lll|lll|lll}
    \hline
    \multicolumn{13}{c}{Panel A: Test Results for the Predicting Power of CP and ${LN1}$}\\
    \hline
     $y_t$ & \multicolumn{3}{c|}{$rx(2)$} & \multicolumn{3}{c|}{$rx(3)$} & \multicolumn{3}{c|}{$rx(4)$} & \multicolumn{3}{c}{$rx(5)$} \\
     \hline
      Predictors     & $Q_m$  & IVX & LN  & $Q_m$  & IVX & LN  & $Q_m$  & IVX & LN  & $Q_m$  & IVX & LN \\
    \hline
    \multicolumn{13}{c}{Sample studied in \cite{ludvigson2009macro}: 1964:01~2003:12} \\
    \hline
    {CP} & 0.7*** & 0.0*** & 0.0*** & 0.1*** & 0.0*** & 0.0*** & 0.3*** & 0.0*** & 0.0*** & 1.6** & 0.0*** & 0.0*** \\
    {{LN1}} & 0.0*** & 0.0*** & 0.0*** & 0.4*** & 0.0*** & 0.0*** & 3.3*** & 0.0*** & 0.0*** & 9.0* & 0.0*** & 0.0*** \\
    \hline
    \multicolumn{13}{c}{Sample during the COVID-19 Epdemic Period: 2020:01~2022:12} \\
    \hline
    {CP} &  63.2 & 0.0*** & 0.4*** & 8.3 & 0.0*** & 1.3** & 60.8 & 0.3*** & 6.4* & 56.7 & 1.3** & 7.7** \\
    {{LN1}} & 84.3 & 63.8 & 1.5** & 84.3 & 20.2 & 0.5*** & 94.7 & 17.1 & 0.3*** & 94.9 & 10.7 & 0.9*** \\
    \hline
    \multicolumn{13}{c}{Panel B: Test Results for the Predicting Power of CP and ${LN2}$}\\
    \hline
    \multicolumn{13}{c}{Sample studied in \cite{ludvigson2009macro} 1964:01~2003:12} \\
    \hline
    {CP} &  1.0** & 0.0*** & 0.0*** & 0.1*** & 0.0*** & 0.0*** & 0.2*** & 0.0*** & 0.0*** & 1.4** & 0.0*** & 0.0*** \\
    {{LN2}} &  0.0*** & 0.0*** & 0.0*** & 0.0*** & 0.0*** & 0.0*** & 2.6** & 0.0*** & 0.0*** & 7.1* & 0.0*** & 0.0*** \\
    \hline
    \multicolumn{13}{c}{Sample during the COVID-19 Epdemic Period: 2020:01~2022:12} \\
    \hline
    {CP} & 48.0  & 0.0*** & 0.8*** & 98.4 & 0.1*** & 2.8** & 73.9 & 0.6*** & 11.8 & 75.1 & 1.7** & 12.3 \\
    {{LN2}} &89.3 & 42.9 & 0.7*** & 90.3 & 25.6 & 0.5*** & 99.2 & 30.1 & 0.4*** & 89.8 & 25.5 & 1.8** \\
    \hline
    \end{tabular}%
        \vspace{-0.2cm}
   \begin{tablenotes}
         \footnotesize
    \item   {\bf{Notes}}: $Q_m$ is our proposed test using the test statistic defined by \eqref{qmeqn} in the main text.
    \item  IVX represents the test using a test statistic proposed by \cite{Phillips2013PredictiveRU}.
    \item LN stands for \cite{ludvigson2009macro} whose inference used the test based on OLS.
    \end{tablenotes}
    \end{threeparttable}}}%
  \label{tab:Example one table2}%
\end{table}%

To further check the robustness of the main results, we also do the tests of the predictability of F1-F5 and M1-M8  from 1964:01 to 2019:12. Results similar to Table \ref{tab:Example two table4} in Section \ref{section6} are observed in Table \ref{new:Example two table4}. Therefore, the robustness of the main results in Section \ref{section6} is confirmed.

\begin{table}[H]
  \centering
  \caption{P Value (\%) of Inference for F1-F5 and M1-M8  (1964:01-2019:12)}
  \resizebox{\textwidth}{!}{
      \begin{tabular}{l|lll|lll|lll|lll}
    \hline
    $y_t$ & \multicolumn{3}{c|}{$rx(2)$} & \multicolumn{3}{c|}{$rx(3)$} & \multicolumn{3}{c|}{$rx(4)$} & \multicolumn{3}{c}{$rx(5)$} \\
    \hline
    Predictors      & $Q_m$  & IVX & OLS & $Q_m$  & IVX & OLS & $Q_m$  & IVX & OLS & $Q_m$  & IVX & OLS \\
    \hline
     F1 & 0.0*** & 0.0*** & 0.0*** & 0.4*** & 0.0*** & 0.0*** & {14.3} & 0.0*** & 0.0*** & {55.2} & 0.0*** & 0.0*** \\

     F2 & 0.0*** & 0.0*** & 0.0*** & 88.7 & 66.5 & 19.1 & 20.0 & 45.1 & 51.0 & 13.5 & 34.5 & 65.5 \\

     F3 & {0.3} & 00.1*** & 23.9 & 3.3** & 0.0*** & 0.0*** & {93.8} & 0.0*** & 17.9 & {26.2} & 0.7*** & 29.0 \\

     F4 & 65.9 & 29.3 & 12.6 & 87.5 & 22.7 & 22.3 & 0.0*** & 0.3*** & 0.0*** & {10.2} & 4.7** & 0.0*** \\

     F5 & 35.0 & 26.8 & 48.9 & 63.0 & 19.7 & 25.0 & 84.7 & 46.4 & 0.7*** & 0.2*** & 1.4** & 0.0*** \\

     M1   & 0.0*** & 0.0*** & 0.0*** & 0.0*** & 0.0*** & 0.0*** & 0.0*** & 1.8** & 0.0*** & 0.0*** & 0.1*** & 0.0*** \\

     M2 & 42.7 & 5.7** & 74.7 & 28.0 & 34.5 & 48.8 & 44.9 & 30.1 & 48.8 & 34.9 & 29.9 & 38.8 \\

     M3 & 0.4*** & 0.4*** & 0.2*** & 1.7** & 2.2** & 0.1*** & 1.6** & 8.4* & 0.1*** & 1.5** & 2.2** & 0.0*** \\

     M4 & 6.4* & 0.2*** & 9.3* & {13.4} & 0.3*** & 4.7* & 7.5* & 0.0*** & 0.2*** & {16.3} & 0.0*** & 1.1** \\

     M5 & 9.0* & 9.1* & 0.9*** & 14.6 & 33.4 & 1.4** &  {3.7**} & 54.2 & 0.2*** &  {1.1**} & 14.4 & 0.0*** \\

     M6 & 71.1 & 12.6 & 16.4 & 88.1 & 20.4 & 80.8 & 61.7 & 81.5 & 67.6 & 88.7 & 85.1 & 86.9 \\

     M7 & 0.0*** & 0.7*** & 0.0*** & 0.2*** & 0.2*** & 0.0*** & 0.0*** & 7.3* & 0.0*** & 0.0*** & 0.6*** & 0.0*** \\

     M8 & 12.8 & 32.3 & 9.5* & 27.0 & 72.0 & 29.9 & 53.5 & 69.7 & 78.0 & 46.6 & 65.7 & 57.7 \\
    \hline
    \end{tabular}%
    }
          \vspace{-0.2cm}
   \begin{tablenotes}
         \footnotesize
    \item    {\bf{Notes}}: $Q_m$ is our proposed test using the test statistic defined by \eqref{qmeqn} in the main text.
    \item  IVX represents the test statistic proposed by \cite{Phillips2013PredictiveRU}.
    \item OLS is the estimation method applied in \cite{ludvigson2009macro}.
    \end{tablenotes}
  \label{new:Example two table4}
\end{table}%

\bibliographystyle{apalike}

\bibliography{Bibliography-MM-MC}
\end{document}